\documentclass[graybox]{svmult}

\usepackage{type1cm}        
\usepackage{makeidx}         
\usepackage{graphicx}        
\usepackage{multicol}        
\usepackage[bottom]{footmisc}

\usepackage{newtxtext}       %
\usepackage[varvw]{newtxmath}       

\makeindex             

\usepackage{natbib}    

\newcommand{\msun}{M_\odot}

\newcommand{\LCDM}{$\Lambda$CDM}
\newcommand{\gureft}{\textsc{gureft}}

\begin{document}

\title*{Galaxy formation in the first billion years}
\author{Rachel S. Somerville}
\institute{Center for Computational Astrophysics, Flatiron Institute,  \email{rsomerville@flatironinstitute.org}}
%
%
\maketitle

\abstract{
These notes present material from lectures given at the 54th Saas-Fee Advanced Course of the Swiss Society of Astrophysics and Astronomy in January 2025, entitled ``Galaxies and Black Holes in the First Billion Years as seen by the JWST'', and are intended for early career researchers or those new to the sub-field. My lectures covered the theory of galaxy formation with a focus on the first billion years of cosmic evolution. In these notes, I discuss cosmological structure formation, properties of dark matter halos at $z\gtrsim 6$, and whether any of the JWST observations to date present a serious and fundamental challenge for the $\Lambda$ Cold Dark Matter Paradigm. I then give an overview of physical processes and modeling techniques, including translating simulation-based quantities to observables, and discuss recent progress and future directions in galaxy formation modeling. The closing section presents a summary of some of the theoretical puzzles and challenges raised by the first three years of high redshift observations with JWST, and how our models of galaxy formation may need to be revised to accommodate them.}


\section{Preamble and Context}
\label{sec:intro}
Origins hold a particular fascination for humans. One of the reasons that we build ever larger and more sophisticated machines is to peer further and further back in time, in the hope that we might someday understand our beginnings. Indeed, this was one of the reasons that the James Webb Space Telescope was built and deployed -- and it has not disappointed. Since the first JWST data became public on July 12, 2022, our knowledge about what happened during the first billion years of cosmic time has expanded dramatically, and many of our expectations have been upended. Thus it seems very timely that the topic of the 54th ``Saas-Fee Advanced Course'' of the Swiss Society for Astrophysics and Astronomy was chosen to be ``Galaxies and Black Holes in the First Billion Years as seen with the JWST''. During this course, held in January 2025 in Saas-Fee, Switzerland, four lecturers covered topics on observational and theoretical aspects of black holes and galaxies at $z \gtrsim 6$. Professor Richard Ellis covered mostly observational aspects of early stars and galaxies revealed by JWST. Professor Eduardo Ba\~{n}ados and Professor Marta Volonteri covered observational and theoretical aspects of supermassive black holes. My lectures covered theory and simulations of galaxy formation in the high redshift Universe, and how to make connections between these theoretical predictions and observations. I have tried to make these notes fairly self-contained, but given the volume of material, I will cover the basic cosmological background very briefly, since there are other excellent pedagogical sources that cover it. Moreover, this is not intended to be a comprehensive review article, and it is inevitably biased towards the work that I am most familiar with. I give suggestions for additional reading at the end of each section. I have tried to remain as faithful as possible to the material that was presented in the lectures, but the field has continued moving quickly, and I include some work that has appeared in the literature since January 2025, although again I do not attempt to be comprehensive. 

The structure of these notes is as follows. In \S\ref{sec:structure}, I review the basics of cosmological structure formation, and discuss the properties of dark matter halos in the first billion years of cosmic evolution and connections between halo properties and the galaxy population observed by JWST. In \S\ref{sec:models}, I give an overview of the physical processes that are thought to be the most important for shaping galaxy formation. Then I discuss various existing approaches for working out the implications of these physical processes via physics-based models and simulations, give a short update on the status of theory and simulations just before JWST launched, and briefly discuss recent advances and future directions in galaxy formation modeling. In \S\ref{sec:synthobs}, I discuss how the intrinsic quantities that are readily predicted by simulations can be translated into observables. I focus on UV-optical galaxy SEDs, and cover stellar population synthesis modeling, and modeling of nebular emission and dust. In \S\ref{sec:results}, I highlight several of the main results from the first three years of JWST observations of the $z\gtrsim 6$ Universe, and how this has modified our theoretical picture of galaxy formation. 

\section{Formation of structure in the first billion years}
\label{sec:structure}
Our modern framework for understanding galaxy formation is built on several axioms, which will be familiar to most readers: 1) the Universe began with a hot singularity (Big Bang), and the early Universe was homogeneous and isotropic on large scales (the Cosmological Principle). 2) The matter density of the Universe is dominated by a substance that does not interact with itself or other matter (or does so only very weakly) except via gravity (dark matter) 3) the early Universe experienced a period of ultra-rapid expansion (inflation), which laid down the seeds of structure formation\footnote{I note here that while there is strong observational evidence for these axioms, they are not definitively proven. In particular, many details about inflationary theory remain unclear.}. One of the beautiful aspects of the paradigm that emerges from these axioms is that, given the initial density perturbation spectrum laid down by inflation, and Newtonian gravity, many aspects of the large scale structures that we observe emerge very naturally and with few additional assumptions required (see e.g. \citealt{MvdBWbook} [hereafter MvdBW], Ch. 3 and references therein).  

In this section, I briefly overview the basics of structure formation within the standard $\Lambda$ Cold Dark Matter ($\Lambda$CDM) paradigm. There are many other excellent presentations of this material, so I keep this part very brief, and suggest other sources where readers can read up on this material. I discuss the properties of dark matter halos in the first billion years of cosmic history, and how they evolved over this time period. I discuss both the number density of halos as a function of their virial mass, and internal properties such as density profile and angular momentum, as predicted by modern dark matter only N-body simulations. I then discuss simple arguments that link observable properties of high redshift galaxies with halo properties, and summarize the implications for interpreting the recent JWST observations.  

\subsection{Structure formation basics}
\label{sec:structure:basics}

In the standard picture, our Universe contains non-relativistic and relativistic matter (or radiation), comprised of ``standard model'' particles (such as quarks, leptons, and bosons; about 5\% of the total matter-energy budget), and a mysterious and as-yet unidentified ``dark matter'' particle\footnote{Dark matter is widely believed to be a particle, but there are other possibilities for its nature.} which is not part of the standard model (about 27\% of the total matter-energy budget). The dark matter dominates over the ``normal'' (standard model) matter by about a factor of 5 to 1, and is distinguished by not interacting with light and thus not experiencing the electromagnetic force (MvdBW Ch. 2). 

The geometry of space-time is described by the Einstein field equation, a tensor equation that describes how space-time curves in response to the matter content of the Universe (see e.g. Eqn.~3.54 in Chapter 3.2 of MvdBW). The Friedmann-Robertson-Walker (FRW) metric describes a spatially homogeneous and isotropic four-dimensional space-time, as is appropriate under the ansatz of the Cosmological Principle (MvdBW 3.1.2). Adopting the FRW metric allows the Einstein equation to be simplified into a single scalar equation called the Friedmann equation, which describes the time dependence of the scale factor $a$ as a function of the matter density $\rho$, curvature $K$, and cosmological constant $\Lambda$ (see Eqn.~3.60 and 3.61 in MvdBW). 

According to our standard theory of cosmology, the Universe was born in a hot Big Bang about 13.8 billion years ago. A fraction of a second later, the Universe entered a period of superluminal expansion known as inflation. Inflation caused microscopic scales to stretch to a size larger than the current horizon, thereby seeding macroscopic inhomogeneities in the matter density field, arising from quantum fluctuations. After the end of inflation at around $10^{-32}$ seconds after the Big Bang, the Universe continued to expand (though at a more moderate pace), and neutrons, protons, and electrons began to form at $\sim 10^{-5}$--$10^{-4}$ s. Nucleosynthesis began a few minutes after the Big Bang, synthesizing protons and neutrons to create primordial D, He, and small quantities of a few other light elements. During this epoch of nucleosynthesis, the Universe was still very hot ($T \sim 10^9$ K), so all of these elements were initially highly ionized. As the Universe continued to expand and cool, at around a few hundred thousand years after the Big Bang, free electrons formed neutral atoms, and the Universe became transparent to photons. These photons are what we observe today as the Cosmic Microwave Background (CMB). The subsequent period of time is often referred to as the \emph{dark ages}, which persisted until the epoch of \emph{first light}, when primordial gas clouds were first able to condense into the \emph{first stars}. The precise timing of this transition remains unknown, but is thought to have occurred at around a redshift of $z \sim 30$--20, or about 100 to 200 million years after the Big Bang (MvdBW 3.3.1). 

The primordial density inhomogeneities that arise from quantum fluctuations amplified by inflation have a very specific property: the power spectrum of the matter density field is expected to be very nearly \emph{scale-free}. In mathematical terms, this means that the primordial power spectrum $P_k \propto k^n$ with $n$ very close to unity, also known as a \emph{Harrison-Zeldovich-Peebles} spectrum (MvdBW Ch. 4.4.4). In more colloquial terms, this means that the Universe is \emph{lumpy on all scales}. Perturbations grow differently during the subsequent radiation-dominated and matter-dominated epochs, resulting in a power spectrum that turns over on a scale of order the size of the horizon at matter-radiation equality. We have direct and very precise constraints on the power spectrum at the surface of last scattering from observations of the CMB (MvdBW Ch. 2.9). 

An alternative way to quantify the matter density field is via the mass variance $\sigma(R)$, defined as follows. Imagine placing spheres with radius $R$ randomly about the Universe and measuring how much mass they contain, then taking the variance among all the spheres. We can write this as:
\begin{equation}
\sigma^2(M) = \left< \left( \frac{M(\mathbf{x};R)-\bar{M}(R)}{\bar{M}(R)} \right)^2 \right>
\end{equation}
where $M(R)$ is the mass contained within a sphere of radius $R$. This can equivalently be expressed as an integral over a window function in $k$ space times the power spectrum (MvdBW 6.1.3). 

The force of gravity is stronger in overdense regions and weaker in underdense ones, so fluctuations continue to grow as time goes on. We define the overdensity $\delta(\mathbf{x}, t) \equiv \rho(\mathbf{x}, t)/\bar{\rho}(t)-1$, where $\rho(\mathbf{x}, t)$ is the density at location $\mathbf{x}$ and time $t$, and $\bar{\rho}(t)$ is the average density of the Universe at time $t$. We can obtain an analytic description of the evolution of the density field using the standard ideal fluid equations (continuity, Euler, Poisson; see MvdBW 4.1.1). Expressing these equations in an expanding FRW space, and making the assumption that both $\delta$ and the velocity $\mathbf{v}$ are small, such that non-linear terms can be neglected, we can obtain a single differential equation for $\delta$ (see Eqn. 4.24 in MvdBW). It is then standard to write down the Fourier transform of this equation (Eqn. 4.26, MvdBW), as the modes decouple and we can solve for the evolution of each individual mode $\delta_{\mathbf{k}}$. Under specific simplifying assumptions (such as a pressureless fluid in an Einstein-de Sitter Universe), there exist analytic solutions for $\delta(t)$ (see MvdBW 4.1.6). It is customary to write $\delta \propto D_{\rm lin} (a)$ where $a$ is the scale factor, and $D_{\rm lin}(a)$ is called the \emph{linear growth rate} (typically defined so that $D_{\rm lin}(z=0)=1$). For the cosmological parameters of the concordance \LCDM\ Universe, there is no analytic solution for $D_{\rm lin}(a)$, but there are standard fitting functions (see MvdBW 4.1.6, Fig. 4.1). $D_{\rm lin}(a)$ also provides a good description of the evolution of the power spectrum $P_k$ or mass variance $\sigma(M)$ in the \emph{linear regime} ($\delta \lesssim 1$). 

As gravity continues to do its work, inevitably, $\delta$ will begin to exceed unity. Soon thereafter, the force of gravity will overcome the pressure of the expansion, and that patch of the Universe will collapse to form a gravitationally bound \emph{dark matter halo}. We can define the quantity $\delta_L(t) =\delta_i \, D_{\rm lin}(t)/D_{\rm lin}(t_i)$, i.e., the linearly extrapolated value of $\delta_i$ at time $t$. In the spherical collapse model, collapse occurs when the linear overdensity extrapolated to the present day is $\delta_{\rm crit}(t) \simeq 1.686/D_{\rm lin}(t)$ (for an Einstein-de Sitter universe, with a weak dependence on $\Omega_m$ and $\Omega_{\Lambda}$ for more general cosmologies;  \citealp{Barkana-Loeb:2001}). We can then define the \emph{characteristic mass} $M_*$ as $\sigma(M_*) = \delta_{\rm crit} (t)$. A simple model of spherical ``tophat" collapse (MvdBW 5.1) predicts that bound virialized halos should have an overdensity $\Delta_{\rm vir} = 18 \pi^2 \simeq 178 \bar{\rho}$ (for an Einstein-de Sitter universe; see \citet{Bryan1998} for more general expressions), and this or a similar criterion is frequently used to define dark matter halos in N-body simulations. Due to the scaling of halo definition with the average density of the Universe, a halo of a fixed mass has a much smaller radius, and higher virial velocity and virial temperature, at high redshift (see \citealt{Loeb-Furlanetto:2013}, Fig. 3.5, 3.6).

A simple analytic model that combines the ideas of the linear growth of perturbations and spherical collapse when $\delta > \delta_{\rm crit}$ provides useful intuition for how the abundances and properties of dark matter halos evolve over cosmic time, and also provide a direct intuitive link between halo abundances and the linear power spectrum (via $\sigma(M)$). The Press-Schechter model \citep{Press-Schechter:1974} provides an expression for the number density of collapsed, virialized halos as a function of halo mass and redshift (halo mass function; see e.g. MvdBW 7.2.1), which qualitatively matches many of the features that were later confirmed with numerical N-body simulations. In particular, this model predicts that halos become exponentially more rare as their mass exceeds the characteristic mass $M_*$ (see e.g. Fig.~10 of \citealt{Barkana-Loeb:2001}). The value of $M_*$ drops very rapidly with increasing redshift, from $M_* \simeq 10^{13} \msun$ at $z=0$ to $\simeq 10^{6} \msun$ at $z=6$, to $\simeq 10^{3} \msun$ at $z=10$ (see Fig.~6 of \citealt{Barkana-Loeb:2001}).

More elaborate versions of this model have been presented by \citet{Sheth1999} and other follow-up works. The same ideas have also been used to derive the \emph{conditional halo mass function}, which is the probability that a halo of mass $M_0$ at time $t_0$ had a progenitor with mass $M_1$ at $t_1$ \citep{Lacey:1993}. These expressions (referred to as ``Extended Press-Schechter'') have been used to construct dark matter ``merger trees'', which represent the full assembly history of a dark matter halo over time \citep{Kauffmann:1993,SK:1999,Cole:2000,Parkinson:2008,Jiang:2014}.

Another consequence of the simple picture of halo formation outlined above is that it predicts how clustered collapsed halos should be relative to the underlying matter density field. This is commonly referred to as \emph{halo bias}, where we can define $\delta_h(\mathbf{x}|M) = b_h(M) \delta(\mathbf{x})$. Here $\delta_h(\mathbf{x}|M)$ is the overdensity of halos as a function of spatial position and halo mass, and $b_h(M)$ is the halo bias (where I have not written the implicit time dependence)\footnote{The definition of halo bias is not unique -- it can also be defined as a ratio of the two-point correlation function or power spectrum of halos and matter.}. The same ideas used by \citet{Press-Schechter:1974} to derive the halo mass function can be used to estimate the halo bias as a function of mass and redshift \citep{Cole:1989,Mo-White:1996,Mo-White:2002,Sheth:2001}:
\begin{equation}
b_h(M, z) = 1 + \left(\frac{\nu^2 - 1}{\delta_c}\right)
\end{equation}
where $\nu \equiv \delta_{\rm crit}(z)/\sigma(M)$; see \citet{Loeb-Furlanetto:2013} Fig. 3.14 for a plot of this function. Multiple studies have found that this analytic model is in good qualitative agreement with halo clustering in N-body simulations \citep[e.g.][]{Sheth:2001}. This model predicts that the bias of halos with mass $< M_*$ is roughly flat and close to or slightly less than unity, while $b_h$ increases very rapidly as the halo mass exceeds $M_*$. Due to the strong evolution of $M_*$ towards higher masses as cosmic time progresses, discussed above, this implies that halos massive enough to form galaxies in the early Universe are very highly biased. Thus in spite of the fact that the overall matter density field at early times is much less clustered than at later epochs, we might still expect early galaxies to be fairly strongly clustered. 

\subsection{Dark Matter halos in the early universe}
\label{sec:structure:halos}
Although the analytic models outlined above provide useful intuition and approximate predictions for important halo properties such as their abundances and clustering, numerical simulations are needed to accurately study non-linear structure formation. There is a large literature on the study of how a Universe of pure dark matter evolves, which of course is a greatly simplified problem because the only physical process that needs to be included is gravity. Dark matter only (sometimes called `dissipationless') cosmological N-body simulations typically adopt periodic boundary conditions in a cubical volume. Mass elements are represented by particles that sample the underlying phase-space distribution function. A realization of the power spectrum in the chosen cosmology and dark matter scenario is generated, typically shortly after recombination. A \emph{gravity solver} then solves Newton's laws within an expanding Universe (typically including dark energy or a cosmological constant), assuming that the particles are collisionless. For an excellent pedagogical exposition of the numerical techniques used in modern N-body simulations, I recommend Volker Springel's notes from the 43rd Saas-Fee course \citep{Springel:2016} and the review article by \citet{Vogelsberger:2020}.    
Solving for the force of every particle on every other particle quickly becomes very expensive for large particle numbers. The so-called ``direct N-body'' approach is only used for extremely dense systems (such as star clusters), where the approximation of matter as collisionless breaks down. Cosmological N-body simulations typically use approximations for the longer range forces, such as tree-based methods (used by the codes Gadget, PKDGrav, Treecode), particle-mesh (PM), or adaptive mesh refinement (AMR; used by the codes ART, RAMSES, AMIGA; \citealp{Springel:2016}). 

\begin{figure}
\includegraphics[width=\textwidth]{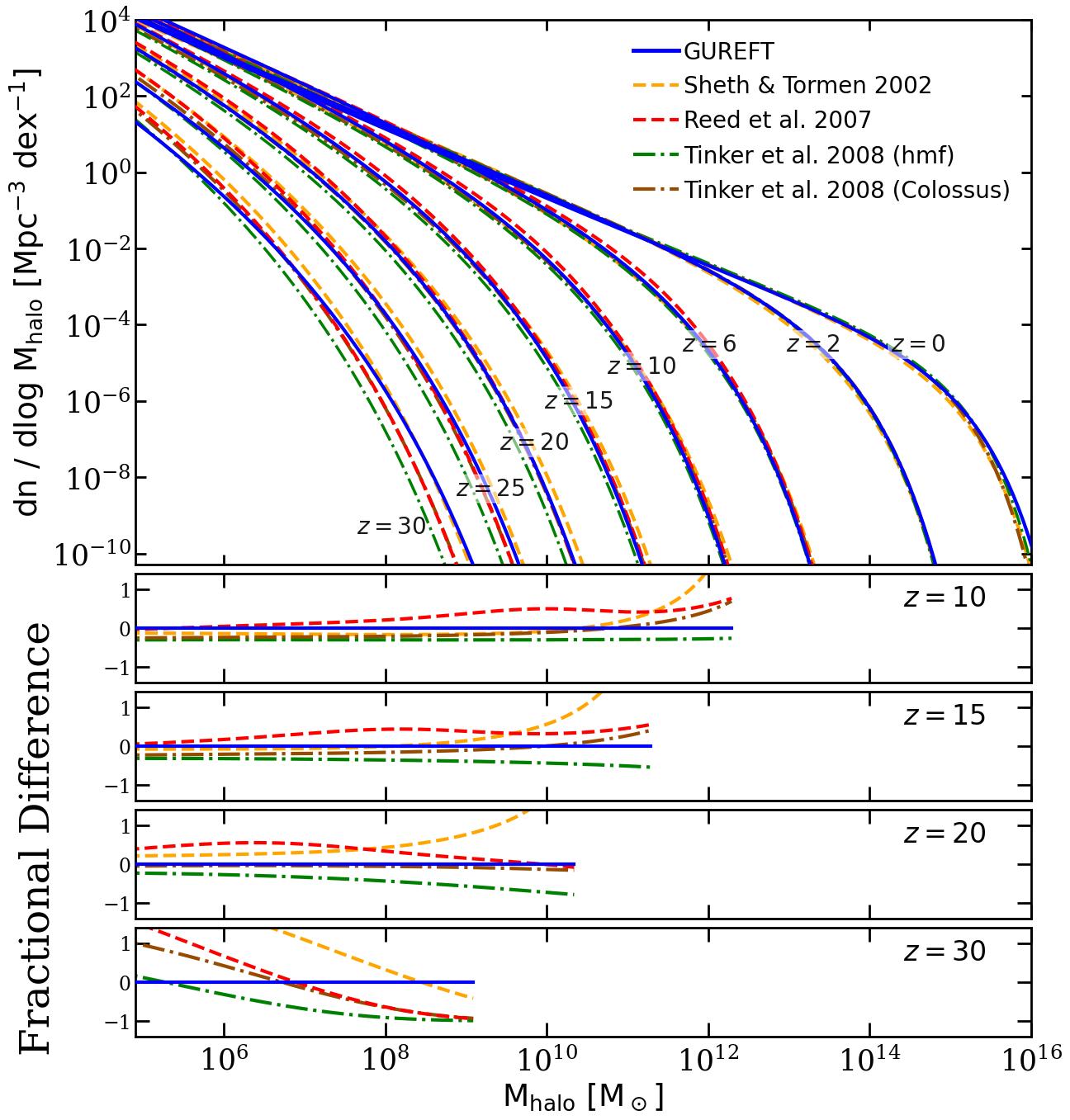}
\caption{Halo mass functions (HMF) from the GUREFT N-body simulation suite shown at $z=0$, 2, 6, 10, 15, 20, 25, and 30, compared with commonly used halo mass functions from analytic models and extrapolations of fitting functions based on lower resolution simulations that were not analyzed at ultra-high redshift (see \citealp{Yung2024a} and \citealp{Yung2025} for details). The bottom panels show the fractional difference of the number density of halos from these HMFs relative to the GUREFT outputs. Halo mass functions that are commonly used in the literature can deviate from these robust N-body based results by up to an order of magnitude at ultra-high redshift. Based on  \citet{Yung2024a}, Fig.~5 and \citet{Yung2025}, Fig.~A1. }
\label{fig:hmfcomp}       
\end{figure}

The first large volume dissipationless cosmological simulations were carried out in the mid-1980's to early 1990's \citep{Davis:1985,Efstathiou:1985,Efstathiou:1988,Klypin:1993} and since then, many groups have carried out such simulations over a very large dynamic range \citep[see][for a review]{Vogelsberger:2020}. In order to interpret the results of these simulations, it is standard practice to run a ``halo finder'', which locates groups of particles that are then identified as ``halos'' \citep[see][for an overview and comparison of different halo finders]{Knebe:2011}. This step is trickier than one might think, since structures in the \LCDM\ model tend to be extremely lumpy and are not, in general, perfectly spherical or even symmetric. Many studies have presented predictions of the halo mass function and other halo properties from a redshift of about 6--8 to the present \citep[e.g.][]{Tinker:2008,Klypin2011,Klypin2016,Rodriguez-Puebla:2016}. However, somewhat surprisingly, very few studies with numerical cosmological N-body simulations have focussed on the ultra-high redshift Universe ($z>10$). 

The recent simulation suite \gureft\ (Gadget at Ultrahigh Redshifts with Extra-fine Timesteps) was designed to characterize the halo populations that are expected to host galaxies from first light ($z\sim 30$) to $z\sim 6$ \citep{Yung2024a}. It consists of four simulation boxes, with volumes ranging from 5$^3$--90$^3$ h$^{-3}$ Mpc$^3$ and particle mass $m_p = 9.92 \times 10^3$ h$^{-1}\, \msun$ (for the smallest box) to $5.78 \times 10^7$ h$^{-1}\, \msun$ (for the largest box), run down to $z\sim 6$. The \gureft\ results show that the predictions of halo number densities at $z\sim 10$ and above from analytic models such as Sheth-Tormen, as well as extrapolations of fitting functions from lower redshift simulation studies, can be quite inaccurate (by up to 1 dex), as illustrated in Fig.~\ref{fig:hmfcomp}. This figure emphasizes how rapidly the number density of massive halos ($M_h \gtrsim 10^{11}-10^{12}\, \msun$) is declining towards early times. \citet{Yung2024a} and \citet{Yung2025} present updated fitting functions for the halo mass function out to $z\sim 30$.  \citet{Yung2024a} also use the \gureft\ results to characterize halo growth rates as a function of halo mass and redshift, and present updated fitting functions. 

\subsection{Linking dark matter halos with galaxies: do the JWST observations challenge $\Lambda$CDM?}
\label{sec:structure:obs}
Under the basic ansatz that galaxies occupy dark matter halos, we can now explore some basic implications and limits obtained from JWST observations of the number densities of ultra-high redshift galaxies. In \S\ref{sec:models}, I discuss detailed physics-based models that incorporate the complex suite of baryonic processes that shape galaxy evolution, and their predictions for the ultra-high redshift Universe. However, before going there, it is instructive to consider what we can learn from much simpler methods for linking galaxies and dark matter halos through empirical models. There are many different variants of these kinds of models (I discuss the landscape of empirical models more generally in \S\ref{sec:models:methods}), but the basic idea is to write down a parameterized mapping between dark matter halo mass and an observable or quasi-observable (such as galaxy luminosity or stellar mass), and then solve for the values of the parameters that match an observed luminosity function or stellar mass function, for an assumed set of cosmological parameters. 

\begin{figure}
\includegraphics[width=\textwidth]{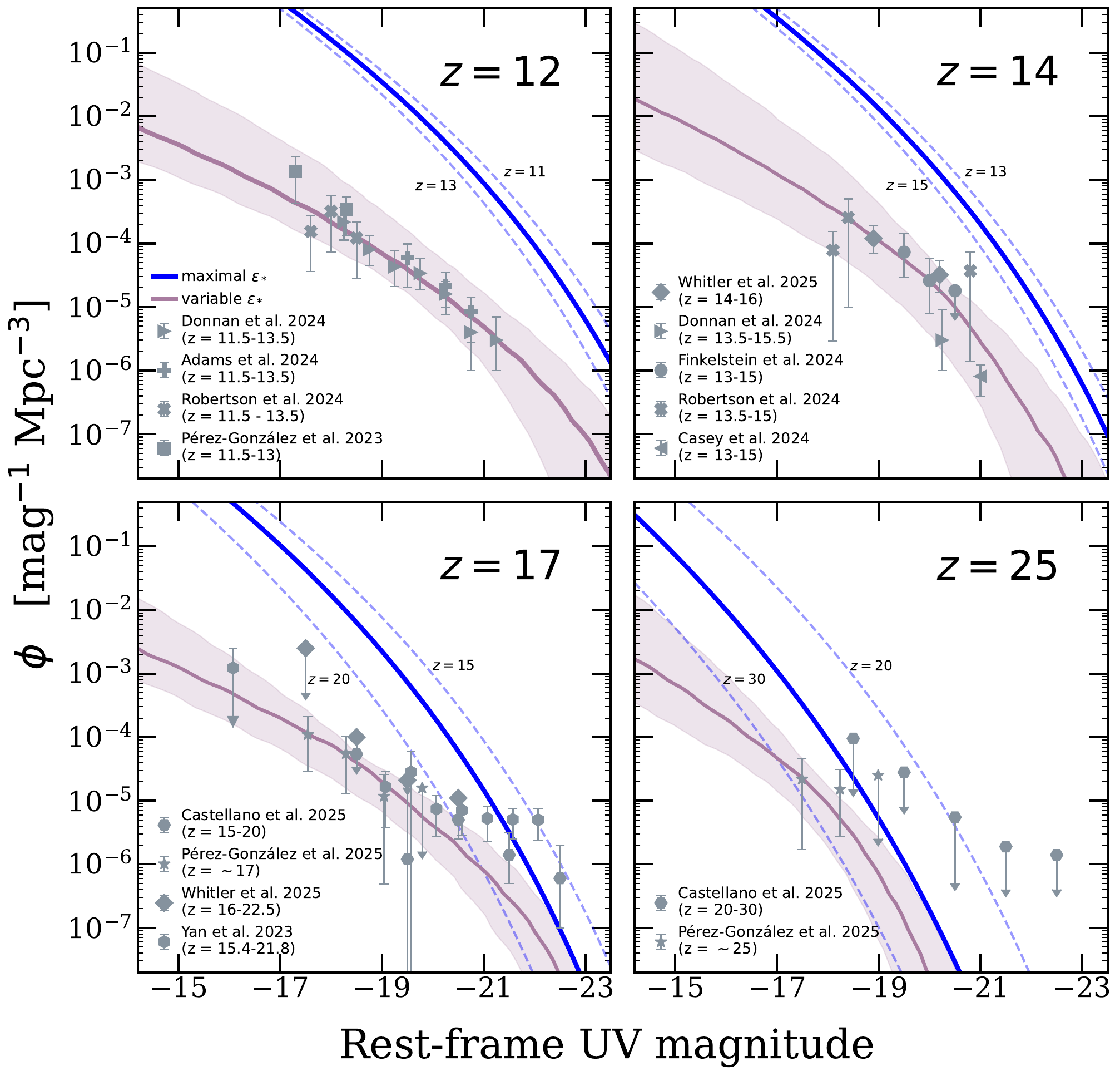}
\caption{UV luminosity functions at $z=12$, 14, 17, and 25, showing a compilation of recent observational estimates (see \citealt{Yung2025} for details). The blue lines show the predicted UVLF from the simple empirical model described in the text, with a maximal baryon conversion efficiency $\epsilon_*=1$. The purple shaded regions show the predicted UVLF for the fitted functional form of $\epsilon_*(M_h,z)$ (Eqn.~\ref{eqn:estaremp}). The fact that the blue lines are everywhere higher than the observations implies that there is no fundamental tension between $\Lambda$CDM and these observations. Reproduced from \citet{Yung2025}, Fig.~4. }
\label{fig:UVLFemp}       
\end{figure}

\begin{figure}
\includegraphics[width=\textwidth]{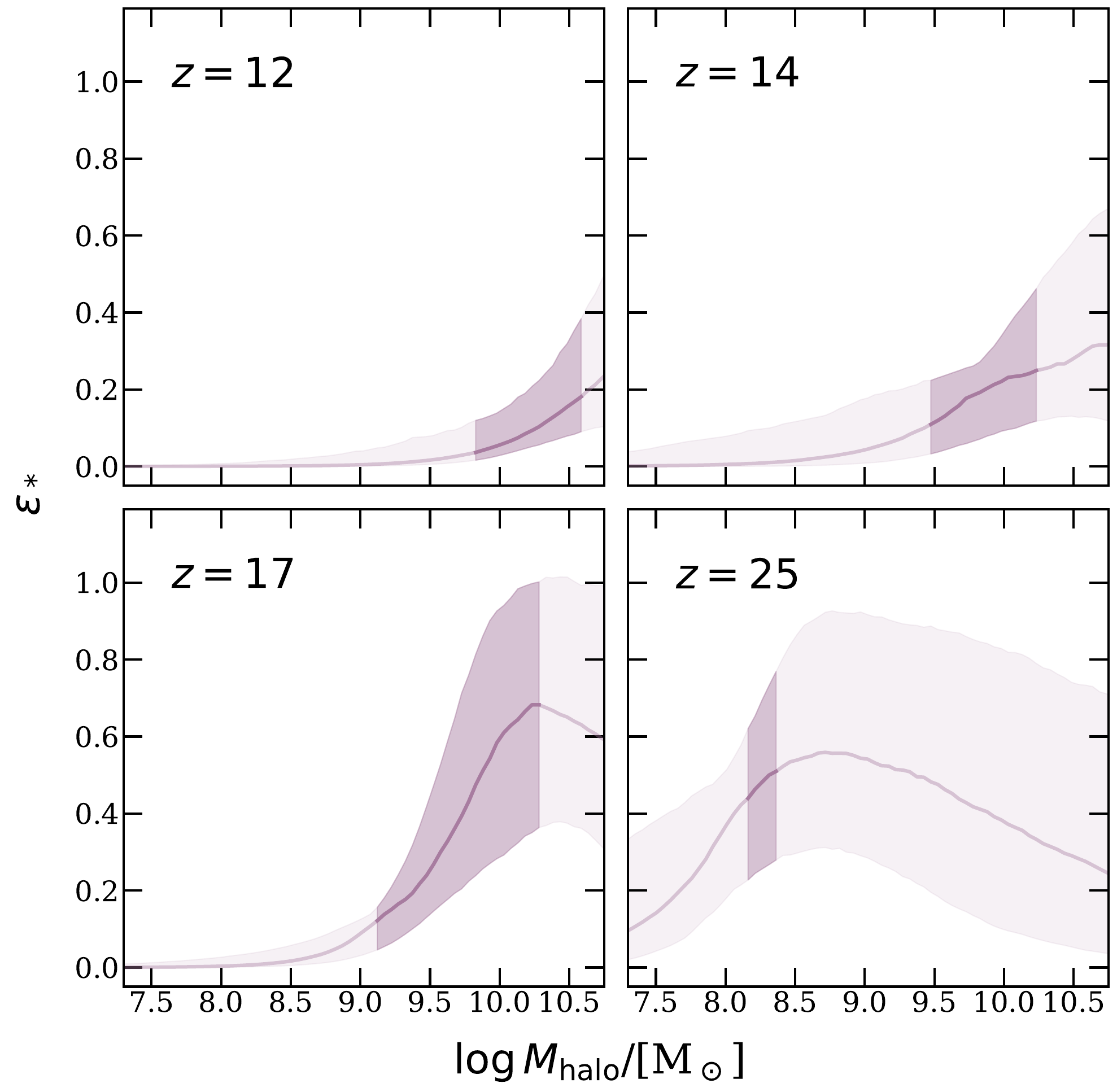}
\caption{The median (solid lines) and 16th and 84th percentiles (shaded regions) baryon conversion efficiency $\epsilon_*(M_h,z)$ obtained from fitting the empirical model described in the text to the observed UVLF. Darker lines and shaded regions show the approximate range of halo mass where there are current observational constraints. The required efficiencies of up to $\sim 20$--50 \% are high, but perhaps not unphysically so. Reproduced from \citet{Yung2025}, Fig.~5. }
\label{fig:SFEemp}       
\end{figure}

Shortly after the first observations from JWST were released, several teams discovered galaxies that seemed to stretch the pre-existing paradigm in various ways (see Ellis Lectures). \citet{Labbe2023} reported galaxy candidates at $z\sim 7$--10 with very large estimated stellar masses, in excess of $10^{10} \msun$ and in a few cases as high as $10^{11} \msun$. \citet{Boylan-Kolchin2023} pointed out that, taken at face value, the total stellar mass density of these few galaxies alone would exceed the entire baryon budget of the Universe at that epoch. Many popular news outlets picked up these results and began to claim that these galaxies could ``break cosmology'' (Scientific American, September 14, 2022)\footnote{https://www.scientificamerican.com/article/jwsts-first-glimpses-of-early-galaxies-could-break-cosmology/}, or even were ``universe breakers'' (The Guardian, February 22, 2023)\footnote{https://www.theguardian.com/science/2023/feb/22/universe-breakers-james-webb-telescope-detects-six-ancient-galaxies}. However, interpreting the cosmological implications of a small number of extreme objects is tricky. \citet{Lovell2022} presented a more rigorous ``extreme value statistics'' (EVS) analysis of the seven \citet{Labbe2023} objects along with a $z\sim 16$ galaxy candidate presented by \citet{Naidu:2022}. Extreme value statistics provide predictions for the greatest (or least) random variable drawn from an underlying distribution (in this case, the DM halo mass function from \LCDM\ along with some assumptions about how to map from DM halo mass to stellar mass). \citet{Lovell2022} concluded that ``either these galaxies are in tension with \LCDM, or there are unaccounted for uncertainties in their stellar mass or redshift estimates''. 

A variety of ``exotic'' resolutions to the purported tension have been proposed, including Early (evolving) Dark Energy \citep{Menci2022,Menci2024}, adding extra small scale power to the primordial power spectrum via a blue spectral index \citep{Parashari2023,Hirano2024}, primordial black holes \citep{Liu2022,Colazo2024} and cosmic strings \citep{Koehler2024}. I believe that we should remain open to these possibilities, and that they are worth exploring, but in the remainder of these notes, I focus on models that retain the context of ``vanilla' $\Lambda$CDM, and explore the implications for baryonic processes.

It is first worth noting that the earliest estimates of stellar masses and redshifts have experienced some revisions. The inflight calibration of JWST's NIRCam instrument resulted in downward revision of the stellar mass estimates overall, and several of the \citet{Labbe2023} objects with anomolously large stellar masses are now thought to be a rather peculiar, previously unknown population of accreting black holes that have come to be known as ``Little Red Dots'' (see Ba\~{n}ados \& Volonteri lectures). Furthermore, spectroscopic follow-up revealed that several of the reported $z\sim 16$ candidates (including the \citet{Naidu:2022} $z=16$ candidate) are actually at much lower redshifts \citep{ArrabalHaro2023}. A re-analysis of JWST massive high redshift galaxy candidates taking into account the updated NIRCam calibration, updated stellar mass and redshift estimates, and with likely AGN removed, found evidence for baryon conversion efficiencies that are somewhat higher than most physics-based models predict, but \emph{did not find any significant fundamental tension with \LCDM} \citep{Chworowsky:2024}.  

An additional, important point to keep in mind when interpreting extreme objects, especially in relatively small fields, is uncertainties and biases arising from galaxy clustering (often referred to as field-to-field variance or cosmic variance). \citet{Jespersen:2025} showed that at the area and volume of existing JWST surveys, field-to-field clustering has a significant impact on sampling statistics, such as the maximum mass of galaxies expected in a given field. They showed that standard EVS analyses, which neglect the impact of clustering and field-to-field variance, can thus significantly overstate the degree of tension. Furthermore, they showed that the galaxy PDF is expected to have significant \emph{skewness}, such that even standard cosmic variance estimates can underestimate the expected difference in object counts from field to field. 

More recently, several groups have reported detections of galaxy candidates at even higher redshifts, $15 \lesssim z \lesssim 30$ \citep{Perez-Gonzalez:2025,Castellano:2025,Gandolfi:2025}, again reopening the question of whether these objects, if really at these extreme redshifts, would present a fundamental challenge for \LCDM. \citet{Yung2025} showed that, with conservative assumptions about mass-to-light ratios, and with accurate halo mass function estimates from the \gureft\ suite of N-body simulations, even these extreme populations can be accounted for within vanilla \LCDM\ without the need for exotic mechanisms (see Fig.~\ref{fig:UVLFemp}). 

Having established that none of the existing JWST observations to date presents a serious fundamental challenge to \LCDM, we can ask what simple empirical analyses can reveal about the physics of these early objects. At $z\gtrsim 6$ and especially at $z\gtrsim 10$, it is difficult to measure the longer wavelengths parts of galaxy spectral energy distributions that are critical for accurate stellar mass estimates. Therefore, many empirical analyses instead focus on the instantaneous star formation efficiency, often using the rest-UV luminosity as a proxy for SFR. A commonly adopted approach is then to assume that the star formation rate is given by SFR = $\epsilon_* f_b\dot{M}_h$, where $f_b$ is the universal baryon fraction and $\dot{M}_h$ is the mass accretion rate of the halo, and to parameterize the SFE as:
\begin{equation}
\epsilon_{\rm *} = \frac{2 \epsilon_0}{(M_h/M_0)^{-\alpha} + (M_h/M_0)^{\beta}}
\label{eqn:estaremp}
\end{equation}
where $\epsilon_0$, $M_0$, $\alpha$, and $\beta$ are free parameters, which may be redshift dependent \citep[e.g.][]{Tacchella:2018}. By adopting a halo mass function, an expression for the halo mass growth rate $\dot{M}_h(M_h, z)$, and a conversion from SFR to rest-UV luminosity, 
one can then fit for the parameter values (or posterior) by matching the observed UV luminosity function. \citet{Shuntov2025} carried out this type of analysis at $z\sim 4$--7 using the FRESCO and CONGRESS JWST surveys, and also presented a compilation of previous analyses of this kind at similar redshifts (see their Figure 7). They find an SFE that is consistent with weak or no evolution over this redshift interval, and which is also consistent with predictions from cosmological hydrodynamics simulations such as First Light \citep{Ceverino2024} and FIREBOX \citep{Feldmann2025}. Similar to integrated SFE in the lower redshift Universe, the values of the SFE peak at around $\epsilon_* \sim 0.2$ and decline to $\epsilon_* \sim 0.03$--0.07 at the lowest halo masses where there are constraints (a few $\times 10^{10} \msun$). 

\citet{Yung2025} quantified the instantaneous star formation $\epsilon_*$ using a similar approach, for the redshift range $z \sim 12$--30. They find evidence for higher values of $\epsilon_*$ at a given halo mass at these epochs, relative to either $z\sim 0$ or $z\sim 4$--7 (see Fig.~\ref{fig:SFEemp}). Assuming that all of the candidates reported by \citet{Castellano:2025} and \citet{Perez-Gonzalez:2025} are in fact galaxies at the estimated photometric redshifts, SFE of $\epsilon_* \sim 0.4$--0.6 would be required at $z\sim 17$--30. These values of $\epsilon_*$ are considerably higher than the typical values at lower redshift at the halo masses where there are constraints ($10^{8.5}$--$10^{10} \msun$). 

However, there are at least two important degeneracies in this kind of analysis. The first is the assumed conversion from SFR to rest-UV luminosity, which depends on the assumed age and metallicity distribution (or star formation and chemical enrichment history), dust content, stellar population model, and stellar initial mass function. Many studies (including \citet{Shuntov2025} and \citet{Yung2025}) assume a fixed conversion factor $\mathcal{K}_\text{UV} = \text{SFR} /L_\text{UV}$, with $\mathcal{K}_\text{UV}$ taken from \citet{Madau2014}. However, \citet{Donnan:2025} showed that an acceptable solution could be found with a non-evolving SFE and an evolving value of $\mathcal{K}_\text{UV}$, which could occur due to stellar populations being younger at earlier times. Evolution in the stellar initial mass function (IMF) towards a more top-heavy IMF at earlier times could also lead to galaxies being brighter in the UV for a given SFR. A second important effect is star formation stochasticity, which could cause scatter in the relationship between  $\dot{M}_h(M_h, z)$ and SFR. I discuss the impact of star formation stochasticity on observed UVLF's further in \S\ref{sec:results:numdens}.

\subsection{Internal structure of early halos: density profiles and spin}
\label{sec:structure:internal}
In the previous section, we focused on the bulk demographics of dark matter halos and the implications for zeroth order observed quantities such as galaxy luminosity functions. However, both dark matter halos and galaxies have internal structure as well, which carries rich information about the physics that shapes their formation. 

\subsubsection{Halo density profiles}
Based on some of the earliest DM-only N-body simulations in the Cold Dark Matter framework, \citet{Navarro:1997} showed that the spherically averaged density profile of dark matter halos could be well described by a functional form in which the density scales with radius as $\rho \propto r^{-1}$ in the inner part, turns over to $\rho \propto r^{-2}$ at a scale radius $r_s$, and steepens to $\rho \propto r^{-3}$ in the outskirts. Thus the density profile could be described to first order via the NFW concentration, defined as $c_{\rm NFW} \equiv R_{\rm vir}/r_s$, where $R_{\rm vir}$ is the virial radius of the halo. Numerous works have quantified how $c_{\rm NFW}$ scales with halo mass and redshift in the concordance \LCDM\ cosmology \citep[e.g.][]{Bullock:2001,Klypin2011,Rodriguez-Puebla:2016}, up to redshifts $z\sim 6$--8. At low to intermediate redshift ($z\sim 0$--2), the average concentration increases with decreasing halo mass. From $z\sim 2$--6, the $c_{\rm NFW}$--$M_{\rm h}$ relation flattens, and $c_{\rm NFW}$ decreases with increasing redshift at fixed halo mass back to $z\sim 6$ (see e.g. Figure 19 of \citealp{Rodriguez-Puebla:2016}). \citet{Yung2024a} used the \gureft\ suite of N-body simulations to probe the evolution of $c_{\rm NFW}$--$M_{\rm h}$ from $z\sim 6$--20. They found that at $z\gtrsim 6$, the sense of the evolution flips, and $c_{\rm NFW}$ starts to slightly increase at fixed halo mass as one looks further back in time. It is well-known that $c_{\rm NFW}$ is strongly correlated with halo formation history, with earlier forming halos having higher values of $c_{\rm NFW}$, reflecting the higher overall density of the Universe at the time that the inner part of the halo collapsed \citep[e.g.][]{Wechsler:2002}.  
\subsubsection{Halo spin distribution}
The collapse process of dark matter halos is not completely symmetric, due to the complex nature of the density perturbation field. Tidal torques during halo collapse in the quasi-linear regime lead to collapsed objects with significant angular momentum \citep[][see MvdBW Ch. 7.5.4 for an overview and more detailed discussion of halo angular momentum and tidal torque theory]{Peebles:1969,Doroshkevich:1970,White:1984,Porciani:2002}. This is traditionally parameterized via a dimensionless spin parameter 
\begin{equation} 
\lambda \equiv \frac{J |E|^{1/2}}{GM^{5/2} }
\end{equation}
\citep{Peebles:1969,Bullock_spin:2001}. Halos can also acquire angular momentum via transfer of orbital angular momentum to internal spin via mergers \citep{Vitvitska:2002,Maller:2002}. Dissipationless numerical N-body simulations have shown that the distribution of spin parameters for a population of halos is well-described by a log-normal distribution (see e.g. MvdBW Eqn. 7.160) with $\bar{\lambda} \simeq 0.035$ and $\sigma_{\ln \lambda}\simeq 0.5$ (see MvdBW Ch. 7.5.4 for references), with only weak dependence of these parameters on halo mass, redshift, and cosmology. However, \citet{Rodriguez-Puebla:2016} reported a moderate decrease of $\bar{\lambda}$ from $\simeq 0.035$ at $z\sim 0$ to $\bar{\lambda} \simeq 0.022$ at $z\sim 8$ (for the Peebles definition of $\lambda$; the Bullock definition shows somewhat weaker evolution). \citet{Yung2024a} extended the study of halo spin distributions from $z \sim 6$--20, finding a continuation of the downward trend, with values of $\bar{\lambda} \simeq 0.02$ or lower at $z\sim 20$, but with a log-normal form still providing a good description of the distribution. 

\subsubsection{Relationship between galaxy size and halo properties}
One of the key structural properties of galaxies is their radial size at a given stellar mass or luminosity. The radial size is often characterized by the 3D or projected radius that contains half of the light (at a given wavelength) or half of the stellar mass, sometimes referred to as the effective radius $r_e$. It is well-known that from the nearby Universe back to $z\sim 6$, galaxies show a correlation between their stellar mass or luminosity and their radius \citep[e.g.][]{Shen:2003,Ravindranath:2004,Ferguson:2004,vanderwel:2014,lange:2015,Oesch:2018}. This correlation has a different slope and evolves differently for late-type (blue or disk-dominated galaxies) and early-type (red or bulge-dominated) galaxies \citep{vanderwel:2014,lange:2015}, though the precise amount of evolution in the stellar mass vs. 3D mass-weighted size relation remains debated due to uncertainties in the corrections from light to mass and from observed projected 2D radius and 3D radius (see \citealt{Somerville2018} and \citealt{Behroozi2022} for a discussion). 

\begin{figure}
\includegraphics[width=\textwidth]{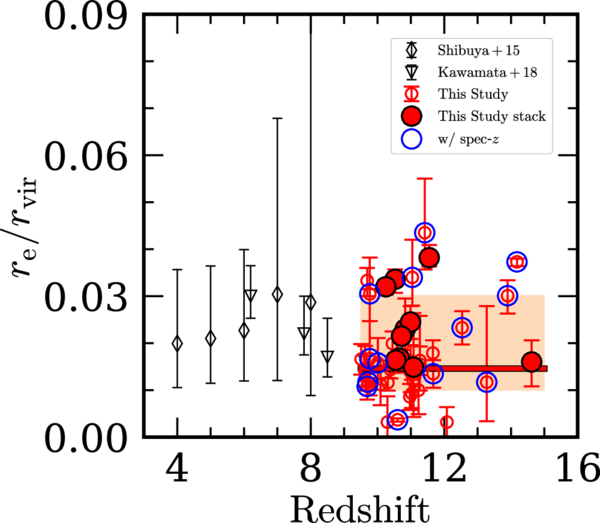}
\caption{Ratio of galaxy UV continuum radius to host dark matter halo radius, $r_e/r_{\rm vir}$, as a function of redshift. The black diamonds and downward triangles indicate the results from pre-JWST studies of $z \sim 4$--8 star-forming galaxies. The red open circles and red filled circles represent the results of individual measurements of relatively luminous galaxies, and results from stacked images of fainter galaxies, respectively, at $z\sim 10$--16 from the analysis of \citet{Ono:2025}. The blue circles indicate spectroscopically confirmed galaxies. It is striking that the ratio of galaxy radius to DM halo virial radius at $z\sim 10$--16 is so similar to that at much lower redshifts. Reproduced from \citet{Ono:2025}, Fig.~20. }
\label{fig:rgalrvir}       
\end{figure}

The physics that shapes galaxy sizes is still poorly understood. There is a standard theory of disk sizes which is adopted in many semi-analytic models (see MvdBW Ch. 11.1-11.2 for a detailed discussion), based on the ``angular momentum partition ansatz''. 
In this picture, described for example in \citet{Blumenthal:1986}, \citet{Dalcanton:1997}, and \citet{Mo1998}, it is assumed that gas is smoothly accreted into halos along with the dark matter, that the diffuse gas has the same angular momentum as the dark matter, and that the gas conserves its specific angular momentum as it cools and collapses to form a disk. It is then generally (though not universally) assumed that the gas settles into a disk with a radially exponential profile, and that the density profile is modified by the gravity of the infalling baryons (``baryonic contraction''; sometimes referred to as ``adiabatic contraction'' due to the common assumption of an adiabatic invariant during disk formation).  In the simplest version of this model, where the halo density profile is an isothermal sphere ($\rho \propto r^{-2}$) and the impact of baryonic contraction is neglected, one can derive an analytic expression for the disk's exponential scale length (e.g. MvdBW Eqn. 11.58; \citealt{Mo1998}): 
\begin{equation}
r_d = \frac{1}{\sqrt 2} \left( \frac{j_d}{m_d}\right) \lambda R_{\rm h}
\end{equation}
where $j_d$ is the fraction of the halo angular momentum that ends up in the disk, and $m_d$ is the baryonic mass of the disk. This simple model can be expanded upon to account for the halo having an NFW density profile and for the baryonic contraction (under the assumption that the disk forms adiabatically) by solving iteratively for a function $f(c_{\rm NFW}, \lambda, f_{\rm d})$ which multiplies the expression above, where $f_{\rm d} = m_{\rm d}/M_{\rm h}$ (MvdBW Ch. 11.2.3; Eqn. 11.64). \citet{Somervillesizev2008} showed that this simple model reproduces the size evolution of disk galaxies quite well out to cosmic noon ($z\sim 2$--3). As pointed out in that work, the observed evolution of galaxy sizes over this period is weaker than the predicted evolution in halo size at fixed mass. However, all else equal, a less concentrated halo is expected to produce a disk with a larger scale length. Thus, the decrease of $c_{\rm NFW}$ towards higher redshift can explain this weaker evolution within the theoretical picture outlined above. 

As discussed above, empirical models can provide constraints on the relationship between global galaxy properties (such as stellar mass) and global halo properties (such as virial mass or velocity). We can take this kind of approach a step further and constrain the relationship between galaxy \emph{structural} properties (such as size) and halo structural properties (such as virial radius). First, one performs standard abundance matching to obtain a relationship between stellar mass and halo mass (often referred to as the SMHM relation). At each redshift, there is a unique relationship between halo virial mass and virial radius, because of the way that halos are defined. Therefore, given an observed relationship between galaxy stellar mass and radius, one can then solve for the relationship between halo virial radius and galaxy radius (sometimes called the SRHR relationship). This kind of analysis was carried out for local galaxies by \citet{Kravtsov:2013} and for galaxies out to cosmic noon ($z\sim 2$--3) by \citet{Huang:2017} and \citet{Somerville2018}. \citet{Shibuya:2015} derived a relationship between galaxy rest-UV size and halo size out to $z\sim 7$, based on observations with HST. 

With JWST, we can now obtain better constraints on the stellar mass weighted radii of galaxies at intermediate redshift ($z\sim 3$--6), and can measure galaxy sizes in the rest-UV out to $z\sim 13$--14 \citep[e.g.][]{Clausen:2025,Allen:2025,Morishita:2024,Ono:2025}.  \citet{Ono:2025} measured sizes for individual galaxies and for a stack of galaxy images at $10 \lesssim z \lesssim 16$, and derived the relationship between galaxy radius and halo virial radius. Intriguingly, they find a median value of $r_{e}/r_{\rm vir} \simeq 0.015$, very similar to the ratio at lower redshift (see Fig.~\ref{fig:rgalrvir}). 

\begin{overview}{Summary}
\begin{itemize}
\item The abundance of dark matter halos of different masses at a given epoch is governed by the primordial power spectrum, the nature of dark matter and early dark energy, and gravitational instability. The characteristic mass that becomes gravitationally unstable and collapses to form a bound object grows rapidly with advancing cosmic time. Therefore, massive halos were much more rare in the past. 

\item Halos that are more rare are also more biased relative to the background dark matter density. Therefore, halos of a given mass were also more biased in the past. 

\item Halos of a given mass had much smaller virial radii and higher virial velocities and temperatures at high redshift, because of the higher overall background density at the time when they formed. 

\item Analytic models for halo abundances and mass functions, such as Press-Schechter and Extended Press-Schechter, as well as halo mass functions extrapolated from fitting functions based on lower redshift simulations, can disagree with state-of-the-art high-resolution N-body simulation results by up to an order of magnitude at $z\sim 10$--30.

\item Adopting accurate N-body based halo mass functions and a simple empirical approach to connect halo mass to UV luminosity, there is no \emph{fundamental} tension between vanilla $\Lambda$CDM and any existing observations of high redshift galaxies. 

\item The halo mass versus concentration relationship at $z\gtrsim 6$ does not follow simple extrapolations from lower redshift N-body simulation based fitting functions. The halo spin distribution remains log-normal, with slightly smaller median spin values at ultra-high redshift.

\item The inferred relationship between galaxy (UV half-light) radius and halo virial radius at $z\sim 10$--14 is very similar to that at lower redshifts.  

\end{itemize}
\end{overview}

\begin{overview}{Additional Reading}
\noindent\citet{MvdBWbook}, Ch. 3, 4, 5.1, 6.1, 7\\
\noindent\citet{Barkana-Loeb:2001}\\
\noindent\citet{Loeb-Furlanetto:2013}\\
\end{overview}

\section{Physical processes and modeling techniques}
\label{sec:models}

I begin this section with an overview of the physical processes that are believed to be important in shaping intrinsic and observable galaxy properties. I then give a brief overview of the methods that are commonly used to model these processes, and present some recent results that are particularly relevant to galaxy formation in the first billion years. I conclude the section with an outlook on new directions and advances in galaxy formation modeling.

\subsection{Physical processes}
\label{sec:models:physics}
\subsubsection{Heating and Cooling}
As we discussed in Section~\ref{sec:structure}, when the force of gravity overcomes the expansion of the Universe in an overdense region, the region 'turns around', separates from the Hubble flow, and forms a gravitationally bound object (a dark matter halo). The gas within this region acquires kinetic (thermal) energy as it loses potential energy. If the cooling time is longer than the dynamical time, then a hot quasi-hydrostatic halo is formed with a temperature close to the virial temperature $kT_{\rm vir} = \frac{1}{2} \mu m_p {V_{\rm vir}}^2 \simeq 35.9 {V_{\rm vir}}^2$, where $\mu$ is the mean molecular weight, $m_p$ is the mass of the proton, and $V_{\rm vir}$ is the virial velocity of the halo (MvdBW Ch. 8.2). If the cooling time is shorter than the dynamical time, the shock energy is rapidly dissipated and gas can cool and fall to the center of the halo on approximately a dynamical time \citep{White:1991,Birnboim:2003,Dekel:2006}.  

Consider an optically thin plasma with primordial composition (hydrogen and helium) in the absence of any radiation field. The relevant atomic cooling processes are: collisional excitation, collisional ionization, recombination, and free-free emission (Bremsstrahlung; see \citealt{Katz:1996}, section 3 for details). In order to calculate cooling rates, we need to know the density of the various ionic species, e.g. $n_e$, $n_{H0}$, $N_{H+}$, $n_{He++}$, etc. It is common to assume \emph{ionization equilibrium}, i.e. that creation and destruction rates are balanced for each species. Since collisional processes drive the ionization balance, this is commonly referred to as \emph{collisional ionization equilibrium}. However, one must think carefully about whether this assumption is appropriate for a given set of conditions.

Under the assumptions of collisional ionization equilibrium, the cooling rate can be written as a function of temperature times the square of the total gas density (or, equivalently, the total hydrogen number density $n_H = X \rho /m_p$, where $X \simeq 0.76$ is the hydrogen mass fraction). Thus the ``cooling function'' can be defined as $\Lambda(T) \equiv \mathcal{C}/{n_H}^2$ (where $\mathcal{C}$ is the cooling rate, and this $\Lambda$ should not be confused with the cosmological constant). Commonly used tabulated cooling functions can be found in \citet{Sutherland:1993}, \citet{Gnat2007}, \citet{Smith:2008}, and \citet{Wiersma:2009}. At temperatures below $\sim 10^4$ K, collisions are not energetic enough to ionize hydrogen atoms or even excite them out of the ground state, so the atomic cooling rate drops to zero below this temperature. Given the expression for the halo virial temperature above, we expect that halos with a virial velocity of less than $\sim 17$ km/s ($M_h \sim 1.8 \times 10^7 \msun$ at $z=30$; $M_h \sim 8.7 \times 10^7 \msun$ at $z=10$, $M_h \sim 1.7 \times 10^8 \msun$ at $z=6$) will be unable to cool via atomic processes. This critical mass or velocity is called the ``atomic cooling limit''.

The first halos to form in the very early Universe have virial temperatures much smaller than the atomic cooling limit (see Fig.~\ref{fig:hmfcomp}), and in the absence of heavy elements, the only available coolant is molecular hydrogen (H$_2$). The lowest energy radiative transition of H$_2$ is the $J = 2 \rightarrow 0$ transition in the rotational ground state, which allows cooling down to a minimum temperature of about 200K \citep{Klessen:2023}. Even at its peak at $\sim 10^4$ K, the H$_2$ cooling function is almost two orders of magnitude lower than the atomic cooling function for a primordial gas (see \citealp{Barkana-Loeb:2001} Fig.~12). At low densities ($n_H \lesssim 10^9$ cm$^{-3}$), and in the absence of dust, there are two primary formation channels for  H$_2$ \citep{Lepp2002,Barkana-Loeb:2001,Klessen:2023}.
The first is \citep{McDowell1961,PeeblesDicke1968}:
\begin{eqnarray}
  H + e^{-} \rightarrow H^{-} + \gamma \\
   H^{-} + H \rightarrow H_2 + e^{-} 
\end{eqnarray}
and the second is \citep{Saslaw1967}:
\begin{eqnarray}
  H + H^{+} \rightarrow {H_2}^{+} + \gamma\\
  {H_2}^{+} + H \rightarrow  H_2 + H^{+}
\end{eqnarray}

Typically, these processes can achieve molecular hydrogen fractions of $10^{-3}$, which is sufficient to lead to cooling and collapse to higher densities. At densities above 10$^9$ cm$^{-3}$, a three-body reaction becomes important:
\begin{equation}
  H + H + H \rightarrow H_2 + H
\end{equation}
This can drive the molecular fractions to values close to unity (see e.g. \citealt{Schauer:2021}). At even higher densities $\gtrsim 10^{11}$ cm$^{-3}$, H$_2$ begins to be collisionally dissociated, and the H$_2$ fraction plateaus or begins to decline a bit \citep{Klessen:2023}. 

Radiation fields play an important role in the chemistry in these primordial halos. Both of the low-density H$_2$ formation channels require free electrons as `catalysts'; the presence of a background X-ray radiation field increases the local supply of free electrons, thereby promoting more efficient H$_2$ formation. On the other hand, UV radiation in the Lyman-Werner bands (11.2 --13.6 eV) photodissociates H$_2$ and suppresses cooling \citep{Sternberg2021}. 

We expect gravitationally bound halos with virial temperatures high enough to allow for H$_2$ cooling, $M_{\rm min, III} \simeq  5 \times 10^4 \msun$,  to begin to form at around $z\sim 30$ \citep{Tegmark:1997,Barkana-Loeb:2001}. However, the relative streaming velocities of baryons and dark matter in the early universe (these relative velocities $\sim 1$ km/s at $z\sim 30$,  are small enough that they only impact very tiny halos) reduce the gas density and delay cooling, pushing this minimum mass up to about 2--4$\times 10^6 \msun$ (see \citealt{Klessen:2023} Fig.~1). As noted above, the presence of a Lyman-Werner radiation field also pushes $M_{\rm min, III}$ to higher values \citep[][and references therein]{Klessen:2023}. 

Once the first stars form and pollute the surrounding gas with metals, metal line cooling becomes important. Even small amounts of metals can greatly enhance the cooling rates. Moreover, metals allow gas to cool down to much lower temperatures (10 K; see \citealt{Drainebook2011}, Ch. 30.4, 34.1). Photo-electric heating (absorption of UV photons by dust grains) and cosmic ray heating can be important heating processes \citep{Bakes1994,Wolfire2003,Drainebook2011}. 

\subsubsection{Cooling in the presence of a photo-ionizing background}
\begin{figure}
\includegraphics[width=\textwidth]{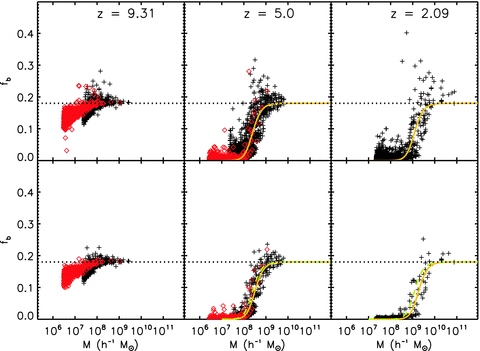}
\caption{Fraction of baryonic mass within halos as a function of total halo mass at three redshifts, for simulations that include a spatially uniform, time-varying meta-galactic UV background. Hydrogen reionization is assumed to start at $z = 9$.  The top row shows all halos, and the bottom row shows isolated halos. After reionization, cooling and accretion is dramatically suppressed in halos below a critical mass, due to photo-ionization ``squelching''. Reproduced from \citet{Okamoto:2008}, Fig.~2. }
\label{fig:squelch}       
\end{figure}

Young massive stars and accreting black holes produce radiation that is energetic enough to ionize hydrogen ($h\nu > 13.6$eV, $\lambda < 912 \mathring{A}$). Neutral hydrogen is opaque to ionizing radiation, but apparently some ionizing photons manage to escape from the ISM of galaxies. We know from observations that most of the IGM was photo-ionized by $z\sim 5.5$--6 (see Ellis lectures and references therein), and thereafter, halos are subjected to a ``meta-galactic'' background of photoionizing radiation. This photoionizing background has two important effects on cooling. First, it removes line excitation and ionization as cooling processes at low densities by destroying H$^0$, He$^0$, and and He$^+$, and second, it heats the gas because the photoelectrons carry away energy \citep{Weinberg:1997}. The heating rate $\cal{H}$ increases as ${n_H}^2$ (where $n_H$ is the hydrogen density) and decreases with increasing temperature. Thus the net rate of radiative energy change per unit volume is $\mathcal{H}-\Lambda$, which scales as ${n_H}^2$ to leading order but has a weak dependence on $n_{H0}$ because of the interplay between photoionization and recombination. At low temperatures (hence in low mass halos), $T \simeq 10^4$--$10^5$ (depending on density),  $\mathcal{H}-\Lambda$ implies \emph{net heating}, i.e., halos below this mass cannot cool \citep{Efstathiou:1992,Thoul:1996,Quinn:1996,Weinberg:1997}. This effect suppresses star formation in halos with masses of $\lesssim 10^{9} \msun$ after the universe is reionized \citep{Gnedin:2000,Kravtsov:2004,Benson:2002}, an effect sometimes referred to as ``photoionization squelching'' \citep{Somerville:2002}. Fig.~\ref{fig:squelch} shows the baryon fraction as a function of halo mass for hydrodynamic simulations with a spatially uniform, time varying UV radiation field \citep{Okamoto2008}, illustrating how cooling and accretion is dramatically suppressed in halos below a critical mass after the Universe is reionized. 

\subsubsection{The first stars: Pop III star formation}
Cooling causes gas to condense to higher density, leading to more rapid cooling and further increase in density, which leads to more rapid cooling, and so on.  Under the right conditions, gas can become cold and dense enough to ignite nuclear fusion: a star is born. The first detailed 3D simulations of the formation of stars from primordial gas (called Pop III stars) were carried out starting in the early 2000's \citep{Bromm:1999,Bromm:2002,Abel:2000,Abel:2002,Nakamura:2001,Nakamura:2002,Yoshida:2003}, and proposed that the first stars may have had typical masses much larger than those born at later times ($\simeq 100$--$1000 \msun$). These early calculations were typically stopped when the density reached $n_H \sim 10^{16}$ cm$^{-3}$, due to the rapidly increasing computational cost. Subsequent simulations in the early 2010's revealed that fragmentation of the gas due to gravitational instabilities is nearly inevitable, leading to multiple proto-stellar clumps or possibly clusters of Pop III stars \citep[e.g.][]{Clark:2011,Greif:2011}. However, the ultimate stellar initial mass function (IMF) of Pop III stars is shaped by a complex mixture of physical processes including turbulence and stellar feedback, and there is not yet a strong consensus on its exact form (see Fig. 6 of \citealp{Klessen:2023} and the review by \citealp{Hennebelle2024}). One thing that all of these predicted Pop III IMFs do have in common, however, is that they tend to be significantly ``flatter'' (more ``top heavy'', i.e. richer in high mass stars) than the IMF in the nearby Universe \citep{Klessen:2023}.

A single Pop III star produces enough metals to self-enrich its host halo to a metallicity of $\sim 10^{-3}$. Additionally, many of these metals may escape the shallow potential wells of their host halos and pollute the IGM, even perhaps contaminating nearby pristine halos with metals. As the metallicity of the ISM increases, simulations suggest that the IMF may grow gradually less flat and less top heavy, until by a metallicity of 0.1 Z$_\odot$ it very closely resembles a \citet{KroupaIMF} or \citet{Chabrier2003} IMF, as is typical in the nearby Universe \citep{Chon2021,Chon2022,Chon2024}. The potential evolution of the IMF has multiple implications for galaxy formation and observability of early galaxies. A stellar population that is enhanced in high-mass stars produces more UV and ionizing radiation per unit mass of stars formed, making galaxies brighter and easier to detect. However, these same high-mass stars produce strong feedback through stellar winds and radiation (as discussed below), perhaps leading to less efficient star formation. Also, stellar populations with a top-heavy IMF will produce more metals, and should have distinctive elemental abundance ratio patterns that may leave an imprint that could be detectable in the oldest stars in our own Galaxy or nearby dwarf galaxies \citep[e.g.][]{Salvadori:2019,Rossi:2025}. 

\subsubsection{Star formation, the interstellar medium, and stellar feedback}
\label{sec:models:physics:sf}
\begin{figure}
\includegraphics[width=\textwidth]{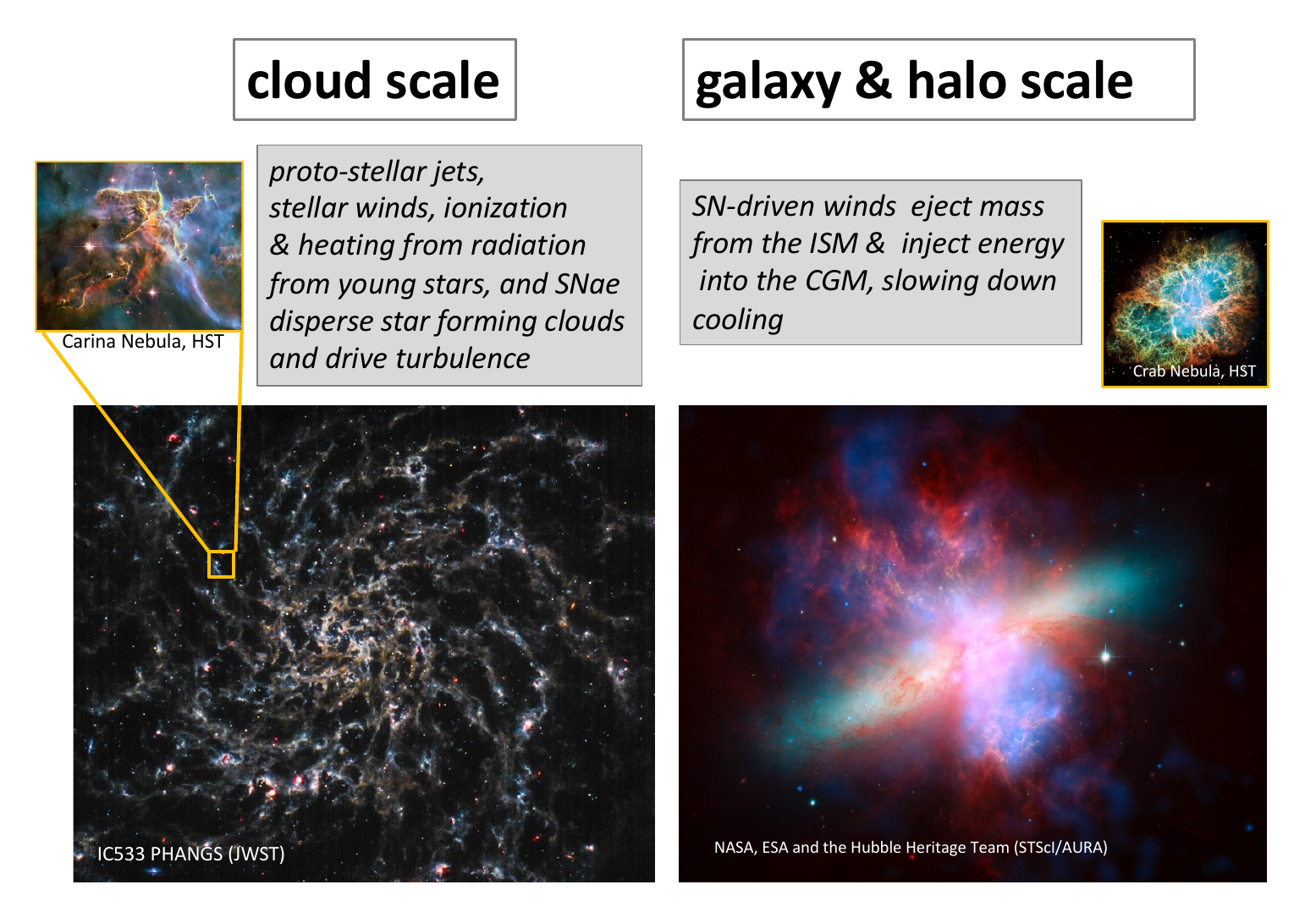}
\caption{`Cloud scale' and `galaxy/halo scale' feedback processes: Proto-stellar jets, stellar winds, ionization and heating from radiation from young massive stars (sometimes called `early stellar feedback'), and supernovae can disperse star forming clouds and drive turbulence in the ISM, reducing the cloud-scale (local) efficiency of star formation. In addition, supernovae explosions can drive galaxy-scale winds, which can eject gas from the ISM and inject thermal energy into the CGM and beyond, slowing cooling and reducing the gas inflow rate into the ISM.}
\label{fig:SFEcartoon}       
\end{figure}

Let's shift our gaze for a moment from the $z\sim 30$ universe to the $z=0$ universe. In our own Galaxy and in nearby galaxies, the ISM is \emph{multi-phase}, with different phases having vastly different temperatures and densities: the coolest/densest phase (the `cold neutral medium' or CNM, which includes molecular clouds) has a density of $\sim 30$--100 cm$^{-3}$ and a temperature $\sim 10$--100 K, and the warmest/lowest density phase (the `hot ionized medium' or HIM) has a density of $\sim 10^{-3}$ cm$^{-3}$ and a temperature $\sim 5 \times 10^5$--10$^7$ K \citep{Drainebook2011}. The standard picture is that thermal instability leads to a three-phase ISM (cold $T\sim 100$K, warm $T\sim 10^4$ K, and hot $T\sim 10^6$ K) in rough pressure equilibrium \citep{McKee:1977}. The formation and destruction of Giant Molecular Clouds (GMC), the sites of star formation, is regulated by a complex interplay between gravitational instability, thermo-chemistry, radiation, and turbulence (see e.g. MvdBW Ch. 9.2). In the presence of metals, cooling to low temperatures ($T< 10^4$K) is dominated by metal ions like [CII] and [OI] \citep{Drainebook2011}. Self-shielding and dust shielding are important to prevent photo-dissociation of molecules. The photo-electric effect is the dominant heating source in low-A$_V$ atomic and molecular gas, and cosmic rays provide the main heating in high-A$_V$ regions \citep{Bakes1994,Wolfire2003,Dalgarno2006}. The importance of photo-electric and cosmic ray heating changes as metallicity decreases, and at sufficiently low metallicities $Z \lesssim 0.01$ Z$_\odot$, the multiphase structure of the ISM may disappear altogether \citep{Bialy:2019}. 

A fundamental timescale for a GMC is the free-fall time: 
\begin{equation}
t_{\rm ff} \equiv \left(\frac{3 \pi}{32 G \bar{\rho}}\right)^{1/2} \simeq 3.6 \times 10^6 {\rm yr} \left(\frac{n_{\rm H_2}}{100 {\rm cm}^{-3}} \right)^{-1/2}
\end{equation}
where $\bar{\rho}$ is the average density and $n_{\rm H_2}$ is the molecular hydrogen number density. The \emph{efficiency} of star formation on sub-galactic scales is often quantified in terms of the fraction of the cloud's mass that turns into stars \emph{per free fall time} (which I will denote $\epsilon_{\rm ff}$). This quantity can be measured in nearby galaxies and out to cosmic noon ($z \sim 2$). The typical values of $\epsilon_{\rm ff}$ are a few percent on scales of GMC to scales of a few hundred pc \citep[e.g.][]{Krumholz2007,Sunphangs2023}. One of the major questions in star formation is \emph{why is star formation in the local universe so inefficient?}

One proposed explanation, popular in the 1980's, is that magnetic fields provide pressure support that prevents GMCs from collapsing under the force of gravity \citep[e.g.][]{Shu:1987}.  However, observational measurements of magnetic field strengths later showed that most GMCs are magnetically super-critical, meaning that the magnetic field cannot effectively regulate star formation. Moreover, the magnetic support theory would predict star formation timescales on the order of the ambipolar diffusion timescale -- but this timescale is about an order of magnitude larger than observationally derived GMC lifetimes (see MvdBW Ch. 9.3.1; \citealp{Maclow:2004}).    

The currently favored picture is that star formation is regulated primarily by a combination of supersonic turbulence and stellar feedback (see \citealt{McKee:2007} for a review). Turbulence is ubiquitous in the ISM across all scales, and is driven by different processes including cosmological accretion, viscous flows within the galaxy, and supernova explosions \citep{Forbes:2023}. The collapse of gas clouds in a turbulent medium is affected in two ways. Turbulence increases the effective velocity dispersion of the gas, thereby \emph{delaying or suppressing} gravitational collapse, but it can also \emph{promote gravitational collapse} by sweeping up and compressing gas in shocks, increasing the density. Self-gravitating supersonic turbulence is thought to be the primary process shaping the probability distribution function (PDF) of cloud masses within galaxies, from GMC scales all the way down to proto-stellar clouds, influencing both the star formation efficiency \citep{Padoan:2012,Federrath:2013} and the stellar IMF \citep{Hennebelle2024}.

The other major factor that influences both star formation efficiency and the IMF is stellar feedback. Once stars start to form, they interact with their environments in a variety of ways, including proto-stellar outflows, stellar winds, and radiation. Radiation can ionize and heat gas, as well as deposit momentum via radiation pressure. These processes start to occur as soon as the first massive stars ignite, and are sometimes collectively referred to as ``early stellar feedback'' or ``massive star feedback''. Fig.~\ref{fig:SFEcartoon} schematically illustrates stellar feedback processes acting on different scales within galaxies.

\begin{figure}    
\includegraphics[width=\textwidth]{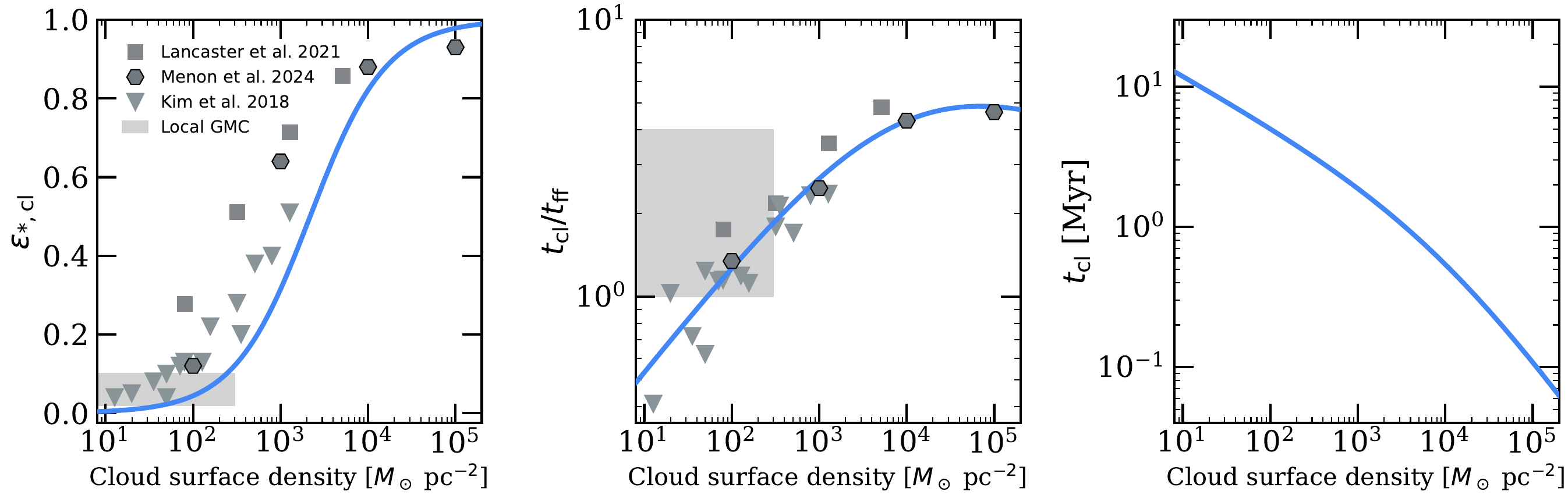}
    \caption{
        Star formation efficiency per star forming cloud (left) and cloud lifetimes (in units of the free fall time (middle), and in Myr (right)) as a function of cloud surface density. The grey shaded regions represent the star formation efficiency (left panel), lifetimes (middle), and surface density range of GMCs in local universe star forming galaxies \citep{Chevance2023}. Symbols show cloud-scale star formation efficiencies (integrated over the cloud lifetime; left) and lifetimes (middle) from cloud-scale simulations by \citet[][squares]{Lancaster2021}, \citet[][hexagons]{Menonfesc2024}, and \citet[][triangles]{Kimjg2018}. The solid blue line in the left panel shows the analytic model of Eqn.~\ref{eqn:estarcloud}. The blue line in the middle panel is an empirical fit to the simulation results. Reproduced from \citet{Somerville:2025}, Fig.~1.    
    }
    \label{fig:estar_tcloud}
\end{figure}

Over the past decade, there have been a large number of numerical studies of how stellar feedback works in individual (idealized) GMC or proto-star clusters (see \citealp{Chevance2023} for a review). These studies have shown that the efficiency of star formation (here characterized by the fraction of the initial cloud mass that is converted to stars by the end of the simulation, which I denote $\epsilon_{\rm *, cl}$) is strongly dependent on the initial surface density of the cloud $\Sigma_{\rm cl}$. The value of $\epsilon_{\rm *, cl}$ increases from a few to ten percent for clouds with $\Sigma_{\rm cl} \simeq 10$--100 M$_\odot$ pc$^{-2}$ (which are typical values for GMCs in nearby galaxies), up to 80--90 \% for clouds with $\Sigma_{\rm cl} \simeq 10^{4}$--$10^5$ M$_\odot$ pc$^{-2}$ (see Fig.~\ref{fig:estar_tcloud} and \citealp{Chevance2023}). In addition, the lifetime of the cloud (in units of the cloud freefall time), defined as the time from the onset of star formation until the cloud is dispersed, is \emph{longer} for denser clouds (Fig.~\ref{fig:estar_tcloud}, middle panel). 

We can understand these results in terms of a simple analytic model \citep{Grudic2018,Grudic2020,Chevance2023}. Let us define $\dot{p}/{m_*}$ as the specific rate of momentum deposition from stellar populations, which here is assumed to include mechanical energy from stellar winds and radiation pressure (but not supernovae). The gas will be unbound, and the cloud dispersed, when the outward momentum is able to overcome the restoring force of gravity. This condition is equivalent to the surface density of the gas exceeding a critical surface density, defined as 
\begin{equation}
    \Sigma_{\rm crit} = \frac{\langle \dot{p}/{m_*} \rangle}{\pi G} 
\end{equation}
The star formation efficiency over the lifetime of the cloud is then
\begin{equation}
\epsilon_{*, \rm cl} = \frac{\Sigma_{\rm cl}/\Sigma_{\rm crit}}{(1.0+\Sigma_{\rm cl}/\Sigma_{\rm crit})}
\label{eqn:estarcloud}
\end{equation}
Fig.~\ref{fig:estar_tcloud} shows the prediction of this model (assuming $\Sigma_{\rm crit}=2176.0$ $\msun$ pc$^{-2}$, based on $\dot{p}/{m_*}$ from Starburst99 \citep{Leitherer1999} models with a \citet{KroupaIMF} IMF; see S25 for details), compared with a compilation of cloud-scale simulations over a range of cloud surface densities. The model does a remarkably good job of qualitatively reproducing the results of these cloud-scale simulations, as seen in Fig.~\ref{fig:estar_tcloud} (left panel). 

A somewhat related picture that holistically summarizes how and why star formation is regulated on galaxy scales is termed the Pressure-Regulated Feedback-Modulated (PRFM) model \citep{Ostriker:2022}. On the one hand, gravity tries to make gas collapse, leading to higher densities, more rapid cooling, and an increase in the potential fuel for star formation. Star formation injects energy and momentum into the gas, providing pressure support that counteracts gravity. The PRFM model posits that the SFR in a galaxy (or a patch of ISM) will adjust until all forms of pressure (e.g. thermal, turbulent, magnetic, cosmic ray) balance the gravitational weight, leading to an equilibrium between gravity and feedback. The PRFM model works extremely well to describe resolved ISM simulations and is supported by observations in nearby galaxies \citep{Ostriker:2022}.  

Supernova driven winds acting on galaxy-wide scales have been invoked for decades to regulate star formation \citep{Dekel:1986,MacLow1999}.
A few million years after the massive stars form, they begin to explode as core collapse supernovae, with each supernova depositing around $10^{51}$ erg of energy into the ISM\footnote{Type I supernovae of course also contribute to feedback, but may be less important in the early Universe due to the longer time delay before they explode.}. The impact depends strongly on the environment in which the supernova explodes. If it explodes within dense gas, much of the thermal energy cools away rapidly, which may delay the next generation of star formation, but it will not convert much of the total energy into momentum, which is able to drive a large scale outflow \citep{Walch:2015b}. If the supernova explodes in lower density, warmer gas, or if clustered star formation leads to a large number of supernovae exploding nearly simultaneously, a large scale outflow may drive gas out of the ISM and perhaps even out of the potential well of the halo altogether (`ejective feedback'; \citealp{Fielding:2018}). This energy may also couple with gas in the circumgalactic medium, leading to reduced or delayed cooling, sometimes termed `preventative feedback' \citep{Lu2015,SomervilleDave2015}. 

\begin{figure}
\includegraphics[width=\textwidth]{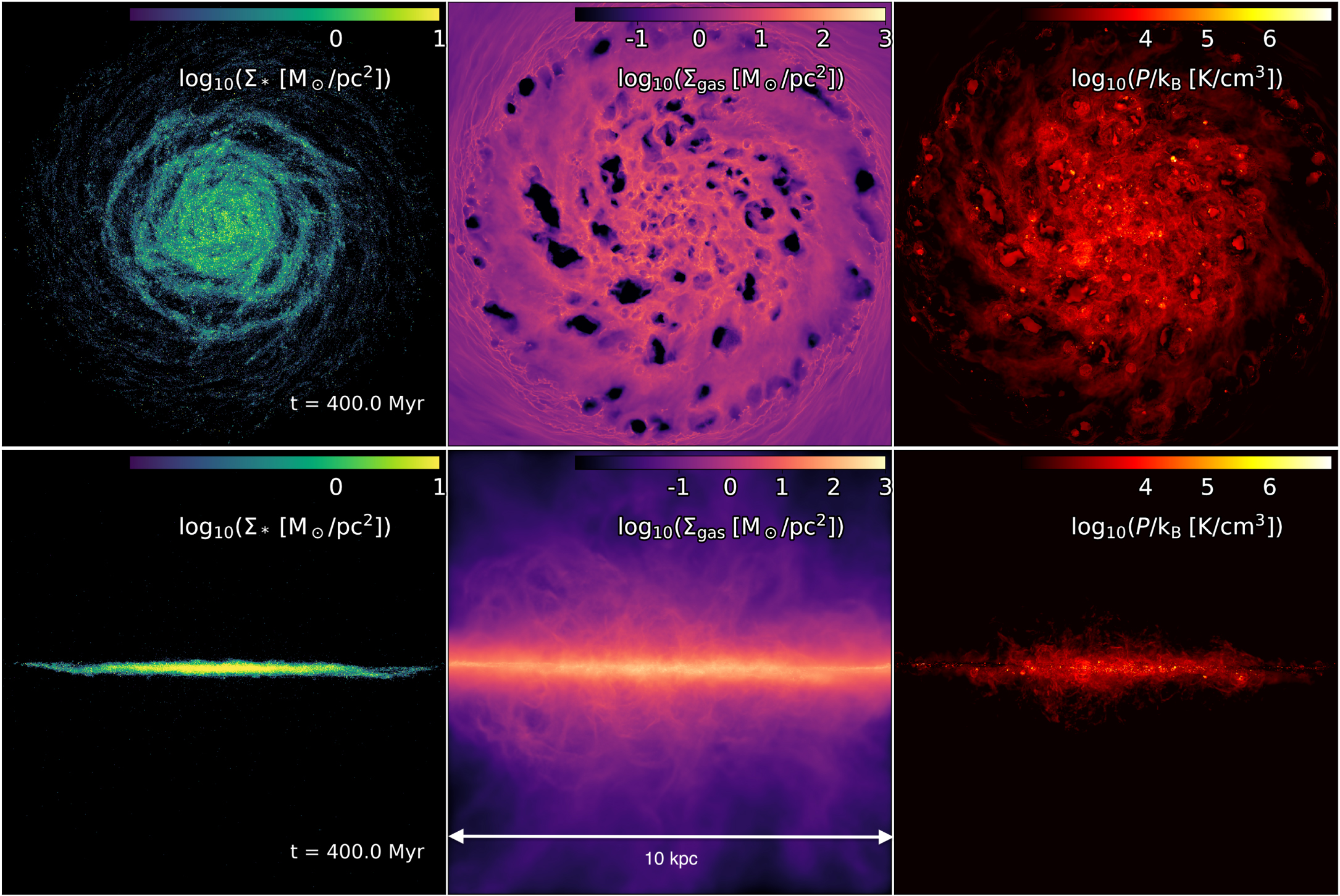}
\caption{Face-on (top) and edge-on (bottom) projections for stellar surface density (left), gas surface density (middle) and thermal pressure (right) for a high resolution, idealized Large Magellanic Cloud analog simulation that resolves the multi-phase ISM and individual supernova explosions. One can see the pressure-driven, SN-inflated bubbles, which drive the large scale outflow visible in the bottom middle panel. Reproduced from \citet{Steinwandel:2024}, Fig.~1.}
\label{fig:steinwandelLMC}       
\end{figure}

Supernova driven galactic-scale winds may be characterized by two quantities --- \emph{mass loading} and \emph{energy loading}. Mass loading is defined as
\begin{equation}
\eta_M \equiv \dot{m}_{\rm out}/\dot{m}_*
\end{equation}
where $\dot{m}_{\rm out}$ is the mass outflow rate of the wind and $\dot{m}_*$ is the star formation rate; and \emph{energy loading} as
\begin{equation}
\eta_E \equiv \dot{E}_{\rm out}/(e_{\rm SN} \dot{m}_*)
\end{equation}
where $\dot{E}_{\rm out}$ is the energy outflow rate of the wind and $e_{\rm SN}$ is the energy deposited by SNae per solar mass of stars formed. The specific energy of the wind is then $\eta_E/\eta_M$. Simulations that resolve individual supernova explosions in sub-galactic slices of ISM (so-called `tall boxes';  \citealp{Martizzi:2015,Kim2015,Walch:2015a,Kimcg2017,Kim:2020}) or idealized galaxies \citep{Hu:2019,Emerick:2019,Gutcke:2021,Steinwandel:2024} are able to predict the emergent mass and energy loading of the supernovae driven winds that arise in these simulations. Fig.~\ref{fig:steinwandelLMC} shows an example of a simulation of an idealized galaxy with resolved stellar feedback from \citet{Steinwandel:2024}, with properties chosen to be similar to that of the Large Magellanic Cloud, with $\sim$ solar mass resolution and sub-pc spatial resolution.

\begin{figure}
\includegraphics[width=\textwidth]{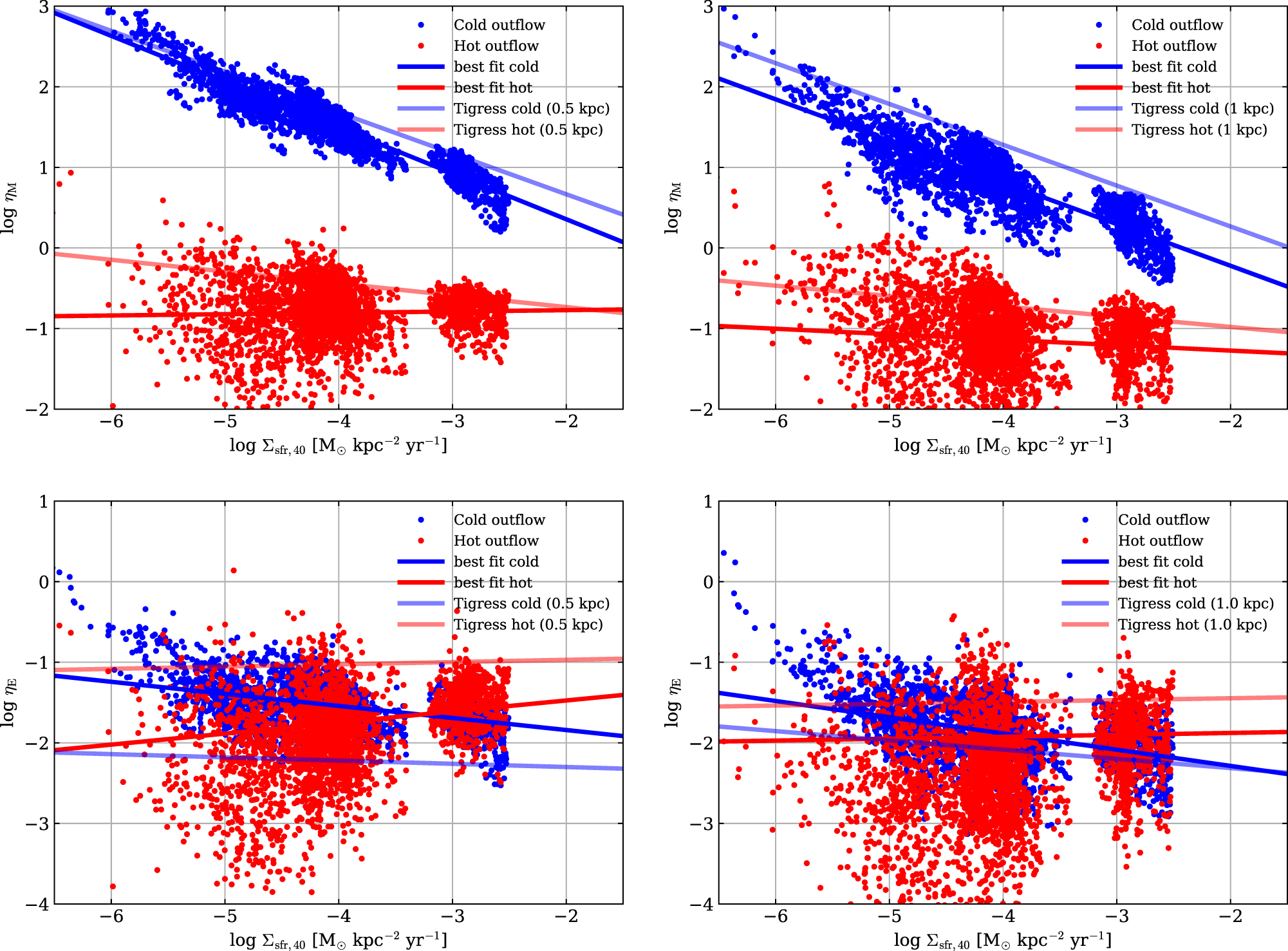}
\caption{Mass loading (top) and energy loading (bottom) as a function of star formation rate surface density for the LMC analog simulation shown in Fig.~\ref{fig:steinwandelLMC}, shown separately for the hot phase ($T > 5 \times 10^{5}$ K; red dots) and the warm phase ($T < 5 \times 10^5$ K; blue dots). The left panels show the mass and energy loading at 0.5 kpc and 1 kpc, respectively. The darker red and blue lines represent a fit to the dots in log–log space, and the lighter lines show fits to the TIGRESS simulations (see text). Simulations with resolved feedback predict that most of the outflowing mass is carried in a cooler, slower moving phase, with a mass loading that decreases with increasing SFR surface density, while most of the outflow energy is carried in a hot, faster wind. Reproduced from \citet{Steinwandel:2024}, Fig.~14. }
\label{fig:massenergyloading}       
\end{figure}

Like the ISM, observed galactic scale outflows are highly multiphase. The simulations mentioned above, with parsec resolution and mass resolution comparable to the masses of individual stars, also naturally produce a multi-phase ISM and multi-phase outflows. One of the key findings of these numerical studies is that the mass loading is dominated by the slower, cold/warm material, while the energy loading is dominated by hot, high velocity outflowing material \citep{Kim:2020,Steinwandel:2024}. Thus, it is important to measure and characterize these key wind parameters separately for these two different phases. When expressed in this way, these studies consistently find that the mass loading $\eta_M$ for the cold phase decreases with increasing gas surface density (or SFR surface density, or disk mid-plane pressure, all of which are strongly correlated), ranging from $\eta_M \sim 100$--1 over three orders of magnitude in surface density. The mass loading of the hot phase is $\eta_M \lesssim 0.1$ with no significant dependence on surface density. The energy loading of the hot phase is $\eta_E \simeq 0.01$--0.1, also with little dependence on density. Fig.~\ref{fig:massenergyloading} shows the mass and energy loadings as a function of $\Sigma_{\rm SFR}$ for the LMC simulation of \citet{Steinwandel:2024} described above, along with results from the TIGRESS simulations \citep{Kim:2020}.

\subsubsection{Chemical enrichment and dust}
\label{sec:models:physics:chemdust}
All elements heavier than helium (aside from boron, beryllium, and a small fraction of lithium) are synthesized in stars or supernovae. Note that I give only the briefest of summaries here, and refer to recent reviews
\citep{Maiolino:2019,Kobayashi:2020,Curti:2025} for a much more complete and detailed account of this subject. Massive stars ($M_* \gtrsim 1.3 \msun$) produce carbon, nitrogen, and oxygen through stellar nucleosynthesis in the so-called CNO cycle. Nitrogen is produced through two channels:  a `primary' channel in massive (or rapidly rotating), low-metallicity stars, when freshly synthesized C and O are mixed into the He burning shell, and a `secondary' channel occurring in Asymptotic Branch Giant (AGB) stars in which pre-existing C and O from the star's birth cloud act as catalysts in the main-sequence CNO cycle. In the primary channel, N production is independent of the initial metallicity, leading to a characteristic flat dependence of N/O on O/H, while in the secondary channel, the nitrogen yield depends on metallicity, leading to an increasing trend of N/O with O/H \citep{Maiolino:2019}.  

In stars more massive than 8 $\msun$, nucleosynthesis continues to create heavier and heavier elements up to iron. Once the stellar core is dominated by iron, nuclear fusion can no longer support the star against gravitational collapse and the star explodes in a core collapse supernova, producing elements heavier than iron through neutron capture of iron-seed nuclei (r-process), followed by $\beta$ decay.  Type Ia supernovae produce a different mix of heavy elements, primarily iron peak, silicon, argon, sulfur, and calcium. Type I supernovae explode after a minimum of about 30 Myr, and most of them likely have considerably longer delay times of up to 1 Gyr. Neutron star mergers are an additional source of elements heavier than iron \citep{Maiolino:2019,Kobayashi:2020}. 

The mass of metals deposited into the ISM per unit mass of stars formed is called the \emph{chemical yield}. Specifically, the yield of a given element $p_i$ is the mass of the newly formed element $i$ relative to the mass of the progenitor star on the main sequence. Alternatively, chemical yields may be expressed for an integrated stellar population with a specific assumed IMF. The theoretical prediction of stellar yields is complex and subject to large uncertainties, and depends on the progenitor star metallicity, assumed mass loss and rotation \citep{Maiolino:2019}. 

As we have seen in \S\ref{sec:models:physics}, the metal content of the gas has a large effect on the efficiency of cooling, with metal enriched gas cooling faster and to lower temperatures. In addition to the uncertainties on the stellar yields, modeling how metals are distributed and mixed into the ISM (and beyond) is extremely challenging. As different elements contribute in different ways to the cooling curves, element abundance ratios also impact cooling rates. Metallicity also impacts the observed spectral energy distributions (SEDs) of stars, through the direct impact on the intrinsic stellar SEDs as well as through the role of metals in the formation of dust, as discussed below. 

Cosmic dust is composed of solid particles with sizes ranging from around 5-250 $\mu$m, and is made up of carbonaceous materials, including graphite and polycyclic
aromatic hydrocarbons (PAHs), and silicates \citep{Draine:2003}. Dust is formed in supernova ejecta and AGB stars, and is released into the ISM. Dust grains can grow in the ISM through the accretion of free metals, and can be destroyed by SN shocks and sputtering. The grain size distribution is modified by shattering and coagulation (see reviews by \citealp{Draine:2003,Schneider:2024}). Dust plays a critical role in the thermo-chemistry of the ISM, and dramatically impacts the observed SEDs of galaxies by absorbing light in the UV-optical and re-radiating it in the IR \citep{Draine:2003}. Dust also provides an important source of opacity for radiation pressure driven winds to act upon. 

\subsubsection{Black Hole feedback}
We have strong evidence from the nearby Universe that most, if not all, massive bulge dominated galaxies contain supermassive black holes \citep{Kormendy:2013}. These black holes must have released a vast amount of energy during their formation, and the effects of black hole or Active Galactic Nuclei (AGN) feedback on galaxy evolution are thought to be fundamental \citep{SomervilleDave2015,NaabOstriker2017}. Black holes can act upon their surroundings via a variety of physical mechanisms and across a very wide range of spatial scales \citep{Heckman:2014}. Up until recently, it was commonly assumed that massive black holes were extremely rare in the very early universe, and therefore that black hole feedback was unlikely to be a dominant physical process in very high redshift ($z\gtrsim 6$) and certainly in ultra-high redshift ($z\gtrsim 10$) galaxies. However, JWST has discovered early SMBH with masses of $\gtrsim 10^6 \msun$ in surprisingly large numbers, with evidence for SMBH at redshifts of $\sim 9$--11 (see Ellis, Ba\~{n}ados, and Volonteri lectures). This suggests that the questions about how, when, and where the first black holes formed, how they grew, and how BH feedback works in early galaxies are likely to be important for understanding galaxy formation in the early Universe. However, I will not discuss this topic further in these notes, as this is covered in the Volonteri lectures. 

\subsection{How to build a galaxy from scratch: methods}
\label{sec:models:methods}
Having discussed the main physical processes that are thought to shape how galaxies form, I briefly discuss how to put these ingredients together into a comprehensive, self-consistent model or simulation of a galaxy or population of galaxies. Note that I have discussed the \emph{conceptual} physical processes and their \emph{implementation} into models and simulations separately for a reason: different implementations of nominally the same physical processes can yield very different predictions. 

Astrophysicists have developed a broad range of tools to model galaxy formation within the framework of hierarchical structure formation in $\Lambda$CDM (or in principle, any alternative cosmological structure formation model)\footnote{Historically, purely empirical models of galaxy populations have also been used, with no grounding in an underlying theoretical cosmological context. Such models are used more infrequently today, and I do not discuss them in these lecture notes.}. I divide such tools into two broad categories: empirical models and physics-based models. Empirical models, which we have already discussed briefly in \S\ref{sec:structure:obs}, derive an empirical \emph{mapping} between dark matter halos as predicted by a cosmological model and observable (e.g. luminosity) or quasi-observable (e.g. stellar mass) properties of galaxies. Such models are generally not specific regarding the physical processes that produce these mappings --- they are more descriptive than predictive. Sub-categories in this class include halo occupation distribution (HOD) models, (sub)-halo abundance matching models (SHAM), and semi-empirical models; see \citet{Wechsler:2018} for a review of these methods. 
In contrast, physics-based models attempt, at varying levels of detail and complexity, to represent physical baryonic processes such as those just outlined (i.e. heating and cooling, star formation, stellar and AGN feedback, chemical enrichment, etc), again within the backbone of cosmological structure formation. Working from the least to the most detailed implementations of physics, we have `gas regulator' models, `semi-analytic models', and numerical (magneto-)(radiation-)hydrodynamic simulations. I briefly summarize each of these techniques in the following sub-sections. 

\subsubsection{Gas regulator models and semi-analytic models}
\label{ref:sec:models:methods:SAM}
In `gas regulator' models, flows of baryons between different reservoirs are tracked by solving ordinary differential equations for global quantities like the mass and metal content of hot (CGM) gas, cold (ISM) gas, and stars \citep{Lilly:2013}. Some recent regulator models also track flows of energy \citep{Carr2023,Voit2024}. There are variants of gas regulator models called `bathtub' or `equilibrium' models, in which it is assumed that gas flows into galaxies are balanced by star formation and outflows, such that the total ISM mass remains roughly constant \citep{Finlator:2008,Dave:2012,Dekel:2014}. 

`Semi-analytic' models (SAMs) are close cousins of gas regulator models, also tracking flows of mass, metals and (in some recent SAMs; see e.g. \citealp{Pandya2023}) energy into and out of the IGM, CGM, ISM, and stellar component of galaxies by solving systems of ODEs \citep[see][for reviews]{Benson:2010,SomervilleDave2015}. There is not a bright line between gas regulator models and SAMs, but SAMs tend to be more detailed and to represent a broader range of physical processes. For example, most SAMs model the full multi-branch merger history of host halos (instead of just the main branch, generally used in gas regulator models), and commonly include modeling of halo and galaxy mergers, separate tracking of stellar bulges and disks (and sometimes stellar halos), and black hole formation and feedback. 

\begin{figure}
\includegraphics[width=\textwidth]{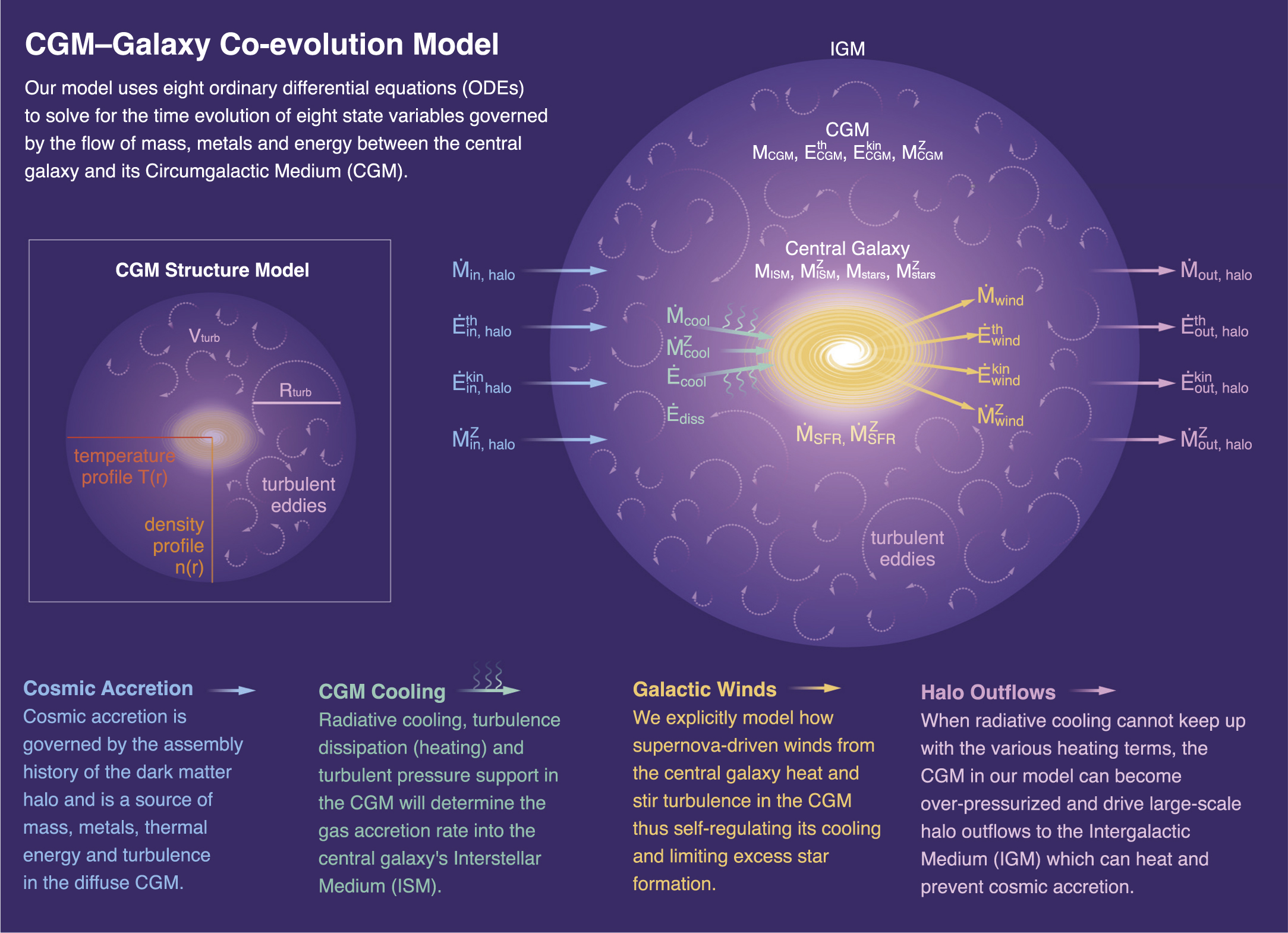}
\caption{Schematic diagram of the ingredients in an example semi-analytic model. Reproduced from \citet{Pandya2023}, Fig.~1. }
\label{fig:SAMschematic}       
\end{figure}

Concrete examples of scaling relations that are adopted in SAMs include:\\
the rate at which mass flows into the CGM: 
\begin{equation}
\dot{M}_{\rm CGM, in} = f_{\rm UV} f_b \dot{M}_{\rm h} + \chi_{\rm re-infall} \frac{M_{\rm ejected}}{t_{\rm dyn}}
\end{equation}
(where $f_{\rm UV}$ is the factor suppressing gas accretion due to photoionization squelching by UV radiation \citep{Kravtsov:2004}, $f_b$ is the universal baryon fraction, $M_{\rm ejected}$ is the mass in an `ejected' gas reservoir, $t_{\rm dyn}$ is the halo dynamical time, and $\chi_{\rm re-infall}$ is a free parameter); \\
the rate that mass cools from the CGM and flows into the ISM:
\begin{equation}
\dot{m}_{\rm ISM, in} = \frac{1}{2} M_{\rm CGM} \frac{R_{\rm cool}}{R_{\rm vir}} \frac{1}{t_{\rm dyn}}
\end{equation}
(where $R_{\rm cool}$ is the radius within which the gas has had time to radiate all of its thermal energy away via cooling \citep{White:1991});\\
and the rate that the ISM gas is converted into stars:
\begin{equation}
\dot{m}_* = \int \Sigma_{\rm SFR} 2\pi r dr 
\end{equation}
where the star formation rate density is generally a variant of a Kennicutt-like relation $\Sigma_{\rm SFR} \propto A \Sigma_{\rm ISM}^N$. Some SAMs adopt a recipe to partition gas into molecular and atomic phases, and use a star formation recipe that is based only on molecular gas \citep[e.g.][]{Bigiel:2008}. 
Similarly, simple scalings are adopted for outflow terms, such as the rate that gas is ejected from the ISM by SNae driven winds:
\begin{equation}
\dot{m}_{\rm ISM, out} = \epsilon_{\rm SN} \left( \frac{V_{\rm disk}}{V_0} \right)^{-\alpha_{\rm rh}} \dot{m}_*
\end{equation}
where $\epsilon_{\rm SN}$ and $\alpha_{\rm rh}$ are free parameters, $V_0$ is a reference parameter with a fixed value, $V_{\rm disk}$ is the circular velocity of the disk, and $\dot{m}_*$ is the star formation rate. Fig.~\ref{fig:SAMschematic} shows a schematic representation of the flow cycle in the semi-analytic model presented in \citet{Pandya2023}.

\subsubsection{Numerical hydrodynamic simulations}
\label{sec:models:methods:hydrosims}
Numerical (magneto)-hydrodynamic simulations explicitly solve the equations of gravity, (magneto)-hydrodynamics, thermodynamics, and chemistry by representing matter as particles or grid cells; for reviews see \citet{Springel:2010,Teyssier:2015,Springel:2016,Vogelsberger:2020,Feldmann:2025}. 

\begin{figure}
\includegraphics[width=\textwidth]{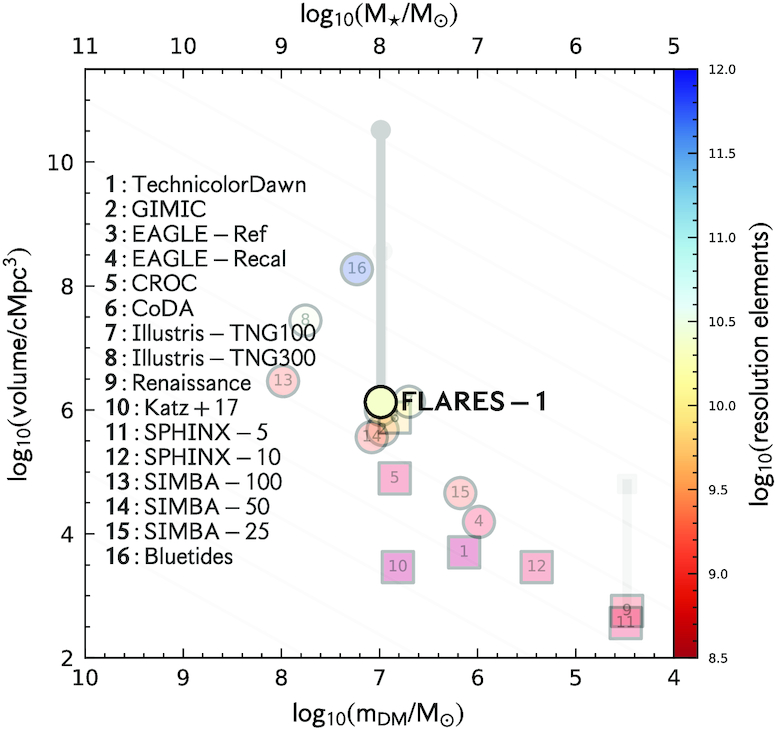}
\caption{Dark matter element resolution versus volume for some recent cosmological simulations. The color of individual points indicates the approximate number of resolution 
elements (dark matter + baryonic gas, excluding stars). Reproduced from \citet{Lovell:2021}, Fig.~1. }
\label{fig:simdynrange}       
\end{figure}

Probably the greatest challenge in numerical simulations of galaxies arises from their limited dynamic range, which is a fundamental consequence of the finite amount of memory and  CPU (or GPU) hours available on computers. This imposes an unavoidable trade-off between simulation volume and resolution (see Fig.~\ref{fig:simdynrange}), and necessitates the adoption of `sub-grid' recipes to model processes that occur at resolutions smaller than those that are able to be explicitly simulated. These considerations also preclude the inclusion of all of the physical processes that are believed to be important in galaxy formation, or necessitate simplified approximate treatments. For example, many/most cosmological simulations of the past decade do not include magnetic fields, on-the-fly radiation transport, non-equilibrium cooling and chemistry, and cosmic rays (though there has been much recent progress in these areas; see Section~\ref{sec:models:outlook}). 

\begin{figure}
\includegraphics[width=\textwidth]{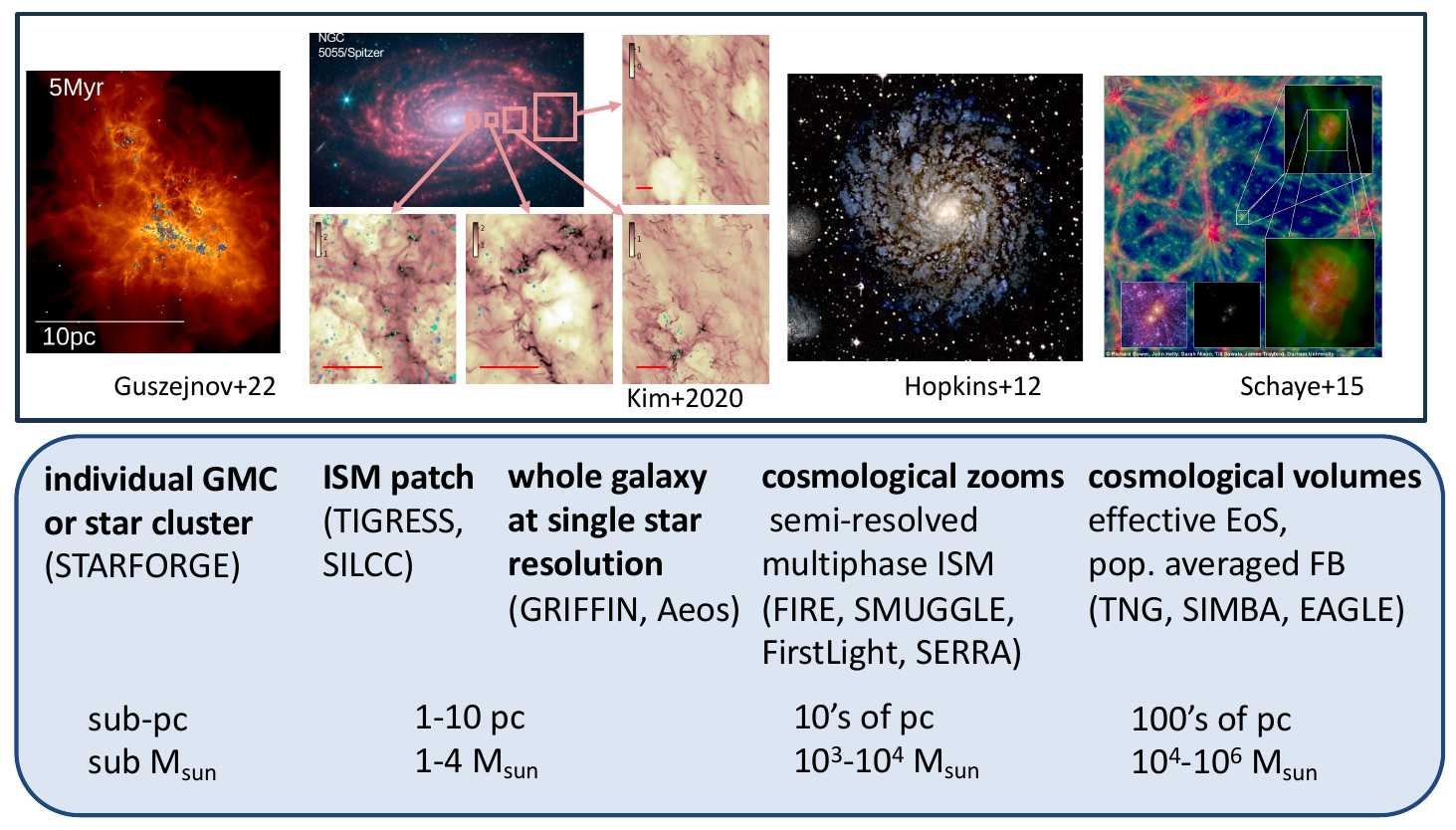}
\caption{Examples of multi-scale simulations from individual GMC or star cluster scales to cosmological volumes. The typical spatial and mass resolution is given at the bottom. The images depicted are from STARFORGE \citep{Grudic:2021,Guszejnov:2022}, TIGRESS \citep{Kim:2020}, FIRE \citep{Hopkins2012}, and EAGLE \citep{Schaye:2015}. See text for a more complete list of simulations. }
\label{fig:simcartoon}       
\end{figure}

One way to overcome this challenge is to adopt initial conditions representing different scales, and to carefully choose the physics that is included and how sub-grid physics is implemented.  By combining the insights from these multi-scale simulations, we can begin to build up a complete, physics-grounded picture of galaxy formation. Fig.~\ref{fig:simcartoon} shows examples of different types of numerical simulations that have been designed to study different scales. These range from simulations of individual GMC and proto-star-clusters\footnote{Of course, simulations of even smaller scale astrophysical objects such as individual stars, protoplanetary disks and planets are also an active field of research} to cosmological volumes spanning thousands of Mpc$^3$ to Gpc$^3$. The smallest scale simulations naturally have the highest spatial and mass resolution (sub-pc and sub-$\msun$) while the largest volume simulations have the coarsest ($\sim 100$'s of pc to kpc; 10$^4$--10$^6 \msun$). 

For large-volume cosmological simulations, many of the key physical processes that we discussed above (including star formation, stellar feedback, chemical enrichment, and black hole formation, growth, and feedback) are modeled with sub-grid recipes. Currently, most groups adopt phenomenological sub-grid recipes that contain somewhat arbitrary choices, and different groups use different implementations of what are notionally the same physical processes. The common practice over the last decade or so has been to parameterize (some of) the uncertainties in the sub-grid recipes, and to calibrate these parameters to reproduce a chosen set of global observable (or quasi-observable) galaxy properties, usually at $z\sim 0$. 
Early cosmological hydro simulations within the $\Lambda$CDM paradigm had difficulty reproducing basic properties of galaxies --- they suffered from a profound `over-cooling' problem, turning far too large a fraction of baryons into stars, as well as an `angular momentum catastrophe', producing galaxies that were too compact and bulge dominated \citep{Navarro:1995,Steinmetz:1999,Sommer-Larsen:1999}. Stellar and black hole feedback are now invoked to solve these problems, but when energy from stars or black holes is injected into gas in a straightforward manner in large-volume, coarse resolution simulations, most of the energy cools away rapidly, and cooling and star formation fail to be regulated to the required degree. As a result, current large volume simulations adopt various ``tricks'' to make feedback efficient enough to reproduce the low star formation efficiencies and baryon conversion efficiencies seen in the local universe. We briefly describe some of the most commonly used approaches for sub-grid modeling in cosmological simulations below, and refer to \citet{SomervilleDave2015} and \citet{NaabOstriker2017} for a more detailed discussion. 

\begin{figure}
\includegraphics[width=\textwidth]{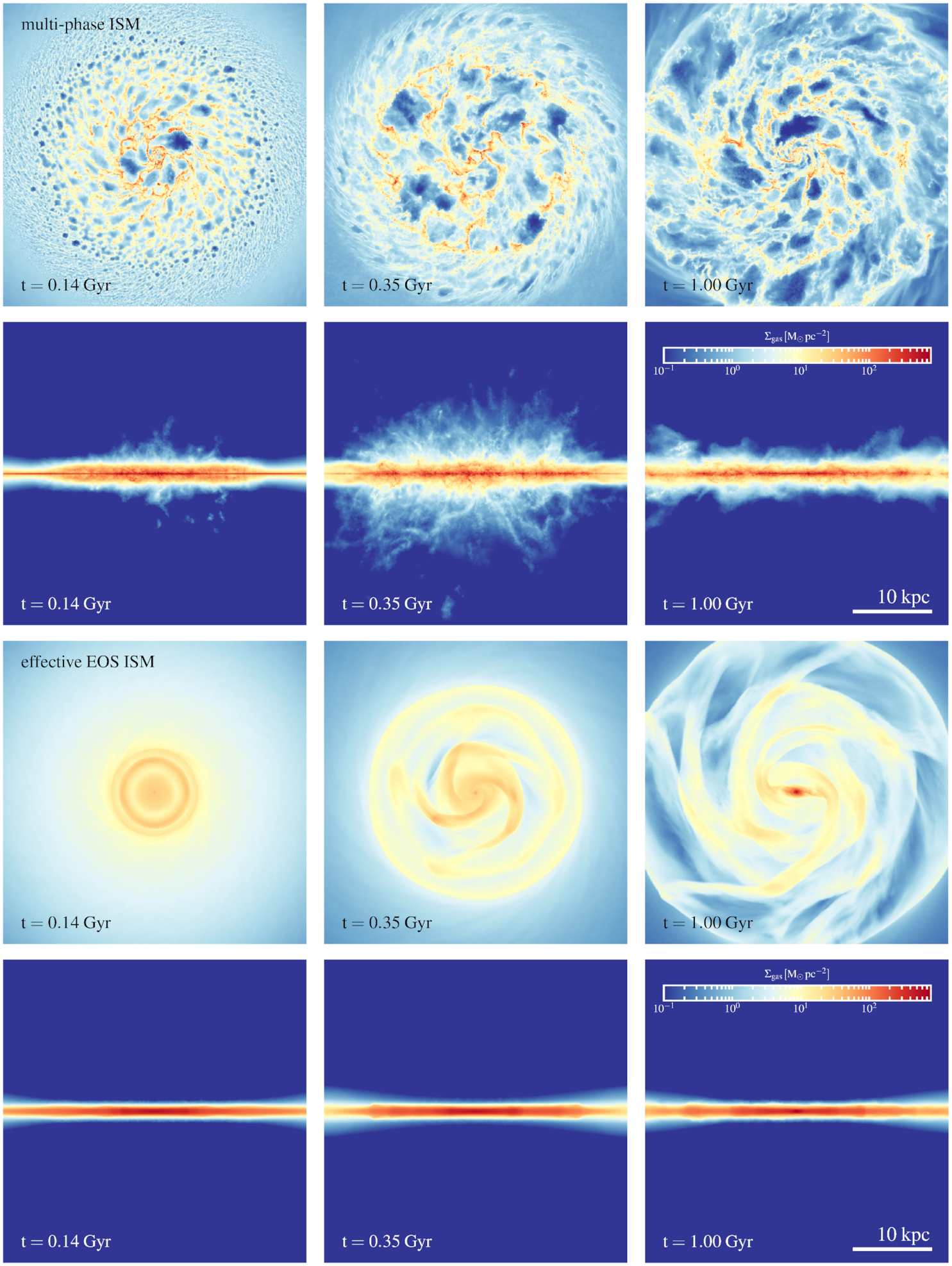}
\caption{Gas column density in a face-on and edge-on projection of a Milky Way-like idealized galaxy, shown at times indicated on the panels. Top: including an explicit, resolved model of the ISM. Bottom: using the \citet{Springel:2003} ISM sub-grid model. The galaxy with the sub-grid ISM model is much smoother, and no outflows are generated. }
\label{fig:image_eEOS}       
\end{figure}

\noindent \textbf{ISM and star formation:} As we discussed in \S\ref{sec:models:physics:sf}, the ISM is multiphase, spanning a vast range of temperatures and densities. Large-volume cosmological simulations generally cannot explicitly resolve this multiphase ISM, and therefore frequently adopt a sub-grid treatment of the dynamics of the unresolved phases of the ISM, often implemented via an `effective equation of state' (eEoS).  A relationship between pressure and density of the form 
\begin{equation}
P = K \rho^\gamma
\end{equation}
is known as a \emph{polytropic} equation of state. Here, $\gamma$ is not generally assumed to be constant as in an ideal gas, but may be a function that depends on other variables. One commonly adopted approach is based on the sub-grid ISM model presented by \citet{Springel:2003}, which is motivated by the picture of cold clouds in pressure equilibrium with an ambient hot medium, as in the two-phase ISM picture of \citet{McKee:1977}. Their model includes star formation, cloud evaporation by SNae, and cloud growth due to cooling. \citet{Springel:2003} write down a set of differential equations describing the masses of the hot and cold phases under the interplay between these processes, and derive analytic solutions that apply under certain conditions and sets of assumptions. In their model, stars form only in the dense cold phase, and star formation leads to supernovae that evaporate these cold clouds, decreasing the star formation rate. Conversely, as star formation decreases, the rate of formation of cold clouds increases, leading to increased star formation and supernovae, etc.  This tends to lead to a self-regulated system that settles into an equilibrium state. In this self-regulated regime, one can derive an effective pressure, and hence an effective polytropic index $\gamma_{\rm eff} \equiv {\rm d\log} P_{\rm eff}/{\rm d\log}\rho$  that is constant in time and is only a function of density. Another approach, adopted for example by \citet{Schaye:2008,Schaye:2015} is to simply assume a polytropic equation of state with a constant effective index $\gamma_{\rm eff}$. Cosmological simulations also frequently adopt a gas temperature and/or pressure floor. Fig.~\ref{fig:image_eEOS} shows images of an idealized Milky Way galaxy, simulated with the same resolution with an explicit ISM model and resolved feedback, and with the \citet{Springel:2003} eEOS ISM model \citep{Marinacci2019}. It is clear that the galaxy simulated with the eEOS model is much smoother and vertically thinner. The eEOS also results in a less bursty star formation history (see \citealp{Marinacci2019} Fig.~3).

Various criteria to determine which gas is eligible to form stars are then adopted. Coarse resolution simulations typically simply adopt a density and temperature threshold. Higher resolution simulations that at least partially resolve the multi-phase ISM sometimes adopt additional criteria, such as requiring the gas to be predominantly molecular, self-gravitating, Jeans unstable, and/or in a region with convergent flow \citep[e.g.][]{Hopkins2014,Hopkins2018}. The star formation rate density in gas that satisfies these conditions is then given by some variant of a Schmidt-type relation, i.e. 
\begin{equation}
\dot{\rho}_* = \frac{\epsilon_{*, \rm ff} \rho_{\rm gas}}{t_{\rm ff}}
\end{equation}
where $\epsilon_{*, \rm ff}$ is generally assumed to be a constant that is calibrated to reproduce observations such as the local Kennicutt-Schmidt relation\footnote{Some simulations assume that the dependence on density steepens above a critical density.}. Following \citet{Schaye:2008}, the EAGLE simulations and their kin cast the SF law as a function of pressure instead of density \citep{Schaye:2015}. 

As discussed in \S\ref{sec:models:physics:sf}, turbulence is thought to be one of the main factors that influences star formation efficiency in the ISM, but the full turbulent cascade is not resolved in galaxy scale or cosmological simulations. Therefore some simulations introduce sub-grid recipes for turbulence and incorporate local gas properties such as turbulent velocity dispersion or Mach number in the star formation criteria and efficiency \citep[e.g.][]{Federrath:2013,Semenov:2016,Kretschmer:2020,Semenov2025a,Semenov:2025b}. For example, some simulations adopt a star formation efficiency $\epsilon_{*, \rm ff}$ that is a function of the virial parameter
\begin{equation}
\alpha_{\rm vir} \equiv \frac{5 \sigma_{\rm 1D}^2 R}{GM}
\end{equation}
where $\sigma_{\rm 1D}$ is the 1D velocity dispersion, $R$ is the length of the grid cell and $M$ is the gas mass within the cell. The typically adopted scalings $\epsilon_{*, \rm ff} \propto \exp -\sqrt{(\alpha_{\rm vir}/0.53)}$ are motivated by analytic models and numerical simulations of star formation in turbulent clouds \citep[see][for a review]{Padoan:2014}.\\

\noindent \textbf{Stellar and SNae feedback}: As alluded to above, early attempts to model stellar feedback by simply injecting thermal energy were unsuccessful, as the energy quickly cooled away without significantly impacting the pressure of the ISM, let alone driving a large scale outflow. There are two broad categories of approaches to mitigate this `overcooling' problem that have been adopted in large volume/coarse resolution simulations: 
\begin{itemize}
\item \textbf{kinetic decoupled winds:} Wind particles are launched with an assumed velocity (which may be a function of other variables), and inherit the properties of the gas cell from which they originate. A parameterized mass loading function determines the probability of launching a wind particle. Hydrodynamics forces are artificially switched off (i.e. wind particles are decoupled) until the particle finds itself in a cell with a density below a critical value. The wind mass loading function (which may be a function of other variables, such as gas phase metallicity) and velocity are tuned to match a set of calibration quantities derived from observations. This approach is used, for example, in the Illustris \citep{Vogelsberger:2014}, IllustrisTNG \citep{Pillepich2018}, and SIMBA \citep{Dave2019} simulations. 

\item \textbf{delayed cooling (blastwave) or stochastic thermal feedback:} In delayed cooling models \citep[e.g.][]{Stinson:2006,Bournaud:2010}, after the thermal energy is injected, radiative cooling is shut off for the lifetime of the SN-driven blastwave as predicted by an analytic model (used for example in the NIHAO \citep{Wang:2015} and ROMULUS \citep{Tremmel:2017} simulations). In stochastic thermal feedback models \citep{DallaVecchia:2012}, the mean energy released by SNae is stored up until a specified temperature boost can be achieved ($\Delta T \sim 10^{7.5}$K). A parameterized function $f_{\rm th}$ determines the probability that a given gas particle will get heated. The minimum temperature boost and the function  $f_{\rm th}$ (which is a function of other variables, such as gas density and metallicity) are tuned to match a set of calibration quantities derived from observations. This approach is used, for example, in the EAGLE simulations \citep{Schaye:2015} and follow-on simulations based on the EAGLE physics model. 
\end{itemize}

These different implementations of the same underlying conceptual physical processes may respond differently to different numerical resolution, and may interact differently with different underlying hydro solvers. Furthermore, up until recently, they have been calibrated and tested most extensively in the relatively low redshift Universe. Various sub-grid implementations may extrapolate differently in the very different conditions that are typical of the high redshift Universe. 

\subsection{Status of theoretical models \& simulations pre-JWST launch}
\label{sec:models:status}
A large number of detailed semi-analytic models and numerical cosmological hydrodynamic simulations have been developed over the past two decades. This topic has been reviewed (relatively) recently by \citet{SomervilleDave2015}, \citet{NaabOstriker2017}, \citet{Vogelsberger2020}, \citet{Crain:2023}, and \citet{Feldmann:2025}. As described there and discussed above, all of these models and simulations contain assumptions about physical scaling laws and/or sub-grid physics, and many of them were calibrated to match global low redshift galaxy observations. Each simulation group chooses a different set of calibration quantities as well as different observational estimates of those quantities, which is an additional factor leading to dispersion between different predictions. Perhaps the one nearly universally used calibration quantity is the $z\sim 0$ stellar mass function (SMHM relation), which probably explains why different simulations yield fairly similar predictions for this quantity. 

These models have yielded some non-trivial successes. They seem to naturally reproduce some observables that were not explicitly calibrated, such as large-scale clustering of galaxies of different stellar masses and colors \citep{Springel:2018,Yung2022}, and qualitative demographics of disk and spheroid dominated galaxies as well as the (qualitative) correlation between morphology or color and large scale environment (see \citealt{SomervilleDave2015} for further discussion and references). 

However, it has been shown that when comparing other quantities that are less widely used in calibration (in some cases because they are difficult to constrain observationally), particularly those related to more diffuse phases of gas, different models and simulations show much larger differences in their predictions. For example, the dispersion in the predicted HI masses \citep{Dave:2020}, CGM masses \citep{Crain:2023}, and IGM column density distributions \citep{Tillman:2023} show much larger differences between simulations with different sub-grid physics implementations. Also, perhaps due to the additional complexity and uncertainties in chemical yields and the detailed implementation of chemical enrichment models, mass-metallicity relationships for both gas and stars also show large differences between models \citep{SomervilleDave2015}.

\begin{figure}
\includegraphics[width=\textwidth]{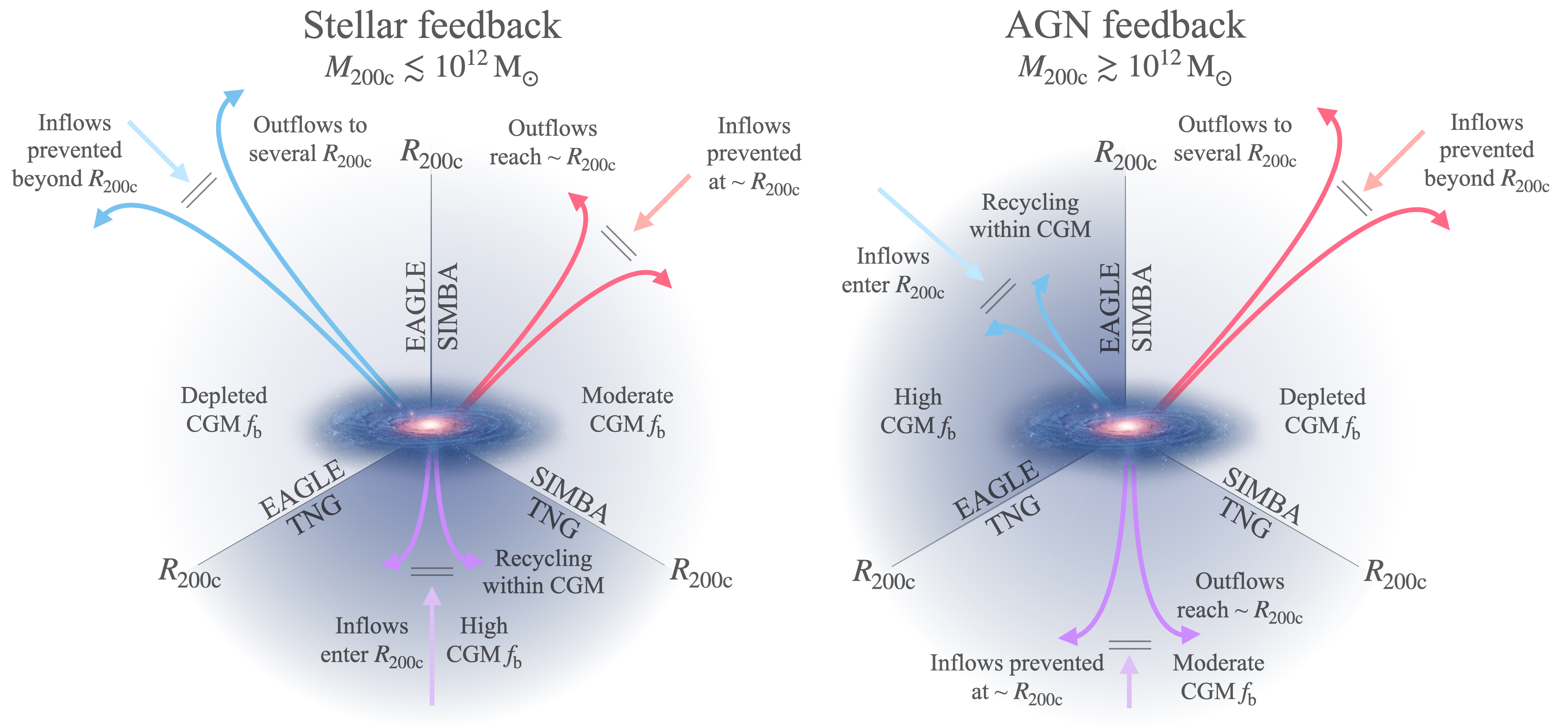}
\includegraphics[width=\textwidth]{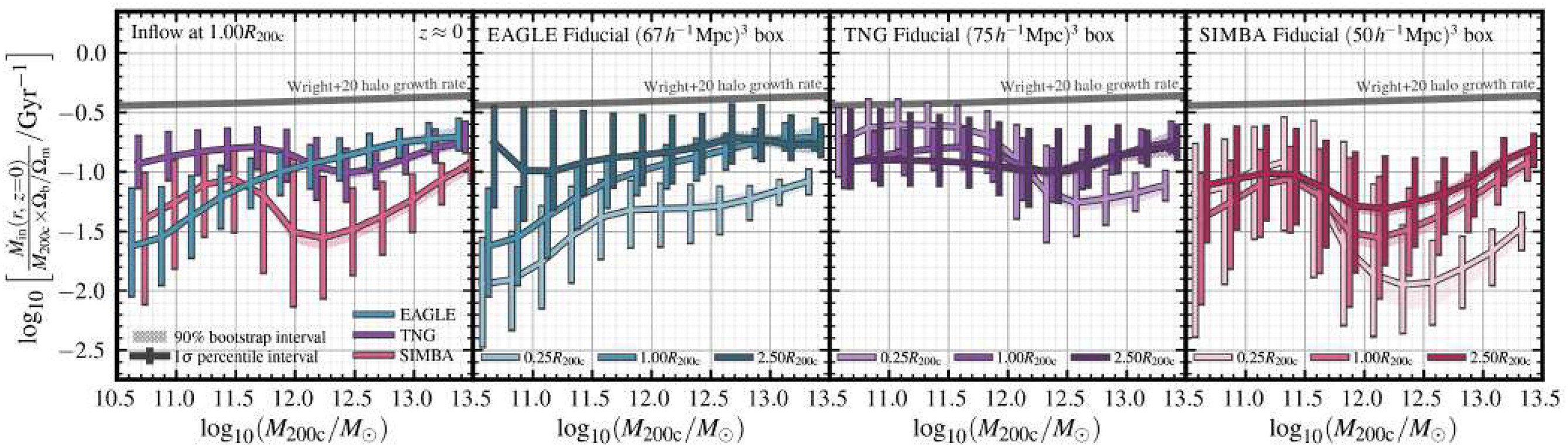}
\includegraphics[width=\textwidth]{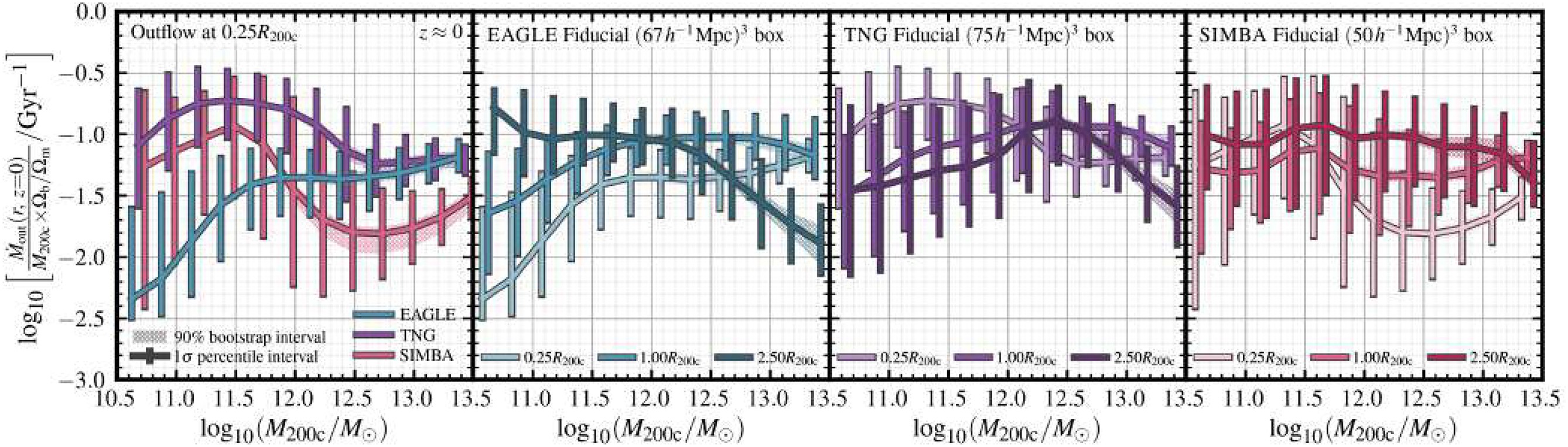}
\caption{Top: Schematic indicating the magnitude and extent of galaxy and halo scale outflows in three cosmological simulations (IllustrisTNG, SIMBA, and EAGLE). 
Middle: Mass inflow rates at the halo scale (left) and at several other scales, as indicated, for the same three simulations. Bottom: mass outflow rates at the ISM scale ($0.25 R_{\rm vir}$) and several other scales, for the same three simulations. Error bars correspond to the 16th–84th percentile range at a given mass, and hatched regions correspond to the bootstrap-generated confidence interval on the medians at a given mass. Although these simulations produce similar results for the median stellar mass as a function of halo mass, they achieve this in very different ways, as the large differences in the inflow and outflow rates reveal. Reproduced from \citet{Wright:2024}, Fig.~8, 5, and 7.  }
\label{fig:wright_flows}       
\end{figure}

Studies of the \emph{flow rates} of gas into and out of halos and galaxies has revealed that a primary reason for these differences is that galaxies regulate star formation in very different manners in these different models. For example, \citet{Pandya:2020} studied halo and galaxy scale inflow and outflow rates in the FIRE-2 zoom-in simulation suite and in the Santa Cruz SAM run within the same DM-only merger trees. They found that at high redshift, turbulence is important for suppressing cooling and star formation, an effect not included in the Santa Cruz SAM. At lower redshift, preventative feedback slows down the accretion and cooling of gas in low mass halos in FIRE, while the Santa Cruz SAM (which does not include preventative SN feedback) relies on very high mass loading factors to instead suppress star formation with strongly ejective feedback. \citet{Wright:2024} carried out a study of gas inflows and outflows on different scales in the EAGLE, IllustrisTNG, and SIMBA simulations, and also showed significant differences between the inflow \emph{and} outflow rates in these simulations, as well as the scales on which they act. For example, in EAGLE and SIMBA, stellar driven outflows could reach well beyond the virial radius of the halo, while in IllustrisTNG, mass outflow rates were higher on ISM scales but ejected material tended to recycle within the CGM without escaping the halo (see Fig.~\ref{fig:wright_flows}).

\begin{figure}
\includegraphics[width=\textwidth]{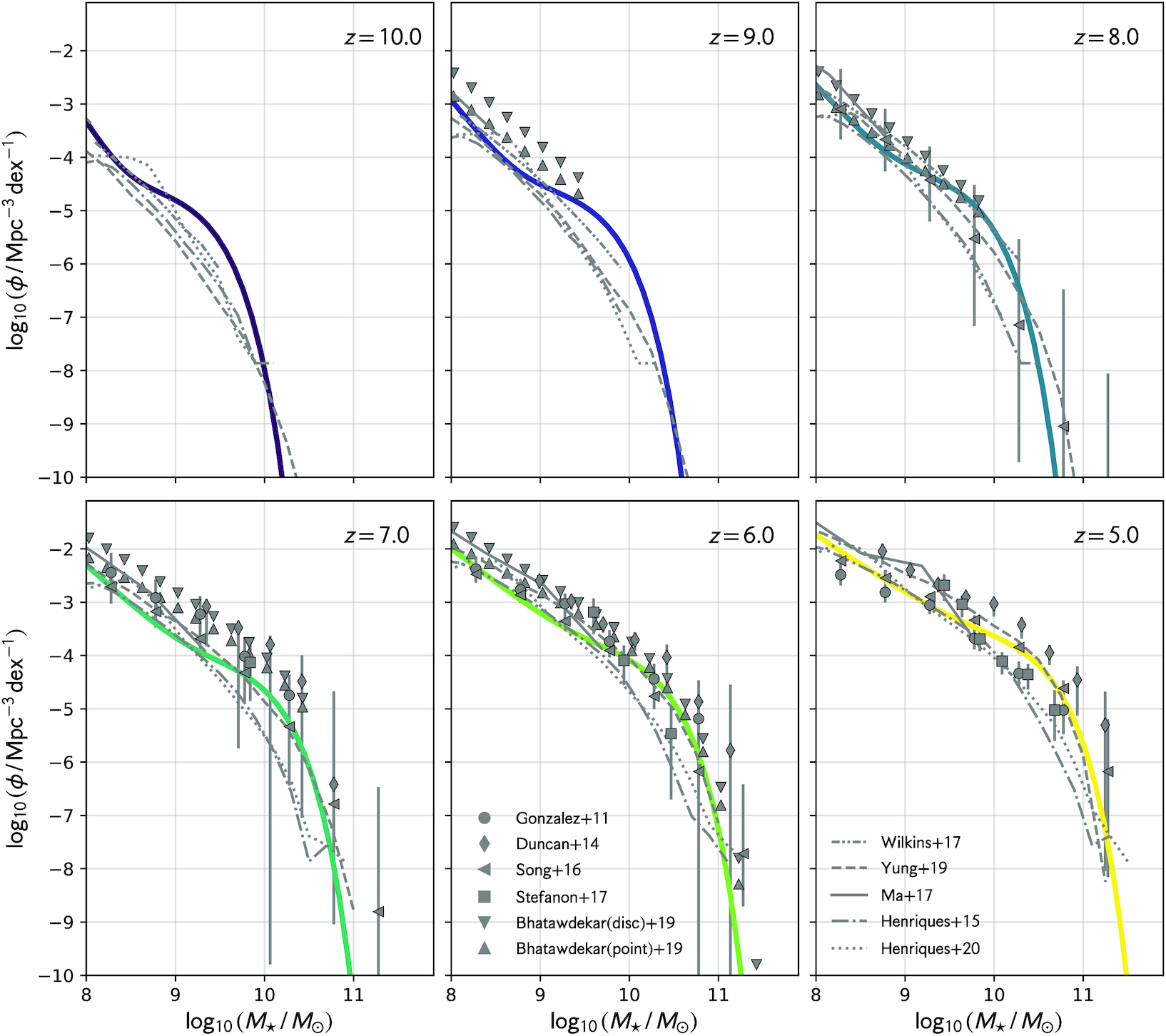}
\caption{Galaxy stellar mass functions predicted by several hydrodynamic simulations and semi-analytic models (colored solid lines show results from the FLARES simulations; other simulations are as listed in the bottom right panel), compared with observational estimates from before the launch of JWST (symbols; see key in bottom middle panel). The predictions of different models and simulations are fairly consistent back to $z\sim 10$. Reproduced from \citet{Lovell:2021}, Fig.~8.  }
\label{fig:lovell21}       
\end{figure}

\begin{figure}
\includegraphics[width=0.9\textwidth]{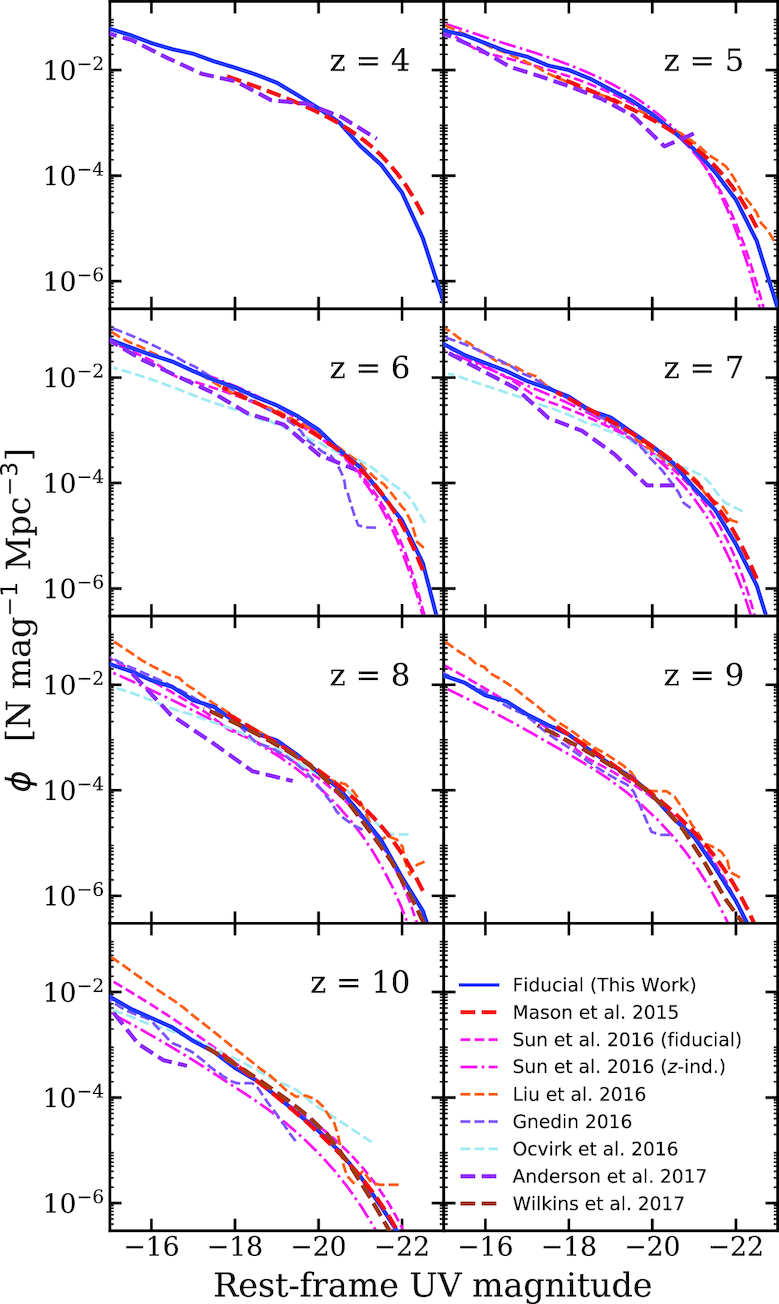}
\caption{Predicted galaxy rest-UV luminosity functions for a compilation of models that were published before the launch of JWST. The model predictions are quite consistent with one another back to $z\sim 9$. Reproduced from \citet{Yung2019a}, Fig.~10.  }
\label{fig:simUVLF}       
\end{figure}

The high redshift Universe clearly represents a very interesting laboratory for testing models, as the conditions at those early epochs were much different from those in the nearby Universe. Before JWST launched, several studies compared results from existing simulations and semi-analytic models back to $z\sim 10$. For example, Fig.~\ref{fig:lovell21} shows a comparison by \citet{Lovell:2021} of the stellar mass function from $z\sim 5$--10 for two semi-analytic models (L-galaxies and SC SAM), a suite of zoom-in simulations (FIRE), and the cosmological hydrodynamic simulations FLARES (which adopt the same physics models as EAGLE). Similarly, \citet{Yung2019a} showed a comparison of UV luminosity functions at $z\sim 4$--10 among different SAMs and hydro simulations (Fig.~\ref{fig:simUVLF}). Overall, considering the large differences discussed above, the models show fairly good agreement even out to the highest redshifts considered, as well as fairly reasonable agreement with the observational estimates of the stellar mass functions available at the time, considering the large observational errors on these pre-JWST estimates. 

I note here in passing that there is an extensive body of work on the reionization of the IGM by galaxies and AGN in empirical models, semi-analytic models, and cosmological hydrodynamic simulations. Although this is an important and fascinating topic, and highly relevant to other subjects discussed during this course, due to time limitations, I did not cover it extensively in my lectures, and do not attempt to cover it here. I recommend that interested readers see the review by \citet{Gnedin:2022}.

\subsection{Outlook: Future directions in galaxy formation modeling}
\label{sec:models:outlook}
Galaxy formation simulations have made enormous progress in the past 5-10 years. One area of progress is in the development of simulations that bridge previous gaps in scale, such as for example, simulations that partially resolve the multi-phase ISM in cosmological zoom-in simulations \citep[e.g.][]{Hopkins2014,Hopkins2018,Hopkins2023,Feldmann2023,Pallottini2022,Rosdahl2022,Ceverino:2017,Marinacci2019,Kannan:2025,Katz:2024,Bhagwat2025}, in some cases even reaching `single star' resolution in a global galaxy model or cosmological zoom \citep{Hu:2019,Emerick:2019,Steinwandel:2024,Gutcke:2021,Brauer:2025}. This enables more explicit implementations of physical processes. Simulations that bridge scales from star clusters to individual stars \citep[e.g.][]{Grudic:2021,Guszejnov:2022} are also extremely exciting. It is also encouraging that some groups are beginning to implement physical processes that have frequently been omitted from galaxy scale/cosmological simulations in the past, such as non-equilibrium cooling and chemistry \citep[e.g.][]{Katz:2022,Sarkar2022,Hu:2023,Steinwandel:2024,Katz:2024}, on-the-fly radiation \citep[e.g.][]{Katz:2024,Kannan2022,Kannan2025,Petersson2025}, cosmic rays \citep[e.g.][]{Chan:2019,Ruszkowski:2023,Girichidis:2024,Bieri:2025}, and live dust models \citep[e.g.][]{Kannan2022,Hu:2023,Choban:2025,Narayanan:2025b}. 

To simulate large cosmological volumes, however, it is still necessary to adopt phenomenological sub-grid recipes with tunable parameters. Due to the high computational expense of running a single cosmological volume, up until recently these parameters were tuned by hand in a very approximate fashion, leaving open the possibility that other points in the high-dimensional parameter space might give similarly good fits to the calibration observations. Ideally, one would use Bayesian Inference to obtain rigorous constraints on the full, multi-dimensional posterior of the parameters.  There have been some studies that used SAMs coupled with Monte Carlo Markov Chains (MCMC) and traditional (explicit likelihood) inference to map out parameter posteriors \citep[e.g.][]{Henriques:2009,Lu:2011}. However, the expense of the forward simulations and the large number of forward executions needed by traditional sampling techniques such as standard MCMC made this impractical for fully numerical simulations. Another concern is that the form of the likelihood function (typically assumed to be Gaussian in the MCMC studies just mentioned) is generally not known. 

Recent work has leveraged machine learning, along with computational speed-ups enabled by a combination of new software and new hardware, to make progress towards overcoming some of these challenges. Machine Learning (ML) based emulators can be used to much more efficiently generate forward simulations (especially summary statistics) \citep[e.g.][]{Jo:2023,Kugel:2023}. These can be coupled with Simulation Based Inference (SBI), in which the relationship between a model and data is learned via neural nets \citep{Ho:2024}. The number of required forward simulations is much smaller in SBI relative to standard MCMC, and the posterior can be learned directly without making any assumptions about the form of the likelihood (implicit likelihood inference). The CAMELS\footnote{https://www.camel-simulations.org/} project \citep{Villaescusa_Navarro:2021} has generated thousands of fully numerical hydrodynamic simulations spanning a broad range of cosmological parameters (matter density $\Omega_0$ and power spectrum normalization $\sigma_8$), and astrophysics parameters controlling the sub-grid physics such as stellar and AGN feedback, providing training data and sandboxes for a large number of promising new ML-based techniques. Furthermore, CAMELS has run the same initial conditions for many of the sub-grid physics model implementations that have been presented in the literature, providing a valuable comparison suite to help understand the impact of different sub-grid implementations. The CAMELS-SAM project has produced thousands of large volume N-body simulations sampling a wide range of values of $\Omega_0$ and $\sigma_8$ and populated them with galaxies using SAMs, again exploring a wide range of astrophysical parameters \citep{Perez:2023}.

Ultimately, it would be preferable to move away from phenomenological, calibrated sub-grid recipes, and towards physically grounded sub-grid recipes. This has been the goal of the SMAUG\footnote{Simulating Multi-scale Astrophysics to Understand Galaxies, https://www.simonsfoundation.org/flatiron/center-for-computational-astrophysics/galaxy-formation/smaug/} project. The SMAUG approach is to use a ladder of numerical simulations on different scales, to attempt to understand how the coarse-grained emergent physics arises from the physical process that is actually driving it on smaller scales. An example is the SMAUG suite of resolved ISM, tall box simulations using the TIGRESS \citep{Kimcg2017} physics model \citep{Kim:2020,Kim:2020b}, which resolve individual star clusters and supernovae explosions and can thus predict the emergent properties of supernova driven winds, such as wind mass loadings and energy loadings, as discussed in \S\ref{sec:models:physics:sf}. Insights from these simulations helped to guide the new Arkenstone scheme for simulating galactic winds in cosmological simulations \citep{Smith:2024a,Smith:2024b,Bennett:2025}. Arkenstone treats the hot and cool phases of the winds separately, allowing for interactions and exchanges of mass, energy, and momentum between the phases. It uses a custom refinement scheme to better resolve the hot, low density, fast moving component of the wind, and embeds an analytic model based on idealized simulations of turbulent radiative mixing layers to represent the evolution of cold clouds interacting with the hot wind \citep{Fielding:2022}. Another example is the adoption of a PRFM-motivated effective equation of state in cosmological simulations, based again on resolved ISM simulations, replacing the commonly used \citet{Springel:2003} ISM sub-grid model \citep{Hassan:2024,Burger:2025}. 

Another important development that came out of SMAUG is a `next generation' semi-analytic modeling framework and code. Traditional semi-analytic models track mass and metal flows, as described in \S\ref{ref:sec:models:methods:SAM}. However, recent studies using a range of techniques including gas regulator models \citep{Carr2023}, analytic models \citep{Voit2024}, and detailed semi-analytic models \citep[][Pandya et al. in prep]{Pandya2023} have shown that tracking \emph{energy flows} in both the CGM and ISM is crucial for understanding how star formation is regulated by stellar and BH feedback. Historically, SAMs needed to adopt high mass loadings for stellar driven winds, especially in dwarf galaxies, in order to obtain reasonable stellar-to-halo mass ratios, which were in apparent tension with observational estimates of mass outflow rates in nearby galaxies as well as predicted mass outflow rates from resolved ISM simulations \citep{Pandya:2020}. However, the energy deposited by winds in the CGM and IGM effectively reduces cooling and inflows (preventative feedback), thus requiring lower mass loadings. This picture has not only been shown to agree with results from high-resolution numerical simulations like FIRE \citep{Pandya2023}, it provides a useful framework for interpreting the results from coarser resolution cosmological simulations with different sub-grid implementations of winds \citep{Voit2024b}.

In addition to this significant new physical aspect, the new semi-analytic model {\sc sapphire}, developed by Pandya and collaborators \citep[][Pandya et al. in prep]{Pandya2023} also incorporates important technical advances. {\sc sapphire} is written in JAX, which enables just in time compilation on either CPU or GPU, leading to substantial decrease in computation time when run on multiple GPUs. Moreover, {\sc sapphire} has been constructed to be a fully differentiable model, and it can output not only the state variables as a function of time, but also the first and second derivatives of each state variable with respect to each parameter (the Jacobian and Hessian matrices). Thus it is straightforward to couple  {\sc sapphire} with efficient gradient descent sampling methods such as Hamiltonian Monte Carlo or SBI, for automated parameter calibration (Pandya et al. in prep). 

The Simons Collaboration Learning the Universe (LtU; https://learning-the-universe.org/) is carrying out an ambitious program to combine several of these techniques. The goals of LtU are to 1) develop physics grounded sub-grid recipes for galaxy formation simulations using the SMAUG approach 2) build machine learning powered emulators to speed up the forward computation of cosmological volumes with realistic baryonic physics  3) Wrap these realistic forward models in a rigorous Bayesian Simulation Based Inference framework, to constrain both cosmological and astrophysical parameters from the next generation of galaxy survey and CMB data. Although the original focus of LtU was the interpretation of low redshift observations, there is considerable promise in applying these techniques at high redshifts.


\begin{overview}{Summary}
\begin{itemize}
\item Large volume cosmological simulations must adopt sub-grid recipes for many key physical processes, including star formation \& stellar feedback. Parameters are calibrated primarily to match local stellar properties of galaxies. Different simulations make very different predictions for uncalibrated quantities.
\item The star formation efficiency on GMC scales is primarily regulated by supersonic turbulence and feedback from massive stars.
\item We might expect GMC scale star formation efficiencies to be higher in higher surface density clouds, as it takes more momentum to unbind the cloud. 
\item In galactic winds in simulations with ‘resolved feedback’, the cold phase carries most of the mass, while the hot phase carries most of the energy.
\item The wind mass loadings adopted in large scale (calibrated) cosmological simulations appear to be inconsistent with those that emerge from simulations with `resolved' feedback. 
\item Galaxy formation simulations are advancing by pushing to higher resolution and incorporating more explicit physics, as well as developing more physically motivated sub-grid recipes and leveraging machine learning and related techniques. 
\end{itemize}
\end{overview}

\begin{overview}{Additional Reading}
\noindent general background on galaxy formation: \citet{MvdBWbook} Ch. 8, 9\\
\noindent First Stars: \citet{Klessen:2023}\\
\noindent IMF: \citet{Hennebelle2024}\\
\noindent Chemical Evolution: \citet{Maiolino:2019}, \citet{Curti:2025}\\
\noindent ISM across cosmic time: \citet{Tacconi2020} \\
\noindent Dust: \citet{Schneider:2024}\\
\noindent Numerical Methods: \citet{Springel:2010}, \citet{Teyssier:2015}, \citet{Springel:2016}\\
\noindent Galaxy Formation Models and Simulations: \citet{SomervilleDave2015}, \citet{NaabOstriker2017}, \citet{Crain:2023}, \citet{Feldmann:2025}\\
\noindent Modeling Cosmic Reionization: \citet{Gnedin:2022}
\end{overview}

\section{Bridging Theory and Observations}
\label{sec:synthobs}
One of the main tools that we have for studying galaxy formation is the panchromatic \emph{spectral energy distributions} (SED) of galaxy populations seen at different cosmic epochs. The SED is the flux of emitted light as a function of wavelength, and it is a superposition of radiation produced by a wide range of physical processes, where different processes are dominant at different wavelengths. As shown in Fig.~2 of \citet{Iyer2025}, the rest-UV through NIR part of the SED (observable by JWST) is dominated by light from stars, nebular continuum and nebular emission lines, and in some cases continuum and line emission from an AGN accretion disk. Rest UV-optical light is absorbed by dust, and re-emitted in the mid to far IR and sub-mm. Emission from molecules (like Polycyclic Aromatic Hydrocarbons (PAHs) in the mid-IR and CO in the sub-mm) and ions (like [CII] and [OIII] in the sub-mm) produces prominent features at longer wavelengths \citep{Carilli:2013,Decarli:2025}. We refer to \citet{Iyer2025} for a more detailed discussion.

In contrast, the most direct predictions from theoretical simulations are masses, ages, and metallicities of star particles, along with gas density and temperature distributions. Galaxy scale simulations do not resolve the ISM on the scales that give rise to the UV-sub-mm emission lines, nor do most of them include non-equilibrium chemistry or on-the-fly radiation transport. Most galaxy scale and cosmological simulations also do not attempt to model the formation and destruction of dust, or the evolution of dust properties, and cannot make direct predictions for how radiation interacts with dust within a highly inhomogeneous ISM. As a result, there is a whole layer of modeling, often done in post-processing, involving additional ``sub-grid'' assumptions, that is needed in order to make direct links between simulations and observations. Taking simulation based quantities to the observational plane is sometimes called `forward modeling'. 

The other approach to try to interpret SEDs in terms of physical properties could be called `backwards modeling', i.e., one tries to deduce underlying physical properties such as stellar mass, star formation histories, metallicity, dust content etc. by fitting the SED. This is also a complex and rich topic, which is covered in the Ellis lectures (Lecture 2, Section 3), as well as in \citet{Iyer2025}, \citet{Conroy2013}, and references therein. 

\subsection{Stellar populations}
\label{sec:synthobs:stellpop}
The study of the interplay between the evolution of stellar populations and the cosmological evolution of galaxies was pioneered by Tinsley \citep[e.g.][]{Tinsley1968,Tinsley1972,Tinsley1980}. In the modern context, both semi-analytic models and numerical simulations record the mass of stars that is formed with a given age and metallicity in all of the progenitor galaxies that have merged to form a galaxy at a given output redshift. In semi-analytic models, these predictions are tabulated in a two dimensional histogram, while in numerical simulations they are represented by discrete `star particles'. We can then couple these predictions with \emph{Stellar Population Synthesis} (SPS) models, which describe the evolution in time of the SED of a stellar population with a single age and metallicity. These models are created by combining stellar evolution theory (in the form of stellar isochrones) with stellar spectral libraries and an assumed stellar initial mass function (IMF). Each of these components may depend on metallicity and elemental abundance ratios. There are a large number of SSP model variants available in the literature, e.g. \citet[][STARBURST99]{Leitherer1999}, \citet{Bruzual:1993,Bruzual:2003,Maraston:2005,Maraston:2009}, FSPS \citep{Conroy:2010}, BPASS \citep{Eldridge2017}. See \citet{Conroy2013} for a discussion of the major uncertainties in these models. Perhaps the largest factors that impact modeling of high redshift stellar populations are 1) uncertainties in the IMF and its evolution 2) uncertainties in the modeling of massive, low metallicity stars (for which there are few low redshift analogues) 3) uncertainties due to the modeling of binary and multiple stars, and the binary/multiple fraction in early stellar populations. 

\subsection{Nebular emission}
\label{sec:synthobs:nebemission}
Nebular emission arises from ionized gas in the ISM, and consists of continuum emission from free-free, free-bound, and two photon emission, and recombination line emission \citep[see][]{Iyer2025}. This gas may be ionized by young, massive stars, post AGB stars, an active galactic nucleus (AGN), or fast radiative shocks \citep[e.g.][]{Hirschmann:2017}. Most models focus on excitation by young stars, which are likely to be dominant in most high redshift galaxies that do not harbor an AGN. A commonly used approach for forward modeling nebular emission is to couple hydrodynamic simulations or semi-analytic models with a photo-ionization code such as {\sc cloudy} \citep{Ferland:1998,Ferland:2013,Ferland:2017} or {\sc mappings} \citep{Dopita:2002,Sutherland:2013,Sutherland:2018} in post-processing. The ionizing radiation field is provided by a star formation and chemical evolution history coupled with an SSP model, as described above. In this type of approach, the ensemble of H$_{\rm II}$ regions and diffuse gas within an entire galaxy are represented by effective parameters \citep{Charlot:2001,Gutkin:2016}. The emergent line emission depends primarily on the following parameters: ISM metallicity, stellar metallicity, ionization parameter, dust-to-metal mass ratios, dust composition and grain size distribution, H$_{\rm II}$-region densities, and carbon-to-oxygen abundance ratio. The ionization parameter is defined as the dimensionless ratio of the number density of 
H-ionizing photons to that of hydrogen. Most of these parameters are not directly predicted by cosmological simulations (especially large volume ones), so assumptions must be adopted for how these H$_{\rm II}$-region properties are related to the coarser scale predictions provided by cosmological simulations, or the even more imprecise estimates available from SAMs \citep{Scharre:2024}. For some of these parameters (e.g. dust-to-metal ratio, C/O ratio), there may not be any information available in most simulations, in which case fixed values are generally used. This approach has been used by \citet{Hirschmann:2017,Hirschmann:2019,Hirschmann:2023}, \citet{Wilkins:2020,Wilkins:2023b} and \citet{Garg:2024}. These studies have demonstrated that many observational quantities, such as emission line ratios and emission line luminosity functions, can be reproduced reasonably well with this approach.

\begin{figure}
\includegraphics[width=\textwidth]{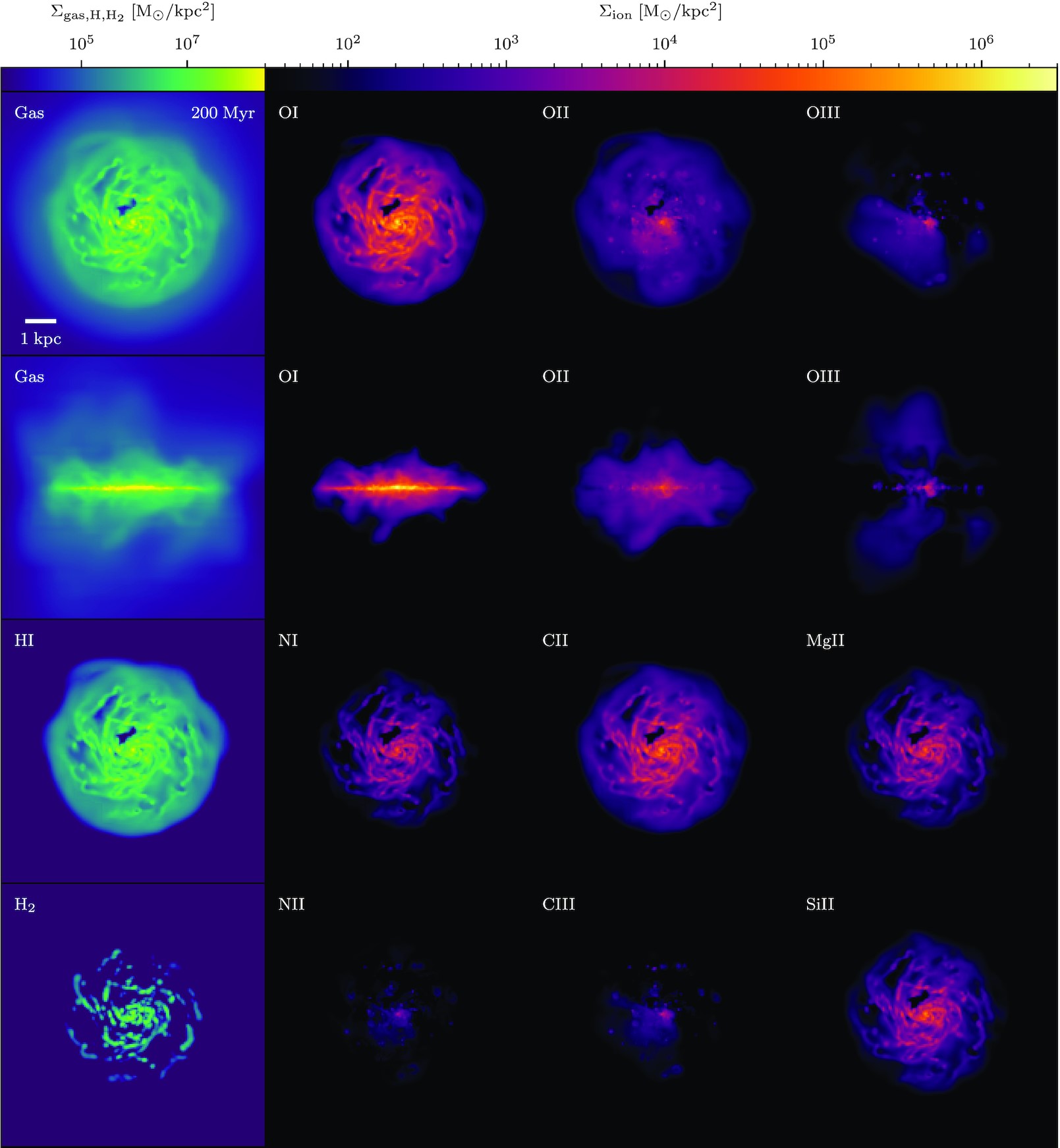}
\caption{Face-on and edge-on images of the gas surface density and the surface density 
of emission from various ions as indicated on the images for a galaxy simulated with RAMSES-RTZ, which includes non-equilibrium cooling and chemistry and on-the-fly radiation transport. The width of each image is 10 kpc. Reproduced from \citet{Katz:2022}, Fig.~10.  }
\label{fig:katz22}       
\end{figure}

The other approach is to carry out very high resolution simulations that can resolve the multi-phase ISM, compute full non-equilibrium chemical networks, and couple the ionization states of individual metals to an on-the-fly radiation hydro solver. This approach has been pioneered by \citet{Katz:2022} with RAMSES-RTZ and the MEGATRON simulations \citep{Katz:2024}. Although it is only feasible to carry out these very expensive calculations for idealized galaxies or a small number of zoom-in simulations, they provide very useful tests of the impact of the simplifying assumptions adopted in equilibrium photo-ionization calculations. Fig.~\ref{fig:katz22} shows an example of the predicted line emission from a RAMSES-RTZ simulation for several lines. 

\subsection{Dust attenuation and emission}
\label{sec:synthobs:dust}
Dust is distributed throughout the ISM, and absorbs and scatters UV-optical light. The dust \emph{extinction curve} is defined as $A_\lambda = 2.5 \log_{10} [F^0_\lambda/F_\lambda]$
\emph{along a specific line of sight to a single source}, where $F_\lambda$ is the observed flux at wavelength $\lambda$ (affected by scattering out of the line of sight and attenuation by dust) and $F^0_\lambda$ is the flux in the absence of extinction \citep[e.g.][]{Salim-Narayanan:2020}. The extinction curve depends only on the composition and grain size distribution of the dust. The \emph{attenuation} curve (also commonly, and confusingly, denoted $A_{\lambda}$) represents the ratio of observed to emitted flux for an \emph{entire galaxy} or region within a galaxy. The attenuation includes the effects of star-dust geometry, as well as scattering back into the line of sight, and unobscured stars (see \citealt{Salim-Narayanan:2020}). It is common to treat extinction curves and attenuation curves interchangeably, but the distinction is quite important. Until recently, it was very common to adopt a single, fixed attenuation curve in all galaxies both in forward modeling of model predictions and in derivation of physical parameters from SED fitting.

Since most SAMs and hydrodynamic simulations do not track the physical processes of dust formation, growth, and destruction with a so-called ``live dust model'', it is common to assume a fixed dust-to-metal ratio, or a simple scaling of dust-to-metal ratio with metallicity, motivated by observations of nearby galaxies \citep{Remy-ruyer:2014,Devis:2019}. The dust composition and grain size distribution are typically assumed to be fixed to be similar to those in the Milky Way. 

Dust modeling in SAMs typically assumes that the dust optical depth $\tau_{\rm dust} \propto Z_{\rm gas} M_{\rm ISM}/r_{\rm dust}^2$. The attenuation at a reference wavelength such as the V-band is then computed by assuming a simple geometry, such as a screen or slab model \citep[e.g.][]{Somerville2012}. A fixed attenuation curve is then generally adopted to compute the wavelength dependence of the attenuation. One can then assume that all of the light that is absorbed from the UV-NIR is re-radiated in the IR. Empirical or theory based templates for the dust emission SED can then be used to predict fluxes at IR-sub-mm wavelengths \citep{Somerville2012,Lacey:2010,Lacey:2016}.  

\begin{figure}
\includegraphics[width=\textwidth]{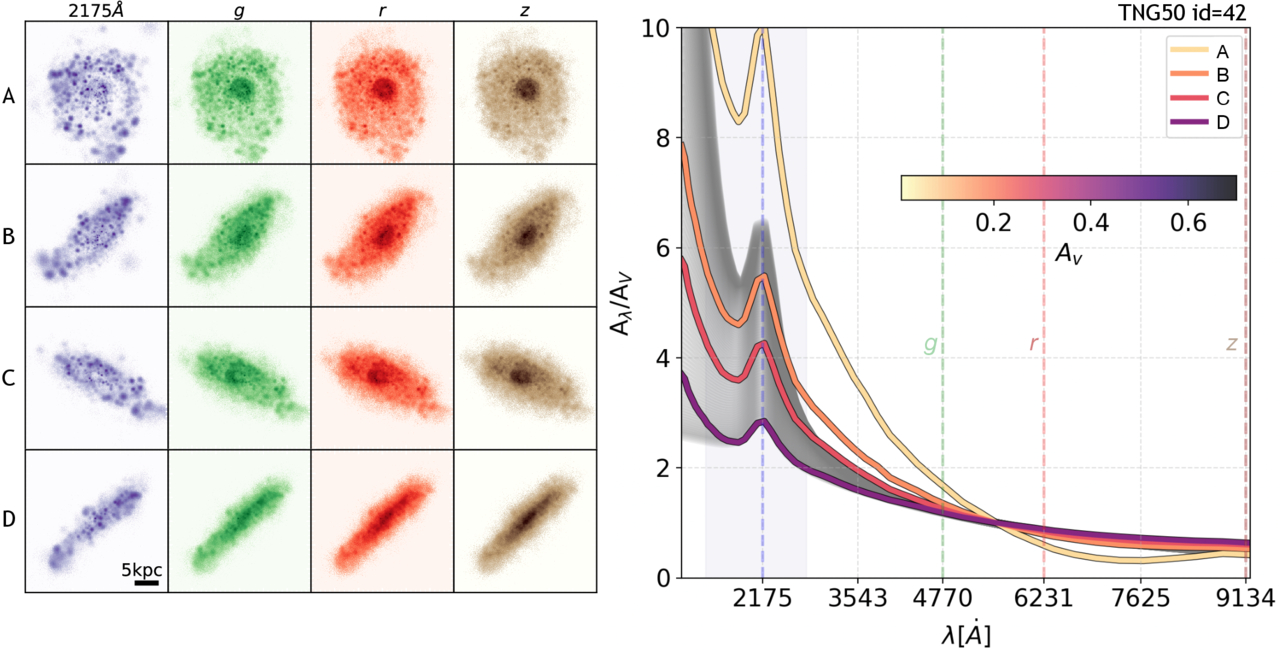}
\caption{Left panel: synthetic images of a galaxy observed along different lines of sight (LOS), from the SKIRT radiative transfer code run in post-processing on galaxies selected from the IllustrisTNG simulation. The value of the visual attenuation (A$_V$) increases from top to bottom. Each column shows the predicted emission at a different wavelength: 2175 Å, and the SDSS g, r, and z bands. Right panel: corresponding attenuation curves for each LOS, color-coded according to A$_V$ (see color bar). The gray shaded area represents the 16th–84th percentile variation of attenuation curves across the entire TNG combined sample, including all sources and LOS. The wavelengths corresponding to the filters shown in the left panel are highlighted by vertical dashed lines. Even with a fixed underlying model for the dust composition and grain size distribution, there can be a very wide dispersion in galaxy attenuation curves, especially in the UV. Reproduced from \citet{Sommovigo:2025}, Fig.~2. }
\label{fig:sommovigo_alambda}       
\end{figure}

One can couple numerical hydrodynamic simulations with radiative transfer codes, such as SKIRT \citep{Camps:2015}, Powderday \citep{Narayanan:2021}, RADMC-3D \citep{Dullemond:2012}, Hyperion \citep{Robitaille:2011} or ART2 \citep{Li:2020} in post-processing. This allows the creation of detailed mock images at different wavelengths, as well as integrated SEDs \citep[e.g.][]{Popping:2022}. Such studies have shown that, even in the nearby Universe and with fixed dust composition and grain size distributions, there is a large galaxy to galaxy variation in the attenuation curves (Fig.~\ref{fig:sommovigo_alambda}, reproduced from \citealt{Sommovigo:2025}; see also \citealt{Salim-Narayanan:2020}), as is also seen in observations \citep{Salim:2018}. This is largely driven by geometrical effects as different lines of sight encounter different structures in the ISM. Moreover, the dust attenuation curve shape is correlated with macroscopic galaxy properties such as the average gas density or optical depth \citep{Salim-Narayanan:2020,Sommovigo2025}. It is important to keep in mind that radiative transfer modeling based on large volume cosmological simulations, which do not resolve the multi-phase ISM, are likely to be missing some of the complexity of the true dust-star geometry of the ISM. Moreover, the dust-to-metal ratio, dust composition, and grain size distribution may vary within galaxies and from galaxy to galaxy, and may evolve over cosmic time. We discuss the implications of these effects further in \S\ref{sec:results:dustev}.

Another common approximation is to assume that the attenuation of the stellar continuum and that of the nebular line emission is the same, or that they are related by a fixed ratio. However, observations of nearby galaxies indicate a large scatter in this ratio as well \citep{Salim:2018}.

\begin{overview}{Summary}
\begin{itemize}
\item The SED contributed by the (unattenuated) stellar continuum can be modeled by combining star formation and chemical enrichment histories from SAMs or hydro sims with Simple Stellar Population (SSP) models. The largest uncertainties in this modeling for high redshift ($z\gtrsim 6$) are due to lack of knowledge about the stellar IMF, evolution and atmospheres of massive, metal poor stars, and the effects of stellar multiplicity. 
\item Nebular emission (line and continuum) is produced by gas that has been ionized by young stars, AGN, or shocks. It can be a very significant component of the UV-optical emission in high redshift galaxies. 
\item Dust \emph{extinction} curves depend on the composition and grain size distribution of dust. Dust \emph{attenuation} curves also depend on star-dust geometry. There can be a broad dispersion in UV-optical attenuation curves even for a fixed dust composition and grain size distribution. 
\end{itemize}
\end{overview}

\begin{overview}{Additional Reading}
\noindent Stellar Population Synthesis: \citet{Conroy2013}\\
\noindent Galaxy SEDs: \citet{Iyer2025} \\ 
\noindent Dust Attenuation curves: \citet{Salim-Narayanan:2020}\\
\end{overview}

\section{Galaxy formation models in the JWST era: insights and puzzles}
\label{sec:results}
The launch of JWST opened a new window onto the high redshift $z\gtrsim 6$ Universe, and provided a view of the ultra-high redshift Universe ($z\gtrsim 10$) for the first time. As discussed in much more detail in the Ellis lectures, before JWST, the wavelength and sensitivity limitations of existing facilities such as the Hubble Space Telescope (HST) and Spitzer Space Telescope precluded the identification of robust galaxy candidates at $z\gtrsim 9$, and spectroscopy was impossible for all but the brightest candidates (see Fig.~23 of Ellis lectures). JWST has not only discovered galaxies at much earlier cosmic times, but the onboard spectrographs have enabled detailed studies of the physical conditions within galaxies at these early epochs. These unprecedented observations have posed a new set of challenges for theory and simulations. In this section I summarize several theoretical puzzles raised by the past three years of JWST observations of galaxies during the first Gyr of cosmic evolution, and provide a status report on recent theoretical ideas that have been proposed to solve them. 

\subsection{Galaxy Evolution at $z\gtrsim10$}
\label{sec:results:numdens}
Almost as soon as the first data became public, JWST broke the previous $z\sim 9$ redshift barrier and various teams began reporting surprisingly large numbers of galaxy candidates at $z\gtrsim 10$ (see Ellis lectures). Soon thereafter, spectroscopic follow-up with NIRSpec showed that, although there were a few catastrophic failures in the photometric redshift estimates (for example, a few $z\sim 16$ candidates turned out to be at $z\sim 4.9$), for the most part the samples of $z \sim 9$--12 candidates turned out to contain few lower redshift interlopers \citep{ArrabalHaro2023,RobertsBorsani2024}. As of this writing, JWST has discovered galaxy candidates out to redshifts as high as $z\sim 30$ \citep{Perez-Gonzalez:2025,Castellano:2025,Gandolfi:2025}, with the highest redshift spectroscopically confirmed object at $z = 14.44$ \citep{Naidu2025}. 

\begin{figure}
\includegraphics[width=\textwidth]{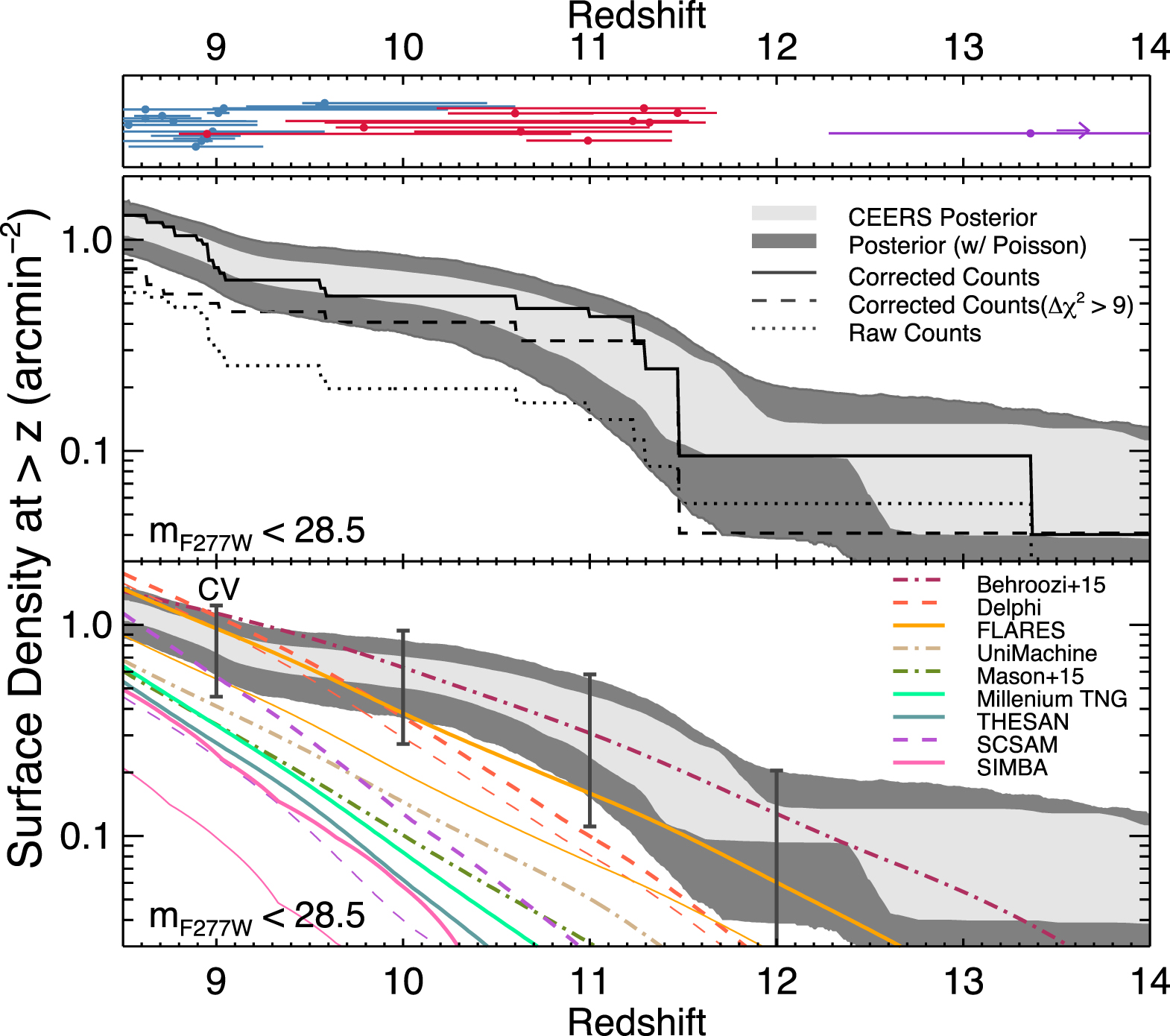}
\caption{Cumulative surface density of observed sources in the CEERS survey (first release) with mF277W $< 28.5$ at redshifts greater than a given x-axis value, starting at $z>8.5$. The top panel shows the redshifts of individual objects. In the middle panel, the solid line shows the observed surface density after applying a correction for incompleteness; the dotted line shows the uncorrected (incomplete) values. The light shaded region shows the posterior on the distribution of the completeness-corrected surface density derived from Monte Carlo simulations marginalizing over the uncertainties in magnitude and photometric redshift; the dark shaded region includes Poisson uncertainty in this marginalization. The bottom panel repeats the shaded region, and shows a comparison with the predictions of pre-JWST-launch models and simulations, shown by the various colored lines (with solid, dotted–dashed, and dashed denoting predictions from hydrodynamical, semi-empirical, and semi-analytic models, respectively). Thicker/thinner lines do not/do include dust attenuation. Nearly all of the models predict a much more rapid decline in the number density of bright galaxies than is seen in the observations. Reproduced from \citet{Finkelstein2023}, Fig.~14. }
\label{fig:cumcounts_ceers}       
\end{figure}

\begin{figure}
\includegraphics[width=\textwidth]{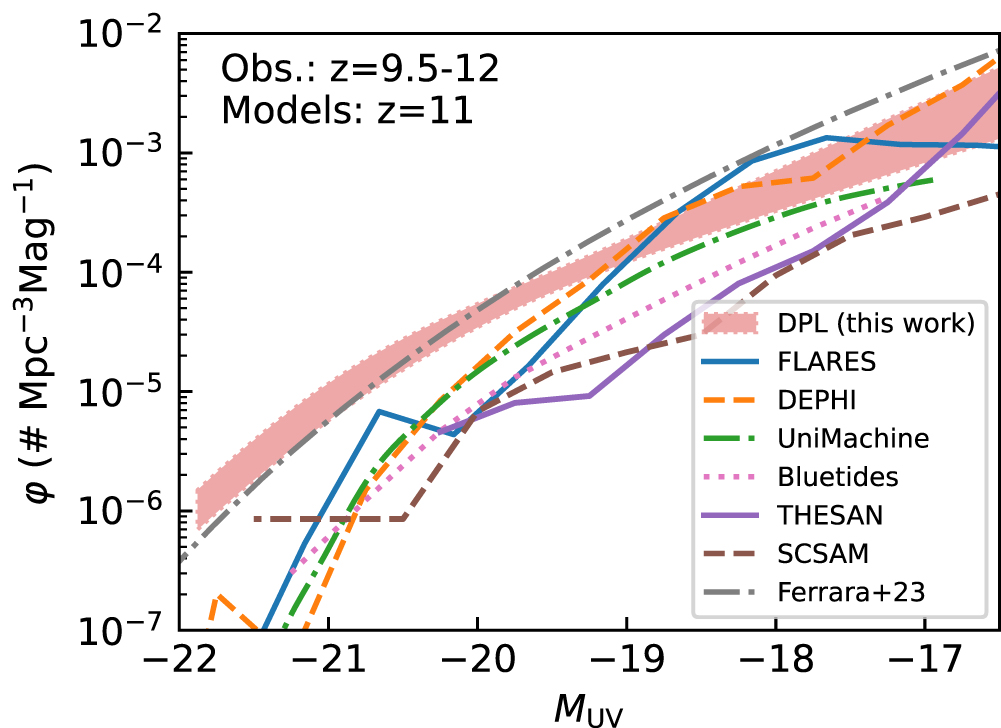}
\caption{The rest-frame UV luminosity function of galaxies at $9.5 < z < 12$. The red shaded region shows the observational estimates from the CEERS and NGDEEP surveys combined. The colored lines show predictions from (mostly) pre-launch theoretical models and simulations. This shows that the tension between model predictions and the observations is stronger at the brighter UV magnitudes probed by current surveys. Reproduced from \citet{Leung2023}, Fig.~6. }
\label{fig:UVLFobs}       
\end{figure}

Many if not most physics-based theoretical models, with a wide variety of different sub-grid physics assumptions and implementations, produced predictions for galaxy stellar mass functions and UV luminosity functions out to $z\sim 9$--10 that were not in dramatic tension with the updated observational estimates from JWST (see Section~\ref{sec:models:status}). Fig.~\ref{fig:cumcounts_ceers} shows the cumulative counts of galaxies from the CEERS Epoch 1 sample \citep{Finkelstein2023} compared with predictions from a large number of pre-launch theoretical models and simulations. The striking take-away from this comparison is that nearly all of these models, regardless of whether they were empirical, semi-analytic, or numerical, predicted a \emph{much steeper decline in the number density of UV-luminous galaxies at $z \gtrsim 10$} than the JWST observations.
Fig.~\ref{fig:UVLFobs} shows a comparison of models and JWST observations of the UV luminosity function at $z\sim 11$, illustrating another point: although some physics-based simulations were able to reproduce enough \emph{relatively low luminosity}  galaxies at this redshift, the high observed number density of \emph{UV luminous} ($M_{\rm UV} \lesssim -19.5$) galaxies seems to be more challenging for models to reproduce (see also \citealt{Adams2024,Whitler2025}).

Large volume hydrodynamic simulations such as IllustrisTNG \citep{Pillepich2018,Nelson2018}, MillenniumTNG \citep{KannanMTNG2023}, THESAN \citep{Kannan2022}, SIMBA and SIMBAEoR \citep{Dave2019,Jones2024}, BlueTides \citep{Feng2016} and FLARES \citep{Wilkins2022} all underproduce UV-luminous galaxies by increasingly large factors at $z\gtrsim 10$, and predict a steeper decline in the comoving number density of UV-bright galaxies at these redshifts than seen in observations \citep{Finkelstein2023,Finkelstein2024,Leung2023,Adams2024}. This may be because these simulations contain sub-grid models for star formation and supernova-driven winds that are calibrated to reproduce the low integrated global galaxy SFE (i.e. $m_*/(f_{\rm b}M_{\rm h})$ relation) at low redshift.  Higher resolution simulations such as FIRE \citep{Hopkins2014,Hopkins2018,Hopkins2023} and FIREBox \citep{Feldmann2023}, SPHYNX \citep{Rosdahl2022}, FirstLight \citep{Ceverino2019,Ceverino2024}, the RAMSES simulations of \citet{Andalman2024} and THESAN-zoom \citep{Kannan2025}, are able to incorporate more physically grounded sub-grid recipes --- but because these simulations are typically zoom-ins or very small volumes, they are limited in their ability to make predictions for the bright end of the UVLF. 

Thinking back to \S\ref{sec:structure:obs} of these lectures, we actually did not even need to run a fancy simulation to realize how surprising the JWST results are. The number density of massive dark matter halos declines dramatically at $z\gtrsim 10$ (see Fig.~1). Therefore, if the efficiency of converting gas to stars and the ratio between SFR and UV light have not changed, it is \emph{inevitable} in the $\Lambda$CDM framework for the number density of UV luminous galaxies to decline rapidly with increasing redshift. 

Unsurprisingly then, the solutions to this puzzle that retain a vanilla $\Lambda$CDM framework generally propose that one or both of the following were the case at earlier epochs:
\begin{enumerate}
\item \textbf{evolving light-to-mass:} higher redshift galaxies were brighter for a given amount of star formation
\item \textbf{evolving SFE:} The star formation efficiency was higher at higher redshifts
\end{enumerate}

\begin{figure}
\includegraphics[width=\textwidth]{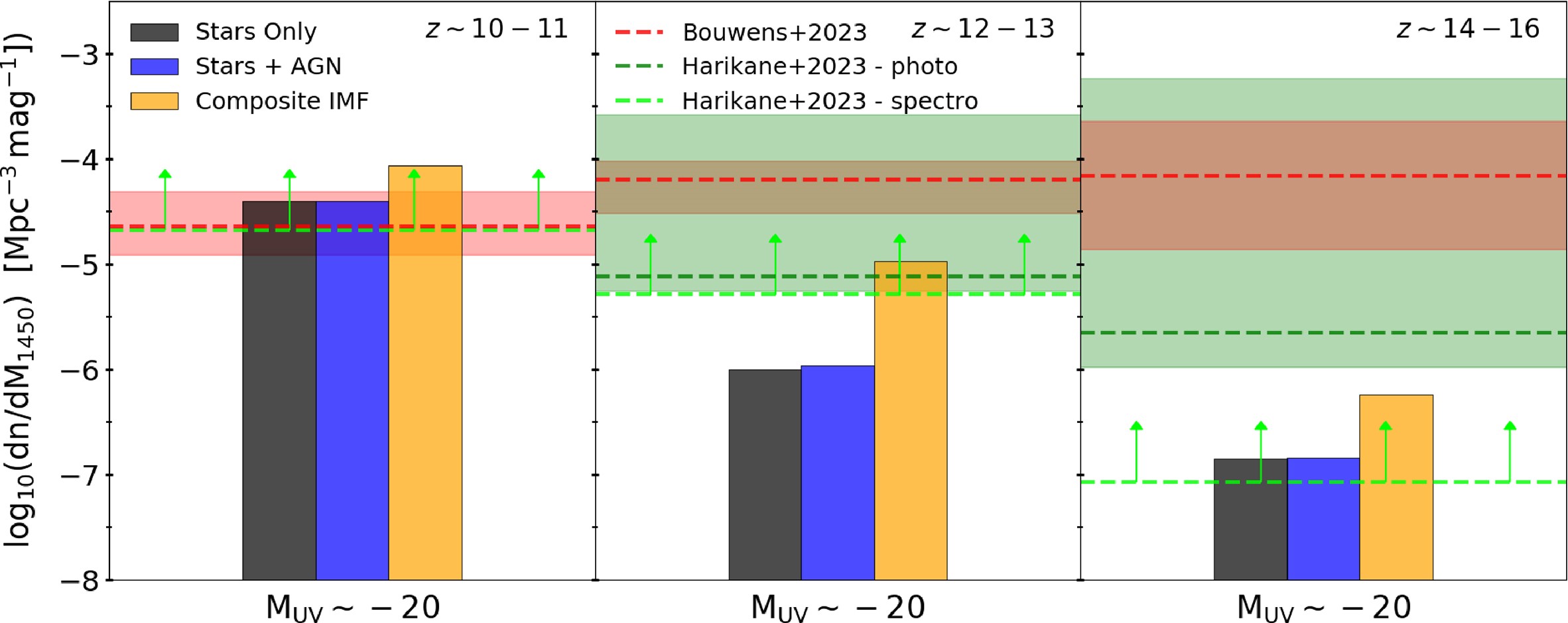}
\caption{The comoving number density of bright galaxies ($M_{\rm UV} \sim -20$) at $10<z<11$, $12<z<13$, and $14<z<16$ (left, middle, right) predicted by the CAT semi-analytic model. Different colored histograms represent the number density from stars only with a standard IMF (black), when including the light from AGN (blue), and when including a metallicity dependent, composite IMF (yellow). The horizontal lines represent observational estimates, with the shaded regions showing the error. The light green horizontal lines show the estimates based only on spectroscopically confirmed galaxies, which are represented here as lower limits of the galaxy number density. The contribution of UV light from AGN has little effect within the context of the CAT model, while the composite IMF can boost the number density of UV bright galaxies by $\sim 0.5$--1 dex. Reproduced from \citet{Trinca2024}, Fig.~12. }
\label{fig:trincauvlf}       
\end{figure}

\noindent \textbf{Evolving light-to-mass}: Several physical explanations for why 1) might be the case have been proposed. Ferrara et al. (\citeyear{Ferrara2023,Ferrara2024,Ferrara2025}) suggested that galaxies at $z\gtrsim 10$ might be brighter because they are less dust attenuated (we discuss this picture in more detail in the next sub-section). However, most of the physics-based theoretical models shown in Fig.~\ref{fig:cumcounts_ceers} fail to reproduce the $z\gtrsim 10$ counts even when attenuation due to dust is neglected. Galaxies would be brighter if the IMF was richer in massive, UV-bright stars \citep{Trinca2024,Cueto:2024}, or if there is a contribution to the UV light from an (unobscured) accreting black hole \citep{Inayoshi2022,Trinca2024}. The work of \citet{Trinca2024}, based on the Cosmic Archeaology Tool (CAT) SAM, suggests that neither of these factors can fully resolve the tension (see Fig.~\ref{fig:trincauvlf}), but these conclusions may be somewhat model dependent and should be investigated further \citep[see also][]{Cueto:2024,Mauerhofer2025}. 

Another explanation is that galaxies could become brighter temporarily due to stochastic, bursty star formation \citep{Mason2023,Shen2023,Sun2023,Kravtsov2024}. In fact, simulations that resolve a multi-phase ISM do show rather bursty star formation \citep[e.g.][]{Iyer2020,Sun2023,Pallottini2023,Basu2025}. As noted in \S\ref{sec:models:methods:hydrosims}, the adoption of an effective equation of state in the ISM, common practice in larger volume cosmological simulations, smooths out the ISM and leads to less bursty star formation, so these simulations likely underestimate burstyness. SAMs also generally have less bursty star formation than high resolution simulations \citep{Iyer2020}. \citet{Sun2023} quantified the burstyness of star formation in the FIRE-2 suite of high resolution simulations over the redshift range $8 \lesssim z \lesssim 12$, and found that the burstyness (expressed as the dispersion in M$_{\rm UV}$ at fixed halo mass, $\sigma_{\rm UV}$) is a strong function of halo mass but does not depend significantly on redshift over this interval. \citet{Gelli2024} implemented this parameterization of burstyness into an empirical model, and showed that this level of stochastic star formation could reconcile a non-evolving underlying SFE with the observations out to $z\sim 11$, but still produced UVLFs that fall short of the observations at higher redshifts. However, the degree of stochasticity differs between different simulations, and \citet{Basu2025} has shown using the SPICE simulations \citep{Bhagwat2025} that the amplitude and slope of the halo mass dependence of $\sigma_{\rm UV}$ is sensitive to the details of the implementation of stellar feedback. 

\noindent \textbf{Evolving SFE}: \citet{Dekel2023} argue that a highly efficient ``feedback free'' burst (FFB) mode of star formation should occur in gas with high densities ($\gtrsim$ few $\times 10^3$ cm$^{-3}$) and low metallicity. At these densities, the free fall time ($\lesssim$ Myr) is short compared with the time before the first core collapse supernovae begin to explode. Furthermore, they argue that stellar winds are weak at metallicities $Z_{\rm gas} \lesssim 0.2$ Z$_{\odot}$, and that radiation pressure cannot disrupt clouds above a critical surface density. They show that these conditions should apply in halos above a redshift dependent critical mass ($M_{\rm FFB} \sim 10^{10.8} \msun$ at $z=9$; see also \citealt{Li2024}). 

\citet[][S25]{Somerville:2025} present another scenario that predicts that the star formation efficiency should increase with increasing redshift, which they call Density Modulated Star Formation Efficiency (DMSFE). They divide regulation of star formation into `local', GMC scale processes and galaxy and halo scale processes like supernova driven winds. The GMC scale SFE determines the rate that gas in the ISM turns into stars per unit time, while supernova driven winds regulate the supply of cold ISM gas that is available for star formation. S25 adopt the analytic model for cloud scale SFE as a function of cloud surface density described in Eqn.~\ref{eqn:estarcloud} (see \S\ref{sec:models:physics:sf}), and swap this into the Santa Cruz semi-analytic model in place of the traditional Kennicutt-Schmidt star formation recipe. They find that if all of the ISM is in dense star forming clouds, the number of UV luminous galaxies at $z\gtrsim 12$ is actually significantly \emph{over-predicted} relative to the observations. They parameterize the fraction of the ISM in dense clouds that are eligible for star formation as $f_{\rm dense}$, and find that $f_{\rm dense}=0.1$ reproduces the observations well at $z\sim 12$, and $f_{\rm dense}=0.5$ works well at $z\sim 14$. However, \emph{no single value} of $f_{\rm dense}$ is able to reproduce the observed shallow evolution of UV-bright galaxies at $z \gtrsim 12$. Thus, either $f_{\rm dense}$ depends on a physical property that changes quite rapidly with cosmic time, or other physical effects are contributing. This highlights that \emph{it is not the high number density of UV-bright galaxies at high redshift that is so difficult to explain theoretically, but rather the shallow evolution}. 

The closely related FFB and DMSFE pictures carry several intriguing implications with potentially observable consequences. In both scenarios, rapid inflows of gas into the ISM make it highly gravitationally unstable, and the gas fragments into large clouds or proto-star clusters ($M_{\rm cl} \simeq 10^4$--$10^5 \msun$). This is supported by images of highly magnified galaxies at $6 \lesssim z \lesssim 10$ behind lensing clusters, which show multiple dense star forming clumps with sizes of less than a parsec, and implied surface densities of $10^4$--$10^5 \msun$ pc$^{-2}$ \citep{Adamo2024,Mowla2024,Fujimotograpes2024}. The combination of rapid gas inflows and highly efficient star formation in these dense clouds is expected to lead to very bursty star formation, contributing to a differential boost in bright galaxies as discussed above. Moreover, high surface density environments may be favorable for hosting stellar populations with a top-heavy IMF \citep{Hennebelle2024}. Ultra-dense star clusters can also develop Very Massive ($M_* \gtrsim 100$), Extremely Massive ($1000 \lesssim M_* \lesssim 10^4 \msun$), and super-massive Stars ($M_* \gtrsim 10^4 \msun$) through runaway collisions in their nuclei \citep[e.g.][]{Rantala2024}. Self-enrichment of the gas by these massive stars could produce elevated N/O as in the ``Nitrogen enhanced'' galaxies that have been observed with JWST \citep{Cameron2023,Topping:2025,Ji2025}.
Furthermore, these dense star clusters may nurture Intermediate Mass Black Holes (IMBH), formed by runaway core collapse \citep{Inayoshi2020}. High redshift galaxies may contain up to several hundred of these star clusters, each hosting an IMBH. If these star clusters and their BH can merge rapidly, this could provide a promising mechanism to create relatively massive early BH, perhaps seeding the $\sim 10^6$--$10^8 \msun$ BH observed by JWST at $5 \lesssim z \lesssim 10$ \citep{Dekel2025}. See the lectures by Marta Volonteri for a more complete discussion of BH seeding models. 
 
\subsection{Evolution of dust in the early Universe}
\label{sec:results:dustev}
As described in \S\ref{sec:models:physics:chemdust}, dust is formed in supernovae, AGB stars, and via grain accretion in the ISM. We have direct evidence that fairly large reservoirs of dust exist by $z\sim 7$--6 ($\sim 300$-500 Myr after the Big Bang) from the observed dust continuum emission seen with ALMA, which imply dust to stellar mass ratios comparable to those seen in local galaxies \citep{Sommovigo:2022,Algera:2025}. However, another surprise delivered by JWST is that the UV spectral slope of most of the ultra-high redshift $z\gtrsim 9$ galaxies is quite blue, leaving little room for reddening by dust \citep{Topping2022,Cullen2023,Morales2024}. Adopting standard SNae dust yields, assuming that the dust is in a screen around the stars, and assuming a standard attenuation curve, the observed UV luminosities of the ``blue monsters'' would seem to imply dust extinctions that are orders of magnitude larger than the observed SEDs suggest \citep{Ziparo2022,Ferrara2023}.

SNae dust yields are highly uncertain (by at least an order of magnitude; see e.g. \citealt{Schneider:2024}), and may be different in the very early SNae that enrich the blue monsters, due to different metallicities, IMF, or other factors. However, if SNae dust yields are significantly lower than is generally assumed, this would require very efficient grain growth in order to build up large enough masses of dust to explain $z\sim7$ dusty galaxies, over the relatively short time period ($\sim$ 285 Myr) between $z\sim 10$ and $z\sim 7$ \citep[e.g.][]{Popping2017}. Timescales for dust growth via grain accretion in the ISM are also quite uncertain \citep{Schneider:2024}.  A wide range of semi-analytic models and numerical simulations that follow dust formation and evolution support a picture in which very early galaxies are dominated by stellar dust, and dust formed via grain accretion becomes increasingly important as time progresses, in most cases contributing significantly to the dust content by $z\sim 6$--7 \citep{Popping2017,Vijayan:2019,Graziani:2020,Esmerian:2022,Lewis:2023,Choban:2025,Narayanan2025}.  Due to the above uncertainties as well as many other uncertainties in dust modeling, the details of when and how this transition is expected to occur are still unclear. However, recent work by \citet{Burgarella2025} suggests that this transition between stellar dust and ISM-grown dust may now have been observed in high redshift galaxies with JWST. 

A key point is that the grain size distribution of stellar dust is dominated by large grains \citep{Nozawa:2007}, in contrast to the grain size distribution adopted in standard Milky Way dust models \citep{Mathis:1977,Draine2001}, which are richer in smaller grains. Large grains are optically thin to UV radiation, and therefore the attenuation curves of galaxies with dust that is primarily composed of unprocessed stellar dust are expected to be ``greyer'', implying less attenuation in the UV. The grain size distribution tends to evolve towards smaller grain sizes over time, due to shattering via collisions in the ISM. Thus one explanation for the  $z\gtrsim 9$--10 ``blue monsters'' is that they already contain significant reservoirs of dust, but due to the large grain dominated nature of this dust, it does not cause significant attenuation or reddening in the UV \citep{Narayanan:2025b}. 

\begin{figure}
\includegraphics[width=0.5\textwidth]{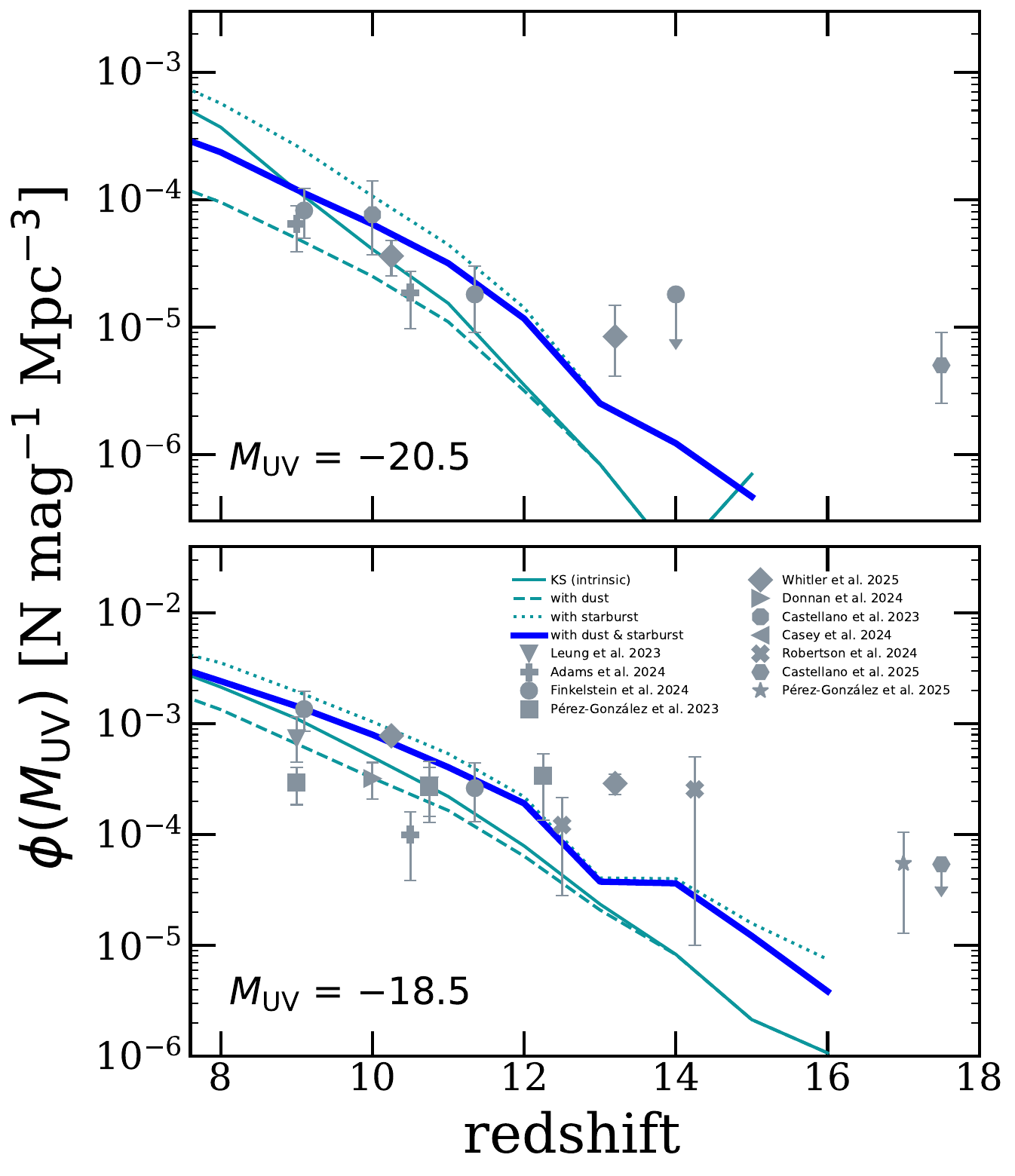}
\includegraphics[width=0.5\textwidth]{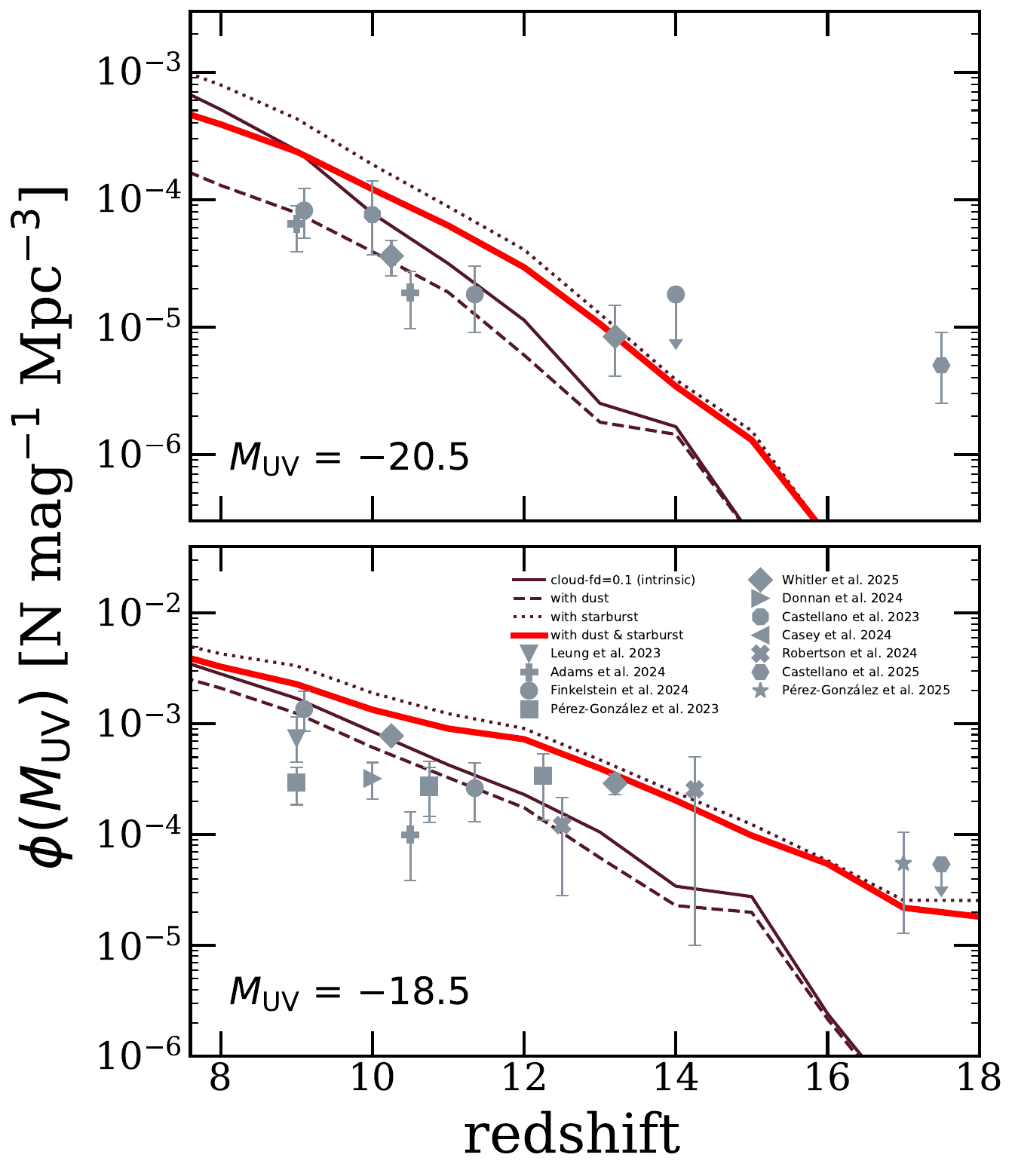}
\caption{Comoving number density of UV bright galaxies ($M_{\rm UV} \sim -20.5$; top) and fainter galaxies ($M_{\rm UV }\sim -18.5$; bottom) as a function of redshift, predicted by the Santa Cruz semi-analytic models. Symbols show observational estimates compiled from the literature. The left panel shows the predictions of a standard Kennicutt-Schmidt (KS) recipe for star formation, calibrated to low redshift observations, and the right panel shows the prediction of the Density Modulated Star Formation Efficiency model, in which the SFE increases with increasing gas density. Light solid lines show the models without dust and without enhanced burstyness (see S25 for details). Dashed lines show the models with dust but no enhanced bursts, and dotted lines show models with bursts but no dust. Thick solid lines show the model predictions with dust and enhanced bursts. In the standard KS model without dust and enhanced bursts, the number density of both bright and fainter galaxies declines much more rapidly than the observations. Adding the dust model brings down the number density especially of UV luminous galaxies at $z\lesssim 10$, leading to better agreement with the observations. Adding bursts preferentially boosts the number density at higher redshifts, but not enough to bring the KS model into agreement with all of the observations. The DMSFE model produces more high redshift galaxies, but the number density still declines more rapidly than the observations in the absence of enhanced bursts. A combination of dust, halo mass dependent bursty star formation, and density modulated SFE may be able to explain the observed trends. Reproduced from \citet{Somerville:2025}, Fig. A3 and 15. }
\label{fig:phiuv_kscloud}       
\end{figure}

Another proposed solution is that radiation pressure from  young stars ejects the dust or puffs it up enough to greatly reduce the optical depth \citep{Ziparo2022,Ferrara2023,Ferrara2024,Ferrara2025}. \citet{Ferrara2023} show that the dust can be ejected by radiation pressure when an Eddington-like limit is exceeded, which translates under certain assumptions into a critical specific SFR (sSFR $ \equiv \dot{m}_*/m_*$). The average sSFR decreases with increasing cosmic time due to the decreasing average halo accretion rate. Thus one expects that most galaxies at $z \gtrsim 10$ will be above the critical sSFR, and will have ejected their dust, while lower redshift galaxies will be below the critical sSFR and will retain their dust and be attenuated \citep{Ferrara2023}. S25 implemented the dust ejection model in post-processing within the Santa Cruz SAMs, using the same critical sSFR suggested by \citet{Ferrara2023}. S25 confirmed that in spite of the many differences in the underlying modeling framework, this leads to a natural transition between galaxies with significant dust attenuation in the UV at $z\sim 6$--7 to most UV-luminous galaxies being nearly attenuation-free at $z\gtrsim 10$. The increasing level of UV attenuation flattens the rise in the number density of UV-bright galaxies from $z\sim 12$--6 compared to the growth rate of dark matter halos, thus helping to resolve some of the tension in this redshift range. However, at $z\gtrsim 12$, most galaxies are already nearly attenuation free, so this does not help with the remaining tension at higher redshifts.

Fig.~\ref{fig:phiuv_kscloud} shows the redshift evolution of the number density of UV-luminous galaxies (at $M_{\rm UV} \sim -20.5$ and $-18.5$) for a suite of models in which several of the physical effects discussed above have been introduced one by one into the Santa Cruz SAM, separately showing the impact of Density Modulated SFE, halo mass dependent enhanced bursty star formation, and the impact of dust. None of these effects alone can reproduce all of the observations, but it appears plausible that the combined effects might be able to do so. 

\subsection{Probing physical properties of galaxies at the epoch of reionization and beyond}
\label{sec:results:physprop}
One of the most exciting things about JWST is that the high resolution onboard spectrograph NIRSpec enables us to go beyond simply counting galaxies or studying their broad band SEDs, and to learn about the detailed physical properties of stellar populations and conditions in the ISM of galaxies in the first billion years of cosmic evolution. NIRSpec observations have already enabled us to begin to probe the burstyness of star formation histories, chemical abundances, chemical abundance ratios, and ISM electron densities in $z\gtrsim 6$ galaxies (see Ellis lectures). Theoretical models are only starting to be able to make predictions that can directly confront these observations, and the available observational samples are still rather small. However, I predict that this will be a particularly exciting area of progress in the coming years. 

\subsubsection{Chemical abundances and abundance ratios}
\label{sec:results:physprop:MZR}
As discussed in more detail in Ellis Lecture 2 Section 4, and the recent review of \citet{Curti:2025}, gas phase metallicities can be estimated using ratios of strong lines like [OIII]4959 and 5007$\mathring{A}$ and Balmer H$\beta$, and there are extensive pre-JWST studies of these metallicity indicators in galaxies at $z\sim 0$--6. However, the calibration between strong line indicators and metallicity is highly uncertain. \citet{Hirschmann:2023b} have shown using cosmological simulations with nebular line emission modeled in post-processing (as described in \S\ref{sec:synthobs:nebemission}) that using locally calibrated relations can yield metallicity estimates that are too low by one order of magnitude. These lines can also be very sensitive to the detailed density and temperature structure of the ISM \citep{Katz:2024}. Thus it may be preferable to forward model from simulations to predict synthetic SEDs and line emission, and compare directly in the space of observed line ratios. 

\begin{figure}
\includegraphics[width=\textwidth]{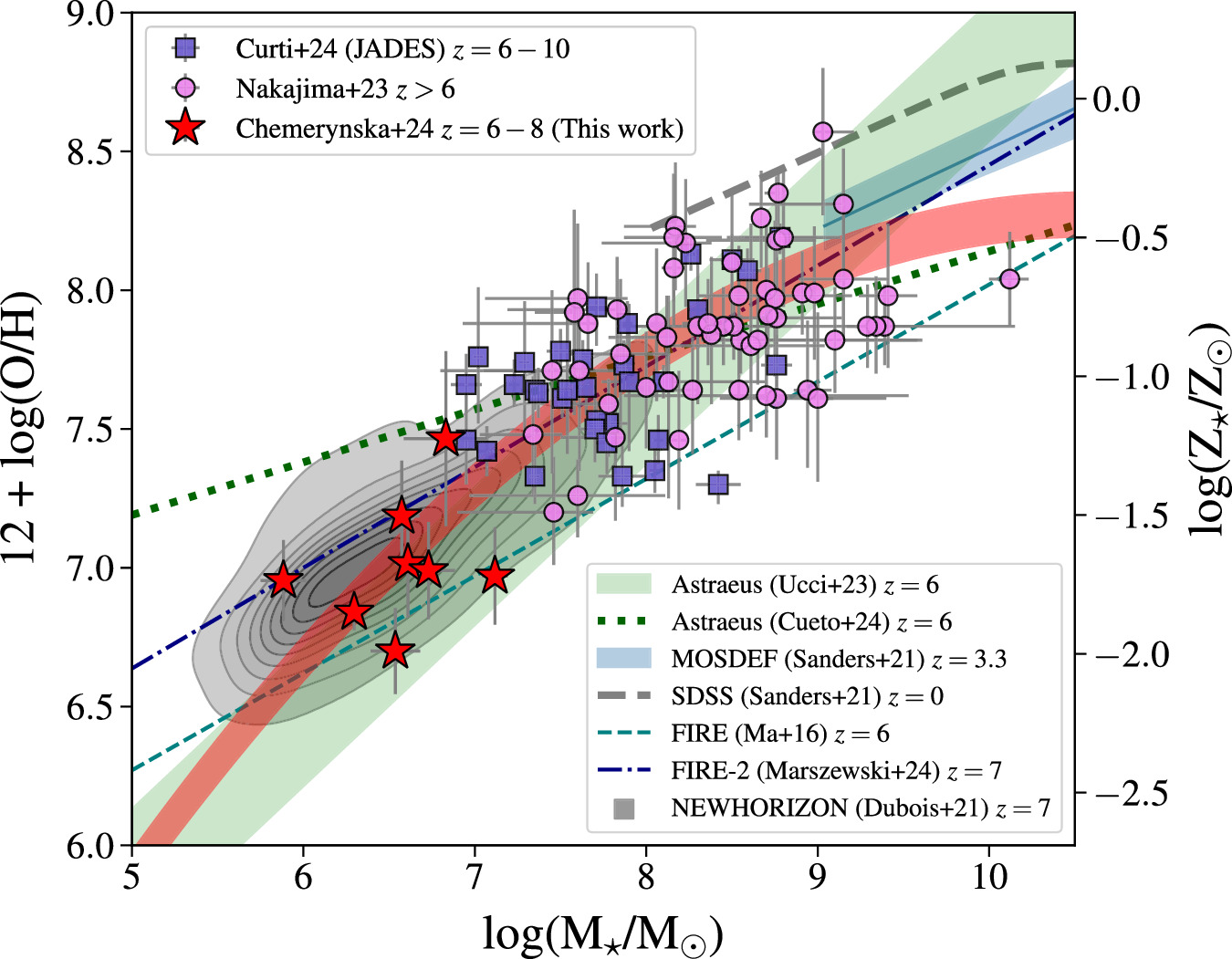}
\caption{Gas phase oxygen abundance vs. stellar mass (MZR) at $6\lesssim z \lesssim 10$. Symbols show a compilation of observations from different studies. The long dashed grey line shows the relation from observations at $z\sim 0$. Other lines and shaded regions show the predictions of various semi-analytic models and numerical simulations. There is a large dispersion in MZR predictions from different models, especially at the low-mass end. Reproduced from \citet{Chemerynska:2024}, Fig.~1. }
\label{fig:MZR}       
\end{figure}

More robust metallicity estimates (though still, of course, subject to calibration uncertainties) can be obtained when weak auroral lines such as [OIII]4363 $\mathring{A}$ can be measured, which provide an estimate of the electron temperature T$_e$ (so-called ``direct temperature'' methods). These measurements are now possible to obtain for $z\gtrsim6$ galaxies with NIRSpec \citep[][and references therein]{Sanders:2025,Curti:2025}. 

Gas phase metallicities have been measured over a large range in stellar mass at $z\sim 6$--8, showing that a correlation between stellar mass and metallicity (MZR relation) is already in place \citep{Nakajima:2023,Curti2023,Chemerynska:2024}. Interestingly, different theoretical models (which agree fairly well in their predictions for galaxy stellar mass functions and luminosity functions at these redshifts; see \S\ref{sec:models:status}) show a wide range of predictions for the MZR evolution at these epochs (see Fig.~\ref{fig:MZR}). This is likely due to different implementations of stellar feedback and winds, suggesting that these observations can provide strong tests of these model ingredients. Bursty star formation (which is connected to stellar feedback) can also impact the MZR and correlations between residuals from the MZR and other properties such as SFR (fundamental mass metallicity relation, FMZR; \citealp{Curti:2025}). 

\begin{figure}
\includegraphics[width=\textwidth]{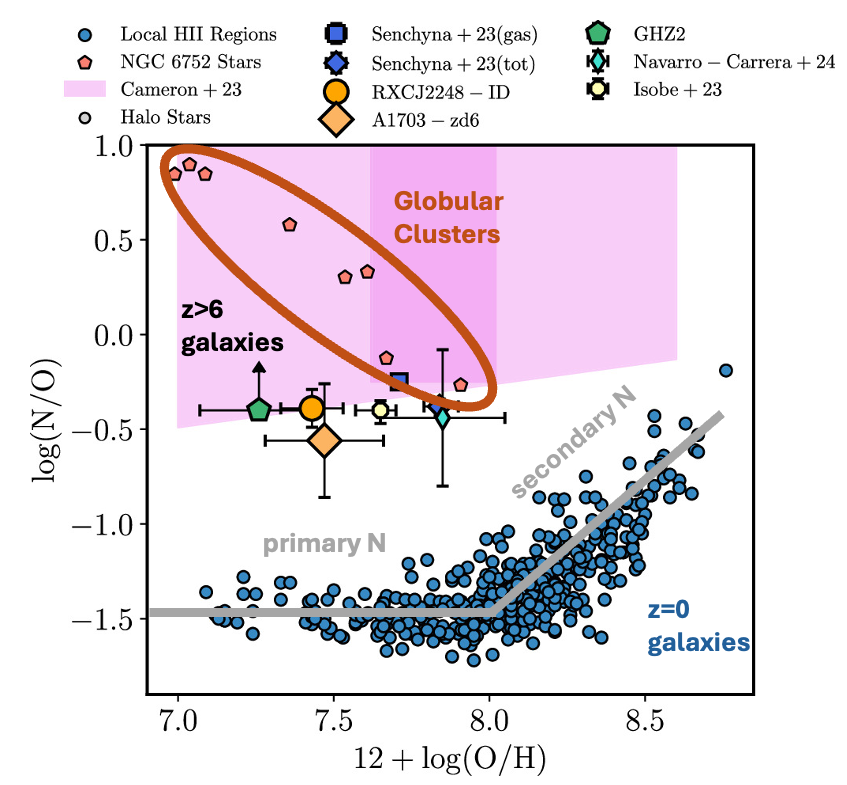}
\caption{Nitrogen to Oxygen ratio (N/O) versus oxygen abundance of interstellar gas. Blue filled circles show the distribution of H$_{\rm II}$ regions in local galaxies, with grey lines schematically highlighting the primary and secondary production sequence of nitrogen. The red outline show the approximate location of globular cluster stars. High redshift ($z\gtrsim 6$) galaxies observed with JWST (larger colored symbols) appear to inhabit the same region of this diagram as globular cluster stars, but a very different region than local star forming galaxies. Adapted from Fig.~13 of \citet{Topping:2025}. }
\label{fig:abundratios}       
\end{figure}

Another intriguing result is the measurement of ratios of different element abundances in $z\sim 6$--10 galaxies. As we alluded to in Section~\ref{sec:models:physics:chemdust}, various elements are produced on different timescales, so abundance ratios are clocks that provide insights into galaxy star formation histories and the physical processes that shape them. For example, [N/O] ratios have been measured in a sample of $z\gtrsim 6$ galaxies as summarized in Fig.~\ref{fig:abundratios}. H$_{\rm II}$ regions in nearby galaxies show a familiar ``hockey stick'' shaped diagram for [N/O] vs. [O/H], with a near constant value of [N/O] at low [O/H] (12+$\log$(O/H) $\lesssim 8$) and [N/O] increasing with increasing [O/H] at higher values, as expected for the primary and secondary nitrogen production channels described in Section~\ref{sec:models:physics:chemdust}.  At least some $z\gtrsim 6$ galaxies have elevated [N/O] at these low [O/H] values compared with the nearby galaxies, more similar to the [N/O] values seen in Milky Way globular cluster stars \citep{Cameron2023,Isobe2023b,Topping:2025,Belokurov2023}. Various mechanisms to explain enhanced N production at low O/H have recently been suggested, including fast rotating stars, very massive stars ($M_*\gtrsim 100 \msun$), Wolf-Rayet stars, and tidal disruption events (TDEs; \citealt{Ji2025} and references therein).  \citet{Kobayashi:2024} suggest that elevated N/O at low O/H could also be produced by multiple bursts of star formation.  In contrast, C/O vs. O/H ratios for the same sample of $z\gtrsim 6$ JWST galaxies appear similar to the distribution for local galaxies. However, C/O vs. O/H measurements are of great interest for further study, as they could potentially uncover evidence for enrichment by Pop III stars or pair instability SNae \citep{Curti:2025}.

\subsubsection{Star formation histories, stellar feedback, and ISM conditions}
\label{sec:results:sfhism}
\begin{figure}
\includegraphics[width=\textwidth]{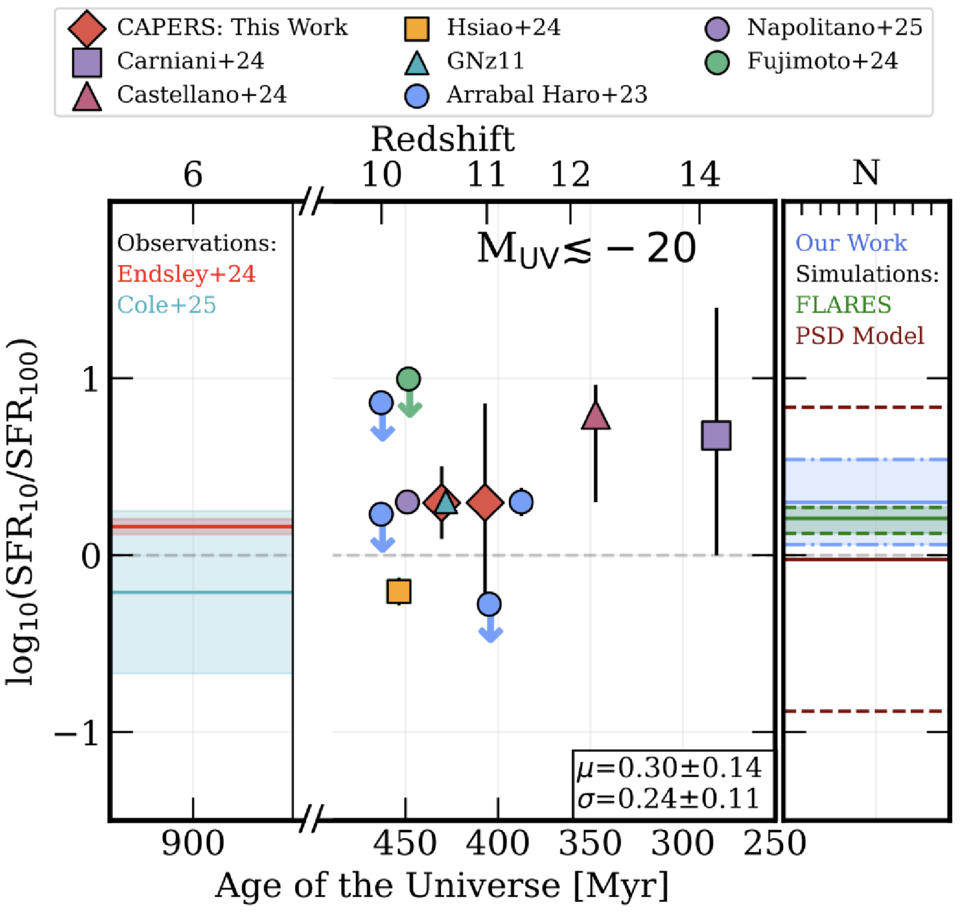}
\caption{The ratio of the SFR averaged over 10 Myr to that averaged over 100 Myr, which is an indicator of `burstyness', as a function of the age of the Universe (or redshift). Observational estimates are shown at $z\sim 6$ and $10 \lesssim z \lesssim 14$. The right panel shows expectations from simulations. A value of $\log({\rm SFR}_{10}/{\rm SFR}_{100}) >0$ indicates that a galaxy is in a `bursting' state; $z\gtrsim 10$ UV-bright galaxies seem to be predominantly in this state, though this could be a selection effect. Reproduced from \citet{Kokorev:2025}, Fig.~4. }
\label{fig:sfr10sfr100}       
\end{figure}

Balmer lines such as H$\alpha$, H$\beta$, etc are signatures of very young stellar populations, with ages of less than a few tens of Myr. The UV continuum is also produced by young massive stars, but has a significant contribution from stars that are a bit older, up to several hundred Myr. Thus ratios of Balmer line strengths to UV continuum can be used to estimate the ratio of the SFR averaged over $\sim10$ Myr to that averaged over $\sim$ 100 Myr (SFR$_{10}$/SFR$_{100}$). This quantity has been used as a measure of ``burstiness'', where SFR$_{10}$/SFR$_{100}>1$ indicates that a galaxy is in a ``bursting'' state while SFR$_{10}$/SFR$_{100}<1$ indicates that a galaxy is in a ``lulling'' or ``napping'' state. Fig.~\ref{fig:sfr10sfr100} shows a recent compilation of estimates of SFR$_{10}$/SFR$_{100}$ at $z\sim 6$ and $z\gtrsim 10$ reproduced from \citet{Kokorev2025}. There are hints that  SFR$_{10}$/SFR$_{100}$ is a bit higher at high redshift, although this could be a bias arising from selecting the most UV-luminous objects, which will naturally pick out objects that are in ``burst'' states. The observational samples are still small, but this seems like a promising approach to constrain physical processes such as stellar feedback and conditions in the ISM.  

(Relatively) large area JWST prism and slitless spectroscopy surveys such as CAPERS \citep{CAPERS:2024} and COSMOS-3D \citep{cosmos3d:2024} are starting to yield emission line measurements for large statistical samples of $z\gtrsim 6$ galaxies, enabling comparisons of distributions of line luminosities and equivalent widths (EW). These can be compared directly with predictions from semi-analytic models and hydrodynamic simulations that either compute line emission in post-processing or on-the-fly (\S\ref{sec:synthobs:nebemission}). For example, \citet{Meyer:2025} present luminosity functions and EW distributions for [OIII] 5008$\mathring{A}$ at $z\sim 7$--8 from COSMOS-3D and FRESCO, and compare with predictions from the Santa Cruz SAM and the THESAN, SPHYNX, and FLARES hydro simulations. Interestingly, the models that provide a closer match to the observed [OIII] luminosity functions (SC SAM and FLARES) are in greater tension with the EW distribution \citep[see also][]{Wilkins:2023b}. 

The blessing and the curse of nebular emission lines is that their luminosities are sensitive to a host of modeling details, including chemical enrichment, the detailed density and temperature distribution in the ISM, and radiation fields. As we have already noted, all of these quantities are in turn very sensitive to the detailed implementation of sub-grid recipes for star formation, the ISM and stellar feedback. \citet{Katz:2024} nicely illustrate this in the MEGATRON simulations. Their Fig.~14 shows how the distribution of the O32 ratio ($\log$ ([OIII] $\lambda$ 5007/[O II] $\lambda \lambda$ 3727) changes when they vary the details of the implementation of their sub-grid recipe for star formation. 

\begin{overview}{Summary}
\begin{itemize}
\item All pre-launch physics based galaxy formation models predicted a stronger decline in the number density of UV-luminous galaxies at $z\gtrsim 10$ than what has been measured by JWST.

\item One class of solutions to this problem involves making galaxies brighter due to a top-heavy IMF, bursty star formation, or contribution from an AGN. The other class of solutions involves making star formation more efficient at early times, perhaps because the high densities and pressures in the ISM lead to weaker stellar feedback. 

\item Another puzzle raised by JWST's first three years of observations is how to reconcile the very blue observed UV slopes at $z \gtrsim 9$ (which leave very little room for dust reddening) with the significant dust reservoirs seen in emission by $z\sim 7$. The $z \gtrsim 9$ galaxies might have significant amounts of dust, but with a size distribution that is skewed towards large grains, which are optically thin to UV radiation. Alternatively, dust may have been ejected or redistributed via radiation pressure, reducing the optical depth. 

\item High resolution spectra with JWST can probe detailed physical properties of high redshift galaxies, including star formation histories and star formation burstyness, chemical abundances and abundance ratios, and ISM electron densities. These have great potential for constraining the most uncertain aspects of theoretical modeling, such as the sub-grid treatment of the ISM, stellar feedback, and chemical evolution. Observational samples are currently small, and theoretical models are just starting to be able to rise to the challenge of making direct comparisons with these observations. 

\end{itemize}
\end{overview}

\section{Concluding Thoughts}
\label{sec:conclusions}

The past three years have been some of the most scientifically exciting of my career, and I know that I am not alone in feeling this way. JWST is working even better than we hoped, and it is more fun when the Universe provides surprises! I would also argue that we actually learn more when our models break. Our current situation is that we have multiple plausible solutions to the puzzles and tensions between theory and JWST observations that I have outlined, and this is in some ways unsatisfying. However, many of the proposed mechanisms (e.g. bursty star formation, weaker feedback, a top-heavy IMF) are actually expected to co-exist for physical reasons. Another piece of good news is that we expect JWST to continue to operate for many more years --- and all of these theoretical ideas have observational consequences that can be directly tested. As theorists, we should dare to make predictions for quantities that cannot \emph{yet} be observed, and try to provide insights into how combinations of observables from next-generation facilities can break some of the current degeneracies in our modeling. As observers, we should push the observational capabilities at our disposal in innovative ways to confront theory. As theorists and observers, we should constantly be asking ourselves and each other: how can a particular theory or theoretical idea be tested by observations?

Many of the comparisons between theory and JWST observations to date have been conducted in the `theoretical' plane, i.e., by estimating intrinsic physical properties from observations to compare with direct theoretical predictions. I believe that it is important for us as a community to work harder to build more accurate `forward models' that can bring theoretical predictions directly to the observational plane. This naturally involves additional assumptions (as well as more work), but `backwards modeling' also involves assumptions. Using both approaches can provide more robust conclusions as well as insights into the validity of these assumptions. 

I am excited to see what the next few years of JWST observations will bring. I'll bet the Universe still has a few surprises in store for us. 

\begin{acknowledgement}
I warmly thank the organizers of the 54th ``Saas-Fee Advanced Course'' of the Swiss Society for Astrophysics and Astronomy: Romain Meyer, Michaela Hirschmann, and Pascal Oesch. Every aspect of the planning and logistics of the school were carried out seamlessly, making it an ideal setting in which to discuss and enjoy the science. I also thank and commend the students --- I was deeply impressed by their level of dedication and engagement. I thank my fellow lecturers, Eduardo Ba\~{n}ados, Richard Ellis, and Marta Volonteri, for their excellent lectures and for many enjoyable discussions.   

I would also like to thank Frank van den Bosch, Benedikt Diemer, Kartheik Iyer, and Raffaella Schneider for pedogogical materials that they provided either publicly or privately, from which these lectures benefited greatly. I am grateful to Aaron Yung for creating Figure 1. I warmly thank my collaborators on JWST observational teams that I have been part of over the years, especially CEERS and its successors. A warm thank you also to Amiel Sternberg, for comments that improved the quality of these notes.

I dedicate these notes to the memory of two of my beloved mentors, Joel Primack and Avishai Dekel, who sadly both passed away in November 2025. Their curiosity about the workings of the Universe and their inexhaustible passion for science were a great inspiration to me and countless other colleagues, students, and postdocs.  They will be dearly missed. 

\end{acknowledgement}

\bibliographystyle{myspbasic}
\bibliography{somerville-saasfee25}

@BOOK{MvdBWbook,
   author = {{Mo}, H. and {van den Bosch}, F.~C. and {White}, S.},
    title = "{Galaxy Formation and Evolution}",
booktitle = {Galaxy Formation and Evolution},
publisher= {Cambridge University Press},
     year = {2010},
    month = may,
   adsurl = {http://adsabs.harvard.edu/abs/2010gfe..book.....M}
}

@ARTICLE{White:1991,
        AUTHOR  = {White, S.~D.~M. and Frenk, C.~S.},
        YEAR    = 1991,
        JOURNAL = {\apj},
        VOLUME  = 379,
        PAGES   = 52
}

@ARTICLE{Somerville:2002,
    author = {{Somerville}, R.~S.},
    journal = {\apjl},
    year = 2002,
    month = jun,
    volume = 572,
    pages = {23}
}

@ARTICLE{Benson:2002,
    author = {{Benson}, A.~J. and {Lacey}, C.~G. and {Baugh}, C.~M. and {Cole}, S. and 
	{Frenk}, C.~S.},
    journal = {\mnras},
    year = 2002,
    month = jun,
    volume = 333,
    pages = {156}
}

@ARTICLE{Efstathiou:1992,
   author = {{Efstathiou}, G.},
  journal = {\mnras},
     year = 1992,
    month = may,
   volume = 256,
    pages = {43}
}

@ARTICLE{Dekel:1986,
   author = {{Dekel}, A. and {Silk}, J.},
  journal = {\apj},
     year = 1986,
    month = apr,
   volume = 303,
    pages = {39}
}

@article{Okamoto:2008,
	Author = {{Okamoto}, T. and {Gao}, L. and {Theuns}, T.},
	Journal = {\mnras},
	Month = nov,
	Pages = {920},
	Volume = 390,
	Year = 2008
}

@ARTICLE{Peebles:1969,
   author = {{Peebles}, P.~J.~E.},
    title = "{Origin of the Angular Momentum of Galaxies}",
  journal = {\apj},
     year = 1969,
    month = feb,
   volume = 155,
    pages = {393},
   adsurl = {http://adsabs.harvard.edu/cgi-bin/nph-bib_query?bibcode=1969ApJ...155..393P&db_key=AST}
}

@ARTICLE{Ji2025,
       author = {{Ji}, Xihan and {Belokurov}, Vasily and {Maiolino}, Roberto and {Monty}, Stephanie and {Isobe}, Yuki and {Kravtsov}, Andrey and {McClymont}, William and {{\"U}bler}, Hannah},
        title = "{Connecting JWST discovered N/O-enhanced galaxies to globular clusters: evidence from chemical imprints}",
      journal = {\mnras},
         year = 2026,
        month = jan,
       volume = {545},
       number = {3},
          eid = {staf2110},
        pages = {staf2110},
          doi = {10.1093/mnras/staf2110},
archivePrefix = {arXiv},
       eprint = {2505.12505},
 primaryClass = {astro-ph.GA},
       adsurl = {https://ui.adsabs.harvard.edu/abs/2026MNRAS.545f2110J}
}

@ARTICLE{Carr2023,
       author = {{Carr}, Christopher and {Bryan}, Greg L. and {Fielding}, Drummond B. and {Pandya}, Viraj and {Somerville}, Rachel S.},
        title = "{Regulation of Star Formation by a Hot Circumgalactic Medium}",
      journal = {\apj},
         year = 2023,
        month = may,
       volume = {949},
       number = {1},
          eid = {21},
        pages = {21},
          doi = {10.3847/1538-4357/acc4c7},
archivePrefix = {arXiv},
       eprint = {2211.05115},
 primaryClass = {astro-ph.GA},
       adsurl = {https://ui.adsabs.harvard.edu/abs/2023ApJ...949...21C}
}

@ARTICLE{Naidu2025,
       author = {{Naidu}, Rohan P. and {Oesch}, Pascal A. and {Brammer}, Gabriel and {Weibel}, Andrea and {Li}, Yijia and {Matthee}, Jorryt and {Chisolm}, John and {Pollock}, Clara L. and {Heintz}, Kasper E. and {Johnson}, Benjamin D. and {Shen}, Xuejian and {Hviding}, Raphael E. and {Leja}, Joel and {Tacchella}, Sandro and {Ganguly}, Arpita and {Witten}, Callum and {Atek}, Hakim and {Belli}, Siro and {Bose}, Sownak and {Bouwens}, Rychard and {Dayal}, Pratika and {Decarli}, Roberto and {de Graaff}, Anna and {Fudamoto}, Yoshinobu and {Giovinazzo}, Emma and {Greene}, Jenny E. and {Illingworth}, Garth and {Inoue}, Akio K. and {Kane}, Sarah G. and {Labbe}, Ivo and {Leonova}, Ecaterina and {Marques-Chaves}, Rui and {Meyer}, Roman A. and {Nelson}, Erica J. and {Roberts-Borsani}, Guido and {Schaerer}, Daniel and {Simcoe}, Robert A. and {Stefanon}, Mauro and {Sugahara}, Yuma and {Toft}, Sune and {van der Wel}, Arjen and {van Dokkum}, Pieter and {Walter}, Fabian and {Watson}, Darrach and {Weaver}, John R. and {Whitaker}, Katherine E.},
        title = "{A Cosmic Miracle: A Remarkably Luminous Galaxy at zspec = 14.44 Confirmed with JWST}",
      journal = {The Open Journal of Astrophysics},
         year = 2026,
        month = jan,
       volume = {9},
        pages = {56033},
          doi = {10.33232/001c.156033},
archivePrefix = {arXiv},
       eprint = {2505.11263},
 primaryClass = {astro-ph.GA},
       adsurl = {https://ui.adsabs.harvard.edu/abs/2026OJAp....956033N}
}

@ARTICLE{Adamo2024,
       author = {{Adamo}, Angela and {Bradley}, Larry D. and {Vanzella}, Eros and {Claeyssens}, Ad{\'e}la{\"\i}de and {Welch}, Brian and {Diego}, Jose M. and {Mahler}, Guillaume and {Oguri}, Masamune and {Sharon}, Keren and {Abdurro'uf} and {Hsiao}, Tiger Yu-Yang and {Xu}, Xinfeng and {Messa}, Matteo and {Lassen}, Augusto E. and {Zackrisson}, Erik and {Brammer}, Gabriel and {Coe}, Dan and {Kokorev}, Vasily and {Ricotti}, Massimo and {Zitrin}, Adi and {Fujimoto}, Seiji and {Inoue}, Akio K. and {Resseguier}, Tom and {Rigby}, Jane R. and {Jim{\'e}nez-Teja}, Yolanda and {Windhorst}, Rogier A. and {Hashimoto}, Takuya and {Tamura}, Yoichi},
        title = "{Bound star clusters observed in a lensed galaxy 460 Myr after the Big Bang}",
      journal = {\nat},
         year = 2024,
        month = aug,
       volume = {632},
       number = {8025},
        pages = {513-516},
          doi = {10.1038/s41586-024-07703-7},
archivePrefix = {arXiv},
       eprint = {2401.03224},
 primaryClass = {astro-ph.GA},
       adsurl = {https://ui.adsabs.harvard.edu/abs/2024Natur.632..513A}
}

@article{Finkelstein2023,
archivePrefix = {arXiv},
arxivId = {2211.05792},
author = {Finkelstein, Steven L. and Bagley, Micaela B. and Ferguson, Henry C. and Wilkins, Stephen M. and Kartaltepe, Jeyhan S. and Papovich, Casey and Yung, L. Y. Aaron and Haro, Pablo Arrabal and Behroozi, Peter and Dickinson, Mark and Kocevski, Dale D. and Koekemoer, Anton M. and Larson, Rebecca L. and {Le Bail}, Aur{\'{e}}lien and Morales, Alexa M. and P{\'{e}}rez-Gonz{\'{a}}lez, Pablo G. and Burgarella, Denis and Dav{\'{e}}, Romeel and Hirschmann, Michaela and Somerville, Rachel S. and Wuyts, Stijn and Bromm, Volker and Casey, Caitlin M. and Fontana, Adriano and Fujimoto, Seiji and Gardner, Jonathan P. and Giavalisco, Mauro and Grazian, Andrea and Grogin, Norman A. and Hathi, Nimish P. and Hutchison, Taylor A. and Jha, Saurabh W. and Jogee, Shardha and Kewley, Lisa J. and Kirkpatrick, Allison and Long, Arianna S. and Lotz, Jennifer M. and Pentericci, Laura and Pierel, Justin D. R. and Pirzkal, Nor and Ravindranath, Swara and Ryan, Russell E. and Trump, Jonathan R. and Yang, Guang and Bhatawdekar, Rachana and Bisigello, Laura and Buat, V{\'{e}}ronique and Calabr{\`{o}}, Antonello and Castellano, Marco and Cleri, Nikko J. and Cooper, M. C. and Croton, Darren and Daddi, Emanuele and Dekel, Avishai and Elbaz, David and Franco, Maximilien and Gawiser, Eric and Holwerda, Benne W. and Huertas-Company, Marc and Jaskot, Anne E. and Leung, Gene C. K. and Lucas, Ray A. and Mobasher, Bahram and Pandya, Viraj and Tacchella, Sandro and Weiner, Benjamin J. and Zavala, Jorge A.},
doi = {10.3847/2041-8213/acade4},
eprint = {2211.05792},
issn = {2041-8205},
journal = {\apjl},
month = {mar},
number = {1},
pages = {L13},
title = {{CEERS Key Paper. I. An Early Look into the First 500 Myr of Galaxy Formation with JWST}},
url = {http://arxiv.org/abs/2211.05792 https://iopscience.iop.org/article/10.3847/2041-8213/acade4},
volume = {946},
year = {2023}
}

@ARTICLE{Finkelstein2024,
       author = {{Finkelstein}, Steven L. and {Leung}, Gene C.~K. and {Bagley}, Micaela B. and {Dickinson}, Mark and {Ferguson}, Henry C. and {Papovich}, Casey and {Akins}, Hollis B. and {Arrabal Haro}, Pablo and {Dav{\'e}}, Romeel and {Dekel}, Avishai and {Kartaltepe}, Jeyhan S. and {Kocevski}, Dale D. and {Koekemoer}, Anton M. and {Pirzkal}, Nor and {Somerville}, Rachel S. and {Yung}, L.~Y. Aaron and {Amor{\'\i}n}, Ricardo O. and {Backhaus}, Bren E. and {Behroozi}, Peter and {Bisigello}, Laura and {Bromm}, Volker and {Casey}, Caitlin M. and {Ch{\'a}vez Ortiz}, {\'O}scar A. and {Cheng}, Yingjie and {Chworowsky}, Katherine and {Cleri}, Nikko J. and {Cooper}, M.~C. and {Davis}, Kelcey and {de la Vega}, Alexander and {Elbaz}, David and {Franco}, Maximilien and {Fontana}, Adriano and {Fujimoto}, Seiji and {Giavalisco}, Mauro and {Grogin}, Norman A. and {Holwerda}, Benne W. and {Huertas-Company}, Marc and {Hirschmann}, Michaela and {Iyer}, Kartheik G. and {Jogee}, Shardha and {Jung}, Intae and {Larson}, Rebecca L. and {Lucas}, Ray A. and {Mobasher}, Bahram and {Morales}, Alexa M. and {Morley}, Caroline V. and {Mukherjee}, Sagnick and {P{\'e}rez-Gonz{\'a}lez}, Pablo G. and {Ravindranath}, Swara and {Rodighiero}, Giulia and {Rowland}, Melanie J. and {Tacchella}, Sandro and {Taylor}, Anthony J. and {Trump}, Jonathan R. and {Wilkins}, Stephen M.},
        title = "{The Complete CEERS Early Universe Galaxy Sample: A Surprisingly Slow Evolution of the Space Density of Bright Galaxies at z {\ensuremath{\sim}} 8.5{\textendash}14.5}",
      journal = {\apjl},
         year = 2024,
        month = jul,
       volume = {969},
       number = {1},
          eid = {L2},
        pages = {L2},
          doi = {10.3847/2041-8213/ad4495},
archivePrefix = {arXiv},
       eprint = {2311.04279},
 primaryClass = {astro-ph.GA},
       adsurl = {https://ui.adsabs.harvard.edu/abs/2024ApJ...969L...2F}
}

@ARTICLE{NaabOstriker2017,
       author = {{Naab}, Thorsten and {Ostriker}, Jeremiah P.},
        title = "{Theoretical Challenges in Galaxy Formation}",
      journal = {\araa},
         year = 2017,
        month = aug,
       volume = {55},
       number = {1},
        pages = {59-109},
          doi = {10.1146/annurev-astro-081913-040019},
archivePrefix = {arXiv},
       eprint = {1612.06891},
 primaryClass = {astro-ph.GA},
       adsurl = {https://ui.adsabs.harvard.edu/abs/2017ARA&A..55...59N}
}

@article{Pallottini2023,
author = {Pallottini, A. and Ferrara, A.},
doi = {10.1051/0004-6361/202347384},
issn = {0004-6361},
journal = {\aap},
month = {sep},
pages = {L4},
title = {{Stochastic star formation in early galaxies: Implications for the James Webb Space Telescope}},
url = {https://www.aanda.org/10.1051/0004-6361/202347384},
volume = {677},
year = {2023}
}

@article{Vogelsberger2020,
archivePrefix = {arXiv},
arxivId = {1904.07238},
author = {Vogelsberger, Mark and Nelson, Dylan and Pillepich, Annalisa and Shen, Xuejian and Marinacci, Federico and Springel, Volker and Pakmor, R{\"{u}}diger and Tacchella, Sandro and Weinberger, Rainer and Torrey, Paul and Hernquist, Lars},
doi = {10.1093/mnras/staa137},
eprint = {1904.07238},
issn = {0035-8711},
journal = {\mnras},
month = {mar},
number = {4},
pages = {5167--5201},
title = {{High-redshift JWST predictions from IllustrisTNG: dust modelling and galaxy luminosity functions}},
url = {http://arxiv.org/abs/1904.07238 https://academic.oup.com/mnras/article/492/4/5167/5707414},
volume = {492},
year = {2020}
}

@article{Yung2019a,
archivePrefix = {arXiv},
arxivId = {1803.09761},
author = {Yung, L. Y. Aaron and Somerville, Rachel S. and Finkelstein, Steven L. and Popping, Gerg{\"{o}} and Dav{\'{e}}, Romeel},
doi = {10.1093/mnras/sty3241},
eprint = {1803.09761},
issn = {0035-8711},
journal = {\mnras},
month = {mar},
number = {3},
pages = {2983--3006},
title = {{Semi-analytic forecasts for JWST – I. UV luminosity functions at z = 4–10}},
url = {http://arxiv.org/abs/1803.09761 https://academic.oup.com/mnras/advance-article/doi/10.1093/mnras/sty3241/5218517 https://academic.oup.com/mnras/article/483/3/2983/5218517},
volume = {483},
year = {2019}
}

@article{Sheth1999,
archivePrefix = {arXiv},
arxivId = {astro-ph/9901122},
author = {Sheth, Ravi K. and Tormen, Giuseppe},
doi = {10.1046/j.1365-8711.1999.02692.x},
eprint = {9901122},
isbn = {doi:10.1046/j.1365-8711.2001.04006.x},
issn = {00358711},
journal = {\mnras},
number = {1},
pages = {119--126},
primaryClass = {astro-ph},
title = {{Large-scale bias and the peak background split}},
volume = {308},
year = {1999}
}

@article{Leung2023,
archivePrefix = {arXiv},
arxivId = {2306.06244},
author = {Leung, Gene C. K. and Bagley, Micaela B. and Finkelstein, Steven L. and Ferguson, Henry C. and Koekemoer, Anton M. and P{\'{e}}rez-Gonz{\'{a}}lez, Pablo G. and Morales, Alexa and Kocevski, Dale D. and {Yang}, Guang and Somerville, Rachel S. and Wilkins, Stephen M. and Yung, L. Y. Aaron and Fujimoto, Seiji and Larson, Rebecca L. and Papovich, Casey and Pirzkal, Nor and Berg, Danielle A. and Lotz, Jennifer M. and Castellano, Marco and {Ch{\'{a}}vez Ortiz}, {\'{O}}scar A. and Cheng, Yingjie and Dickinson, Mark and Giavalisco, Mauro and Hathi, Nimish P. and Hutchison, Taylor A. and Jung, Intae and Kartaltepe, Jeyhan S. and Natarajan, Priyamvada and Rothberg, Barry},
doi = {10.3847/2041-8213/acf365},
eprint = {2306.06244},
issn = {2041-8205},
journal = {\apjl},
month = {sep},
number = {2},
pages = {L46},
title = {{NGDEEP Epoch 1: The Faint End of the Luminosity Function at z ∼ 9–12 from Ultradeep JWST Imaging}},
url = {http://arxiv.org/abs/2306.06244 https://iopscience.iop.org/article/10.3847/2041-8213/acf365},
volume = {954},
year = {2023}
}

@article{Feng2016,
archivePrefix = {arXiv},
arxivId = {1504.06619},
author = {Feng, Yu and Di-Matteo, Tiziana and Croft, Rupert A. and Bird, Simeon and Battaglia, Nicholas and Wilkins, Stephen},
doi = {10.1093/mnras/stv2484},
eprint = {1504.06619},
isbn = {1523-7060},
issn = {0035-8711},
journal = {\mnras},
month = {jan},
number = {3},
pages = {2778--2791},
title = {{The BlueTides simulation: first galaxies and reionization}},
url = {https://academic.oup.com/mnras/article-lookup/doi/10.1093/mnras/stv2484},
volume = {455},
year = {2016}
}

@article{Shen2023,
archivePrefix = {arXiv},
arxivId = {2305.05679},
author = {Shen, Xuejian and Vogelsberger, Mark and Boylan-Kolchin, Michael and Tacchella, Sandro and Kannan, Rahul},
doi = {10.1093/mnras/stad2508},
eprint = {2305.05679},
issn = {0035-8711},
journal = {\mnras},
month = {sep},
number = {3},
pages = {3254--3261},
title = {{The impact of UV variability on the abundance of bright galaxies at z ≥ 9}},
url = {http://arxiv.org/abs/2305.05679 https://academic.oup.com/mnras/article/525/3/3254/7246904},
volume = {525},
year = {2023}
}

@article{SomervilleDave2015,
archivePrefix = {arXiv},
arxivId = {1412.2712v1},
author = {Somerville, Rachel S. and Dav{\'{e}}, Romeel},
doi = {10.1146/annurev-astro-082812-140951},
eprint = {1412.2712v1},
issn = {0066-4146},
journal = {\araa},
pages = {31--113},
title = {{Physical Models of Galaxy Formation in a Cosmological Framework}},
url = {http://www.annualreviews.org/doi/abs/10.1146/annurev-astro-082812-140951},
volume = {53},
year = {2015}
}

@article{Mason2023,
archivePrefix = {arXiv},
arxivId = {2207.14808},
author = {Mason, Charlotte A. and Trenti, Michele and Treu, Tommaso},
doi = {10.1093/mnras/stad035},
eprint = {2207.14808},
issn = {0035-8711},
journal = {\mnras},
month = {mar},
number = {1},
pages = {497--503},
title = {{The brightest galaxies at cosmic dawn}},
url = {http://arxiv.org/abs/2207.14808 https://academic.oup.com/mnras/article/521/1/497/6979826},
volume = {521},
year = {2023}
}

@article{ArrabalHaro2023,
archivePrefix = {arXiv},
arxivId = {2304.05378},
author = {{Arrabal Haro}, Pablo and Dickinson, Mark and Finkelstein, Steven L. and Fujimoto, Seiji and Fern{\'{a}}ndez, Vital and Kartaltepe, Jeyhan S. and Jung, Intae and Cole, Justin W. and Burgarella, Denis and Chworowsky, Katherine and Hutchison, Taylor A. and Morales, Alexa M. and Papovich, Casey and Simons, Raymond C. and Amor{\'{i}}n, Ricardo O. and Backhaus, Bren E. and Bagley, Micaela B. and Bisigello, Laura and Calabr{\`{o}}, Antonello and Castellano, Marco and Cleri, Nikko J. and Dav{\'{e}}, Romeel and Dekel, Avishai and Ferguson, Henry C. and Fontana, Adriano and Gawiser, Eric and Giavalisco, Mauro and Harish, Santosh and Hathi, Nimish P. and Hirschmann, Michaela and Holwerda, Benne W. and Huertas-Company, Marc and Koekemoer, Anton M. and Larson, Rebecca L. and Lucas, Ray A. and Mobasher, Bahram and P{\'{e}}rez-Gonz{\'{a}}lez, Pablo G. and Pirzkal, Nor and Rose, Caitlin and Santini, Paola and Trump, Jonathan R. and de la Vega, Alexander and Wang, Xin and Weiner, Benjamin J. and Wilkins, Stephen M. and Yang, Guang and Yung, L. Y. Aaron and Zavala, Jorge A.},
doi = {10.3847/2041-8213/acdd54},
eprint = {2304.05378},
issn = {2041-8205},
journal = {\apjl},
month = {jul},
number = {1},
pages = {L22},
title = {{Spectroscopic Confirmation of CEERS NIRCam-selected Galaxies at z ≃ 8–10}},
url = {http://arxiv.org/abs/2304.05378 https://iopscience.iop.org/article/10.3847/2041-8213/acdd54},
volume = {951},
year = {2023}
}

@article{Kannan2022,
archivePrefix = {arXiv},
arxivId = {2110.00584},
author = {Kannan, R. and Garaldi, E. and Smith, A. and Pakmor, R. and Springel, V. and Vogelsberger, M. and Hernquist, L.},
doi = {10.1093/mnras/stab3710},
eprint = {2110.00584},
issn = {0035-8711},
journal = {\mnras},
month = {feb},
number = {3},
pages = {4005--4030},
title = {{Introducing the <scp>thesan</scp> project: radiation-magnetohydrodynamic simulations of the epoch of reionization}},
url = {http://arxiv.org/abs/2110.00584 https://academic.oup.com/mnras/article/511/3/4005/6484814},
volume = {511},
year = {2022}
}

@article{Wilkins2022,
archivePrefix = {arXiv},
arxivId = {2204.09431},
author = {Wilkins, Stephen M. and Vijayan, Aswin P. and Lovell, Christopher C. and Roper, William J and Irodotou, Dimitrios and Caruana, Joseph and Seeyave, Louise T. C. and Kuusisto, Jussi K. and Thomas, Peter A. and Parris, Shedeur A. K.},
doi = {10.1093/mnras/stac3280},
eprint = {2204.09431},
issn = {0035-8711},
journal = {\mnras},
month = {dec},
number = {2},
pages = {3118--3128},
title = {{First light and reionization epoch simulations (FLARES) V: the redshift frontier}},
url = {http://arxiv.org/abs/2204.09431 https://academic.oup.com/mnras/article/519/2/3118/6832298},
volume = {519},
year = {2022}
}

@article{Ferrara2023,
archivePrefix = {arXiv},
arxivId = {2208.00720},
author = {Ferrara, Andrea and Pallottini, Andrea and Dayal, Pratika},
doi = {10.1093/mnras/stad1095},
eprint = {2208.00720},
issn = {0035-8711},
journal = {\mnras},
month = {may},
number = {3},
pages = {3986--3991},
title = {{On the stunning abundance of super-early, luminous galaxies revealed by JWST}},
url = {http://arxiv.org/abs/2208.00720 http://dx.doi.org/10.1093/mnras/stad1095 https://academic.oup.com/mnras/article/522/3/3986/7156962},
volume = {522},
year = {2023}
}

@article{Boylan-Kolchin2023,
archivePrefix = {arXiv},
arxivId = {2208.01611},
author = {Boylan-Kolchin, Michael},
doi = {10.1038/s41550-023-01937-7},
eprint = {2208.01611},
issn = {2397-3366},
journal = {Nat. Astron.},
month = {apr},
number = {6},
pages = {731--735},
title = {{Stress testing $\Lambda$CDM with high-redshift galaxy candidates}},
url = {http://arxiv.org/abs/2208.01611 http://dx.doi.org/10.1038/s41550-023-01937-7 https://www.nature.com/articles/s41550-023-01937-7},
volume = {7},
year = {2023}
}

@article{Topping2022,
archivePrefix = {arXiv},
arxivId = {2208.01610},
author = {Topping, Michael W. and Stark, Daniel P. and Endsley, Ryan and Plat, Adele and Whitler, Lily and Chen, Zuyi and Charlot, St{\'{e}}phane},
doi = {10.3847/1538-4357/aca522},
eprint = {2208.01610},
issn = {0004-637X},
journal = {\apj},
month = {dec},
number = {2},
pages = {153},
title = {{Searching for Extremely Blue UV Continuum Slopes at z = 7–11 in JWST/NIRCam Imaging: Implications for Stellar Metallicity and Ionizing Photon Escape in Early Galaxies}},
url = {http://arxiv.org/abs/2208.01610 http://dx.doi.org/10.3847/1538-4357/aca522 https://iopscience.iop.org/article/10.3847/1538-4357/aca522},
volume = {941},
year = {2022}
}

@article{Bryan1998,
archivePrefix = {arXiv},
arxivId = {astro-ph/9710107},
author = {Bryan, Greg L. and Norman, Michael L.},
doi = {10.1086/305262},
eprint = {9710107},
issn = {0004-637X},
journal = {\apj},
number = {1},
pages = {80--99},
primaryClass = {astro-ph},
title = {{Statistical Properties of X‐Ray Clusters: Analytic and Numerical Comparisons}},
url = {http://stacks.iop.org/0004-637X/495/i=1/a=80 http://adsabs.harvard.edu/abs/1998ApJ...495...80B},
volume = {495},
year = {1998}
}

@article{Popping2017,
archivePrefix = {arXiv},
arxivId = {1609.08622},
author = {Popping, Gerg{\"{o}} and Somerville, Rachel S. and Galametz, Maud},
doi = {10.1093/mnras/stx1545},
eprint = {1609.08622},
issn = {0035-8711},
journal = {\mnras},
month = {nov},
number = {3},
pages = {3152--3185},
title = {{The dust content of galaxies from z = 0 to z = 9}},
url = {http://arxiv.org/abs/1609.08622 http://academic.oup.com/mnras/article/471/3/3152/3873954/The-dust-content-of-galaxies-from-z0-to-z9},
volume = {471},
year = {2017}
}

@article{Menci2022,
archivePrefix = {arXiv},
arxivId = {2208.11471},
author = {Menci, N. and Castellano, M. and Santini, P. and Merlin, E. and Fontana, A. and Shankar, F.},
doi = {10.3847/2041-8213/ac96e9},
eprint = {2208.11471},
issn = {2041-8205},
journal = {\apjl},
number = {1},
pages = {L5},
publisher = {IOP Publishing},
title = {{High-redshift Galaxies from Early JWST Observations: Constraints on Dark Energy Models}},
url = {http://dx.doi.org/10.3847/2041-8213/ac96e9},
volume = {938},
year = {2022}
}

@ARTICLE{Menci2024,
       author = {{Menci}, N. and {Sen}, A.~A. and {Castellano}, M.},
        title = "{The Excess of JWST Bright Galaxies: A Possible Origin in the Ground State of Dynamical Dark Energy in the Light of DESI 2024 Data}",
      journal = {\apj},
         year = 2024,
        month = dec,
       volume = {976},
       number = {2},
          eid = {227},
        pages = {227},
          doi = {10.3847/1538-4357/ad8d5b},
archivePrefix = {arXiv},
       eprint = {2410.22940},
 primaryClass = {astro-ph.CO},
       adsurl = {https://ui.adsabs.harvard.edu/abs/2024ApJ...976..227M}
}

@article{KannanMTNG2023,
archivePrefix = {arXiv},
arxivId = {2210.10066},
author = {Kannan, Rahul and Springel, Volker and Hernquist, Lars and Pakmor, R{\"{u}}diger and Delgado, Ana Maria and Hadzhiyska, Boryana and Hern{\'{a}}ndez-Aguayo, C{\'{e}}sar and Barrera, Monica and Ferlito, Fulvio and Bose, Sownak and White, Simon D M and Frenk, Carlos and Smith, Aaron and Garaldi, Enrico},
doi = {10.1093/mnras/stac3743},
eprint = {2210.10066},
issn = {0035-8711},
journal = {\mnras},
month = {jul},
number = {2},
pages = {2594--2605},
title = {{The MillenniumTNG project: the galaxy population at z ≥ 8}},
url = {http://arxiv.org/abs/2210.10066 http://dx.doi.org/10.1093/mnras/stac3743 https://academic.oup.com/mnras/article/524/2/2594/7226462},
volume = {524},
year = {2023}
}

@article{Dekel2023,
archivePrefix = {arXiv},
arxivId = {2303.04827},
author = {Dekel, Avishai and Sarkar, Kartick C and Birnboim, Yuval and Mandelker, Nir and Li, Zhaozhou},
doi = {10.1093/mnras/stad1557},
eprint = {2303.04827},
issn = {0035-8711},
journal = {\mnras},
month = {jun},
number = {3},
pages = {3201--3218},
title = {{Efficient formation of massive galaxies at cosmic dawn by feedback-free starbursts}},
url = {http://arxiv.org/abs/2303.04827 http://dx.doi.org/10.1093/mnras/stad1557 https://academic.oup.com/mnras/article/523/3/3201/7179993},
volume = {523},
year = {2023}
}

@article{Pillepich2018,
archivePrefix = {arXiv},
arxivId = {1703.02970},
author = {Pillepich, Annalisa and Springel, Volker and Nelson, Dylan and Genel, Shy and Naiman, Jill and Pakmor, R{\"{u}}diger and Hernquist, Lars and Torrey, Paul and Vogelsberger, Mark and Weinberger, Rainer and Marinacci, Federico},
doi = {10.1093/mnras/stx2656},
eprint = {1703.02970},
issn = {0035-8711},
journal = {\mnras},
month = {jan},
number = {3},
pages = {4077--4106},
title = {{Simulating galaxy formation with the IllustrisTNG model}},
url = {http://arxiv.org/abs/1703.02970 http://academic.oup.com/mnras/article/473/3/4077/4494369},
volume = {473},
year = {2018}
}

@article{Madau2014,
archivePrefix = {arXiv},
arxivId = {1403.0007},
author = {Madau, Piero and Dickinson, Mark},
doi = {10.1146/annurev-astro-081811-125615},
eprint = {1403.0007},
issn = {0066-4146},
journal = {\araa},
month = {aug},
number = {1},
pages = {415--486},
title = {{Cosmic Star-Formation History}},
url = {http://arxiv.org/abs/1403.0007 http://dx.doi.org/10.1146/annurev-astro-081811-125615 http://www.annualreviews.org/doi/10.1146/annurev-astro-081811-125615},
volume = {52},
year = {2014}
}

@article{Yung2022,
archivePrefix = {arXiv},
arxivId = {2206.13521},
author = {Yung, L. Y. Aaron and Somerville, Rachel S. and Ferguson, Henry C. and Finkelstein, Steven L. and Gardner, Jonathan P. and Dav{\'{e}}, Romeel and Bagley, Micaela B and Popping, Gerg{\"{o}} and Behroozi, Peter},
doi = {10.1093/mnras/stac2139},
eprint = {2206.13521},
issn = {0035-8711},
journal = {\mnras},
month = {aug},
number = {4},
pages = {5416--5436},
title = {{Semi-analytic forecasts for JWST – VI. Simulated light-cones and galaxy clustering predictions}},
url = {https://arxiv.org/pdf/2206.13521.pdf http://arxiv.org/abs/2206.13521 https://academic.oup.com/mnras/article/515/4/5416/6652501},
volume = {515},
year = {2022}
}

@article{Cullen2023,
archivePrefix = {arXiv},
arxivId = {2208.04914},
author = {Cullen, Fergus and McLure, R. J. and McLeod, D. J. and Dunlop, J S and Donnan, C T and Carnall, A C and Bowler, R A A and Begley, R and Hamadouche, M L and Stanton, T. M.},
doi = {10.1093/mnras/stad073},
eprint = {2208.04914},
issn = {0035-8711},
journal = {\mnras},
month = {jan},
number = {1},
pages = {14--23},
title = {{The ultraviolet continuum slopes ( $\beta$ ) of galaxies at z ≃ 8-16 from JWST and ground-based near-infrared imaging}},
url = {http://arxiv.org/abs/2208.04914 http://dx.doi.org/10.1093/mnras/stad073 https://academic.oup.com/mnras/article/520/1/14/6982919},
volume = {520},
year = {2023}
}

@article{Lovell2022,
archivePrefix = {arXiv},
arxivId = {2208.10479},
author = {Lovell, Christopher C. and Harrison, Ian and Harikane, Yuichi and Tacchella, Sandro and Wilkins, Stephen M.},
doi = {10.1093/mnras/stac3224},
eprint = {2208.10479},
issn = {0035-8711},
journal = {\mnras},
month = {nov},
number = {2},
pages = {2511--2520},
title = {{Extreme value statistics of the halo and stellar mass distributions at high redshift: are JWST results in tension with $\Lambda$CDM?}},
url = {http://arxiv.org/abs/2208.10479 http://dx.doi.org/10.1093/mnras/stac3224 https://academic.oup.com/mnras/article/518/2/2511/6823705},
volume = {518},
year = {2022}
}

@article{Labbe2023,
archivePrefix = {arXiv},
arxivId = {2207.12446},
author = {Labb{\'{e}}, Ivo and van Dokkum, Pieter and Nelson, Erica and Bezanson, Rachel and Suess, Katherine A. and Leja, Joel and Brammer, Gabriel and Whitaker, Katherine and Mathews, Elijah and Stefanon, Mauro and Wang, Bingjie},
doi = {10.1038/s41586-023-05786-2},
eprint = {2207.12446},
issn = {0028-0836},
journal = {Nature},
month = {apr},
number = {7956},
pages = {266--269},
title = {{A population of red candidate massive galaxies $\sim$600 Myr after the Big Bang}},
url = {http://arxiv.org/abs/2207.12446 https://www.nature.com/articles/s41586-023-05786-2},
volume = {616},
year = {2023}
}

@article{Eldridge2017,
archivePrefix = {arXiv},
arxivId = {1710.02154},
author = {Eldridge, J. J. and Stanway, E. R. and Xiao, L. and McClelland, L. A. S. and Taylor, G. and Ng, M. and Greis, S. M. L. and Bray, J. C.},
doi = {10.1017/pasa.2017.51},
eprint = {1710.02154},
issn = {1323-3580},
journal = {PASA},
month = {nov},
pages = {e058},
title = {{Binary Population and Spectral Synthesis Version 2.1: Construction, Observational Verification, and New Results}},
url = {https://www.cambridge.org/core/product/identifier/S1323358017000510/type/journal_article},
volume = {34},
year = {2017}
}

@article{Nelson2018,
archivePrefix = {arXiv},
arxivId = {1707.03395},
author = {Nelson, Dylan and Pillepich, Annalisa and Springel, Volker and Weinberger, Rainer and Hernquist, Lars and Pakmor, R{\"{u}}diger and Genel, Shy and Torrey, Paul and Vogelsberger, Mark and Kauffmann, Guinevere and Marinacci, Federico and Naiman, Jill},
doi = {10.1093/mnras/stx3040},
eprint = {1707.03395},
issn = {13652966},
journal = {\mnras},
number = {1},
pages = {624--647},
title = {{First results from the IllustrisTNG simulations: The galaxy colour bimodality}},
volume = {475},
year = {2018}
}

@article{Klypin2016,
archivePrefix = {arXiv},
arxivId = {1411.4001},
author = {Klypin, Anatoly and Yepes, Gustavo and Gottl{\"{o}}ber, Stefan and Prada, Francisco and He{\ss}, Steffen},
doi = {10.1093/mnras/stw248},
eprint = {1411.4001},
isbn = {0035-8711},
issn = {0035-8711},
journal = {\mnras},
month = {apr},
number = {4},
pages = {4340--4359},
title = {{MultiDark simulations: the story of dark matter halo concentrations and density profiles}},
url = {https://academic.oup.com/mnras/article-lookup/doi/10.1093/mnras/stw248},
volume = {457},
year = {2016}
}

@article{Sun2023,
archivePrefix = {arXiv},
arxivId = {2307.15305},
author = {Sun, Guochao and Faucher-Gigu{\`{e}}re, Claude-Andr{\'{e}} and Hayward, Christopher C. and Shen, Xuejian and Wetzel, Andrew and Cochrane, Rachel K.},
doi = {10.3847/2041-8213/acf85a},
eprint = {2307.15305},
issn = {2041-8205},
journal = {\apjl},
month = {oct},
number = {2},
pages = {L35},
title = {{Bursty Star Formation Naturally Explains the Abundance of Bright Galaxies at Cosmic Dawn}},
url = {http://arxiv.org/abs/2307.15305 https://iopscience.iop.org/article/10.3847/2041-8213/acf85a},
volume = {955},
year = {2023}
}

@article{Chabrier2003,
archivePrefix = {arXiv},
arxivId = {astro-ph/0304382},
author = {Chabrier, Gilles},
doi = {10.1086/376392},
eprint = {0304382},
isbn = {10.1086/376392},
issn = {0004-6280},
journal = {PASP},
month = {jul},
number = {809},
pages = {763--795},
pmid = {26438271},
primaryClass = {astro-ph},
title = {{Galactic Stellar and Substellar Initial Mass Function}},
url = {http://arxiv.org/abs/astro-ph/0304382%0Ahttp://dx.doi.org/10.1086/376392 http://iopscience.iop.org/article/10.1086/376392},
volume = {115},
year = {2003}
}

@article{Somerville2012,
archivePrefix = {arXiv},
arxivId = {1104.0669},
author = {Somerville, Rachel S. and Gilmore, Rudy C. and Primack, Joel R. and Dom{\'{i}}nguez, Alberto},
doi = {10.1111/j.1365-2966.2012.20490.x},
eprint = {1104.0669},
issn = {00358711},
journal = {\mnras},
month = {jul},
number = {3},
pages = {1992--2015},
title = {{Galaxy properties from the ultraviolet to the far-infrared: $\Lambda$ cold dark matter models confront observations}},
url = {https://academic.oup.com/mnras/article-lookup/doi/10.1111/j.1365-2966.2012.20490.x},
volume = {423},
year = {2012}
}

@ARTICLE{Shuntov2025,
       author = {{Shuntov}, Marko and {Oesch}, Pascal A. and {Toft}, Sune and {Meyer}, Romain A. and {Covelo-Paz}, Alba and {Paquereau}, Louise and {Bouwens}, Rychard and {Brammer}, Gabriel and {Gelli}, Viola and {Giovinazzo}, Emma and {Herard-Demanche}, Thomas and {Illingworth}, Garth D. and {Mason}, Charlotte and {Naidu}, Rohan P. and {Weibel}, Andrea and {Xiao}, Mengyuan},
        title = "{Constraints on the early Universe star formation efficiency from galaxy clustering and halo modeling of H{\ensuremath{\alpha}} and [O III] emitters}",
      journal = {\aap},
         year = 2025,
        month = jul,
       volume = {699},
          eid = {A231},
        pages = {A231},
          doi = {10.1051/0004-6361/202554618},
archivePrefix = {arXiv},
       eprint = {2503.14280},
 primaryClass = {astro-ph.GA},
       adsurl = {https://ui.adsabs.harvard.edu/abs/2025A&A...699A.231S}
}

@article{Kokorev2025,
archivePrefix = {arXiv},
arxivId = {2504.12504},
author = {Kokorev, Vasily and Ortiz, {\'{O}}scar A. Ch{\'{a}}vez and Taylor, Anthony J. and Finkelstein, Steven L. and Haro, Pablo Arrabal and Dickinson, Mark and Chisholm, John and Fujimoto, Seiji and Mu{\~{n}}oz, Julian B. and Endsley, Ryan and Hu, Weida and Napolitano, Lorenzo and Wilkins, Stephen M. and Akins, Hollis B. and Amori{\'{i}}n, Ricardo and Casey, Caitlin M. and Cheng, Yingjie and Cleri, Nikko J. and Cole, Justin and Cullen, Fergus and Daddi, Emanuele and Davis, Kelcey and Donnan, Callum T. and Dunlop, James S. and Fern{\'{a}}ndez, Vital and Giavalisco, Mauro and Grogin, Norman A. and Hathi, Nimish and Hirschmann, Michaela and Kartaltepe, Jeyhan S. and Koekemoer, Anton M. and Leung, Ho-Hin and Lucas, Ray A. and McLeod, Derek and Papovich, Casey and Pentericci, Laura and P{\'{e}}rez-Gonz{\'{a}}lez, Pablo G. and Somerville, Rachel S. and Wang, Xin and Yung, L. Y. Aaron and Zavala, Jorge A.},
eprint = {2504.12504},
journal = {arXiv:2504.12504},
title = {{CAPERS Observations of Two UV-Bright Galaxies at z>10. More Evidence for Bursting Star Formation in the Early Universe}},
url = {http://arxiv.org/abs/2504.12504},
year = {2025}
}

@article{Morales2024,
archivePrefix = {arXiv},
arxivId = {2311.04294},
author = {Morales, Alexa M and Finkelstein, Steven L and Leung, Gene C K and Bagley, Micaela B and Cleri, Nikko J and Dave, Romeel and Dickinson, Mark and Ferguson, Henry C and Hathi, Nimish P and Jones, Ewan and Koekemoer, Anton M and Papovich, Casey and P{\'{e}}rez-Gonz{\'{a}}lez, Pablo G. and Pirzkal, Nor and Smith, Britton and Wilkins, Stephen M and Yung, L. Y. Aaron},
doi = {10.3847/2041-8213/ad2de4},
eprint = {2311.04294},
issn = {2041-8205},
journal = {\apjl},
month = {apr},
number = {2},
pages = {L24},
title = {{Rest-frame UV Colors for Faint Galaxies at z ∼ 9–16 with the JWST NGDEEP Survey}},
url = {http://arxiv.org/abs/2311.04294 https://iopscience.iop.org/article/10.3847/2041-8213/ad2de4},
volume = {964},
year = {2024}
}

@article{Adams2024,
archivePrefix = {arXiv},
arxivId = {2304.13721},
author = {Adams, Nathan J and Conselice, Christopher J and Austin, Duncan and Harvey, Thomas and Ferreira, Leonardo and Trussler, James and Juod{\v{z}}balis, Ignas and Li, Qiong and Windhorst, Rogier and Cohen, Seth H and Jansen, Rolf A. and Summers, Jake and Tompkins, Scott and Driver, Simon P and Robotham, Aaron and D'Silva, Jordan C. J. and Yan, Haojing and Coe, Dan and Frye, Brenda and Grogin, Norman A and Koekemoer, Anton M and Marshall, Madeline A and Pirzkal, Nor and Ryan, Russell E and Maksym, W Peter and Rutkowski, Michael J and Willmer, Christopher N A and Hammel, Heidi B and Nonino, Mario and Bhatawdekar, Rachana and Wilkins, Stephen M. and Bradley, Larry D. and Broadhurst, Tom and Cheng, Cheng and Dole, Herv{\'{e}} and Hathi, Nimish P. and Zitrin, Adi},
doi = {10.3847/1538-4357/ad2a7b},
eprint = {2304.13721},
issn = {0004-637X},
journal = {\apj},
month = {apr},
number = {2},
pages = {169},
title = {{EPOCHS. II. The Ultraviolet Luminosity Function from 7.5 < z < 13.5 Using 180 arcmin 2 of Deep, Blank Fields from the PEARLS Survey and Public JWST Data}},
url = {https://iopscience.iop.org/article/10.3847/1538-4357/ad2a7b},
volume = {965},
year = {2024}
}

@article{Trinca2024,
archivePrefix = {arXiv},
arxivId = {2305.04944},
author = {Trinca, Alessandro and Schneider, Raffaella and Valiante, Rosa and Graziani, Luca and Ferrotti, Arianna and Omukai, Kazuyuki and Chon, Sunmyon},
doi = {10.1093/mnras/stae651},
eprint = {2305.04944},
issn = {0035-8711},
journal = {\mnras},
month = {mar},
number = {4},
pages = {3563--3581},
title = {{Exploring the nature of UV-bright z ≳ 10 galaxies detected by JWST : star formation, black hole accretion, or a non-universal IMF?}},
url = {http://arxiv.org/abs/2305.04944 https://academic.oup.com/mnras/article/529/4/3563/7623041},
volume = {529},
year = {2024}
}

@ARTICLE{Whitler2025,
       author = {{Whitler}, Lily and {Stark}, Daniel P. and {Topping}, Michael W. and {Robertson}, Brant and {Rieke}, Marcia and {Hainline}, Kevin N. and {Endsley}, Ryan and {Chen}, Zuyi and {Baker}, William M. and {Bhatawdekar}, Rachana and {Bunker}, Andrew J. and {Carniani}, Stefano and {Charlot}, St{\'e}phane and {Chevallard}, Jacopo and {Curtis-Lake}, Emma and {Egami}, Eiichi and {Eisenstein}, Daniel J. and {Helton}, Jakob M. and {Ji}, Zhiyuan and {Johnson}, Benjamin D. and {P{\'e}rez-Gonz{\'a}lez}, Pablo G. and {Rinaldi}, Pierluigi and {Tacchella}, Sandro and {Williams}, Christina C. and {Willmer}, Christopher N.~A. and {Willott}, Chris and {Witstok}, Joris},
        title = "{The z {\ensuremath{\gtrsim}} 9 Galaxy UV Luminosity Function from the JWST Advanced Deep Extragalactic Survey: Insights into Early Galaxy Evolution and Reionization}",
      journal = {\apj},
         year = 2025,
        month = oct,
       volume = {992},
       number = {1},
          eid = {63},
        pages = {63},
          doi = {10.3847/1538-4357/adfddc},
archivePrefix = {arXiv},
       eprint = {2501.00984},
 primaryClass = {astro-ph.GA},
       adsurl = {https://ui.adsabs.harvard.edu/abs/2025ApJ...992...63W}
}

@article{Krumholz2007,
archivePrefix = {arXiv},
arxivId = {astro-ph/0606277},
author = {Krumholz, Mark R. and Tan, Jonathan C.},
doi = {10.1086/509101},
eprint = {0606277},
issn = {0004-637X},
journal = {\apj},
number = {1},
pages = {304--315},
primaryClass = {astro-ph},
title = {{Slow Star Formation in Dense Gas: Evidence and Implications}},
volume = {654},
year = {2007}
}

@article{Conroy2013,
archivePrefix = {arXiv},
arxivId = {1301.7095},
author = {Conroy, Charlie},
doi = {10.1146/annurev-astro-082812-141017},
eprint = {1301.7095},
issn = {0066-4146},
journal = {\araa},
number = {1},
pages = {393--455},
title = {{Modeling the Panchromatic Spectral Energy Distributions of Galaxies}},
volume = {51},
year = {2013}
}

@article{Chevance2023,
archivePrefix = {arXiv},
arxivId = {2203.09570},
author = {Chevance, M{\'{e}}lanie and Krumholz, Mark R. and McLeod, Anna F. and Ostriker, Eve C. and Rosolowsky, Erik W. and Sternberg, Amiel},
doi = {10.48550/arXiv.2203.09570},
eprint = {2203.09570},
journal = {Protostars Planets VII},
pages = {1},
title = {{The Life and Times of Giant Molecular Clouds}},
url = {http://arxiv.org/abs/2203.09570},
volume = {534},
year = {2023}
}

@article{Yung2024a,
archivePrefix = {arXiv},
arxivId = {2309.14408},
author = {Yung, L. Y. Aaron and Somerville, Rachel S. and Nguyen, Tri and Behroozi, Peter and Modi, Chirag and Gardner, Jonathan P.},
doi = {10.1093/mnras/stae1188},
eprint = {2309.14408},
issn = {0035-8711},
journal = {\mnras},
month = {may},
number = {4},
pages = {4868--4886},
title = {{Characterizing ultra-high-redshift dark matter halo demographics and assembly histories with the <scp>gureft</scp> simulations}},
url = {http://arxiv.org/abs/2309.14408 https://academic.oup.com/mnras/article/530/4/4868/7663579},
volume = {530},
year = {2024}
}

@article{Chon2024,
author = {Chon, Sunmyon and Hosokawa, Takashi and Omukai, Kazuyuki and Schneider, Raffaella},
doi = {10.1093/mnras/stae1027},
issn = {0035-8711},
journal = {\mnras},
month = {apr},
number = {3},
pages = {2453--2474},
title = {{Impact of radiative feedback on the initial mass function of metal-poor stars}},
url = {https://academic.oup.com/mnras/article/530/3/2453/7646872},
volume = {530},
year = {2024}
}

@article{Gelli2024,
archivePrefix = {arXiv},
arxivId = {2405.13108},
author = {Gelli, Viola and Mason, Charlotte and Hayward, Christopher C.},
doi = {10.3847/1538-4357/ad7b36},
eprint = {2405.13108},
issn = {0004-637X},
journal = {\apj},
month = {nov},
number = {2},
pages = {192},
title = {{The Impact of Mass-dependent Stochasticity at Cosmic Dawn}},
url = {http://arxiv.org/abs/2405.13108 https://iopscience.iop.org/article/10.3847/1538-4357/ad7b36},
volume = {975},
year = {2024}
}

@ARTICLE{Li2024,
       author = {{Li}, Zhaozhou and {Dekel}, Avishai and {Sarkar}, Kartick C. and {Aung}, Han and {Giavalisco}, Mauro and {Mandelker}, Nir and {Tacchella}, Sandro},
        title = "{Feedback-free starbursts at cosmic dawn: Observable predictions for JWST}",
      journal = {\aap},
         year = 2024,
        month = oct,
       volume = {690},
          eid = {A108},
        pages = {A108},
          doi = {10.1051/0004-6361/202348727},
archivePrefix = {arXiv},
       eprint = {2311.14662},
 primaryClass = {astro-ph.GA},
       adsurl = {https://ui.adsabs.harvard.edu/abs/2024A&A...690A.108L}
}

@article{Ceverino2024,
author = {Ceverino, D and Nakazato, Y. and Yoshida, N. and Klessen, R. S. and Glover, S. C. O.},
doi = {10.1051/0004-6361/202450224},
issn = {0004-6361},
journal = {\aap},
month = {sep},
pages = {A244},
title = {{Redshift-dependent galaxy formation efficiency at z = 5 − 13 in the FirstLight Simulations}},
url = {https://www.aanda.org/10.1051/0004-6361/202450224},
volume = {689},
year = {2024}
}

@ARTICLE{Ceverino:2017,
       author = {{Ceverino}, Daniel and {Glover}, Simon C.~O. and {Klessen}, Ralf S.},
        title = "{Introducing the FirstLight project: UV luminosity function and scaling relations of primeval galaxies}",
      journal = {\mnras},
         year = 2017,
        month = sep,
       volume = {470},
       number = {3},
        pages = {2791-2798},
          doi = {10.1093/mnras/stx1386},
archivePrefix = {arXiv},
       eprint = {1703.02913},
 primaryClass = {astro-ph.GA},
       adsurl = {https://ui.adsabs.harvard.edu/abs/2017MNRAS.470.2791C}
}

@ARTICLE{Ceverino2019,
       author = {{Ceverino}, Daniel and {Klessen}, Ralf S. and {Glover}, Simon C.~O.},
        title = "{FirstLight III: rest-frame UV-optical spectral energy distributions of simulated galaxies at cosmic dawn}",
      journal = {\mnras},
         year = 2019,
        month = mar,
       volume = {484},
       number = {1},
        pages = {1366-1377},
          doi = {10.1093/mnras/stz079},
archivePrefix = {arXiv},
       eprint = {1810.09754},
 primaryClass = {astro-ph.GA},
       adsurl = {https://ui.adsabs.harvard.edu/abs/2019MNRAS.484.1366C}
}

@article{Hopkins2012,
archivePrefix = {arXiv},
arxivId = {1110.4638},
author = {Hopkins, Philip F. and Quataert, Eliot and Murray, Norman},
doi = {10.1111/j.1365-2966.2012.20593.x},
eprint = {1110.4638},
issn = {00358711},
journal = {\mnras},
number = {4},
pages = {3522--3537},
title = {{Stellar feedback in galaxies and the origin of galaxy-scale winds}},
volume = {421},
year = {2012}
}

@article{Mauerhofer2025,
archivePrefix = {arXiv},
arxivId = {2502.02647},
author = {Mauerhofer, Valentin and Dayal, Pratika and Haehnelt, Martin G. and Kimm, Taysun and Rosdahl, Joakim and Teyssier, Romain},
doi = {10.1051/0004-6361/202554042},
eprint = {2502.02647},
issn = {0004-6361},
journal = {\aap},
month = {apr},
pages = {A157},
title = {{Synergising semi-analytical models and hydrodynamical simulations to interpret JWST data from the first billion years}},
url = {http://arxiv.org/abs/2502.02647 https://www.aanda.org/10.1051/0004-6361/202554042},
volume = {696},
year = {2025}
}

@article{Lancaster2021,
archivePrefix = {arXiv},
arxivId = {2104.07722},
author = {Lancaster, Lachlan and Ostriker, Eve C. and Kim, Jeong-Gyu and Kim, Chang-Goo},
doi = {10.3847/1538-4357/abf8ac},
eprint = {2104.07722},
issn = {0004-637X},
journal = {\apj},
number = {2},
pages = {90},
publisher = {IOP Publishing},
title = {{Efficiently Cooled Stellar Wind Bubbles in Turbulent Clouds. II. Validation of Theory with Hydrodynamic Simulations}},
url = {http://dx.doi.org/10.3847/1538-4357/abf8ac},
volume = {914},
year = {2021}
}

@ARTICLE{Ferrara2025,
       author = {{Ferrara}, A. and {Pallottini}, A. and {Sommovigo}, L.},
        title = "{Blue monsters at z > 10: Where all their dust has gone}",
      journal = {\aap},
         year = 2025,
        month = feb,
       volume = {694},
          eid = {A286},
        pages = {A286},
          doi = {10.1051/0004-6361/202452707},
archivePrefix = {arXiv},
       eprint = {2410.19042},
 primaryClass = {astro-ph.GA},
       adsurl = {https://ui.adsabs.harvard.edu/abs/2025A&A...694A.286F}
}

@ARTICLE{Kravtsov2024,
       author = {{Kravtsov}, Andrey and {Belokurov}, Vasily},
        title = "{Stochastic star formation and the abundance of $z>10$ UV-bright galaxies}",
      journal = {arXiv e-prints},
         year = 2024,
        month = may,
          eid = {arXiv:2405.04578},
        pages = {arXiv:2405.04578},
          doi = {10.48550/arXiv.2405.04578},
archivePrefix = {arXiv},
       eprint = {2405.04578},
 primaryClass = {astro-ph.GA},
       adsurl = {https://ui.adsabs.harvard.edu/abs/2024arXiv240504578K}
}

@ARTICLE{RobertsBorsani2024,
       author = {{Roberts-Borsani}, Guido and {Treu}, Tommaso and {Shapley}, Alice and {Fontana}, Adriano and {Pentericci}, Laura and {Castellano}, Marco and {Morishita}, Takahiro and {Bergamini}, Pietro and {Rosati}, Piero},
        title = "{Between the Extremes: A JWST Spectroscopic Benchmark for High-redshift Galaxies Using {\ensuremath{\sim}}500 Confirmed Sources at z {\ensuremath{\geq}} 5}",
      journal = {\apj},
         year = 2024,
        month = dec,
       volume = {976},
       number = {2},
          eid = {193},
        pages = {193},
          doi = {10.3847/1538-4357/ad85d3},
archivePrefix = {arXiv},
       eprint = {2403.07103},
 primaryClass = {astro-ph.GA},
       adsurl = {https://ui.adsabs.harvard.edu/abs/2024ApJ...976..193R}
}

@ARTICLE{Mowla2024,
       author = {{Mowla}, Lamiya and {Iyer}, Kartheik and {Asada}, Yoshihisa and {Desprez}, Guillaume and {Tan}, Vivian Yun Yan and {Martis}, Nicholas and {Sarrouh}, Ghassan and {Strait}, Victoria and {Abraham}, Roberto and {Brada{\v{c}}}, Maru{\v{s}}a and {Brammer}, Gabriel and {Muzzin}, Adam and {Pacifici}, Camilla and {Ravindranath}, Swara and {Sawicki}, Marcin and {Willott}, Chris and {Estrada-Carpenter}, Vince and {Jahan}, Nusrath and {Noirot}, Ga{\"e}l and {Matharu}, Jasleen and {Rihtar{\v{s}}i{\v{c}}}, Gregor and {Zabl}, Johannes},
        title = "{Formation of a low-mass galaxy from star clusters in a 600-million-year-old Universe}",
      journal = {\nat},
         year = 2024,
        month = dec,
       volume = {636},
       number = {8042},
        pages = {332-336},
          doi = {10.1038/s41586-024-08293-0},
archivePrefix = {arXiv},
       eprint = {2402.08696},
 primaryClass = {astro-ph.GA},
       adsurl = {https://ui.adsabs.harvard.edu/abs/2024Natur.636..332M}
}

@ARTICLE{Cameron2023,
       author = {{Cameron}, Alex J. and {Katz}, Harley and {Rey}, Martin P. and {Saxena}, Aayush},
        title = "{Nitrogen enhancements 440 Myr after the big bang: supersolar N/O, a tidal disruption event, or a dense stellar cluster in GN-z11?}",
      journal = {\mnras},
         year = 2023,
        month = aug,
       volume = {523},
       number = {3},
        pages = {3516-3525},
          doi = {10.1093/mnras/stad1579},
archivePrefix = {arXiv},
       eprint = {2302.10142},
 primaryClass = {astro-ph.GA},
       adsurl = {https://ui.adsabs.harvard.edu/abs/2023MNRAS.523.3516C}
}

@ARTICLE{Ferrara2024,
       author = {{Ferrara}, A.},
        title = "{Super-early JWST galaxies, outflows, and Ly{\ensuremath{\alpha}} visibility in the Epoch of Reionization}",
      journal = {\aap},
         year = 2024,
        month = apr,
       volume = {684},
          eid = {A207},
        pages = {A207},
          doi = {10.1051/0004-6361/202348321},
archivePrefix = {arXiv},
       eprint = {2310.12197},
 primaryClass = {astro-ph.GA},
       adsurl = {https://ui.adsabs.harvard.edu/abs/2024A&A...684A.207F}
}

@ARTICLE{Isobe2023b,
       author = {{Isobe}, Yuki and {Ouchi}, Masami and {Tominaga}, Nozomu and {Watanabe}, Kuria and {Nakajima}, Kimihiko and {Umeda}, Hiroya and {Yajima}, Hidenobu and {Harikane}, Yuichi and {Fukushima}, Hajime and {Xu}, Yi and {Ono}, Yoshiaki and {Zhang}, Yechi},
        title = "{JWST Identification of Extremely Low C/N Galaxies with [N/O] {\ensuremath{\gtrsim}} 0.5 at z 6-10 Evidencing the Early CNO-cycle Enrichment and a Connection with Globular Cluster Formation}",
      journal = {\apj},
         year = 2023,
        month = dec,
       volume = {959},
       number = {2},
          eid = {100},
        pages = {100},
          doi = {10.3847/1538-4357/ad09be},
archivePrefix = {arXiv},
       eprint = {2307.00710},
 primaryClass = {astro-ph.GA},
       adsurl = {https://ui.adsabs.harvard.edu/abs/2023ApJ...959..100I}
}

@ARTICLE{Ziparo2022,
       author = {{Ziparo}, Francesco and {Ferrara}, Andrea and {Sommovigo}, Laura and {Kohandel}, Mahsa},
        title = "{Blue monsters. Why are JWST super-early, massive galaxies so blue?}",
      journal = {\mnras},
         year = 2023,
        month = apr,
       volume = {520},
       number = {2},
        pages = {2445-2450},
          doi = {10.1093/mnras/stad125},
archivePrefix = {arXiv},
       eprint = {2209.06840},
 primaryClass = {astro-ph.GA},
       adsurl = {https://ui.adsabs.harvard.edu/abs/2023MNRAS.520.2445Z}
}

@ARTICLE{Curti2023,
       author = {{Curti}, Mirko and {Maiolino}, Roberto and {Curtis-Lake}, Emma and {Chevallard}, Jacopo and {Carniani}, Stefano and {D'Eugenio}, Francesco and {Looser}, Tobias J. and {Scholtz}, Jan and {Charlot}, Stephane and {Cameron}, Alex and {{\"U}bler}, Hannah and {Witstok}, Joris and {Boyett}, Kristian and {Laseter}, Isaac and {Sandles}, Lester and {Arribas}, Santiago and {Bunker}, Andrew and {Giardino}, Giovanna and {Maseda}, Michael V. and {Rawle}, Tim and {Rodr{\'\i}guez Del Pino}, Bruno and {Smit}, Renske and {Willott}, Chris J. and {Eisenstein}, Daniel J. and {Hausen}, Ryan and {Johnson}, Benjamin and {Rieke}, Marcia and {Robertson}, Brant and {Tacchella}, Sandro and {Williams}, Christina C. and {Willmer}, Christopher and {Baker}, William M. and {Bhatawdekar}, Rachana and {Egami}, Eiichi and {Helton}, Jakob M. and {Ji}, Zhiyuan and {Kumari}, Nimisha and {Perna}, Michele and {Shivaei}, Irene and {Sun}, Fengwu},
        title = "{JADES: Insights on the low-mass end of the mass--metallicity--star-formation rate relation at $3 < z < 10$ from deep JWST/NIRSpec spectroscopy}",
      journal = {arXiv e-prints},
         year = 2023,
        month = apr,
          eid = {arXiv:2304.08516},
        pages = {arXiv:2304.08516},
          doi = {10.48550/arXiv.2304.08516},
archivePrefix = {arXiv},
       eprint = {2304.08516},
 primaryClass = {astro-ph.GA},
       adsurl = {https://ui.adsabs.harvard.edu/abs/2023arXiv230408516C}
}

@ARTICLE{Grudic2018,
       author = {{Grudi{\'c}}, Michael Y. and {Hopkins}, Philip F. and {Faucher-Gigu{\`e}re}, Claude-Andr{\'e} and {Quataert}, Eliot and {Murray}, Norman and {Kere{\v{s}}}, Du{\v{s}}an},
        title = "{When feedback fails: the scaling and saturation of star formation efficiency}",
      journal = {\mnras},
         year = 2018,
        month = apr,
       volume = {475},
       number = {3},
        pages = {3511-3528},
          doi = {10.1093/mnras/sty035},
archivePrefix = {arXiv},
       eprint = {1612.05635},
 primaryClass = {astro-ph.GA},
       adsurl = {https://ui.adsabs.harvard.edu/abs/2018MNRAS.475.3511G}
}

@ARTICLE{Kimcg2017,
       author = {{Kim}, Chang-Goo and {Ostriker}, Eve C.},
        title = "{Three-phase Interstellar Medium in Galaxies Resolving Evolution with Star Formation and Supernova Feedback (TIGRESS): Algorithms, Fiducial Model, and Convergence}",
      journal = {\apj},
         year = 2017,
        month = sep,
       volume = {846},
       number = {2},
          eid = {133},
        pages = {133},
          doi = {10.3847/1538-4357/aa8599},
archivePrefix = {arXiv},
       eprint = {1612.03918},
 primaryClass = {astro-ph.GA},
       adsurl = {https://ui.adsabs.harvard.edu/abs/2017ApJ...846..133K}
}

@ARTICLE{Kimjg2018,
       author = {{Kim}, Jeong-Gyu and {Kim}, Woong-Tae and {Ostriker}, Eve C.},
        title = "{Modeling UV Radiation Feedback from Massive Stars. II. Dispersal of Star-forming Giant Molecular Clouds by Photoionization and Radiation Pressure}",
      journal = {\apj},
         year = 2018,
        month = may,
       volume = {859},
       number = {1},
          eid = {68},
        pages = {68},
          doi = {10.3847/1538-4357/aabe27},
archivePrefix = {arXiv},
       eprint = {1804.04664},
 primaryClass = {astro-ph.GA},
       adsurl = {https://ui.adsabs.harvard.edu/abs/2018ApJ...859...68K}
}

@ARTICLE{Kim2015,
       author = {{Kim}, Chang-Goo and {Ostriker}, Eve C.},
        title = "{Momentum Injection by Supernovae in the Interstellar Medium}",
      journal = {\apj},
         year = 2015,
        month = apr,
       volume = {802},
       number = {2},
          eid = {99},
        pages = {99},
          doi = {10.1088/0004-637X/802/2/99},
archivePrefix = {arXiv},
       eprint = {1410.1537},
 primaryClass = {astro-ph.GA},
       adsurl = {https://ui.adsabs.harvard.edu/abs/2015ApJ...802...99K}
}

@ARTICLE{Inayoshi2022,
       author = {{Inayoshi}, Kohei and {Harikane}, Yuichi and {Inoue}, Akio K. and {Li}, Wenxiu and {Ho}, Luis C.},
        title = "{A Lower Bound of Star Formation Activity in Ultra-high-redshift Galaxies Detected with JWST: Implications for Stellar Populations and Radiation Sources}",
      journal = {\apjl},
         year = 2022,
        month = oct,
       volume = {938},
       number = {2},
          eid = {L10},
        pages = {L10},
          doi = {10.3847/2041-8213/ac9310},
archivePrefix = {arXiv},
       eprint = {2208.06872},
 primaryClass = {astro-ph.GA},
       adsurl = {https://ui.adsabs.harvard.edu/abs/2022ApJ...938L..10I}
}

@ARTICLE{Chon2022,
       author = {{Chon}, Sunmyon and {Ono}, Haruka and {Omukai}, Kazuyuki and {Schneider}, Raffaella},
        title = "{Impact of the cosmic background radiation on the initial mass function of metal-poor stars}",
      journal = {\mnras},
         year = 2022,
        month = aug,
       volume = {514},
       number = {3},
        pages = {4639-4654},
          doi = {10.1093/mnras/stac1549},
archivePrefix = {arXiv},
       eprint = {2205.15328},
 primaryClass = {astro-ph.GA},
       adsurl = {https://ui.adsabs.harvard.edu/abs/2022MNRAS.514.4639C}
}

@ARTICLE{Klypin2011,
       author = {{Klypin}, Anatoly A. and {Trujillo-Gomez}, Sebastian and {Primack}, Joel},
        title = "{Dark Matter Halos in the Standard Cosmological Model: Results from the Bolshoi Simulation}",
      journal = {\apj},
         year = 2011,
        month = oct,
       volume = {740},
       number = {2},
          eid = {102},
        pages = {102},
          doi = {10.1088/0004-637X/740/2/102},
archivePrefix = {arXiv},
       eprint = {1002.3660},
 primaryClass = {astro-ph.CO},
       adsurl = {https://ui.adsabs.harvard.edu/abs/2011ApJ...740..102K}
}

@ARTICLE{Colazo2024,
       author = {{Colazo}, P.~E. and {Stasyszyn}, F. and {Padilla}, N.},
        title = "{Structure formation with primordial black holes to alleviate early star formation tension revealed by JWST}",
      journal = {\aap},
         year = 2024,
        month = may,
       volume = {685},
          eid = {L8},
        pages = {L8},
          doi = {10.1051/0004-6361/202449565},
archivePrefix = {arXiv},
       eprint = {2404.13110},
 primaryClass = {astro-ph.CO},
       adsurl = {https://ui.adsabs.harvard.edu/abs/2024A&A...685L...8C}
}

@ARTICLE{Liu2022,
       author = {{Liu}, Boyuan and {Bromm}, Volker},
        title = "{Accelerating Early Massive Galaxy Formation with Primordial Black Holes}",
      journal = {\apjl},
         year = 2022,
        month = oct,
       volume = {937},
       number = {2},
          eid = {L30},
        pages = {L30},
          doi = {10.3847/2041-8213/ac927f},
archivePrefix = {arXiv},
       eprint = {2208.13178},
 primaryClass = {astro-ph.CO},
       adsurl = {https://ui.adsabs.harvard.edu/abs/2022ApJ...937L..30L}
}

@ARTICLE{Hirano2024,
       author = {{Hirano}, Shingo and {Yoshida}, Naoki},
        title = "{Early Structure Formation from Primordial Density Fluctuations with a Blue, Tilted Power Spectrum: High-redshift Galaxies}",
      journal = {\apj},
         year = 2024,
        month = mar,
       volume = {963},
       number = {1},
          eid = {2},
        pages = {2},
          doi = {10.3847/1538-4357/ad22e0},
archivePrefix = {arXiv},
       eprint = {2306.11993},
 primaryClass = {astro-ph.GA},
       adsurl = {https://ui.adsabs.harvard.edu/abs/2024ApJ...963....2H}
}

@article{Koehler2024,
archivePrefix = {arXiv},
arxivId = {2412.00182},
author = {Koehler, Sonja M. and Jiao, Hao and Kannan, Rahul},
eprint = {2412.00182},
journal = {arXiv:2412.00182},
title = {{Investigating cosmic strings using large-volume hydrodynamical simulations in the context of JWST's massive UV-bright galaxies}},
url = {http://arxiv.org/abs/2412.00182},
year = {2024}
}

@ARTICLE{Narayanan2025,
       author = {{Narayanan}, Desika and {Stark}, Daniel P. and {Finkelstein}, Steven L. and {Torrey}, Paul and {Li}, Qi and {Cullen}, Fergus and {Topping}, Micheal W. and {Marinacci}, Federico and {Sales}, Laura V. and {Shen}, Xuejian and {Vogelsberger}, Mark},
        title = "{The Ultraviolet Slopes of Early Universe Galaxies: The Impact of Bursty Star Formation, Dust, and Nebular Continuum Emission}",
      journal = {\apj},
         year = 2025,
        month = mar,
       volume = {982},
       number = {1},
          eid = {7},
        pages = {7},
          doi = {10.3847/1538-4357/adb41c},
archivePrefix = {arXiv},
       eprint = {2408.13312},
 primaryClass = {astro-ph.GA},
       adsurl = {https://ui.adsabs.harvard.edu/abs/2025ApJ...982....7N}
}

@ARTICLE{Narayanan:2025b,
       author = {{Narayanan}, Desika and {Torrey}, Paul and {Stark}, Daniel and {Chisholm}, John and {Finkelstein}, Steven and {Garcia}, Alex and {Kelley-Derzon}, Jessica and {Marinacci}, Federico and {Sales}, Laura and {Savitch}, Ethan and {Vogelsberger}, Mark and {Zimmerman}, Dhruv},
        title = "{The Growth of Dust in Galaxies in the First Billion Years with Applications to Blue Monsters}",
      journal = {arXiv e-prints},
         year = 2025,
        month = sep,
          eid = {arXiv:2509.18266},
        pages = {arXiv:2509.18266},
          doi = {10.48550/arXiv.2509.18266},
archivePrefix = {arXiv},
       eprint = {2509.18266},
 primaryClass = {astro-ph.GA},
       adsurl = {https://ui.adsabs.harvard.edu/abs/2025arXiv250918266N}
}

@ARTICLE{Fujimotograpes2024,
       author = {{Fujimoto}, S. and {Ouchi}, M. and {Kohno}, K. and {Valentino}, F. and {Gim{\'e}nez-Arteaga}, C. and {Brammer}, G.~B. and {Furtak}, L.~J. and {Kohandel}, M. and {Oguri}, M. and {Pallottini}, A. and {Richard}, J. and {Zitrin}, A. and {Bauer}, F.~E. and {Boylan-Kolchin}, M. and {Dessauges-Zavadsky}, M. and {Egami}, E. and {Finkelstein}, S.~L. and {Ma}, Z. and {Smail}, I. and {Watson}, D. and {Hutchison}, T.~A. and {Rigby}, J.~R. and {Welch}, B.~D. and {Ao}, Y. and {Bradley}, L.~D. and {Caminha}, G.~B. and {Caputi}, K.~I. and {Espada}, D. and {Endsley}, R. and {Fudamoto}, Y. and {Gonz{\'a}lez-L{\'o}pez}, J. and {Hatsukade}, B. and {Koekemoer}, A.~M. and {Kokorev}, V. and {Laporte}, N. and {Lee}, M. and {Magdis}, G.~E. and {Ono}, Y. and {Rizzo}, F. and {Shibuya}, T. and {Shimasaku}, K. and {Sun}, F. and {Toft}, S. and {Umehata}, H. and {Wang}, T. and {Yajima}, H.},
        title = "{Primordial rotating disk composed of at least 15 dense star-forming clumps at cosmic dawn}",
      journal = {Nature Astronomy},
         year = 2025,
        month = aug,
       volume = {9},
        pages = {1553-1567},
          doi = {10.1038/s41550-025-02592-w},
archivePrefix = {arXiv},
       eprint = {2402.18543},
 primaryClass = {astro-ph.GA},
       adsurl = {https://ui.adsabs.harvard.edu/abs/2025NatAs...9.1553F}
}

@ARTICLE{Somervillesizev2008,
       author = {{Somerville}, Rachel S. and {Barden}, Marco and {Rix}, Hans-Walter and {Bell}, Eric F. and {Beckwith}, Steven V.~W. and {Borch}, Andrea and {Caldwell}, John A.~R. and {H{\"a}u{\ss}ler}, Boris and {Heymans}, Catherine and {Jahnke}, Knud and {Jogee}, Shardha and {McIntosh}, Daniel H. and {Meisenheimer}, Klaus and {Peng}, Chien Y. and {S{\'a}nchez}, Sebastian F. and {Wisotzki}, Lutz and {Wolf}, Christian},
        title = "{An Explanation for the Observed Weak Size Evolution of Disk Galaxies}",
      journal = {\apj},
         year = 2008,
        month = jan,
       volume = {672},
       number = {2},
        pages = {776-786},
          doi = {10.1086/523661},
archivePrefix = {arXiv},
       eprint = {astro-ph/0612428},
 primaryClass = {astro-ph},
       adsurl = {https://ui.adsabs.harvard.edu/abs/2008ApJ...672..776S}
}

@ARTICLE{Iyer2020,
       author = {{Iyer}, Kartheik G. and {Tacchella}, Sandro and {Genel}, Shy and {Hayward}, Christopher C. and {Hernquist}, Lars and {Brooks}, Alyson M. and {Caplar}, Neven and {Dav{\'e}}, Romeel and {Diemer}, Benedikt and {Forbes}, John C. and {Gawiser}, Eric and {Somerville}, Rachel S. and {Starkenburg}, Tjitske K.},
        title = "{The diversity and variability of star formation histories in models of galaxy evolution}",
      journal = {\mnras},
         year = 2020,
        month = oct,
       volume = {498},
       number = {1},
        pages = {430-463},
          doi = {10.1093/mnras/staa2150},
archivePrefix = {arXiv},
       eprint = {2007.07916},
 primaryClass = {astro-ph.GA},
       adsurl = {https://ui.adsabs.harvard.edu/abs/2020MNRAS.498..430I}
}

@ARTICLE{Yung2025,
       author = {{Yung}, L.~Y. Aaron and {Somerville}, Rachel S. and {Iyer}, Kartheik G.},
        title = "{{\ensuremath{\Lambda}}CDM is still not broken: empirical constraints on the star formation efficiency at z {\ensuremath{\sim}}12─30}",
      journal = {\mnras},
         year = 2025,
        month = nov,
       volume = {543},
       number = {4},
        pages = {3802-3813},
          doi = {10.1093/mnras/staf1699},
archivePrefix = {arXiv},
       eprint = {2504.18618},
 primaryClass = {astro-ph.GA},
       adsurl = {https://ui.adsabs.harvard.edu/abs/2025MNRAS.543.3802Y}
}

@ARTICLE{Pallottini2022,
       author = {{Pallottini}, A. and {Ferrara}, A. and {Gallerani}, S. and {Behrens}, C. and {Kohandel}, M. and {Carniani}, S. and {Vallini}, L. and {Salvadori}, S. and {Gelli}, V. and {Sommovigo}, L. and {D'Odorico}, V. and {Di Mascia}, F. and {Pizzati}, E.},
        title = "{A survey of high-z galaxies: SERRA simulations}",
      journal = {\mnras},
         year = 2022,
        month = jul,
       volume = {513},
       number = {4},
        pages = {5621-5641},
          doi = {10.1093/mnras/stac1281},
archivePrefix = {arXiv},
       eprint = {2201.02636},
 primaryClass = {astro-ph.GA},
       adsurl = {https://ui.adsabs.harvard.edu/abs/2022MNRAS.513.5621P}
}

@ARTICLE{Draine2001,
       author = {{Draine}, B.~T. and {Li}, Aigen},
        title = "{Infrared Emission from Interstellar Dust. I. Stochastic Heating of Small Grains}",
      journal = {\apj},
         year = 2001,
        month = apr,
       volume = {551},
       number = {2},
        pages = {807-824},
          doi = {10.1086/320227},
archivePrefix = {arXiv},
       eprint = {astro-ph/0011318},
 primaryClass = {astro-ph},
       adsurl = {https://ui.adsabs.harvard.edu/abs/2001ApJ...551..807D}
}

@ARTICLE{Sommovigo2025,
       author = {{Sommovigo}, L. and {Cochrane}, R.~K. and {Somerville}, R.~S. and {Hayward}, C.~C. and {Lovell}, C.~C. and {Starkenburg}, T. and {Popping}, G. and {Iyer}, K. and {Gabrielpillai}, A. and {Ho}, M. and {Steinwandel}, U.~P. and {Perez}, L.~A.},
        title = "{Learning the Universe: Physically Motivated Priors for Dust Attenuation Curves}",
      journal = {\apj},
         year = 2025,
        month = sep,
       volume = {990},
       number = {2},
          eid = {114},
        pages = {114},
          doi = {10.3847/1538-4357/addec1},
archivePrefix = {arXiv},
       eprint = {2502.13240},
 primaryClass = {astro-ph.GA},
       adsurl = {https://ui.adsabs.harvard.edu/abs/2025ApJ...990..114S}
}

@ARTICLE{Belokurov2023,
       author = {{Belokurov}, Vasily and {Kravtsov}, Andrey},
        title = "{Nitrogen enrichment and clustered star formation at the dawn of the Galaxy}",
      journal = {\mnras},
         year = 2023,
        month = nov,
       volume = {525},
       number = {3},
        pages = {4456-4473},
          doi = {10.1093/mnras/stad2241},
archivePrefix = {arXiv},
       eprint = {2306.00060},
 primaryClass = {astro-ph.GA},
       adsurl = {https://ui.adsabs.harvard.edu/abs/2023MNRAS.525.4456B}
}

@ARTICLE{Inayoshi2020,
       author = {{Inayoshi}, Kohei and {Visbal}, Eli and {Haiman}, Zolt{\'a}n},
        title = "{The Assembly of the First Massive Black Holes}",
      journal = {\araa},
         year = 2020,
        month = aug,
       volume = {58},
        pages = {27-97},
          doi = {10.1146/annurev-astro-120419-014455},
archivePrefix = {arXiv},
       eprint = {1911.05791},
 primaryClass = {astro-ph.GA},
       adsurl = {https://ui.adsabs.harvard.edu/abs/2020ARA&A..58...27I}
}

@ARTICLE{Rantala2024,
       author = {{Rantala}, Antti and {Naab}, Thorsten and {Lah{\'e}n}, Natalia},
        title = "{FROST-CLUSTERS - I. Hierarchical star cluster assembly boosts intermediate-mass black hole formation}",
      journal = {\mnras},
         year = 2024,
        month = jul,
       volume = {531},
       number = {3},
        pages = {3770-3799},
          doi = {10.1093/mnras/stae1413},
archivePrefix = {arXiv},
       eprint = {2403.10602},
 primaryClass = {astro-ph.GA},
       adsurl = {https://ui.adsabs.harvard.edu/abs/2024MNRAS.531.3770R}
}

@ARTICLE{Dekel2025,
       author = {{Dekel}, Avishai and {Stone}, Nicholas C. and {Chowdhury}, Dhruba Dutta and {Gilbaum}, Shmuel and {Li}, Zhaozhou and {Mandelker}, Nir and {van den Bosch}, Frank C.},
        title = "{Growth of massive black holes in FFB galaxies at cosmic dawn}",
      journal = {\aap},
         year = 2025,
        month = mar,
       volume = {695},
          eid = {A97},
        pages = {A97},
          doi = {10.1051/0004-6361/202452393},
archivePrefix = {arXiv},
       eprint = {2409.18605},
 primaryClass = {astro-ph.GA},
       adsurl = {https://ui.adsabs.harvard.edu/abs/2025A&A...695A..97D}
}

@article{Iyer2025,
archivePrefix = {arXiv},
arxivId = {2502.17680},
author = {Iyer, Kartheik G. and Pacifici, Camilla and Calistro-Rivera, Gabriela and Lovell, Christopher C.},
eprint = {2502.17680},
journal = {arXiv:2502.17680},
title = {{The Spectral Energy Distributions of Galaxies}},
url = {http://arxiv.org/abs/2502.17680},
year = {2025}
}

@ARTICLE{Grudic2020,
       author = {{Grudi{\'c}}, Michael Y. and {Boylan-Kolchin}, Michael and {Faucher-Gigu{\`e}re}, Claude-Andr{\'e} and {Hopkins}, Philip F.},
        title = "{The universal acceleration scale from stellar feedback}",
      journal = {\mnras},
         year = 2020,
        month = jul,
       volume = {496},
       number = {1},
        pages = {L127-L132},
          doi = {10.1093/mnrasl/slaa103},
archivePrefix = {arXiv},
       eprint = {1910.06345},
 primaryClass = {astro-ph.GA},
       adsurl = {https://ui.adsabs.harvard.edu/abs/2020MNRAS.496L.127G}
}

@ARTICLE{Hennebelle2024,
       author = {{Hennebelle}, P. and {Grudi{\'c}}, M.~Y.},
        title = "{The Physical Origin of the Stellar Initial Mass Function}",
      journal = {\araa},
         year = 2024,
        month = sep,
       volume = {62},
       number = {1},
        pages = {63-111},
          doi = {10.1146/annurev-astro-052622-031748},
archivePrefix = {arXiv},
       eprint = {2404.07301},
 primaryClass = {astro-ph.GA},
       adsurl = {https://ui.adsabs.harvard.edu/abs/2024ARA&A..62...63H}
}

@ARTICLE{Pandya2023,
       author = {{Pandya}, Viraj and {Fielding}, Drummond B. and {Bryan}, Greg L. and {Carr}, Christopher and {Somerville}, Rachel S. and {Stern}, Jonathan and {Faucher-Gigu{\`e}re}, Claude-Andr{\'e} and {Hafen}, Zachary and {Angl{\'e}s-Alc{\'a}zar}, Daniel and {Forbes}, John C.},
        title = "{A Unified Model for the Coevolution of Galaxies and Their Circumgalactic Medium: The Relative Roles of Turbulence and Atomic Cooling Physics}",
      journal = {\apj},
         year = 2023,
        month = oct,
       volume = {956},
       number = {2},
          eid = {118},
        pages = {118},
          doi = {10.3847/1538-4357/acf3ea},
archivePrefix = {arXiv},
       eprint = {2211.09755},
 primaryClass = {astro-ph.GA},
       adsurl = {https://ui.adsabs.harvard.edu/abs/2023ApJ...956..118P}
}

@ARTICLE{Voit2024,
       author = {{Voit}, G. Mark and {Pandya}, Viraj and {Fielding}, Drummond B. and {Bryan}, Greg L. and {Carr}, Christopher and {Donahue}, Megan and {Oppenheimer}, Benjamin D. and {Somerville}, Rachel S.},
        title = "{Equilibrium States of Galactic Atmospheres. I. The Flip Side of Mass Loading}",
      journal = {\apj},
         year = 2024,
        month = dec,
       volume = {976},
       number = {2},
          eid = {150},
        pages = {150},
          doi = {10.3847/1538-4357/ad81d6},
archivePrefix = {arXiv},
       eprint = {2406.07631},
 primaryClass = {astro-ph.GA},
       adsurl = {https://ui.adsabs.harvard.edu/abs/2024ApJ...976..150V}
}

@ARTICLE{Jones2024,
       author = {{Jones}, E. and {Smith}, B. and {Dav{\'e}}, R. and {Narayanan}, D. and {Li}, Q.},
        title = "{SIMBA-EOR: early galaxy formation in the SIMBA simulation including a new sub-grid interstellar medium model}",
      journal = {\mnras},
         year = 2024,
        month = dec,
       volume = {535},
       number = {2},
        pages = {1293-1314},
          doi = {10.1093/mnras/stae2445},
archivePrefix = {arXiv},
       eprint = {2402.06728},
 primaryClass = {astro-ph.GA},
       adsurl = {https://ui.adsabs.harvard.edu/abs/2024MNRAS.535.1293J}
}

@ARTICLE{Dave2019,
       author = {{Dav{\'e}}, Romeel and {Angl{\'e}s-Alc{\'a}zar}, Daniel and {Narayanan}, Desika and {Li}, Qi and {Rafieferantsoa}, Mika H. and {Appleby}, Sarah},
        title = "{SIMBA: Cosmological simulations with black hole growth and feedback}",
      journal = {\mnras},
         year = 2019,
        month = jun,
       volume = {486},
       number = {2},
        pages = {2827-2849},
          doi = {10.1093/mnras/stz937},
archivePrefix = {arXiv},
       eprint = {1901.10203},
 primaryClass = {astro-ph.GA},
       adsurl = {https://ui.adsabs.harvard.edu/abs/2019MNRAS.486.2827D}
}

@ARTICLE{Marinacci2019,
       author = {{Marinacci}, Federico and {Sales}, Laura V. and {Vogelsberger}, Mark and {Torrey}, Paul and {Springel}, Volker},
        title = "{Simulating the interstellar medium and stellar feedback on a moving mesh: implementation and isolated galaxies}",
      journal = {\mnras},
         year = 2019,
        month = nov,
       volume = {489},
       number = {3},
        pages = {4233-4260},
          doi = {10.1093/mnras/stz2391},
archivePrefix = {arXiv},
       eprint = {1905.08806},
 primaryClass = {astro-ph.GA},
       adsurl = {https://ui.adsabs.harvard.edu/abs/2019MNRAS.489.4233M}
}

@ARTICLE{Hopkins2014,
       author = {{Hopkins}, Philip F. and {Kere{\v{s}}}, Du{\v{s}}an and {O{\~n}orbe}, Jos{\'e} and {Faucher-Gigu{\`e}re}, Claude-Andr{\'e} and {Quataert}, Eliot and {Murray}, Norman and {Bullock}, James S.},
        title = "{Galaxies on FIRE (Feedback In Realistic Environments): stellar feedback explains cosmologically inefficient star formation}",
      journal = {\mnras},
         year = 2014,
        month = nov,
       volume = {445},
       number = {1},
        pages = {581-603},
          doi = {10.1093/mnras/stu1738},
archivePrefix = {arXiv},
       eprint = {1311.2073},
 primaryClass = {astro-ph.CO},
       adsurl = {https://ui.adsabs.harvard.edu/abs/2014MNRAS.445..581H}
}

@ARTICLE{Hopkins2018,
       author = {{Hopkins}, Philip F. and {Wetzel}, Andrew and {Kere{\v{s}}}, Du{\v{s}}an and {Faucher-Gigu{\`e}re}, Claude-Andr{\'e} and {Quataert}, Eliot and {Boylan-Kolchin}, Michael and {Murray}, Norman and {Hayward}, Christopher C. and {Garrison-Kimmel}, Shea and {Hummels}, Cameron and {Feldmann}, Robert and {Torrey}, Paul and {Ma}, Xiangcheng and {Angl{\'e}s-Alc{\'a}zar}, Daniel and {Su}, Kung-Yi and {Orr}, Matthew and {Schmitz}, Denise and {Escala}, Ivanna and {Sanderson}, Robyn and {Grudi{\'c}}, Michael Y. and {Hafen}, Zachary and {Kim}, Ji-Hoon and {Fitts}, Alex and {Bullock}, James S. and {Wheeler}, Coral and {Chan}, T.~K. and {Elbert}, Oliver D. and {Narayanan}, Desika},
        title = "{FIRE-2 simulations: physics versus numerics in galaxy formation}",
      journal = {\mnras},
         year = 2018,
        month = oct,
       volume = {480},
       number = {1},
        pages = {800-863},
          doi = {10.1093/mnras/sty1690},
archivePrefix = {arXiv},
       eprint = {1702.06148},
 primaryClass = {astro-ph.GA},
       adsurl = {https://ui.adsabs.harvard.edu/abs/2018MNRAS.480..800H}
}

@ARTICLE{Hopkins2023,
       author = {{Hopkins}, Philip F. and {Wetzel}, Andrew and {Wheeler}, Coral and {Sanderson}, Robyn and {Grudi{\'c}}, Michael Y. and {Sameie}, Omid and {Boylan-Kolchin}, Michael and {Orr}, Matthew and {Ma}, Xiangcheng and {Faucher-Gigu{\`e}re}, Claude-Andr{\'e} and {Kere{\v{s}}}, Du{\v{s}}an and {Quataert}, Eliot and {Su}, Kung-Yi and {Moreno}, Jorge and {Feldmann}, Robert and {Bullock}, James S. and {Loebman}, Sarah R. and {Angl{\'e}s-Alc{\'a}zar}, Daniel and {Stern}, Jonathan and {Necib}, Lina and {Choban}, Caleb R. and {Hayward}, Christopher C.},
        title = "{FIRE-3: updated stellar evolution models, yields, and microphysics and fitting functions for applications in galaxy simulations}",
      journal = {\mnras},
         year = 2023,
        month = feb,
       volume = {519},
       number = {2},
        pages = {3154-3181},
          doi = {10.1093/mnras/stac3489},
archivePrefix = {arXiv},
       eprint = {2203.00040},
 primaryClass = {astro-ph.GA},
       adsurl = {https://ui.adsabs.harvard.edu/abs/2023MNRAS.519.3154H}
}

@ARTICLE{Feldmann2023,
       author = {{Feldmann}, Robert and {Quataert}, Eliot and {Faucher-Gigu{\`e}re}, Claude-Andr{\'e} and {Hopkins}, Philip F. and {{\c{C}}atmabacak}, Onur and {Kere{\v{s}}}, Du{\v{s}}an and {Bassini}, Luigi and {Bernardini}, Mauro and {Bullock}, James S. and {Cenci}, Elia and {Gensior}, Jindra and {Liang}, Lichen and {Moreno}, Jorge and {Wetzel}, Andrew},
        title = "{FIREbox: simulating galaxies at high dynamic range in a cosmological volume}",
      journal = {\mnras},
         year = 2023,
        month = jul,
       volume = {522},
       number = {3},
        pages = {3831-3860},
          doi = {10.1093/mnras/stad1205},
archivePrefix = {arXiv},
       eprint = {2205.15325},
 primaryClass = {astro-ph.GA},
       adsurl = {https://ui.adsabs.harvard.edu/abs/2023MNRAS.522.3831F}
}

@ARTICLE{Feldmann2025,
       author = {{Feldmann}, Robert and {Boylan-Kolchin}, Michael and {Bullock}, James S. and {{\c{C}}atmabacak}, Onur and {Faucher-Gigu{\`e}re}, Claude-Andr{\'e} and {Hayward}, Christopher C. and {Kere{\v{s}}}, Du{\v{s}}an and {Lazar}, Alexandres and {Liang}, Lichen and {Moreno}, Jorge and {Oesch}, Pascal A. and {Quataert}, Eliot and {Shen}, Xuejian and {Sun}, Guochao},
        title = "{Elevated UV luminosity density at Cosmic Dawn explained by non-evolving, weakly mass-dependent star formation efficiency}",
      journal = {\mnras},
         year = 2025,
        month = jan,
       volume = {536},
       number = {1},
        pages = {988-1016},
          doi = {10.1093/mnras/stae2633},
archivePrefix = {arXiv},
       eprint = {2407.02674},
 primaryClass = {astro-ph.CO},
       adsurl = {https://ui.adsabs.harvard.edu/abs/2025MNRAS.536..988F}
}

@ARTICLE{Bhagwat2025,
       author = {{Bhagwat}, Aniket and {Napolitano}, Lorenzo and {Pentericci}, Laura and {Ciardi}, Benedetta and {Costa}, Tiago},
        title = "{Ly {\ensuremath{\alpha}} with SPICE: interpreting Ly {\ensuremath{\alpha}} emission at z > 5}",
      journal = {\mnras},
         year = 2025,
        month = sep,
       volume = {542},
       number = {1},
        pages = {128-135},
          doi = {10.1093/mnras/staf1121},
archivePrefix = {arXiv},
       eprint = {2408.16063},
 primaryClass = {astro-ph.GA},
       adsurl = {https://ui.adsabs.harvard.edu/abs/2025MNRAS.542..128B}
}

@ARTICLE{Basu2025,
       author = {{Basu}, Arghyadeep and {Bhagwat}, Aniket and {Ciardi}, Benedetta and {Costa}, Tiago},
        title = "{Variability of the UV luminosity function with SPICE}",
      journal = {\mnras},
         year = 2026,
        month = jan,
       volume = {545},
       number = {3},
          eid = {staf2240},
        pages = {staf2240},
          doi = {10.1093/mnras/staf2240},
archivePrefix = {arXiv},
       eprint = {2501.18559},
 primaryClass = {astro-ph.GA},
       adsurl = {https://ui.adsabs.harvard.edu/abs/2026MNRAS.545f2240B}
}

@ARTICLE{Kannan2025,
       author = {{Kannan}, Rahul and {Puchwein}, Ewald and {Smith}, Aaron and {Borrow}, Josh and {Garaldi}, Enrico and {Keating}, Laura and {Vogelsberger}, Mark and {Zier}, Oliver and {McClymont}, William and {Shen}, Xuejian and {Popovic}, Filip and {Tacchella}, Sandro and {Hernquist}, Lars and {Springel}, Volker},
        title = "{Introducing the THESAN-ZOOM project: radiation-hydrodynamic simulations of high-redshift galaxies with a multi-phase interstellar medium}",
      journal = {The Open Journal of Astrophysics},
         year = 2025,
        month = oct,
       volume = {8},
          eid = {153},
        pages = {153},
          doi = {10.33232/001c.145804},
archivePrefix = {arXiv},
       eprint = {2502.20437},
 primaryClass = {astro-ph.GA},
       adsurl = {https://ui.adsabs.harvard.edu/abs/2025OJAp....8E.153K}
}

@ARTICLE{Rosdahl2022,
       author = {{Rosdahl}, Joakim and {Blaizot}, J{\'e}r{\'e}my and {Katz}, Harley and {Kimm}, Taysun and {Garel}, Thibault and {Haehnelt}, Martin and {Keating}, Laura C. and {Martin-Alvarez}, Sergio and {Michel-Dansac}, L{\'e}o and {Ocvirk}, Pierre},
        title = "{LyC escape from SPHINX galaxies in the Epoch of Reionization}",
      journal = {\mnras},
         year = 2022,
        month = sep,
       volume = {515},
       number = {2},
        pages = {2386-2414},
          doi = {10.1093/mnras/stac1942},
archivePrefix = {arXiv},
       eprint = {2207.03232},
 primaryClass = {astro-ph.GA},
       adsurl = {https://ui.adsabs.harvard.edu/abs/2022MNRAS.515.2386R}
}

@ARTICLE{Andalman2024,
       author = {{Andalman}, Zachary L. and {Teyssier}, Romain and {Dekel}, Avishai},
        title = "{On the origin of the high star formation efficiency in massive galaxies at Cosmic Dawn}",
      journal = {\mnras},
         year = 2025,
        month = jul,
       volume = {540},
       number = {4},
        pages = {3350-3383},
          doi = {10.1093/mnras/staf930},
archivePrefix = {arXiv},
       eprint = {2410.20530},
 primaryClass = {astro-ph.GA},
       adsurl = {https://ui.adsabs.harvard.edu/abs/2025MNRAS.540.3350A}
}

@ARTICLE{Chon2021,
       author = {{Chon}, Sunmyon and {Omukai}, Kazuyuki and {Schneider}, Raffaella},
        title = "{Transition of the initial mass function in the metal-poor environments}",
      journal = {\mnras},
         year = 2021,
        month = dec,
       volume = {508},
       number = {3},
        pages = {4175-4192},
          doi = {10.1093/mnras/stab2497},
archivePrefix = {arXiv},
       eprint = {2103.04997},
 primaryClass = {astro-ph.GA},
       adsurl = {https://ui.adsabs.harvard.edu/abs/2021MNRAS.508.4175C}
}

@article{Okamoto2008,
archivePrefix = {arXiv},
arxivId = {0806.0378},
author = {Okamoto, Takashi and Gao, Liang and Theuns, Tom},
doi = {10.1111/j.1365-2966.2008.13830.x},
eprint = {0806.0378},
issn = {00358711},
journal = {\mnras},
number = {3},
pages = {920--928},
title = {{Mass loss of galaxies due to an ultraviolet background}},
url = {http://arxiv.org/abs/0806.0378%5Cnhttp://doi.wiley.com/10.1111/j.1365-2966.2008.13830.x},
volume = {390},
year = {2008}
}

@ARTICLE{Mo1998,
       author = {{Mo}, H.~J. and {Mao}, Shude and {White}, Simon D.~M.},
        title = "{The formation of galactic discs}",
      journal = {\mnras},
         year = 1998,
        month = apr,
       volume = {295},
       number = {2},
        pages = {319-336},
          doi = {10.1046/j.1365-8711.1998.01227.x},
archivePrefix = {arXiv},
       eprint = {astro-ph/9707093},
 primaryClass = {astro-ph},
       adsurl = {https://ui.adsabs.harvard.edu/abs/1998MNRAS.295..319M}
}

@ARTICLE{Leitherer1999,
       author = {{Leitherer}, Claus and {Schaerer}, Daniel and {Goldader}, Jeffrey D. and {Delgado}, Rosa M. Gonz{\'a}lez and {Robert}, Carmelle and {Kune}, Denis Foo and {de Mello}, Du{\'\i}lia F. and {Devost}, Daniel and {Heckman}, Timothy M.},
        title = "{Starburst99: Synthesis Models for Galaxies with Active Star Formation}",
      journal = {ApJs},
         year = 1999,
        month = jul,
       volume = {123},
       number = {1},
        pages = {3-40},
          doi = {10.1086/313233},
archivePrefix = {arXiv},
       eprint = {astro-ph/9902334},
 primaryClass = {astro-ph},
       adsurl = {https://ui.adsabs.harvard.edu/abs/1999ApJS..123....3L}
}

@ARTICLE{Behroozi2022,
       author = {{Behroozi}, Peter and {Hearin}, Andrew and {Moster}, Benjamin P.},
        title = "{Observational measures of halo properties beyond mass}",
      journal = {\mnras},
         year = 2022,
        month = jan,
       volume = {509},
       number = {2},
        pages = {2800-2824},
          doi = {10.1093/mnras/stab3193},
archivePrefix = {arXiv},
       eprint = {2101.05280},
 primaryClass = {astro-ph.GA},
       adsurl = {https://ui.adsabs.harvard.edu/abs/2022MNRAS.509.2800B}
}

@ARTICLE{Somerville2018,
       author = {{Somerville}, Rachel S. and {Behroozi}, Peter and {Pandya}, Viraj and {Dekel}, Avishai and {Faber}, S.~M. and {Fontana}, Adriano and {Koekemoer}, Anton M. and {Koo}, David C. and {P{\'e}rez-Gonz{\'a}lez}, P.~G. and {Primack}, Joel R. and {Santini}, Paola and {Taylor}, Edward N. and {van der Wel}, Arjen},
        title = "{The relationship between galaxy and dark matter halo size from z {\ensuremath{\sim}} 3 to the present}",
      journal = {\mnras},
         year = 2018,
        month = jan,
       volume = {473},
       number = {2},
        pages = {2714-2736},
          doi = {10.1093/mnras/stx2040},
archivePrefix = {arXiv},
       eprint = {1701.03526},
 primaryClass = {astro-ph.GA},
       adsurl = {https://ui.adsabs.harvard.edu/abs/2018MNRAS.473.2714S}
}

@ARTICLE{Sunphangs2023,
       author = {{Sun}, Jiayi and {Leroy}, Adam K. and {Ostriker}, Eve C. and {Meidt}, Sharon and {Rosolowsky}, Erik and {Schinnerer}, Eva and {Wilson}, Christine D. and {Utomo}, Dyas and {Belfiore}, Francesco and {Blanc}, Guillermo A. and {Emsellem}, Eric and {Faesi}, Christopher and {Groves}, Brent and {Hughes}, Annie and {Koch}, Eric W. and {Kreckel}, Kathryn and {Liu}, Daizhong and {Pan}, Hsi-An and {Pety}, J{\'e}r{\^o}me and {Querejeta}, Miguel and {Razza}, Alessandro and {Saito}, Toshiki and {Sardone}, Amy and {Usero}, Antonio and {Williams}, Thomas G. and {Bigiel}, Frank and {Bolatto}, Alberto D. and {Chevance}, M{\'e}lanie and {Dale}, Daniel A. and {Gensior}, Jindra and {Glover}, Simon C.~O. and {Grasha}, Kathryn and {Henshaw}, Jonathan D. and {Jim{\'e}nez-Donaire}, Mar{\'\i}a J. and {Klessen}, Ralf S. and {Kruijssen}, J.~M. Diederik and {Murphy}, Eric J. and {Neumann}, Lukas and {Teng}, Yu-Hsuan and {Thilker}, David A.},
        title = "{Star Formation Laws and Efficiencies across 80 Nearby Galaxies}",
      journal = {\apjl},
         year = 2023,
        month = mar,
       volume = {945},
       number = {2},
          eid = {L19},
        pages = {L19},
          doi = {10.3847/2041-8213/acbd9c},
archivePrefix = {arXiv},
       eprint = {2302.12267},
 primaryClass = {astro-ph.GA},
       adsurl = {https://ui.adsabs.harvard.edu/abs/2023ApJ...945L..19S}
}

@ARTICLE{Menonfesc2024,
       author = {{Menon}, Shyam H. and {Burkhart}, Blakesley and {Somerville}, Rachel S. and {Thompson}, Todd A. and {Sternberg}, Amiel},
        title = "{Bursts of Star Formation and Radiation-driven Outflows Produce Efficient LyC Leakage from Dense Compact Star Clusters}",
      journal = {\apj},
         year = 2025,
        month = jul,
       volume = {987},
       number = {1},
          eid = {12},
        pages = {12},
          doi = {10.3847/1538-4357/add2f9},
archivePrefix = {arXiv},
       eprint = {2408.14591},
 primaryClass = {astro-ph.GA},
       adsurl = {https://ui.adsabs.harvard.edu/abs/2025ApJ...987...12M}
}

@ARTICLE{Burgarella2025,
       author = {{Burgarella}, Denis and {Buat}, V{\'e}ronique and {Theul{\'e}}, Patrice and {Zavala}, Jorge and {Dickinson}, Mark and {Arrabal Haro}, Pablo and {Bagley}, Micaela B. and {Boquien}, M{\'e}d{\'e}ric and {Cleri}, Nikko and {Dewachter}, Tim and {Ferguson}, Henry C. and {Fern{\`a}ndez}, Vital and {Finkelstein}, Steven L. and {Gawiser}, Eric and {Grazian}, Andrea and {Grogin}, Norman and {Holwerda}, Benne W. and {Kartaltepe}, Jeyhan S. and {Kewley}, Lisa and {Kirkpatrick}, Allison and {Kocevski}, Dale and {Koekemoer}, Anton M. and {Long}, Arianna and {Lotz}, Jennifer and {Lucas}, Ray A. and {Mobasher}, Bahram and {Papovich}, Casey and {P{\'e}rez-Gonz{\`a}lez}, Pablo G. and {Pirzkal}, Nor and {Ravindranath}, Swara and {Rodighiero}, Giulia and {Roehlly}, Yannick and {Rose}, Caitlin and {Seill{\'e}}, Lise-Marie and {Somerville}, Rachel and {Wilkins}, Steve and {Yang}, Guang and {Yung}, L.~Y. Aaron},
        title = "{CEERS: Possibly forging the first dust grains in the universe: A population of galaxies with spectroscopically derived extremely low dust attenuation (GELDA) at 4.0 < z {\ensuremath{\lesssim}} 11.4}",
      journal = {\aap},
         year = 2025,
        month = jul,
       volume = {699},
          eid = {A336},
        pages = {A336},
          doi = {10.1051/0004-6361/202554231},
archivePrefix = {arXiv},
       eprint = {2504.13118},
 primaryClass = {astro-ph.GA},
       adsurl = {https://ui.adsabs.harvard.edu/abs/2025A&A...699A.336B}
}

@ARTICLE{Semenov:2025b,
       author = {{Semenov}, Vadim A.},
        title = "{Capturing Turbulence with Numerical Dissipation: A Simple Dynamical Model for Unresolved Turbulence in Hydrodynamic Simulations}",
      journal = {ApJs},
         year = 2025,
        month = dec,
       volume = {281},
       number = {2},
          eid = {37},
        pages = {37},
          doi = {10.3847/1538-4365/ae0cc6},
archivePrefix = {arXiv},
       eprint = {2410.23339},
 primaryClass = {astro-ph.GA},
       adsurl = {https://ui.adsabs.harvard.edu/abs/2025ApJS..281...37S}
}

@ARTICLE{Semenov2025a,
       author = {{Semenov}, Vadim A. and {Conroy}, Charlie and {Hernquist}, Lars},
        title = "{From UV-bright Galaxies to Early Disks: The Importance of Turbulent Star Formation in the Early Universe}",
      journal = {\apj},
         year = 2025,
        month = aug,
       volume = {989},
       number = {2},
          eid = {219},
        pages = {219},
          doi = {10.3847/1538-4357/ade22d},
archivePrefix = {arXiv},
       eprint = {2410.09205},
 primaryClass = {astro-ph.GA},
       adsurl = {https://ui.adsabs.harvard.edu/abs/2025ApJ...989..219S}
}

@ARTICLE{Parashari2023,
       author = {{Parashari}, Priyank and {Laha}, Ranjan},
        title = "{Primordial power spectrum in light of JWST observations of high redshift galaxies}",
      journal = {\mnras},
         year = 2023,
        month = nov,
       volume = {526},
       number = {1},
        pages = {L63-L69},
          doi = {10.1093/mnrasl/slad107},
archivePrefix = {arXiv},
       eprint = {2305.00999},
 primaryClass = {astro-ph.CO},
       adsurl = {https://ui.adsabs.harvard.edu/abs/2023MNRAS.526L..63P}
}

@ARTICLE{KroupaIMF,
       author = {{Kroupa}, Pavel},
        title = "{On the variation of the initial mass function}",
      journal = {\mnras},
         year = 2001,
        month = apr,
       volume = {322},
       number = {2},
        pages = {231-246},
          doi = {10.1046/j.1365-8711.2001.04022.x},
archivePrefix = {arXiv},
       eprint = {astro-ph/0009005},
 primaryClass = {astro-ph},
       adsurl = {https://ui.adsabs.harvard.edu/abs/2001MNRAS.322..231K}
}

@BOOK{Loeb-Furlanetto:2013,
       author = {{Loeb}, Abraham and {Furlanetto}, Steven R.},
        title = "{The First Galaxies in the Universe}",
         year = 2013,
       adsurl = {https://ui.adsabs.harvard.edu/abs/2013fgu..book.....L}
}

@ARTICLE{Press-Schechter:1974,
   author = {{Press}, W.~H. and {Schechter}, P.},
    title = "{Formation of Galaxies and Clusters of Galaxies by Self-Similar Gravitational Condensation}",
  journal = {\apj},
     year = 1974,
    month = feb,
   volume = 187,
    pages = {425-438},
      doi = {10.1086/152650},
   adsurl = {http://adsabs.harvard.edu/abs/1974ApJ...187..425P}
}

@ARTICLE{Lacey:1993,
   author = {{Lacey}, C. and {Cole}, S.},
    title = "{Merger rates in hierarchical models of galaxy formation}",
  journal = {\mnras},
     year = 1993,
    month = jun,
   volume = 262,
    pages = {627},
   adsurl = {http://adsabs.harvard.edu/abs/1993MNRAS.262..627L}
}

@ARTICLE{SK:1999,
   author = {{Somerville}, R.~S. and {Kolatt}, T.~S.},
    title = "{How to plant a merger tree}",
  journal = {\mnras},
     year = 1999,
    month = may,
   volume = 305,
    pages = {1},
   adsurl = {http://adsabs.harvard.edu/abs/1999MNRAS.305....1S}
}

@ARTICLE{Parkinson:2008,
   author = {{Parkinson}, H. and {Cole}, S. and {Helly}, J.},
    title = "{Generating dark matter halo merger trees}",
  journal = {\mnras},
archivePrefix = "arXiv",
   eprint = {0708.1382},
     year = 2008,
    month = jan,
   volume = 383,
    pages = {557},
      doi = {10.1111/j.1365-2966.2007.12517.x},
   adsurl = {http://adsabs.harvard.edu/abs/2008MNRAS.383..557P}
}

@ARTICLE{Cole:1989,
       author = {{Cole}, Shaun and {Kaiser}, Nick},
        title = "{Biased clustering in the cold dark matter cosmogony.}",
      journal = {\mnras},
         year = 1989,
        month = apr,
       volume = {237},
        pages = {1127-1146},
          doi = {10.1093/mnras/237.4.1127},
       adsurl = {https://ui.adsabs.harvard.edu/abs/1989MNRAS.237.1127C}
}

@ARTICLE{Sheth:2001,
       author = {{Sheth}, Ravi K. and {Mo}, H.~J. and {Tormen}, Giuseppe},
        title = "{Ellipsoidal collapse and an improved model for the number and spatial distribution of dark matter haloes}",
      journal = {\mnras},
         year = 2001,
        month = may,
       volume = {323},
       number = {1},
        pages = {1-12},
          doi = {10.1046/j.1365-8711.2001.04006.x},
archivePrefix = {arXiv},
       eprint = {astro-ph/9907024},
 primaryClass = {astro-ph},
       adsurl = {https://ui.adsabs.harvard.edu/abs/2001MNRAS.323....1S}
}

@ARTICLE{Mo-White:1996,
       author = {{Mo}, H.~J. and {White}, S.~D.~M.},
        title = "{An analytic model for the spatial clustering of dark matter haloes}",
      journal = {\mnras},
         year = 1996,
        month = sep,
       volume = {282},
       number = {2},
        pages = {347-361},
          doi = {10.1093/mnras/282.2.347},
archivePrefix = {arXiv},
       eprint = {astro-ph/9512127},
 primaryClass = {astro-ph},
       adsurl = {https://ui.adsabs.harvard.edu/abs/1996MNRAS.282..347M}
}

@ARTICLE{Mo-White:2002,
       author = {{Mo}, H.~J. and {White}, S.~D.~M.},
        title = "{The abundance and clustering of dark haloes in the standard {\ensuremath{\Lambda}}CDM cosmogony}",
      journal = {\mnras},
         year = 2002,
        month = oct,
       volume = {336},
       number = {1},
        pages = {112-118},
          doi = {10.1046/j.1365-8711.2002.05723.x},
archivePrefix = {arXiv},
       eprint = {astro-ph/0202393},
 primaryClass = {astro-ph},
       adsurl = {https://ui.adsabs.harvard.edu/abs/2002MNRAS.336..112M}
}

@Inbook{Springel:2016,
author="Springel, Volker",
editor="Revaz, Yves
and Jablonka, Pascale
and Teyssier, Romain
and Mayer, Lucio",
title="High Performance Computing and Numerical Modelling",
bookTitle="Star Formation in Galaxy Evolution: Connecting Numerical Models to Reality: Saas-Fee Advanced Course 43. Swiss Society for Astrophysics and Astronomy",
year="2016",
publisher="Springer Berlin Heidelberg",
address="Berlin, Heidelberg",
pages="251--358",
isbn="978-3-662-47890-5",
doi="10.1007/978-3-662-47890-5_3",
url="https://doi.org/10.1007/978-3-662-47890-5_3"
}

@ARTICLE{Cueto:2024,
       author = {{Cueto}, Elie R. and {Hutter}, Anne and {Dayal}, Pratika and {Gottl{\"o}ber}, Stefan and {Heintz}, Kasper E. and {Mason}, Charlotte and {Trebitsch}, Maxime and {Yepes}, Gustavo},
        title = "{ASTRAEUS. IX. Impact of an evolving stellar initial mass function on early galaxies and reionisation}",
      journal = {\aap},
         year = 2024,
        month = jun,
       volume = {686},
          eid = {A138},
        pages = {A138},
          doi = {10.1051/0004-6361/202349017},
archivePrefix = {arXiv},
       eprint = {2312.12109},
 primaryClass = {astro-ph.GA},
       adsurl = {https://ui.adsabs.harvard.edu/abs/2024A&A...686A.138C}
}

@ARTICLE{Curti:2025,
       author = {{Curti}, Mirko},
        title = "{The Chemical Evolution of Galaxies}",
      journal = {arXiv e-prints},
         year = 2025,
        month = apr,
          eid = {arXiv:2504.08933},
        pages = {arXiv:2504.08933},
          doi = {10.48550/arXiv.2504.08933},
archivePrefix = {arXiv},
       eprint = {2504.08933},
 primaryClass = {astro-ph.GA},
       adsurl = {https://ui.adsabs.harvard.edu/abs/2025arXiv250408933C}
}

@ARTICLE{Kobayashi:2024,
       author = {{Kobayashi}, Chiaki and {Ferrara}, Andrea},
        title = "{Rapid Chemical Enrichment by Intermittent Star Formation in GN-z11}",
      journal = {\apjl},
         year = 2024,
        month = feb,
       volume = {962},
       number = {1},
          eid = {L6},
        pages = {L6},
          doi = {10.3847/2041-8213/ad1de1},
archivePrefix = {arXiv},
       eprint = {2308.15583},
 primaryClass = {astro-ph.GA},
       adsurl = {https://ui.adsabs.harvard.edu/abs/2024ApJ...962L...6K}
}

@ARTICLE{Kokorev:2025,
       author = {{Kokorev}, Vasily and {Ch{\'a}vez Ortiz}, {\'O}scar A. and {Taylor}, Anthony J. and {Finkelstein}, Steven L. and {Arrabal Haro}, Pablo and {Dickinson}, Mark and {Chisholm}, John and {Fujimoto}, Seiji and {noz}, Julian B. Mu and {Endsley}, Ryan and {Hu}, Weida and {Napolitano}, Lorenzo and {Wilkins}, Stephen M. and {Akins}, Hollis B. and {Amori{\'\i}n}, Ricardo and {Casey}, Caitlin M. and {Cheng}, Yingjie and {Cleri}, Nikko J. and {Cole}, Justin and {Cullen}, Fergus and {Daddi}, Emanuele and {Davis}, Kelcey and {Donnan}, Callum T. and {Dunlop}, James S. and {Fern{\'a}ndez}, Vital and {Giavalisco}, Mauro and {Grogin}, Norman A. and {Hathi}, Nimish and {Hirschmann}, Michaela and {Kartaltepe}, Jeyhan S. and {Koekemoer}, Anton M. and {Leung}, Ho-Hin and {Lucas}, Ray A. and {McLeod}, Derek and {Papovich}, Casey and {Pentericci}, Laura and {P{\'e}rez-Gonz{\'a}lez}, Pablo G. and {Somerville}, Rachel S. and {Wang}, Xin and {Yung}, L.~Y. Aaron and {Zavala}, Jorge A.},
        title = "{CAPERS Observations of Two UV-bright Galaxies at z > 10. More Evidence for Bursting Star Formation in the Early Universe}",
      journal = {\apjl},
         year = 2025,
        month = jul,
       volume = {988},
       number = {1},
          eid = {L10},
        pages = {L10},
          doi = {10.3847/2041-8213/ade8f5},
archivePrefix = {arXiv},
       eprint = {2504.12504},
 primaryClass = {astro-ph.GA},
       adsurl = {https://ui.adsabs.harvard.edu/abs/2025ApJ...988L..10K}
}

@ARTICLE{Meyer:2025,
       author = {{Meyer}, Romain A. and {Wang}, Feige and {Kakiichi}, Koki and {Brammer}, Gabe and {Champagne}, Jackie and {Jurk}, Katharina and {Li}, Zihao and {Li}, Zijian and {Musin}, Marat and {Satyavolu}, Sindhu and {Schindler}, Jan-Torge and {Shuntov}, Marko and {Xu}, Yi and {Zou}, Siwei and {Bian}, Fuyan and {Casey}, Caitlin and {Egami}, Eiichi and {Fan}, Xiaohui and {Jiang}, Danyang and {Laporte}, Nicolas and {Liu}, Weizhe and {Oesch}, Pascal and {Tasca}, Lidia and {Yang}, Jinyi and {Zhang}, Zijian and {Akins}, Hollis and {Cai}, Zheng and {Coulter}, Dave A. and {Huang}, Jiamu and {Li}, Mingyu and {Liu}, Weizhe and {Liang}, Yongming and {Jin}, Xiangyu and {Kartaltepe}, Jeyhan and {Matharu}, Jasleen and {Pudoka}, Maria and {Tee}, Wei-Leong and {Witten}, Callum and {Zhang}, Haowen and {Zhu}, Yongda},
        title = "{JWST COSMOS-3D: Spectroscopic Census and Luminosity Function of [O III] Emitters at 6.75<z<9.05 in COSMOS}",
      journal = {arXiv e-prints},
         year = 2025,
        month = oct,
          eid = {arXiv:2510.11373},
        pages = {arXiv:2510.11373},
          doi = {10.48550/arXiv.2510.11373},
archivePrefix = {arXiv},
       eprint = {2510.11373},
 primaryClass = {astro-ph.GA},
       adsurl = {https://ui.adsabs.harvard.edu/abs/2025arXiv251011373M}
}

@ARTICLE{Gutkin:2016,
       author = {{Gutkin}, Julia and {Charlot}, St{\'e}phane and {Bruzual}, Gustavo},
        title = "{Modelling the nebular emission from primeval to present-day star-forming galaxies}",
      journal = {\mnras},
         year = 2016,
        month = oct,
       volume = {462},
       number = {2},
        pages = {1757-1774},
          doi = {10.1093/mnras/stw1716},
archivePrefix = {arXiv},
       eprint = {1607.06086},
 primaryClass = {astro-ph.GA},
       adsurl = {https://ui.adsabs.harvard.edu/abs/2016MNRAS.462.1757G}
}

@misc{Sutherland:2013,
       author = {{Sutherland}, Ralph and {Dopita}, Mike and {Binette}, Luc and {Groves}, Brent},
        title = "{MAPPINGS III: Modelling And Prediction in PhotoIonized Nebulae and Gasdynamical Shocks}",
 howpublished = {Astrophysics Source Code Library, record ascl:1306.008},
         year = 2013,
        month = jun,
          eid = {ascl:1306.008},
archivePrefix = {ascl},
       eprint = {1306.008},
       adsurl = {https://ui.adsabs.harvard.edu/abs/2013ascl.soft06008S}
}

@misc{Sutherland:2018,
       author = {{Sutherland}, Ralph and {Dopita}, Mike and {Binette}, Luc and {Groves}, Brent},
        title = "{MAPPINGS V: Astrophysical plasma modeling code}",
 howpublished = {Astrophysics Source Code Library, record ascl:1807.005},
         year = 2018,
        month = jul,
          eid = {ascl:1807.005},
archivePrefix = {ascl},
       eprint = {1807.005},
       adsurl = {https://ui.adsabs.harvard.edu/abs/2018ascl.soft07005S}
}

@ARTICLE{Ferland:2013,
       author = {{Ferland}, G.~J. and {Porter}, R.~L. and {van Hoof}, P.~A.~M. and {Williams}, R.~J.~R. and {Abel}, N.~P. and {Lykins}, M.~L. and {Shaw}, G. and {Henney}, W.~J. and {Stancil}, P.~C.},
        title = "{The 2013 Release of Cloudy}",
      journal = {RMxAA},
         year = 2013,
        month = apr,
       volume = {49},
        pages = {137-163},
          doi = {10.48550/arXiv.1302.4485},
archivePrefix = {arXiv},
       eprint = {1302.4485},
 primaryClass = {astro-ph.GA},
       adsurl = {https://ui.adsabs.harvard.edu/abs/2013RMxAA..49..137F}
}

@ARTICLE{Ferland:2017,
       author = {{Ferland}, G.~J. and {Chatzikos}, M. and {Guzm{\'a}n}, F. and {Lykins}, M.~L. and {van Hoof}, P.~A.~M. and {Williams}, R.~J.~R. and {Abel}, N.~P. and {Badnell}, N.~R. and {Keenan}, F.~P. and {Porter}, R.~L. and {Stancil}, P.~C.},
        title = "{The 2017 Release Cloudy}",
      journal = {RMxAA},
         year = 2017,
        month = oct,
       volume = {53},
        pages = {385-438},
          doi = {10.48550/arXiv.1705.10877},
archivePrefix = {arXiv},
       eprint = {1705.10877},
 primaryClass = {astro-ph.GA},
       adsurl = {https://ui.adsabs.harvard.edu/abs/2017RMxAA..53..385F}
}

@ARTICLE{Hirschmann:2017,
       author = {{Hirschmann}, Michaela and {Charlot}, Stephane and {Feltre}, Anna and {Naab}, Thorsten and {Choi}, Ena and {Ostriker}, Jeremiah P. and {Somerville}, Rachel S.},
        title = "{Synthetic nebular emission from massive galaxies - I: origin of the cosmic evolution of optical emission-line ratios}",
      journal = {\mnras},
         year = 2017,
        month = dec,
       volume = {472},
       number = {2},
        pages = {2468-2495},
          doi = {10.1093/mnras/stx2180},
archivePrefix = {arXiv},
       eprint = {1706.00010},
 primaryClass = {astro-ph.GA},
       adsurl = {https://ui.adsabs.harvard.edu/abs/2017MNRAS.472.2468H}
}

@ARTICLE{Salim-Narayanan:2020,
       author = {{Salim}, Samir and {Narayanan}, Desika},
        title = "{The Dust Attenuation Law in Galaxies}",
      journal = {\araa},
         year = 2020,
        month = aug,
       volume = {58},
        pages = {529-575},
          doi = {10.1146/annurev-astro-032620-021933},
archivePrefix = {arXiv},
       eprint = {2001.03181},
 primaryClass = {astro-ph.GA},
       adsurl = {https://ui.adsabs.harvard.edu/abs/2020ARA&A..58..529S}
}

@ARTICLE{Ono:2025,
       author = {{Ono}, Yoshiaki and {Ouchi}, Masami and {Harikane}, Yuichi and {Yajima}, Hidenobu and {Nakajima}, Kimihiko and {Fujimoto}, Seiji and {Nakane}, Minami and {Xu}, Yi},
        title = "{Morphological Demographics of Galaxies at z {\ensuremath{\sim}} 10{\textendash}16: Log-normal Size Distribution and Exponential Profiles Consistent with the Disk Formation Scenario}",
      journal = {\apj},
         year = 2025,
        month = oct,
       volume = {991},
       number = {2},
          eid = {222},
        pages = {222},
          doi = {10.3847/1538-4357/adfc4d},
archivePrefix = {arXiv},
       eprint = {2502.08885},
 primaryClass = {astro-ph.GA},
       adsurl = {https://ui.adsabs.harvard.edu/abs/2025ApJ...991..222O}
}

@ARTICLE{Steinwandel:2024,
       author = {{Steinwandel}, Ulrich P. and {Kim}, Chang-Goo and {Bryan}, Greg L. and {Ostriker}, Eve C. and {Somerville}, Rachel S. and {Fielding}, Drummond B.},
        title = "{The Structure and Composition of Multiphase Galactic Winds in a Large Magellanic Cloud Mass Simulated Galaxy}",
      journal = {\apj},
         year = 2024,
        month = jan,
       volume = {960},
       number = {2},
          eid = {100},
        pages = {100},
          doi = {10.3847/1538-4357/ad09e1},
archivePrefix = {arXiv},
       eprint = {2212.03898},
 primaryClass = {astro-ph.GA},
       adsurl = {https://ui.adsabs.harvard.edu/abs/2024ApJ...960..100S}
}

@ARTICLE{Pandya:2020,
       author = {{Pandya}, Viraj and {Somerville}, Rachel S. and {Angl{\'e}s-Alc{\'a}zar}, Daniel and {Hayward}, Christopher C. and {Bryan}, Greg L. and {Fielding}, Drummond B. and {Forbes}, John C. and {Burkhart}, Blakesley and {Genel}, Shy and {Hernquist}, Lars and {Kim}, Chang-Goo and {Tonnesen}, Stephanie and {Starkenburg}, Tjitske},
        title = "{First Results from SMAUG: The Need for Preventative Stellar Feedback and Improved Baryon Cycling in Semianalytic Models of Galaxy Formation}",
      journal = {\apj},
         year = 2020,
        month = dec,
       volume = {905},
       number = {1},
          eid = {4},
        pages = {4},
          doi = {10.3847/1538-4357/abc3c1},
archivePrefix = {arXiv},
       eprint = {2006.16317},
 primaryClass = {astro-ph.GA},
       adsurl = {https://ui.adsabs.harvard.edu/abs/2020ApJ...905....4P}
}

@ARTICLE{Wright:2024,
       author = {{Wright}, Ruby J. and {Somerville}, Rachel S. and {Lagos}, Claudia del P. and {Schaller}, Matthieu and {Dav{\'e}}, Romeel and {Angl{\'e}s-Alc{\'a}zar}, Daniel and {Genel}, Shy},
        title = "{The baryon cycle in modern cosmological hydrodynamical simulations}",
      journal = {\mnras},
         year = 2024,
        month = aug,
       volume = {532},
       number = {3},
        pages = {3417-3440},
          doi = {10.1093/mnras/stae1688},
archivePrefix = {arXiv},
       eprint = {2402.08408},
 primaryClass = {astro-ph.GA},
       adsurl = {https://ui.adsabs.harvard.edu/abs/2024MNRAS.532.3417W}
}

@ARTICLE{Lovell:2021,
       author = {{Lovell}, Christopher C. and {Vijayan}, Aswin P. and {Thomas}, Peter A. and {Wilkins}, Stephen M. and {Barnes}, David J. and {Irodotou}, Dimitrios and {Roper}, Will},
        title = "{First Light And Reionization Epoch Simulations (FLARES) - I. Environmental dependence of high-redshift galaxy evolution}",
      journal = {\mnras},
         year = 2021,
        month = jan,
       volume = {500},
       number = {2},
        pages = {2127-2145},
          doi = {10.1093/mnras/staa3360},
archivePrefix = {arXiv},
       eprint = {2004.07283},
 primaryClass = {astro-ph.GA},
       adsurl = {https://ui.adsabs.harvard.edu/abs/2021MNRAS.500.2127L}
}

@ARTICLE{Katz:2024,
       author = {{Katz}, Harley and {Rey}, Martin P. and {Cadiou}, Corentin and {Kimm}, Taysun and {Agertz}, Oscar},
        title = "{The Impact of Star Formation and Feedback Recipes on the Stellar Mass and Interstellar Medium of High-Redshift Galaxies}",
      journal = {arXiv e-prints},
         year = 2024,
        month = nov,
          eid = {arXiv:2411.07282},
        pages = {arXiv:2411.07282},
          doi = {10.48550/arXiv.2411.07282},
archivePrefix = {arXiv},
       eprint = {2411.07282},
 primaryClass = {astro-ph.GA},
       adsurl = {https://ui.adsabs.harvard.edu/abs/2024arXiv241107282K}
}

@ARTICLE{Katz:2022,
       author = {{Katz}, Harley},
        title = "{RAMSES-RTZ: non-equilibrium metal chemistry and cooling coupled to on-the-fly radiation hydrodynamics}",
      journal = {\mnras},
         year = 2022,
        month = may,
       volume = {512},
       number = {1},
        pages = {348-365},
          doi = {10.1093/mnras/stac423},
archivePrefix = {arXiv},
       eprint = {2202.04083},
 primaryClass = {astro-ph.GA},
       adsurl = {https://ui.adsabs.harvard.edu/abs/2022MNRAS.512..348K}
}

@ARTICLE{Popping:2022,
       author = {{Popping}, Gerg{\"o} and {Pillepich}, Annalisa and {Calistro Rivera}, Gabriela and {Schulz}, Sebastian and {Hernquist}, Lars and {Kaasinen}, Melanie and {Marinacci}, Federico and {Nelson}, Dylan and {Vogelsberger}, Mark},
        title = "{The dust-continuum size of TNG50 galaxies at z = 1-5: a comparison with the distribution of stellar light, stars, dust, and H$_{2}$}",
      journal = {\mnras},
         year = 2022,
        month = mar,
       volume = {510},
       number = {3},
        pages = {3321-3334},
          doi = {10.1093/mnras/stab3312},
archivePrefix = {arXiv},
       eprint = {2101.12218},
 primaryClass = {astro-ph.GA},
       adsurl = {https://ui.adsabs.harvard.edu/abs/2022MNRAS.510.3321P}
}

@ARTICLE{Sommovigo:2025,
       author = {{Sommovigo}, L. and {Cochrane}, R.~K. and {Somerville}, R.~S. and {Hayward}, C.~C. and {Lovell}, C.~C. and {Starkenburg}, T. and {Popping}, G. and {Iyer}, K. and {Gabrielpillai}, A. and {Ho}, M. and {Steinwandel}, U.~P. and {Perez}, L.~A.},
        title = "{Learning the Universe: Physically Motivated Priors for Dust Attenuation Curves}",
      journal = {\apj},
         year = 2025,
        month = sep,
       volume = {990},
       number = {2},
          eid = {114},
        pages = {114},
          doi = {10.3847/1538-4357/addec1},
archivePrefix = {arXiv},
       eprint = {2502.13240},
 primaryClass = {astro-ph.GA},
       adsurl = {https://ui.adsabs.harvard.edu/abs/2025ApJ...990..114S}
}

@ARTICLE{Chemerynska:2024,
       author = {{Chemerynska}, Iryna and {Atek}, Hakim and {Dayal}, Pratika and {Furtak}, Lukas J. and {Feldmann}, Robert and {Greene}, Jenny E. and {Maseda}, Michael V. and {Nanayakkara}, Themiya and {Oesch}, Pascal A. and {Fujimoto}, Seiji and {Labb{\'e}}, Ivo and {Bezanson}, Rachel and {Brammer}, Gabriel and {Cutler}, Sam E. and {Leja}, Joel and {Pan}, Richard and {Price}, Sedona H. and {Wang}, Bingjie and {Weaver}, John R. and {Whitaker}, Katherine E.},
        title = "{The Extreme Low-mass End of the Mass{\textendash}Metallicity Relation at z {\ensuremath{\sim}} 7}",
      journal = {\apjl},
         year = 2024,
        month = nov,
       volume = {976},
       number = {1},
          eid = {L15},
        pages = {L15},
          doi = {10.3847/2041-8213/ad8dc9},
archivePrefix = {arXiv},
       eprint = {2407.17110},
 primaryClass = {astro-ph.GA},
       adsurl = {https://ui.adsabs.harvard.edu/abs/2024ApJ...976L..15C}
}

@ARTICLE{Sanders:2025,
       author = {{Sanders}, Ryan L. and {Shapley}, Alice E. and {Topping}, Michael W. and {Reddy}, Naveen A. and {Berg}, Danielle A. and {Khostovan}, Ali Ahmad and {Bouwens}, Rychard J. and {Brammer}, Gabriel and {Carnall}, Adam C. and {Cullen}, Fergus and {Dav{\'e}}, Romeel and {Dunlop}, James S. and {Ellis}, Richard S. and {F{\"o}rster Schreiber}, N.~M. and {Furlanetto}, Steven R. and {Glazebrook}, Karl and {Illingworth}, Garth D. and {Jones}, Tucker and {Kriek}, Mariska and {McLeod}, Derek J. and {McLure}, Ross J. and {Narayanan}, Desika and {Oesch}, Pascal A. and {Pahl}, Anthony J. and {Pettini}, Max and {Schaerer}, Daniel and {Stark}, Daniel P. and {Steidel}, Charles C. and {Tang}, Mengtao and {Clarke}, Leonardo and {Donnan}, Callum T. and {Kehoe}, Emily},
        title = "{The AURORA Survey: High-Redshift Empirical Metallicity Calibrations from Electron Temperature Measurements at z=2-10}",
      journal = {arXiv e-prints},
         year = 2025,
        month = aug,
          eid = {arXiv:2508.10099},
        pages = {arXiv:2508.10099},
          doi = {10.48550/arXiv.2508.10099},
archivePrefix = {arXiv},
       eprint = {2508.10099},
 primaryClass = {astro-ph.GA},
       adsurl = {https://ui.adsabs.harvard.edu/abs/2025arXiv250810099S}
}

@article{Barkana-Loeb:2001,
	title = {In the beginning: the first sources of light and the reionization of the universe},
	volume = {349},
	issn = {0370-1573},
	shorttitle = {In the beginning},
	url = {https://www.sciencedirect.com/science/article/pii/S0370157301000199},
	doi = {10.1016/S0370-1573(01)00019-9},
	number = {2},
	urldate = {2024-10-12},
	journal = {Physics Reports},
	author = {Barkana, Rennan and Loeb, Abraham},
	month = jul,
	year = {2001},
	pages = {125--238},
}

@ARTICLE{Jiang:2014,
       author = {{Jiang}, Fangzhou and {van den Bosch}, Frank C.},
        title = "{Generating merger trees for dark matter haloes: a comparison of methods}",
      journal = {\mnras},
         year = 2014,
        month = may,
       volume = {440},
       number = {1},
        pages = {193-207},
          doi = {10.1093/mnras/stu280},
archivePrefix = {arXiv},
       eprint = {1311.5225},
 primaryClass = {astro-ph.CO},
       adsurl = {https://ui.adsabs.harvard.edu/abs/2014MNRAS.440..193J}
}

@ARTICLE{Kauffmann:1993,
       author = {{Kauffmann}, G. and {White}, S.~D.~M.},
        title = "{The merging history of dark matter haloes in a hierarchical universe.}",
      journal = {\mnras},
         year = 1993,
        month = apr,
       volume = {261},
        pages = {921-928},
          doi = {10.1093/mnras/261.4.921},
       adsurl = {https://ui.adsabs.harvard.edu/abs/1993MNRAS.261..921K}
}

@ARTICLE{Cole:2000,
       author = {{Cole}, Shaun and {Lacey}, Cedric G. and {Baugh}, Carlton M. and {Frenk}, Carlos S.},
        title = "{Hierarchical galaxy formation}",
      journal = {\mnras},
         year = 2000,
        month = nov,
       volume = {319},
       number = {1},
        pages = {168-204},
          doi = {10.1046/j.1365-8711.2000.03879.x},
archivePrefix = {arXiv},
       eprint = {astro-ph/0007281},
 primaryClass = {astro-ph},
       adsurl = {https://ui.adsabs.harvard.edu/abs/2000MNRAS.319..168C}
}

@ARTICLE{Davis:1985,
       author = {{Davis}, M. and {Efstathiou}, G. and {Frenk}, C.~S. and {White}, S.~D.~M.},
        title = "{The evolution of large-scale structure in a universe dominated by cold dark matter}",
      journal = {\apj},
         year = 1985,
        month = may,
       volume = {292},
        pages = {371-394},
          doi = {10.1086/163168},
       adsurl = {https://ui.adsabs.harvard.edu/abs/1985ApJ...292..371D}
}

@ARTICLE{Efstathiou:1985,
       author = {{Efstathiou}, G. and {Davis}, M. and {White}, S.~D.~M. and {Frenk}, C.~S.},
        title = "{Numerical techniques for large cosmological N-body simulations}",
      journal = {ApJs},
         year = 1985,
        month = feb,
       volume = {57},
        pages = {241-260},
          doi = {10.1086/191003},
       adsurl = {https://ui.adsabs.harvard.edu/abs/1985ApJS...57..241E}
}

@ARTICLE{Efstathiou:1988,
       author = {{Efstathiou}, George and {Frenk}, Carlos S. and {White}, Simon D.~M. and {Davis}, Marc},
        title = "{Gravitational clustering from scale-free initial conditions.}",
      journal = {\mnras},
         year = 1988,
        month = dec,
       volume = {235},
        pages = {715-748},
          doi = {10.1093/mnras/235.3.715},
       adsurl = {https://ui.adsabs.harvard.edu/abs/1988MNRAS.235..715E}
}

@ARTICLE{Klypin:1993,
       author = {{Klypin}, Anatoly and {Holtzman}, Jon and {Primack}, Joel and {Regos}, Eniko},
        title = "{Structure Formation with Cold plus Hot Dark Matter}",
      journal = {\apj},
         year = 1993,
        month = oct,
       volume = {416},
        pages = {1},
          doi = {10.1086/173210},
archivePrefix = {arXiv},
       eprint = {astro-ph/9305011},
 primaryClass = {astro-ph},
       adsurl = {https://ui.adsabs.harvard.edu/abs/1993ApJ...416....1K}
}

@ARTICLE{Vogelsberger:2020,
       author = {{Vogelsberger}, Mark and {Marinacci}, Federico and {Torrey}, Paul and {Puchwein}, Ewald},
        title = "{Cosmological simulations of galaxy formation}",
      journal = {Nature Reviews Physics},
         year = 2020,
        month = jan,
       volume = {2},
       number = {1},
        pages = {42-66},
          doi = {10.1038/s42254-019-0127-2},
archivePrefix = {arXiv},
       eprint = {1909.07976},
 primaryClass = {astro-ph.GA},
       adsurl = {https://ui.adsabs.harvard.edu/abs/2020NatRP...2...42V}
}

@ARTICLE{Rodriguez-Puebla:2016,
       author = {{Rodr{\'\i}guez-Puebla}, Aldo and {Behroozi}, Peter and {Primack}, Joel and {Klypin}, Anatoly and {Lee}, Christoph and {Hellinger}, Doug},
        title = "{Halo and subhalo demographics with Planck cosmological parameters: Bolshoi-Planck and MultiDark-Planck simulations}",
      journal = {\mnras},
         year = 2016,
        month = oct,
       volume = {462},
       number = {1},
        pages = {893-916},
          doi = {10.1093/mnras/stw1705},
archivePrefix = {arXiv},
       eprint = {1602.04813},
 primaryClass = {astro-ph.CO},
       adsurl = {https://ui.adsabs.harvard.edu/abs/2016MNRAS.462..893R}
}

@ARTICLE{Tinker:2008,
       author = {{Tinker}, Jeremy and {Kravtsov}, Andrey V. and {Klypin}, Anatoly and {Abazajian}, Kevork and {Warren}, Michael and {Yepes}, Gustavo and {Gottl{\"o}ber}, Stefan and {Holz}, Daniel E.},
        title = "{Toward a Halo Mass Function for Precision Cosmology: The Limits of Universality}",
      journal = {\apj},
         year = 2008,
        month = dec,
       volume = {688},
       number = {2},
        pages = {709-728},
          doi = {10.1086/591439},
archivePrefix = {arXiv},
       eprint = {0803.2706},
 primaryClass = {astro-ph},
       adsurl = {https://ui.adsabs.harvard.edu/abs/2008ApJ...688..709T}
}

@ARTICLE{Naidu:2022,
       author = {{Naidu}, Rohan P. and {Oesch}, Pascal A. and {Setton}, David J. and {Matthee}, Jorryt and {Conroy}, Charlie and {Johnson}, Benjamin D. and {Weaver}, John R. and {Bouwens}, Rychard J. and {Brammer}, Gabriel B. and {Dayal}, Pratika and {Illingworth}, Garth D. and {Barrufet}, Laia and {Belli}, Sirio and {Bezanson}, Rachel and {Bose}, Sownak and {Heintz}, Kasper E. and {Leja}, Joel and {Leonova}, Ecaterina and {Marques-Chaves}, Rui and {Stefanon}, Mauro and {Toft}, Sune and {van der Wel}, Arjen and {van Dokkum}, Pieter and {Weibel}, Andrea and {Whitaker}, Katherine E.},
        title = "{Schrodinger's Galaxy Candidate: Puzzlingly Luminous at $z\approx17$, or Dusty/Quenched at $z\approx5$?}",
      journal = {arXiv e-prints},
         year = 2022,
        month = aug,
          eid = {arXiv:2208.02794},
        pages = {arXiv:2208.02794},
          doi = {10.48550/arXiv.2208.02794},
archivePrefix = {arXiv},
       eprint = {2208.02794},
 primaryClass = {astro-ph.GA},
       adsurl = {https://ui.adsabs.harvard.edu/abs/2022arXiv220802794N}
}

@ARTICLE{Chworowsky:2024,
       author = {{Chworowsky}, Katherine and {Finkelstein}, Steven L. and {Boylan-Kolchin}, Michael and {McGrath}, Elizabeth J. and {Iyer}, Kartheik G. and {Papovich}, Casey and {Dickinson}, Mark and {Taylor}, Anthony J. and {Yung}, L.~Y. Aaron and {Arrabal Haro}, Pablo and {Bagley}, Micaela B. and {Backhaus}, Bren E. and {Bhatawdekar}, Rachana and {Cheng}, Yingjie and {Cleri}, Nikko J. and {Cole}, Justin W. and {Cooper}, M.~C. and {Costantin}, Luca and {Dekel}, Avishai and {Franco}, Maximilien and {Fujimoto}, Seiji and {Hayward}, Christopher C. and {Holwerda}, Benne W. and {Huertas-Company}, Marc and {Hirschmann}, Michaela and {Hutchison}, Taylor A. and {Koekemoer}, Anton M. and {Larson}, Rebecca L. and {Li}, Zhaozhou and {Long}, Arianna S. and {Lucas}, Ray A. and {Pirzkal}, Nor and {Rodighiero}, Giulia and {Somerville}, Rachel S. and {Vanderhoof}, Brittany N. and {de la Vega}, Alexander and {Wilkins}, Stephen M. and {Yang}, Guang and {Zavala}, Jorge A.},
        title = "{Evidence for a Shallow Evolution in the Volume Densities of Massive Galaxies at z = 4{\textendash}8 from CEERS}",
      journal = {\aj},
         year = 2024,
        month = sep,
       volume = {168},
       number = {3},
          eid = {113},
        pages = {113},
          doi = {10.3847/1538-3881/ad57c1},
archivePrefix = {arXiv},
       eprint = {2311.14804},
 primaryClass = {astro-ph.GA},
       adsurl = {https://ui.adsabs.harvard.edu/abs/2024AJ....168..113C}
}

@ARTICLE{Jespersen:2025,
       author = {{Jespersen}, Christian Kragh and {Steinhardt}, Charles L. and {Somerville}, Rachel S. and {Lovell}, Christopher C.},
        title = "{On the Significance of Rare Objects at High Redshift: The Impact of Cosmic Variance}",
      journal = {\apj},
         year = 2025,
        month = mar,
       volume = {982},
       number = {1},
          eid = {23},
        pages = {23},
          doi = {10.3847/1538-4357/adb422},
archivePrefix = {arXiv},
       eprint = {2403.00050},
 primaryClass = {astro-ph.GA},
       adsurl = {https://ui.adsabs.harvard.edu/abs/2025ApJ...982...23J}
}

@ARTICLE{Perez-Gonzalez:2025,
       author = {{P{\'e}rez-Gonz{\'a}lez}, Pablo G. and {{\"O}stlin}, G{\"o}ran and {Costantin}, Luca and {Melinder}, Jens and {Finkelstein}, Steven L. and {Somerville}, Rachel S. and {Annunziatella}, Marianna and {{\'A}lvarez-M{\'a}rquez}, Javier and {Colina}, Luis and {Dekel}, Avishai and {Ferguson}, Henry C. and {Li}, Zhaozhou and {Yung}, L.~Y. Aaron and {Bagley}, Micaela B. and {Boogaard}, Leindert A. and {Burgarella}, Denis and {Calabr{\`o}}, Antonello and {Caputi}, Karina I. and {Cheng}, Yingjie and {Dickinson}, Mark and {Eckart}, Andreas and {Giavalisco}, Mauro and {Gillman}, Steven and {Greve}, Thomas R. and {Hamed}, Mahmoud and {Hathi}, Nimish P. and {Hjorth}, Jens and {Huertas-Company}, Marc and {Kartaltepe}, Jeyhan S. and {Koekemoer}, Anton M. and {Kokorev}, Vasily and {Labiano}, {\'A}lvaro and {Langeroodi}, Danial and {Leung}, Gene C.~K. and {Natarajan}, Priyamvada and {Papovich}, Casey and {Peissker}, Florian and {Pentericci}, Laura and {Pirzkal}, Nor and {Rinaldi}, Pierluigi and {van der Werf}, Paul and {Walter}, Fabian},
        title = "{The Rise of the Galactic Empire: Ultraviolet Luminosity Functions at z {\ensuremath{\sim}} 17 and z {\ensuremath{\sim}} 25 Estimated with the MIDIS+NGDEEP Ultra-deep JWST/NIRCam Data Set}",
      journal = {\apj},
         year = 2025,
        month = oct,
       volume = {991},
       number = {2},
          eid = {179},
        pages = {179},
          doi = {10.3847/1538-4357/adf8c9},
archivePrefix = {arXiv},
       eprint = {2503.15594},
 primaryClass = {astro-ph.GA},
       adsurl = {https://ui.adsabs.harvard.edu/abs/2025ApJ...991..179P}
}

@ARTICLE{Castellano:2025,
       author = {{Castellano}, M. and {Fontana}, A. and {Merlin}, E. and {Santini}, P. and {Napolitano}, L. and {Menci}, N. and {P{\'e}rez-Gonz{\'a}lez}, P.~G. and {Calabr{\`o}}, A. and {Paris}, D. and {Pentericci}, L. and {Zavala}, J.~A. and {Dickinson}, M. and {Finkelstein}, S.~L. and {Treu}, T. and {Amorin}, R.~O. and {Arrabal Haro}, P. and {Bergamini}, P. and {Bisigello}, L. and {Catone}, M. and {Daddi}, E. and {Dayal}, P. and {Dekel}, A. and {Ferrara}, A. and {Fortuni}, F. and {Gandolfi}, G. and {Giavalisco}, M. and {Grillo}, C. and {Guida}, S.~T. and {Hathi}, N.~P. and {Holwerda}, B.~W. and {Koekemoer}, A.~M. and {Kokorev}, V. and {Li}, Z. and {Llerena}, M. and {Lucas}, R.~A. and {Mascia}, S. and {Metha}, B. and {Morishita}, T. and {Nanayakkara}, T. and {Pacucci}, F. and {Roberts-Borsani}, G. and {Rodighiero}, G. and {Rosati}, P. and {Salazar}, V. and {Schneider}, R. and {Somerville}, R.~S. and {Taylor}, A. and {Trenti}, M. and {Trinca}, A. and {Wang}, X. and {Watson}, P.~J. and {Yang}, L. and {Yung}, L.~Y.~A.},
        title = "{Pushing JWST to the extremes: Search and scrutiny of bright galaxy candidates at z ≃ 15─30}",
      journal = {\aap},
         year = 2025,
        month = dec,
       volume = {704},
          eid = {A158},
        pages = {A158},
          doi = {10.1051/0004-6361/202555082},
archivePrefix = {arXiv},
       eprint = {2504.05893},
 primaryClass = {astro-ph.GA},
       adsurl = {https://ui.adsabs.harvard.edu/abs/2025A&A...704A.158C}
}

@ARTICLE{Gandolfi:2025,
       author = {{Gandolfi}, G. and {Rodighiero}, G. and {Castellano}, M. and {Fontana}, A. and {Santini}, P. and {Dickinson}, M. and {Finkelstein}, S. and {Catone}, M. and {Calabr{\`o}}, A. and {Merlin}, E. and {Pentericci}, L. and {Bisigello}, L. and {Grazian}, A. and {Napolitano}, L. and {Vulcani}, B. and {Taylor}, A.~J. and {Arrabal Haro}, P. and {Kirkpatrick}, A. and {Backhaus}, B.~E. and {Holwerda}, B.~W. and {Giulietti}, M. and {Bianchetti}, A. and {Cassata}, P. and {Cleri}, N.~J. and {Daddi}, E. and {Ferguson}, H.~C. and {Girardi}, G. and {Hirschmann}, M. and {Koekemoer}, A.~M. and {Lapi}, A. and {Pacucci}, F. and {P{\'e}rez-Gonz{\'a}lez}, P.~G. and {de la Vega}, A. and {Vietri}, A. and {Wilkins}, S. and {Yung}, L.~Y.~A. and {Bagley}, M. and {Bhatawdekar}, R. and {Kartaltepe}, J. and {Papovich}, C. and {Pirzkal}, N.},
        title = "{Mysteries of Capotauro: Investigating the puzzling nature of an extreme F356W-dropout}",
      journal = {\aap},
         year = 2026,
        month = feb,
       volume = {706},
          eid = {A364},
        pages = {A364},
          doi = {10.1051/0004-6361/202557061},
archivePrefix = {arXiv},
       eprint = {2509.01664},
 primaryClass = {astro-ph.GA},
       adsurl = {https://ui.adsabs.harvard.edu/abs/2026A&A...706A.364G}
}

@ARTICLE{Tacchella:2018,
       author = {{Tacchella}, Sandro and {Bose}, Sownak and {Conroy}, Charlie and {Eisenstein}, Daniel J. and {Johnson}, Benjamin D.},
        title = "{A Redshift-independent Efficiency Model: Star Formation and Stellar Masses in Dark Matter Halos at z {\ensuremath{\gtrsim}} 4}",
      journal = {\apj},
         year = 2018,
        month = dec,
       volume = {868},
       number = {2},
          eid = {92},
        pages = {92},
          doi = {10.3847/1538-4357/aae8e0},
archivePrefix = {arXiv},
       eprint = {1806.03299},
 primaryClass = {astro-ph.GA},
       adsurl = {https://ui.adsabs.harvard.edu/abs/2018ApJ...868...92T}
}

@ARTICLE{Donnan:2025,
       author = {{Donnan}, C.~T. and {Dunlop}, J.~S. and {McLure}, R.~J. and {McLeod}, D.~J. and {Cullen}, F.},
        title = "{No evidence (yet) for increased star-formation efficiency at early times}",
      journal = {\mnras},
         year = 2025,
        month = may,
       volume = {539},
       number = {3},
        pages = {2409-2423},
          doi = {10.1093/mnras/staf641},
archivePrefix = {arXiv},
       eprint = {2501.03217},
 primaryClass = {astro-ph.GA},
       adsurl = {https://ui.adsabs.harvard.edu/abs/2025MNRAS.539.2409D}
}

@ARTICLE{Bullock:2001,
       author = {{Bullock}, J.~S. and {Kolatt}, T.~S. and {Sigad}, Y. and {Somerville}, R.~S. and {Kravtsov}, A.~V. and {Klypin}, A.~A. and {Primack}, J.~R. and {Dekel}, A.},
        title = "{Profiles of dark haloes: evolution, scatter and environment}",
      journal = {\mnras},
         year = 2001,
        month = mar,
       volume = {321},
       number = {3},
        pages = {559-575},
          doi = {10.1046/j.1365-8711.2001.04068.x},
archivePrefix = {arXiv},
       eprint = {astro-ph/9908159},
 primaryClass = {astro-ph},
       adsurl = {https://ui.adsabs.harvard.edu/abs/2001MNRAS.321..559B}
}

@ARTICLE{Wechsler:2002,
       author = {{Wechsler}, Risa H. and {Bullock}, James S. and {Primack}, Joel R. and {Kravtsov}, Andrey V. and {Dekel}, Avishai},
        title = "{Concentrations of Dark Halos from Their Assembly Histories}",
      journal = {\apj},
         year = 2002,
        month = mar,
       volume = {568},
       number = {1},
        pages = {52-70},
          doi = {10.1086/338765},
archivePrefix = {arXiv},
       eprint = {astro-ph/0108151},
 primaryClass = {astro-ph},
       adsurl = {https://ui.adsabs.harvard.edu/abs/2002ApJ...568...52W}
}

@ARTICLE{Navarro:1997,
       author = {{Navarro}, Julio F. and {Frenk}, Carlos S. and {White}, Simon D.~M.},
        title = "{A Universal Density Profile from Hierarchical Clustering}",
      journal = {\apj},
         year = 1997,
        month = dec,
       volume = {490},
       number = {2},
        pages = {493-508},
          doi = {10.1086/304888},
archivePrefix = {arXiv},
       eprint = {astro-ph/9611107},
 primaryClass = {astro-ph},
       adsurl = {https://ui.adsabs.harvard.edu/abs/1997ApJ...490..493N}
}

@ARTICLE{Doroshkevich:1970,
    author = {Doroshkevich, A. G.},
    journal = {Astrofizika},
    year = 1970,
    volume = 6,
    pages = {581},
}

@ARTICLE{Maller:2002,
    author = {{Maller}, A.~H. and {Dekel}, A. and {Somerville}, R.S.},
    title = "{Modelling angular-momentum history in dark-matter haloes}",
    journal = {\mnras},
    year = 2002,
    month = jan,
    volume = 329,
    pages = {423-430},
    adsurl = {http://adsabs.harvard.edu/cgi-bin/nph-bib_query?bibcode=2002MNRAS.329..423M&amp;db_key=AST}
}

@ARTICLE{Bullock_spin:2001,
       author = {{Bullock}, J.~S. and {Dekel}, A. and {Kolatt}, T.~S. and {Kravtsov}, A.~V. and {Klypin}, A.~A. and {Porciani}, C. and {Primack}, J.~R.},
        title = "{A Universal Angular Momentum Profile for Galactic Halos}",
      journal = {\apj},
         year = 2001,
        month = jul,
       volume = {555},
       number = {1},
        pages = {240-257},
          doi = {10.1086/321477},
archivePrefix = {arXiv},
       eprint = {astro-ph/0011001},
 primaryClass = {astro-ph},
       adsurl = {https://ui.adsabs.harvard.edu/abs/2001ApJ...555..240B}
}

@ARTICLE{Vitvitska:2002,
       author = {{Vitvitska}, Maya and {Klypin}, Anatoly A. and {Kravtsov}, Andrey V. and {Wechsler}, Risa H. and {Primack}, Joel R. and {Bullock}, James S.},
        title = "{The Origin of Angular Momentum in Dark Matter Halos}",
      journal = {\apj},
         year = 2002,
        month = dec,
       volume = {581},
       number = {2},
        pages = {799-809},
          doi = {10.1086/344361},
archivePrefix = {arXiv},
       eprint = {astro-ph/0105349},
 primaryClass = {astro-ph},
       adsurl = {https://ui.adsabs.harvard.edu/abs/2002ApJ...581..799V}
}

@ARTICLE{Shen:2003,
   author = {{Shen}, S. and others
	},
    title = "{The size distribution of galaxies in the Sloan Digital Sky Survey}",
  journal = {\mnras},
     year = 2003,
    month = aug,
   volume = 343,
    pages = {978-994},
   adsurl = {http://adsabs.harvard.edu/cgi-bin/nph-bib_query?bibcode=2003MNRAS.343..978S&db_key=AST}
}

@ARTICLE{Ravindranath:2004,
   author = {{Ravindranath}, S. and others},
    title = "{The Evolution of Disk Galaxies in the GOODS-South Field: Number Densities and Size Distribution}",
  journal = {\apjl},
     year = 2004,
    month = mar,
   volume = 604,
    pages = {L9-L12},
   adsurl = {http://adsabs.harvard.edu/cgi-bin/nph-bib_query?bibcode=2004ApJ...604L...9R&db_key=AST}
}

@ARTICLE{Ferguson:2004,
   author = {{Ferguson}, H.~C. and others},
    title = "{The Size Evolution of High-Redshift Galaxies}",
  journal = {\apjl},
     year = 2004,
    month = jan,
   volume = 600,
    pages = {L107-L110},
   adsurl = {http://adsabs.harvard.edu/cgi-bin/nph-bib_query?bibcode=2004ApJ...600L.107F&db_key=AST}
}

@ARTICLE{vanderwel:2014,
   author = {{van der Wel}, A. and others},
    title = "{3D-HST+CANDELS: The Evolution of the Galaxy Size-Mass Distribution since z = 3}",
  journal = {\apj},
archivePrefix = "arXiv",
   eprint = {1404.2844},
     year = 2014,
    month = jun,
   volume = 788,
      eid = {28},
    pages = {28},
      doi = {10.1088/0004-637X/788/1/28},
   adsurl = {http://adsabs.harvard.edu/abs/2014ApJ...788...28V}
}

@ARTICLE{lange:2015,
   author = {{Lange}, R. and others
	},
    title = "{Galaxy And Mass Assembly (GAMA): mass-size relations of z $\lt$ 0.1 galaxies subdivided by S{\'e}rsic index, colour and morphology}",
  journal = {\mnras},
archivePrefix = "arXiv",
   eprint = {1411.6355},
     year = 2015,
    month = mar,
   volume = 447,
    pages = {2603-2630},
      doi = {10.1093/mnras/stu2467},
   adsurl = {http://adsabs.harvard.edu/abs/2015MNRAS.447.2603L}
}

@ARTICLE{White:1984,
   author = {{White}, S.~D.~M.},
    title = "{Angular momentum growth in protogalaxies}",
  journal = {\apj},
     year = 1984,
    month = nov,
   volume = 286,
    pages = {38-41},
      doi = {10.1086/162573},
   adsurl = {http://adsabs.harvard.edu/abs/1984ApJ...286...38W}
}

@ARTICLE{Porciani:2002,
   author = {{Porciani}, C. and {Dekel}, A. and {Hoffman}, Y.},
    title = "{Testing tidal-torque theory - I. Spin amplitude and direction}",
  journal = {\mnras},
   eprint = {astro-ph/0105123},
     year = 2002,
    month = may,
   volume = 332,
    pages = {325-338},
      doi = {10.1046/j.1365-8711.2002.05305.x},
   adsurl = {http://adsabs.harvard.edu/abs/2002MNRAS.332..325P}
}

@ARTICLE{Shibuya:2015,
   author = {{Shibuya}, T. and {Ouchi}, M. and {Harikane}, Y.},
    title = "{Morphologies of {\tilde}190,000 Galaxies at z = 0-10 Revealed with HST Legacy Data. I. Size Evolution}",
  journal = {ApJs},
archivePrefix = "arXiv",
   eprint = {1503.07481},
     year = 2015,
    month = aug,
   volume = 219,
      eid = {15},
    pages = {15},
      doi = {10.1088/0067-0049/219/2/15},
   adsurl = {http://adsabs.harvard.edu/abs/2015ApJS..219...15S}
}

@ARTICLE{Kravtsov:2013,
   author = {{Kravtsov}, A.~V.},
    title = "{The Size-Virial Radius Relation of Galaxies}",
  journal = {\apjl},
archivePrefix = "arXiv",
   eprint = {1212.2980},
     year = 2013,
    month = feb,
   volume = 764,
      eid = {L31},
    pages = {L31},
      doi = {10.1088/2041-8205/764/2/L31},
   adsurl = {http://adsabs.harvard.edu/abs/2013ApJ...764L..31K}
}

@ARTICLE{Blumenthal:1986,
        AUTHOR  = {Blumenthal, G. and Faber, S.M. and Flores, R. and 
                   Primack, J.R.},
        YEAR    = 1986,
        JOURNAL = {\apj},
        VOLUME  = 301,
        PAGES = 27 
}

@ARTICLE{Dalcanton:1997,
        AUTHOR  = {Dalcanton, J.J. and Spergel, D.N. and Summers, F.J.},
        YEAR    = 1997,
        JOURNAL = {\apj},
        VOLUME  = 482,
        PAGES = 659
}

@ARTICLE{Huang:2017,
       author = {{Huang}, Kuang-Han and {Fall}, S. Michael and {Ferguson}, Henry C. and {van der Wel}, Arjen and {Grogin}, Norman and {Koekemoer}, Anton and {Lee}, Seong-Kook and {P{\'e}rez-Gonz{\'a}lez}, Pablo G. and {Wuyts}, Stijn},
        title = "{Relations between the Sizes of Galaxies and Their Dark Matter Halos at Redshifts 0 < z < 3}",
      journal = {\apj},
         year = 2017,
        month = mar,
       volume = {838},
       number = {1},
          eid = {6},
        pages = {6},
          doi = {10.3847/1538-4357/aa62a6},
archivePrefix = {arXiv},
       eprint = {1701.04001},
 primaryClass = {astro-ph.GA},
       adsurl = {https://ui.adsabs.harvard.edu/abs/2017ApJ...838....6H}
}

@ARTICLE{Morishita:2024,
       author = {{Morishita}, Takahiro and {Stiavelli}, Massimo and {Chary}, Ranga-Ram and {Trenti}, Michele and {Bergamini}, Pietro and {Chiaberge}, Marco and {Leethochawalit}, Nicha and {Roberts-Borsani}, Guido and {Shen}, Xuejian and {Treu}, Tommaso},
        title = "{Enhanced Subkiloparsec-scale Star Formation: Results from a JWST Size Analysis of 341 Galaxies at 5 < z < 14}",
      journal = {\apj},
         year = 2024,
        month = mar,
       volume = {963},
       number = {1},
          eid = {9},
        pages = {9},
          doi = {10.3847/1538-4357/ad1404},
archivePrefix = {arXiv},
       eprint = {2308.05018},
 primaryClass = {astro-ph.GA},
       adsurl = {https://ui.adsabs.harvard.edu/abs/2024ApJ...963....9M}
}

@ARTICLE{Allen:2025,
       author = {{Allen}, Natalie and {Oesch}, Pascal A. and {Toft}, Sune and {Matharu}, Jasleen and {McPartland}, Conor J.~R. and {Weibel}, Andrea and {Brammer}, Gabe and {Bowler}, Rebecca A.~A. and {Ito}, Kei and {Gottumukkala}, Rashmi and {Rizzo}, Francesca and {Valentino}, Francesco and {Varadaraj}, Rohan G. and {Weaver}, John R. and {Whitaker}, Katherine E.},
        title = "{Galaxy size and mass build-up in the first 2 Gyr of cosmic history from multi-wavelength JWST NIRCam imaging}",
      journal = {\aap},
         year = 2025,
        month = jun,
       volume = {698},
          eid = {A30},
        pages = {A30},
          doi = {10.1051/0004-6361/202452690},
archivePrefix = {arXiv},
       eprint = {2410.16354},
 primaryClass = {astro-ph.GA},
       adsurl = {https://ui.adsabs.harvard.edu/abs/2025A&A...698A..30A}
}

@ARTICLE{Oesch:2018,
       author = {{Oesch}, P.~A. and {Bouwens}, R.~J. and {Illingworth}, G.~D. and {Labb{\'e}}, I. and {Stefanon}, M.},
        title = "{The Dearth of z {\ensuremath{\sim}} 10 Galaxies in All HST Legacy Fields{\textemdash}The Rapid Evolution of the Galaxy Population in the First 500 Myr}",
      journal = {\apj},
         year = 2018,
        month = mar,
       volume = {855},
       number = {2},
          eid = {105},
        pages = {105},
          doi = {10.3847/1538-4357/aab03f},
archivePrefix = {arXiv},
       eprint = {1710.11131},
 primaryClass = {astro-ph.GA},
       adsurl = {https://ui.adsabs.harvard.edu/abs/2018ApJ...855..105O}
}

@ARTICLE{Clausen:2025,
       author = {{Clausen}, Maike and {Momcheva}, Ivelina G. and {Whitaker}, Katherine E. and {Cutler}, Sam E. and {Bezanson}, Rachel S. and {Dunlop}, James S. and {Grogin}, Norman A. and {Koekemoer}, Anton M. and {McLeod}, Derek J. and {McLure}, Ross J. and {Miller}, Tim B. and {Nelson}, Erica J. and {van der Wel}, Arjen and {Wake}, David A. and {Wuyts}, Stijn},
        title = "{The Evolution of Half-mass Radii and Color Gradients for Young and Old Quiescent Galaxies at 0.5 < z < 3 with JWST/PRIMER}",
      journal = {\apj},
         year = 2025,
        month = nov,
       volume = {993},
       number = {1},
          eid = {106},
        pages = {106},
          doi = {10.3847/1538-4357/ae03aa},
archivePrefix = {arXiv},
       eprint = {2501.04788},
 primaryClass = {astro-ph.GA},
       adsurl = {https://ui.adsabs.harvard.edu/abs/2025ApJ...993..106C}
}

@ARTICLE{Birnboim:2003,
       author = {{Birnboim}, Yuval and {Dekel}, Avishai},
        title = "{Virial shocks in galactic haloes?}",
      journal = {\mnras},
         year = 2003,
        month = oct,
       volume = {345},
       number = {1},
        pages = {349-364},
          doi = {10.1046/j.1365-8711.2003.06955.x},
archivePrefix = {arXiv},
       eprint = {astro-ph/0302161},
 primaryClass = {astro-ph},
       adsurl = {https://ui.adsabs.harvard.edu/abs/2003MNRAS.345..349B}
}

@ARTICLE{Dekel:2006,
       author = {{Dekel}, Avishai and {Birnboim}, Yuval},
        title = "{Galaxy bimodality due to cold flows and shock heating}",
      journal = {\mnras},
         year = 2006,
        month = may,
       volume = {368},
       number = {1},
        pages = {2-20},
          doi = {10.1111/j.1365-2966.2006.10145.x},
archivePrefix = {arXiv},
       eprint = {astro-ph/0412300},
 primaryClass = {astro-ph},
       adsurl = {https://ui.adsabs.harvard.edu/abs/2006MNRAS.368....2D}
}

@ARTICLE{Katz:1996,
   author = {{Katz}, N. and {Weinberg}, D.~H. and {Hernquist}, L.},
    title = "{Cosmological Simulations with TreeSPH}",
  journal = {ApJs},
   eprint = {astro-ph/9509107},
     year = 1996,
    month = jul,
   volume = 105,
    pages = {19},
      doi = {10.1086/192305},
   adsurl = {http://adsabs.harvard.edu/abs/1996ApJS..105...19K}
}

@ARTICLE{Sutherland:1993,
   author = {{Sutherland}, R.~S. and {Dopita}, M.~A.},
    title = "{Cooling functions for low-density astrophysical plasmas}",
  journal = {ApJs},
     year = 1993,
    month = sep,
   volume = 88,
    pages = {253},
      doi = {10.1086/191823},
   adsurl = {http://adsabs.harvard.edu/abs/1993ApJS...88..253S}
}

@ARTICLE{Wiersma:2009,
   author = {{Wiersma}, R.~P.~C. and {Schaye}, J. and {Smith}, B.~D.},
    title = "{The effect of photoionization on the cooling rates of enriched, astrophysical plasmas}",
  journal = {\mnras},
archivePrefix = "arXiv",
   eprint = {0807.3748},
     year = 2009,
    month = feb,
   volume = 393,
    pages = {99},
      doi = {10.1111/j.1365-2966.2008.14191.x},
   adsurl = {http://adsabs.harvard.edu/abs/2009MNRAS.393...99W}
}

@ARTICLE{Smith:2008,
       author = {{Smith}, Britton and {Sigurdsson}, Steinn and {Abel}, Tom},
        title = "{Metal cooling in simulations of cosmic structure formation}",
      journal = {\mnras},
         year = 2008,
        month = apr,
       volume = {385},
       number = {3},
        pages = {1443-1454},
          doi = {10.1111/j.1365-2966.2008.12922.x},
archivePrefix = {arXiv},
       eprint = {0706.0754},
 primaryClass = {astro-ph},
       adsurl = {https://ui.adsabs.harvard.edu/abs/2008MNRAS.385.1443S}
}

@ARTICLE{Klessen:2023,
       author = {{Klessen}, Ralf S. and {Glover}, Simon C.~O.},
        title = "{The First Stars: Formation, Properties, and Impact}",
      journal = {\araa},
         year = 2023,
        month = aug,
       volume = {61},
        pages = {65-130},
          doi = {10.1146/annurev-astro-071221-053453},
archivePrefix = {arXiv},
       eprint = {2303.12500},
 primaryClass = {astro-ph.CO},
       adsurl = {https://ui.adsabs.harvard.edu/abs/2023ARA&A..61...65K}
}

@ARTICLE{Schauer:2021,
       author = {{Schauer}, Anna T.~P. and {Glover}, Simon C.~O. and {Klessen}, Ralf S. and {Clark}, Paul},
        title = "{The influence of streaming velocities and Lyman-Werner radiation on the formation of the first stars}",
      journal = {\mnras},
         year = 2021,
        month = oct,
       volume = {507},
       number = {2},
        pages = {1775-1787},
          doi = {10.1093/mnras/stab1953},
archivePrefix = {arXiv},
       eprint = {2008.05663},
 primaryClass = {astro-ph.GA},
       adsurl = {https://ui.adsabs.harvard.edu/abs/2021MNRAS.507.1775S}
}

@ARTICLE{Tegmark:1997,
       author = {{Tegmark}, Max and {Silk}, Joseph and {Rees}, Martin J. and {Blanchard}, Alain and {Abel}, Tom and {Palla}, Francesco},
        title = "{How Small Were the First Cosmological Objects?}",
      journal = {\apj},
         year = 1997,
        month = jan,
       volume = {474},
        pages = {1},
          doi = {10.1086/303434},
archivePrefix = {arXiv},
       eprint = {astro-ph/9603007},
 primaryClass = {astro-ph},
       adsurl = {https://ui.adsabs.harvard.edu/abs/1997ApJ...474....1T}
}

@ARTICLE{Draine:2003,
       author = {{Draine}, B.~T.},
        title = "{Interstellar Dust Grains}",
      journal = {\araa},
         year = 2003,
        month = jan,
       volume = {41},
        pages = {241-289},
          doi = {10.1146/annurev.astro.41.011802.094840},
archivePrefix = {arXiv},
       eprint = {astro-ph/0304489},
 primaryClass = {astro-ph},
       adsurl = {https://ui.adsabs.harvard.edu/abs/2003ARA&A..41..241D}
}

@book{Drainebook2011,
    author = {Draine, Bruce T.},
    title = {Physics of the Interstellar and Intergalactic Medium},
    publisher = {Princeton University Press},
    address = {Princeton, N.J.},
    year = {2011},
    series = {Princeton Series in Astrophysics},
    note = {Bibcode: 2011piim.book.....D},
    isbn = {978-0691122144}
}

@ARTICLE{Bialy:2019,
       author = {{Bialy}, Shmuel and {Sternberg}, Amiel},
        title = "{Thermal Phases of the Neutral Atomic Interstellar Medium from Solar Metallicity to Primordial Gas}",
      journal = {\apj},
         year = 2019,
        month = aug,
       volume = {881},
       number = {2},
          eid = {160},
        pages = {160},
          doi = {10.3847/1538-4357/ab2fd1},
archivePrefix = {arXiv},
       eprint = {1902.06764},
 primaryClass = {astro-ph.GA},
       adsurl = {https://ui.adsabs.harvard.edu/abs/2019ApJ...881..160B}
}

@ARTICLE{Thoul:1996,
    author = {{Thoul}, A.~A. and {Weinberg}, D.~H.},
    title = "{Hydrodynamic Simulations of Galaxy Formation. II. Photoionization and the Formation of Low-Mass Galaxies}",
    journal = {\apj},
    year = 1996,
    month = jul,
    volume = 465,
    pages = {608},
    adsurl = {http://adsabs.harvard.edu/cgi-bin/nph-bib_query?bibcode=1996ApJ...465..608T&amp;db_key=AST}
}

@ARTICLE{Quinn:1996,
    author = {{Quinn}, T. and {Katz}, N. and {Efstathiou}, G.},
    title = "{Photoionization and the formation of dwarf galaxies}",
    journal = {\mnras},
    year = 1996,
    month = feb,
    volume = 278,
    pages = {49},
    adsurl = {http://adsabs.harvard.edu/cgi-bin/nph-bib_query?bibcode=1996MNRAS.278L..49Q&amp;db_key=AST}
}

@ARTICLE{Gnedin:2000,
    author = {{Gnedin}, N.~Y.},
    title = "{Effect of Reionization on Structure Formation in the Universe}",
    journal = {\apj},
    year = 2000,
    month = oct,
    volume = 542,
    pages = {535},
    adsurl = {http://adsabs.harvard.edu/cgi-bin/nph-bib_query?bibcode=2000ApJ...542..535G&amp;db_key=AST}
}

@ARTICLE{Kravtsov:2004,
   author = {{Kravtsov}, A.~V. and {Gnedin}, O.~Y. and {Klypin}, A.~A.},
    title = "{The Tumultuous Lives of Galactic Dwarfs and the Missing Satellites Problem}",
  journal = {\apj},
   eprint = {arXiv:astro-ph/0401088},
     year = 2004,
    month = jul,
   volume = 609,
    pages = {482},
      doi = {10.1086/421322},
   adsurl = {http://adsabs.harvard.edu/abs/2004ApJ...609..482K}
}

@ARTICLE{Bromm:1999,
       author = {{Bromm}, Volker and {Coppi}, Paolo S. and {Larson}, Richard B.},
        title = "{Forming the First Stars in the Universe: The Fragmentation of Primordial Gas}",
      journal = {\apjl},
         year = 1999,
        month = dec,
       volume = {527},
       number = {1},
        pages = {L5-L8},
          doi = {10.1086/312385},
archivePrefix = {arXiv},
       eprint = {astro-ph/9910224},
 primaryClass = {astro-ph},
       adsurl = {https://ui.adsabs.harvard.edu/abs/1999ApJ...527L...5B}
}

@ARTICLE{Bromm:2002,
       author = {{Bromm}, Volker and {Coppi}, Paolo S. and {Larson}, Richard B.},
        title = "{The Formation of the First Stars. I. The Primordial Star-forming Cloud}",
      journal = {\apj},
         year = 2002,
        month = jan,
       volume = {564},
       number = {1},
        pages = {23-51},
          doi = {10.1086/323947},
archivePrefix = {arXiv},
       eprint = {astro-ph/0102503},
 primaryClass = {astro-ph},
       adsurl = {https://ui.adsabs.harvard.edu/abs/2002ApJ...564...23B}
}

@ARTICLE{Abel:2000,
       author = {{Abel}, Tom and {Bryan}, Greg L. and {Norman}, Michael L.},
        title = "{The Formation and Fragmentation of Primordial Molecular Clouds}",
      journal = {\apj},
         year = 2000,
        month = sep,
       volume = {540},
       number = {1},
        pages = {39-44},
          doi = {10.1086/309295},
archivePrefix = {arXiv},
       eprint = {astro-ph/0002135},
 primaryClass = {astro-ph},
       adsurl = {https://ui.adsabs.harvard.edu/abs/2000ApJ...540...39A}
}

@ARTICLE{Abel:2002,
       author = {{Abel}, Tom and {Bryan}, Greg L. and {Norman}, Michael L.},
        title = "{The Formation of the First Star in the Universe}",
      journal = {Science},
         year = 2002,
        month = jan,
       volume = {295},
       number = {5552},
        pages = {93-98},
          doi = {10.1126/science.1063991},
archivePrefix = {arXiv},
       eprint = {astro-ph/0112088},
 primaryClass = {astro-ph},
       adsurl = {https://ui.adsabs.harvard.edu/abs/2002Sci...295...93A}
}

@ARTICLE{Nakamura:2002,
       author = {{Nakamura}, Fumitaka and {Umemura}, Masayuki},
        title = "{The Stellar Initial Mass Function in Primordial Galaxies}",
      journal = {\apj},
         year = 2002,
        month = apr,
       volume = {569},
       number = {2},
        pages = {549-557},
          doi = {10.1086/339392},
archivePrefix = {arXiv},
       eprint = {astro-ph/0201497},
 primaryClass = {astro-ph},
       adsurl = {https://ui.adsabs.harvard.edu/abs/2002ApJ...569..549N}
}

@ARTICLE{Nakamura:2001,
       author = {{Nakamura}, Fumitaka and {Umemura}, Masayuki},
        title = "{On the Initial Mass Function of Population III Stars}",
      journal = {\apj},
         year = 2001,
        month = feb,
       volume = {548},
       number = {1},
        pages = {19-32},
          doi = {10.1086/318663},
archivePrefix = {arXiv},
       eprint = {astro-ph/0010464},
 primaryClass = {astro-ph},
       adsurl = {https://ui.adsabs.harvard.edu/abs/2001ApJ...548...19N}
}

@ARTICLE{Yoshida:2003,
       author = {{Yoshida}, Naoki and {Abel}, Tom and {Hernquist}, Lars and {Sugiyama}, Naoshi},
        title = "{Simulations of Early Structure Formation: Primordial Gas Clouds}",
      journal = {\apj},
         year = 2003,
        month = aug,
       volume = {592},
       number = {2},
        pages = {645-663},
          doi = {10.1086/375810},
archivePrefix = {arXiv},
       eprint = {astro-ph/0301645},
 primaryClass = {astro-ph},
       adsurl = {https://ui.adsabs.harvard.edu/abs/2003ApJ...592..645Y}
}

@ARTICLE{Clark:2011,
       author = {{Clark}, Paul C. and {Glover}, Simon C.~O. and {Klessen}, Ralf S. and {Bromm}, Volker},
        title = "{Gravitational Fragmentation in Turbulent Primordial Gas and the Initial Mass Function of Population III Stars}",
      journal = {\apj},
         year = 2011,
        month = feb,
       volume = {727},
       number = {2},
          eid = {110},
        pages = {110},
          doi = {10.1088/0004-637X/727/2/110},
archivePrefix = {arXiv},
       eprint = {1006.1508},
 primaryClass = {astro-ph.GA},
       adsurl = {https://ui.adsabs.harvard.edu/abs/2011ApJ...727..110C}
}

@ARTICLE{Greif:2011,
       author = {{Greif}, Thomas H. and {Springel}, Volker and {White}, Simon D.~M. and {Glover}, Simon C.~O. and {Clark}, Paul C. and {Smith}, Rowan J. and {Klessen}, Ralf S. and {Bromm}, Volker},
        title = "{Simulations on a Moving Mesh: The Clustered Formation of Population III Protostars}",
      journal = {\apj},
         year = 2011,
        month = aug,
       volume = {737},
       number = {2},
          eid = {75},
        pages = {75},
          doi = {10.1088/0004-637X/737/2/75},
archivePrefix = {arXiv},
       eprint = {1101.5491},
 primaryClass = {astro-ph.CO},
       adsurl = {https://ui.adsabs.harvard.edu/abs/2011ApJ...737...75G}
}

@ARTICLE{Rossi:2025,
       author = {{Rossi}, Martina and {Salvadori}, Stefania and {Sk{\'u}lad{\'o}ttir}, {\'A}sa and {Vanni}, Irene and {Koutsouridou}, Ioanna},
        title = "{Hidden Population III Descendants in Ultrafaint Dwarf Galaxies}",
      journal = {\apj},
         year = 2025,
        month = jul,
       volume = {987},
       number = {2},
          eid = {121},
        pages = {121},
          doi = {10.3847/1538-4357/add5e9},
archivePrefix = {arXiv},
       eprint = {2406.12960},
 primaryClass = {astro-ph.GA},
       adsurl = {https://ui.adsabs.harvard.edu/abs/2025ApJ...987..121R}
}

@ARTICLE{Salvadori:2019,
       author = {{Salvadori}, S. and {Bonifacio}, P. and {Caffau}, E. and {Korotin}, S. and {Andreevsky}, S. and {Spite}, M. and {Sk{\'u}lad{\'o}ttir}, {\'A}.},
        title = "{Probing the existence of very massive first stars}",
      journal = {\mnras},
         year = 2019,
        month = aug,
       volume = {487},
       number = {3},
        pages = {4261-4284},
          doi = {10.1093/mnras/stz1464},
archivePrefix = {arXiv},
       eprint = {1906.00994},
 primaryClass = {astro-ph.GA},
       adsurl = {https://ui.adsabs.harvard.edu/abs/2019MNRAS.487.4261S}
}

@ARTICLE{McKee:1977,
   author = {{McKee}, C.~F. and {Ostriker}, J.~P.},
    title = "{A theory of the interstellar medium - Three components regulated by supernova explosions in an inhomogeneous substrate}",
  journal = {\apj},
     year = 1977,
    month = nov,
   volume = 218,
    pages = {148},
      doi = {10.1086/155667},
   adsurl = {http://adsabs.harvard.edu/abs/1977ApJ...218..148M}
}

@ARTICLE{Shu:1987,
       author = {{Shu}, Frank H. and {Adams}, Fred C. and {Lizano}, Susana},
        title = "{Star formation in molecular clouds: observation and theory.}",
      journal = {\araa},
         year = 1987,
        month = jan,
       volume = {25},
        pages = {23-81},
          doi = {10.1146/annurev.aa.25.090187.000323},
       adsurl = {https://ui.adsabs.harvard.edu/abs/1987ARA&A..25...23S}
}

@ARTICLE{Maclow:2004,
       author = {{Mac Low}, Mordecai-Mark and {Klessen}, Ralf S.},
        title = "{Control of star formation by supersonic turbulence}",
      journal = {Reviews of Modern Physics},
         year = 2004,
        month = jan,
       volume = {76},
       number = {1},
        pages = {125-194},
          doi = {10.1103/RevModPhys.76.125},
archivePrefix = {arXiv},
       eprint = {astro-ph/0301093},
 primaryClass = {astro-ph},
       adsurl = {https://ui.adsabs.harvard.edu/abs/2004RvMP...76..125M}
}

@article{McKee:2007,
	Author = {Christopher F McKee and Eve C Ostriker},
	Journal = {ARA\&A},
	Month = {Sep},
	Pages = {565},
	Pmid = {2007ARA&A..45..565M},
	Title = {Theory of Star Formation},
	Volume = {45},
	Year = {2007}
}

@ARTICLE{Forbes:2023,
       author = {{Forbes}, John C. and {Emami}, Razieh and {Somerville}, Rachel S. and {Genel}, Shy and {Nelson}, Dylan and {Pillepich}, Annalisa and {Burkhart}, Blakesley and {Bryan}, Greg L. and {Krumholz}, Mark R. and {Hernquist}, Lars and {Tonnesen}, Stephanie and {Torrey}, Paul and {Pandya}, Viraj and {Hayward}, Christopher C.},
        title = "{Gas Accretion Can Drive Turbulence in Galaxies}",
      journal = {\apj},
         year = 2023,
        month = may,
       volume = {948},
       number = {2},
          eid = {107},
        pages = {107},
          doi = {10.3847/1538-4357/acb53e},
archivePrefix = {arXiv},
       eprint = {2204.05344},
 primaryClass = {astro-ph.GA},
       adsurl = {https://ui.adsabs.harvard.edu/abs/2023ApJ...948..107F}
}

@ARTICLE{Padoan:2012,
       author = {{Padoan}, Paolo and {Haugb{\o}lle}, Troels and {Nordlund}, {\r{A}}ke},
        title = "{A Simple Law of Star Formation}",
      journal = {\apjl},
         year = 2012,
        month = nov,
       volume = {759},
       number = {2},
          eid = {L27},
        pages = {L27},
          doi = {10.1088/2041-8205/759/2/L27},
archivePrefix = {arXiv},
       eprint = {1208.3758},
 primaryClass = {astro-ph.GA},
       adsurl = {https://ui.adsabs.harvard.edu/abs/2012ApJ...759L..27P}
}

@ARTICLE{Federrath:2013,
       author = {{Federrath}, Christoph and {Klessen}, Ralf S.},
        title = "{On the Star Formation Efficiency of Turbulent Magnetized Clouds}",
      journal = {\apj},
         year = 2013,
        month = jan,
       volume = {763},
       number = {1},
          eid = {51},
        pages = {51},
          doi = {10.1088/0004-637X/763/1/51},
archivePrefix = {arXiv},
       eprint = {1211.6433},
 primaryClass = {astro-ph.SR},
       adsurl = {https://ui.adsabs.harvard.edu/abs/2013ApJ...763...51F}
}

@ARTICLE{Somerville:2025,
       author = {{Somerville}, Rachel S. and {Yung}, L.~Y. Aaron and {Lancaster}, Lachlan and {Menon}, Shyam and {Sommovigo}, Laura and {Finkelstein}, Steven L.},
        title = "{Density-modulated star formation efficiency: implications for the observed abundance of ultraviolet luminous galaxies at z > 10}",
      journal = {\mnras},
         year = 2025,
        month = dec,
       volume = {544},
       number = {4},
        pages = {3774-3798},
          doi = {10.1093/mnras/staf1824},
archivePrefix = {arXiv},
       eprint = {2505.05442},
 primaryClass = {astro-ph.GA},
       adsurl = {https://ui.adsabs.harvard.edu/abs/2025MNRAS.544.3774S}
}

@ARTICLE{Kim:2020,
       author = {{Kim}, Chang-Goo and {Ostriker}, Eve C. and {Somerville}, Rachel S. and {Bryan}, Greg L. and {Fielding}, Drummond B. and {Forbes}, John C. and {Hayward}, Christopher C. and {Hernquist}, Lars and {Pandya}, Viraj},
        title = "{First Results from SMAUG: Characterization of Multiphase Galactic Outflows from a Suite of Local Star-forming Galactic Disk Simulations}",
      journal = {\apj},
         year = 2020,
        month = sep,
       volume = {900},
       number = {1},
          eid = {61},
        pages = {61},
          doi = {10.3847/1538-4357/aba962},
archivePrefix = {arXiv},
       eprint = {2006.16315},
 primaryClass = {astro-ph.GA},
       adsurl = {https://ui.adsabs.harvard.edu/abs/2020ApJ...900...61K}
}

@ARTICLE{Hu:2019,
       author = {{Hu}, Chia-Yu},
        title = "{Supernova-driven winds in simulated dwarf galaxies}",
      journal = {\mnras},
         year = 2019,
        month = mar,
       volume = {483},
       number = {3},
        pages = {3363-3381},
          doi = {10.1093/mnras/sty3252},
archivePrefix = {arXiv},
       eprint = {1805.06614},
 primaryClass = {astro-ph.GA},
       adsurl = {https://ui.adsabs.harvard.edu/abs/2019MNRAS.483.3363H}
}

@ARTICLE{Fielding:2018,
       author = {{Fielding}, Drummond and {Quataert}, Eliot and {Martizzi}, Davide},
        title = "{Clustered supernovae drive powerful galactic winds after superbubble breakout}",
      journal = {\mnras},
         year = 2018,
        month = dec,
       volume = {481},
       number = {3},
        pages = {3325-3347},
          doi = {10.1093/mnras/sty2466},
archivePrefix = {arXiv},
       eprint = {1807.08758},
 primaryClass = {astro-ph.GA},
       adsurl = {https://ui.adsabs.harvard.edu/abs/2018MNRAS.481.3325F}
}

@ARTICLE{Walch:2015a,
       author = {{Walch}, S. and {Girichidis}, P. and {Naab}, T. and {Gatto}, A. and {Glover}, S.~C.~O. and {W{\"u}nsch}, R. and {Klessen}, R.~S. and {Clark}, P.~C. and {Peters}, T. and {Derigs}, D. and {Baczynski}, C.},
        title = "{The SILCC (SImulating the LifeCycle of molecular Clouds) project - I. Chemical evolution of the supernova-driven ISM}",
      journal = {\mnras},
         year = 2015,
        month = nov,
       volume = {454},
       number = {1},
        pages = {238-268},
          doi = {10.1093/mnras/stv1975},
archivePrefix = {arXiv},
       eprint = {1412.2749},
 primaryClass = {astro-ph.GA},
       adsurl = {https://ui.adsabs.harvard.edu/abs/2015MNRAS.454..238W}
}

@ARTICLE{Walch:2015b,
       author = {{Walch}, Stefanie and {Naab}, Thorsten},
        title = "{The energy and momentum input of supernova explosions in structured and ionized molecular clouds}",
      journal = {\mnras},
         year = 2015,
        month = aug,
       volume = {451},
       number = {3},
        pages = {2757-2771},
          doi = {10.1093/mnras/stv1155},
archivePrefix = {arXiv},
       eprint = {1410.0011},
 primaryClass = {astro-ph.GA},
       adsurl = {https://ui.adsabs.harvard.edu/abs/2015MNRAS.451.2757W}
}

@ARTICLE{Emerick:2019,
       author = {{Emerick}, Andrew and {Bryan}, Greg L. and {Mac Low}, Mordecai-Mark},
        title = "{Simulating an isolated dwarf galaxy with multichannel feedback and chemical yields from individual stars}",
      journal = {\mnras},
         year = 2019,
        month = jan,
       volume = {482},
       number = {1},
        pages = {1304-1329},
          doi = {10.1093/mnras/sty2689},
archivePrefix = {arXiv},
       eprint = {1807.07182},
 primaryClass = {astro-ph.GA},
       adsurl = {https://ui.adsabs.harvard.edu/abs/2019MNRAS.482.1304E}
}

@ARTICLE{Martizzi:2015,
       author = {{Martizzi}, Davide and {Faucher-Gigu{\`e}re}, Claude-Andr{\'e} and {Quataert}, Eliot},
        title = "{Supernova feedback in an inhomogeneous interstellar medium}",
      journal = {\mnras},
         year = 2015,
        month = jun,
       volume = {450},
       number = {1},
        pages = {504-522},
          doi = {10.1093/mnras/stv562},
archivePrefix = {arXiv},
       eprint = {1409.4425},
 primaryClass = {astro-ph.GA},
       adsurl = {https://ui.adsabs.harvard.edu/abs/2015MNRAS.450..504M}
}

@ARTICLE{Gutcke:2021,
       author = {{Gutcke}, Thales A. and {Pakmor}, R{\"u}diger and {Naab}, Thorsten and {Springel}, Volker},
        title = "{LYRA - I. Simulating the multiphase ISM of a dwarf galaxy with variable energy supernovae from individual stars}",
      journal = {\mnras},
         year = 2021,
        month = mar,
       volume = {501},
       number = {4},
        pages = {5597-5615},
          doi = {10.1093/mnras/staa3875},
archivePrefix = {arXiv},
       eprint = {2010.07311},
 primaryClass = {astro-ph.GA},
       adsurl = {https://ui.adsabs.harvard.edu/abs/2021MNRAS.501.5597G}
}

@ARTICLE{Maiolino:2019,
       author = {{Maiolino}, R. and {Mannucci}, F.},
        title = "{De re metallica: the cosmic chemical evolution of galaxies}",
      journal = {\aapr},
         year = 2019,
        month = feb,
       volume = {27},
       number = {1},
          eid = {3},
        pages = {3},
          doi = {10.1007/s00159-018-0112-2},
archivePrefix = {arXiv},
       eprint = {1811.09642},
 primaryClass = {astro-ph.GA},
       adsurl = {https://ui.adsabs.harvard.edu/abs/2019A&ARv..27....3M}
}

@ARTICLE{Kobayashi:2020,
       author = {{Kobayashi}, Chiaki and {Karakas}, Amanda I. and {Lugaro}, Maria},
        title = "{The Origin of Elements from Carbon to Uranium}",
      journal = {\apj},
         year = 2020,
        month = sep,
       volume = {900},
       number = {2},
          eid = {179},
        pages = {179},
          doi = {10.3847/1538-4357/abae65},
archivePrefix = {arXiv},
       eprint = {2008.04660},
 primaryClass = {astro-ph.GA},
       adsurl = {https://ui.adsabs.harvard.edu/abs/2020ApJ...900..179K}
}

@ARTICLE{Schneider:2024,
       author = {{Schneider}, Raffaella and {Maiolino}, Roberto},
        title = "{The formation and cosmic evolution of dust in the early Universe: I. Dust sources}",
      journal = {\aapr},
         year = 2024,
        month = apr,
       volume = {32},
       number = {1},
          eid = {2},
        pages = {2},
          doi = {10.1007/s00159-024-00151-2},
archivePrefix = {arXiv},
       eprint = {2310.00053},
 primaryClass = {astro-ph.GA},
       adsurl = {https://ui.adsabs.harvard.edu/abs/2024A&ARv..32....2S}
}

@article{Kormendy:2013,
	Adsurl = {http://adsabs.harvard.edu/abs/2013ARA%26A..51..511K},
	Archiveprefix = {arXiv},
	Author = {{Kormendy}, J. and {Ho}, L.~C.},
	Doi = {10.1146/annurev-astro-082708-101811},
	Eprint = {1304.7762},
	Journal = {\araa},
	Month = aug,
	Pages = {511},
	Primaryclass = {astro-ph.CO},
	Title = {{Coevolution (Or Not) of Supermassive Black Holes and Host Galaxies}},
	Volume = 51,
	Year = 2013
}

@ARTICLE{Heckman:2014,
   author = {{Heckman}, T. and {Best}, P.},
    title = "{The Co-Evolution of Galaxies and Supermassive Black Holes: Insights from Surveys of the Contemporary Universe}",
  journal = {ArXiv e-prints},
archivePrefix = "arXiv",
   eprint = {1403.4620},
 primaryClass = "astro-ph.GA",
     year = 2014,
    month = mar,
   adsurl = {http://adsabs.harvard.edu/abs/2014arXiv1403.4620H}
}

@ARTICLE{Wechsler:2018,
       author = {{Wechsler}, Risa H. and {Tinker}, Jeremy L.},
        title = "{The Connection Between Galaxies and Their Dark Matter Halos}",
      journal = {\araa},
         year = 2018,
        month = sep,
       volume = {56},
        pages = {435-487},
          doi = {10.1146/annurev-astro-081817-051756},
archivePrefix = {arXiv},
       eprint = {1804.03097},
 primaryClass = {astro-ph.GA},
       adsurl = {https://ui.adsabs.harvard.edu/abs/2018ARA&A..56..435W}
}

@ARTICLE{Dave:2012,
   author = {{Dav{\'e}}, R. and {Finlator}, K. and {Oppenheimer}, B.~D.},
    title = "{An analytic model for the evolution of the stellar, gas and metal content of galaxies}",
  journal = {\mnras},
archivePrefix = "arXiv",
   eprint = {1108.0426},
 primaryClass = "astro-ph.CO",
     year = 2012,
    month = mar,
   volume = 421,
    pages = {98},
      doi = {10.1111/j.1365-2966.2011.20148.x},
   adsurl = {http://adsabs.harvard.edu/abs/2012MNRAS.421...98D}
}

@ARTICLE{Dekel:2014,
       author = {{Dekel}, Avishai and {Mandelker}, Nir},
        title = "{An analytic solution for the minimal bathtub toy model: challenges in the star formation history of high-z galaxies}",
      journal = {\mnras},
         year = 2014,
        month = nov,
       volume = {444},
       number = {3},
        pages = {2071-2084},
          doi = {10.1093/mnras/stu1427},
archivePrefix = {arXiv},
       eprint = {1402.2283},
 primaryClass = {astro-ph.CO},
       adsurl = {https://ui.adsabs.harvard.edu/abs/2014MNRAS.444.2071D}
}

@ARTICLE{Lilly:2013,
   author = {{Lilly}, S.~J. and {Carollo}, C.~M. and {Pipino}, A. and {Renzini}, A. and 
	{Peng}, Y.},
    title = "{Gas Regulation of Galaxies: The Evolution of the Cosmic Specific Star Formation Rate, the Metallicity-Mass-Star-formation Rate Relation, and the Stellar Content of Halos}",
  journal = {\apj},
archivePrefix = "arXiv",
   eprint = {1303.5059},
 primaryClass = "astro-ph.CO",
     year = 2013,
    month = aug,
   volume = 772,
      eid = {119},
    pages = {119},
      doi = {10.1088/0004-637X/772/2/119},
   adsurl = {http://adsabs.harvard.edu/abs/2013ApJ...772..119L}
}

@ARTICLE{Finlator:2008,
   author = {{Finlator}, K. and {Dav{\'e}}, R.},
    title = "{The origin of the galaxy mass-metallicity relation and implications for galactic outflows}",
  journal = {\mnras},
archivePrefix = "arXiv",
   eprint = {0704.3100},
     year = 2008,
    month = apr,
   volume = 385,
    pages = {2181},
      doi = {10.1111/j.1365-2966.2008.12991.x},
   adsurl = {http://adsabs.harvard.edu/abs/2008MNRAS.385.2181F}
}

@ARTICLE{Benson:2010,
   author = {{Benson}, A.~J.},
    title = "{Galaxy formation theory}",
  journal = {Physics Reports},
archivePrefix = "arXiv",
   eprint = {1006.5394},
 primaryClass = "astro-ph.CO",
     year = 2010,
    month = oct,
   volume = 495,
    pages = {33},
      doi = {10.1016/j.physrep.2010.06.001},
   adsurl = {http://adsabs.harvard.edu/abs/2010PhR...495...33B}
}

@article{Bigiel:2008,
	Adsurl = {http://adsabs.harvard.edu/abs/2008AJ....136.2846B},
	Archiveprefix = {arXiv},
	Author = {{Bigiel}, F. and {Leroy}, A. and {Walter}, F. and {Brinks}, E. and {de Blok}, W.~J.~G. and {Madore}, B. and {Thornley}, M.~D.},
	Doi = {10.1088/0004-6256/136/6/2846},
	Eprint = {0810.2541},
	Journal = {\aj},
	Month = dec,
	Pages = {2846},
	Title = {{The Star Formation Law in Nearby Galaxies on Sub-Kpc Scales}},
	Volume = 136,
	Year = 2008
}

@ARTICLE{Feldmann:2025,
       author = {{Feldmann}, Robert and {Bieri}, Rebekka},
        title = "{Cosmological Simulations of Galaxies}",
      journal = {arXiv e-prints},
         year = 2025,
        month = jul,
          eid = {arXiv:2507.08925},
        pages = {arXiv:2507.08925},
          doi = {10.48550/arXiv.2507.08925},
archivePrefix = {arXiv},
       eprint = {2507.08925},
 primaryClass = {astro-ph.GA},
       adsurl = {https://ui.adsabs.harvard.edu/abs/2025arXiv250708925F}
}

@ARTICLE{Teyssier:2015,
       author = {{Teyssier}, Romain},
        title = "{Grid-Based Hydrodynamics in Astrophysical Fluid Flows}",
      journal = {\araa},
         year = 2015,
        month = aug,
       volume = {53},
        pages = {325-364},
          doi = {10.1146/annurev-astro-082214-122309},
       adsurl = {https://ui.adsabs.harvard.edu/abs/2015ARA&A..53..325T}
}

@ARTICLE{Navarro:1995,
   author = {{Navarro}, J.~F. and {Frenk}, C.~S. and {White}, S.~D.~M.},
    title = "{The assembly of galaxies in a hierarchically clustering universe}",
  journal = {\mnras},
   eprint = {astro-ph/9408067},
     year = 1995,
    month = jul,
   volume = 275,
    pages = {56},
   adsurl = {http://adsabs.harvard.edu/abs/1995MNRAS.275...56N}
}

@ARTICLE{Steinmetz:1999,
    author = {{Steinmetz}, M.},
    title = "{Numerical Simulations of Galaxy Formation}",
    journal = {Astrophysics and Space Science},
    year = 1999,
    volume = 269,
    pages = {513},
    url = {http://adsabs.harvard.edu/cgi-bin/nph-bib_query?bibcode=1999Ap%26SS.269..513S&db_key=AST}
}

@ARTICLE{Sommer-Larsen:1999,
    author = {{Sommer-Larsen}, J. and {Gelato}, S. and {Vedel}, H.},
    title = "{Formation of Disk Galaxies: Feedback and the Angular Momentum Problem}",
    journal = {\apj},
    year = 1999,
    month = jul,
    volume = 519,
    pages = {501},
    url = {http://adsabs.harvard.edu/cgi-bin/nph-bib_query?bibcode=1999ApJ...519..501S&db_key=AST}
}

@ARTICLE{Springel:2003,
    author = {{Springel}, V. and {Hernquist}, L.},
    title = "{Cosmological smoothed particle hydrodynamics simulations: a hybrid multiphase model for star formation}",
    journal = {\mnras},
    year = 2003,
    month = feb,
    volume = 339,
    pages = {289},
    adsurl = {http://adsabs.harvard.edu/cgi-bin/nph-bib_query?bibcode=2003MNRAS.339..289S&amp;db_key=AST}
}

@ARTICLE{Schaye:2008,
   author = {{Schaye}, J. and {Dalla Vecchia}, C.},
    title = "{On the relation between the Schmidt and Kennicutt-Schmidt star formation laws and its implications for numerical simulations}",
  journal = {\mnras},
archivePrefix = "arXiv",
   eprint = {0709.0292},
     year = 2008,
    month = jan,
   volume = 383,
    pages = {1210},
      doi = {10.1111/j.1365-2966.2007.12639.x},
   adsurl = {http://adsabs.harvard.edu/abs/2008MNRAS.383.1210S}
}

@ARTICLE{Schaye:2015,
       author = {{Schaye}, Joop and {Crain}, Robert A. and {Bower}, Richard G. and {Furlong}, Michelle and {Schaller}, Matthieu and {Theuns}, Tom and {Dalla Vecchia}, Claudio and {Frenk}, Carlos S. and {McCarthy}, I.~G. and {Helly}, John C. and {Jenkins}, Adrian and {Rosas-Guevara}, Y.~M. and {White}, Simon D.~M. and {Baes}, Maarten and {Booth}, C.~M. and {Camps}, Peter and {Navarro}, Julio F. and {Qu}, Yan and {Rahmati}, Alireza and {Sawala}, Till and {Thomas}, Peter A. and {Trayford}, James},
        title = "{The EAGLE project: simulating the evolution and assembly of galaxies and their environments}",
      journal = {\mnras},
         year = 2015,
        month = jan,
       volume = {446},
       number = {1},
        pages = {521-554},
          doi = {10.1093/mnras/stu2058},
archivePrefix = {arXiv},
       eprint = {1407.7040},
 primaryClass = {astro-ph.GA},
       adsurl = {https://ui.adsabs.harvard.edu/abs/2015MNRAS.446..521S}
}

@INPROCEEDINGS{Padoan:2014,
       author = {{Padoan}, P. and {Federrath}, C. and {Chabrier}, G. and {Evans}, II, N.~J. and {Johnstone}, D. and {J{\o}rgensen}, J.~K. and {McKee}, C.~F. and {Nordlund}, {\r{A}}.},
        title = "{The Star Formation Rate of Molecular Clouds}",
    booktitle = {Protostars and Planets VI},
         year = 2014,
       editor = {{Beuther}, Henrik and {Klessen}, Ralf S. and {Dullemond}, Cornelis P. and {Henning}, Thomas},
        month = jan,
        pages = {77-100},
          doi = {10.2458/azu_uapress_9780816531240-ch004},
archivePrefix = {arXiv},
       eprint = {1312.5365},
 primaryClass = {astro-ph.GA},
       adsurl = {https://ui.adsabs.harvard.edu/abs/2014prpl.conf...77P}
}

@ARTICLE{Kretschmer:2020,
       author = {{Kretschmer}, Michael and {Teyssier}, Romain},
        title = "{Forming early-type galaxies without AGN feedback: a combination of merger-driven outflows and inefficient star formation}",
      journal = {\mnras},
         year = 2020,
        month = feb,
       volume = {492},
       number = {1},
        pages = {1385-1398},
          doi = {10.1093/mnras/stz3495},
archivePrefix = {arXiv},
       eprint = {1906.11836},
 primaryClass = {astro-ph.GA},
       adsurl = {https://ui.adsabs.harvard.edu/abs/2020MNRAS.492.1385K}
}

@ARTICLE{Semenov:2016,
       author = {{Semenov}, Vadim A. and {Kravtsov}, Andrey V. and {Gnedin}, Nickolay Y.},
        title = "{Nonuniversal Star Formation Efficiency in Turbulent ISM}",
      journal = {\apj},
         year = 2016,
        month = aug,
       volume = {826},
       number = {2},
          eid = {200},
        pages = {200},
          doi = {10.3847/0004-637X/826/2/200},
archivePrefix = {arXiv},
       eprint = {1512.03101},
 primaryClass = {astro-ph.GA},
       adsurl = {https://ui.adsabs.harvard.edu/abs/2016ApJ...826..200S}
}

@ARTICLE{Vogelsberger:2014,
       author = {{Vogelsberger}, Mark and {Genel}, Shy and {Springel}, Volker and {Torrey}, Paul and {Sijacki}, Debora and {Xu}, Dandan and {Snyder}, Greg and {Nelson}, Dylan and {Hernquist}, Lars},
        title = "{Introducing the Illustris Project: simulating the coevolution of dark and visible matter in the Universe}",
      journal = {\mnras},
         year = 2014,
        month = oct,
       volume = {444},
       number = {2},
        pages = {1518-1547},
          doi = {10.1093/mnras/stu1536},
archivePrefix = {arXiv},
       eprint = {1405.2921},
 primaryClass = {astro-ph.CO},
       adsurl = {https://ui.adsabs.harvard.edu/abs/2014MNRAS.444.1518V}
}

@ARTICLE{Stinson:2006,
   author = {{Stinson}, G. and {Seth}, A. and {Katz}, N. and {Wadsley}, J. and 
	{Governato}, F. and {Quinn}, T.},
    title = "{Star formation and feedback in smoothed particle hydrodynamic simulations - I. Isolated galaxies}",
  journal = {\mnras},
   eprint = {astro-ph/0602350},
     year = 2006,
    month = dec,
   volume = 373,
    pages = {1074},
      doi = {10.1111/j.1365-2966.2006.11097.x},
   adsurl = {http://adsabs.harvard.edu/abs/2006MNRAS.373.1074S}
}

@ARTICLE{Bournaud:2010,
   author = {{Bournaud}, F. and {Elmegreen}, B.~G. and {Teyssier}, R. and 
	{Block}, D.~L. and {Puerari}, I.},
    title = "{ISM properties in hydrodynamic galaxy simulations: turbulence cascades, cloud formation, role of gravity and feedback}",
  journal = {\mnras},
archivePrefix = "arXiv",
   eprint = {1007.2566},
 primaryClass = "astro-ph.CO",
     year = 2010,
    month = dec,
   volume = 409,
    pages = {1088},
      doi = {10.1111/j.1365-2966.2010.17370.x},
   adsurl = {http://adsabs.harvard.edu/abs/2010MNRAS.409.1088B}
}

@ARTICLE{Wang:2015,
       author = {{Wang}, Liang and {Dutton}, Aaron A. and {Stinson}, Gregory S. and {Macci{\`o}}, Andrea V. and {Penzo}, Camilla and {Kang}, Xi and {Keller}, Ben W. and {Wadsley}, James},
        title = "{NIHAO project - I. Reproducing the inefficiency of galaxy formation across cosmic time with a large sample of cosmological hydrodynamical simulations}",
      journal = {\mnras},
         year = 2015,
        month = nov,
       volume = {454},
       number = {1},
        pages = {83-94},
          doi = {10.1093/mnras/stv1937},
archivePrefix = {arXiv},
       eprint = {1503.04818},
 primaryClass = {astro-ph.GA},
       adsurl = {https://ui.adsabs.harvard.edu/abs/2015MNRAS.454...83W}
}

@ARTICLE{Tremmel:2017,
       author = {{Tremmel}, M. and {Karcher}, M. and {Governato}, F. and {Volonteri}, M. and {Quinn}, T.~R. and {Pontzen}, A. and {Anderson}, L. and {Bellovary}, J.},
        title = "{The Romulus cosmological simulations: a physical approach to the formation, dynamics and accretion models of SMBHs}",
      journal = {\mnras},
         year = 2017,
        month = sep,
       volume = {470},
       number = {1},
        pages = {1121-1139},
          doi = {10.1093/mnras/stx1160},
archivePrefix = {arXiv},
       eprint = {1607.02151},
 primaryClass = {astro-ph.GA},
       adsurl = {https://ui.adsabs.harvard.edu/abs/2017MNRAS.470.1121T}
}

@ARTICLE{DallaVecchia:2012,
   author = {{Dalla Vecchia}, C. and {Schaye}, J.},
    title = "{Simulating galactic outflows with thermal supernova feedback}",
  journal = {\mnras},
archivePrefix = "arXiv",
   eprint = {1203.5667},
 primaryClass = "astro-ph.GA",
     year = 2012,
    month = oct,
   volume = 426,
    pages = {140},
      doi = {10.1111/j.1365-2966.2012.21704.x},
   adsurl = {http://adsabs.harvard.edu/abs/2012MNRAS.426..140D}
}

@ARTICLE{Crain:2023,
       author = {{Crain}, Robert A. and {van de Voort}, Freeke},
        title = "{Hydrodynamical Simulations of the Galaxy Population: Enduring Successes and Outstanding Challenges}",
      journal = {\araa},
         year = 2023,
        month = aug,
       volume = {61},
        pages = {473-515},
          doi = {10.1146/annurev-astro-041923-043618},
archivePrefix = {arXiv},
       eprint = {2309.17075},
 primaryClass = {astro-ph.GA},
       adsurl = {https://ui.adsabs.harvard.edu/abs/2023ARA&A..61..473C}
}

@ARTICLE{Springel:2010,
   author = {{Springel}, V.},
    title = "{Smoothed Particle Hydrodynamics in Astrophysics}",
  journal = {\araa},
archivePrefix = "arXiv",
   eprint = {1109.2219},
 primaryClass = "astro-ph.CO",
     year = 2010,
    month = sep,
   volume = 48,
    pages = {391},
      doi = {10.1146/annurev-astro-081309-130914},
   adsurl = {http://adsabs.harvard.edu/abs/2010ARA%26A..48..391S}
}

@ARTICLE{Springel:2018,
       author = {{Springel}, Volker and {Pakmor}, R{\"u}diger and {Pillepich}, Annalisa and {Weinberger}, Rainer and {Nelson}, Dylan and {Hernquist}, Lars and {Vogelsberger}, Mark and {Genel}, Shy and {Torrey}, Paul and {Marinacci}, Federico and {Naiman}, Jill},
        title = "{First results from the IllustrisTNG simulations: matter and galaxy clustering}",
      journal = {\mnras},
         year = 2018,
        month = mar,
       volume = {475},
       number = {1},
        pages = {676-698},
          doi = {10.1093/mnras/stx3304},
archivePrefix = {arXiv},
       eprint = {1707.03397},
 primaryClass = {astro-ph.GA},
       adsurl = {https://ui.adsabs.harvard.edu/abs/2018MNRAS.475..676S}
}

@ARTICLE{Dave:2020,
       author = {{Dav{\'e}}, Romeel and {Crain}, Robert A. and {Stevens}, Adam R.~H. and {Narayanan}, Desika and {Saintonge}, Amelie and {Catinella}, Barbara and {Cortese}, Luca},
        title = "{Galaxy cold gas contents in modern cosmological hydrodynamic simulations}",
      journal = {\mnras},
         year = 2020,
        month = sep,
       volume = {497},
       number = {1},
        pages = {146-166},
          doi = {10.1093/mnras/staa1894},
archivePrefix = {arXiv},
       eprint = {2002.07226},
 primaryClass = {astro-ph.GA},
       adsurl = {https://ui.adsabs.harvard.edu/abs/2020MNRAS.497..146D}
}

\end{document}